%% file: review.tex
\documentclass{ar-1col}

\usepackage[comma]{natbib}
\usepackage{subcaption}
\usepackage{graphicx}
\usepackage[colorlinks,citecolor=blue]{hyperref}
\hypersetup{
    colorlinks=true,
    linkcolor=red,
    }

\newcommand{\begit}{\begin{itemize}}
\newcommand{\enit}{\end{itemize}}
\newcommand{\beq}{\begin{equation}} 
\newcommand{\eeq}{\end{equation}} 
\newcommand{\beqa}{\begin{eqnarray}} 
\newcommand{\eeqa}{\end{eqnarray}} 
\newcommand{\p}{\partial}

\newcommand{\begen}{\begin{enumerate}}
\newcommand{\enen}{\end{enumerate}}

\def\lesssim{\mathrel{\hbox{\rlap{\hbox{\lower4pt\hbox{$\sim$}}}\raise2pt\hbox{$<$}}}}
\def\gtrsim{\mathrel{\hbox{\rlap{\hbox{\lower4pt\hbox{$\sim$}}}\raise2pt\hbox{$>$}}}}

\def\lesssim{\mathrel{\hbox{\rlap{\hbox{\lower4pt\hbox{$\sim$}}}\raise2pt\hbox{$<$}}}}
\def\gtrsim{\mathrel{\hbox{\rlap{\hbox{\lower4pt\hbox{$\sim$}}}\raise2pt\hbox{$>$}}}}

\jname{Xxxx. Xxx. Xxx. Xxx.}
\jvol{AA}
\jyear{2020}
\doi{10.1146/((please add article doi))}

\begin{document}

\markboth{Thompson \& Heckman}{Galactic Winds}

\title{Theory and Observation of Winds from Star-Forming Galaxies}

\author{Todd A.\ Thompson$^{1}$ \& Timothy M.\  Heckman$^{2,3}$ 
\affil{$^1$Department of Astronomy, Department of Physics, Center for Cosmology \& Astro-Particle Physics, The Ohio State University; thompson.1847@osu.edu}
\affil{$^2$The William H. Miller III Department of Physics and Astronomy, The Johns Hopkins University; theckma1@jhu.edu}
\affil{$^3$School of Earth \& Space Exploration, Arizona State University, Tempe, Arizona}}

\begin{abstract}
Galactic winds shape the stellar, gas, and metal content of galaxies. To quantify their impact, we must understand their physics. We review potential wind-driving mechanisms and observed wind properties, with a focus on the warm ionized and hot X-ray-emitting gas. Energy and momentum injection by supernovae (SNe), cosmic rays, radiation pressure, and magnetic fields are considered in the light of observations:

\vspace*{0.2cm}
\begin{minipage}[l]{0.75\textwidth}
\begin{itemize}
\item  Emission and absorption line measurements of cool/warm gas provide our best physical diagnostics of galactic outflows. 
\item The critical unsolved problem is how to accelerate cool gas to the high velocities observed. Although conclusive evidence for no one mechanism exists, the momentum, energy, and mass-loading budgets observed compare well with theory. 
\item A model where star formation provides a force $\sim L/c$, where $L$ is the bolometric luminosity, and cool gas is pushed out of the galaxy's gravitational potential, compares well with available data. The wind power is $\sim0.1$ that provided by SNe.
\item The very hot X-ray emitting phase, may be a (or the) prime mover. Momentum and energy exchange between the hot and cooler phases is critical to the gas dynamics. 
\item Gaps in our observational knowledge include the hot gas kinematics and the size and structure of the outflows probed with UV absorption lines. 
\end{itemize}
\end{minipage}
\vspace*{0.2cm}

Simulations are needed to more fully understand mixing, cloud-radiation, cloud-cosmic ray, and cloud-hot wind interactions, the collective effects of star clusters, and both distributed and clustered SNe. Observational works should seek secondary correlations in the wind data that provide evidence for specific mechanisms and compare spectroscopy with the column density-velocity results from theory. 
\end{abstract}


\maketitle

\tableofcontents

\section{Prologue}
\label{section:prologue}

In the beginning, there was light, and gas. 14 billion years later and, my God, it's full of stars! We wish to understand the transformation of a Universe filled with a nearly uniform distribution of gas, radiation, and dark matter, into the beautiful Universe we see before us, adorned with galaxies, and replete with the ingredients for and engines of life. Galactic winds are critical to this metamorphosis. By ejecting gas and metals into the surrounding medium, they delay, prolong, and sometimes terminate star formation, leaving ancient stellar populations to evolve until today and long beyond. Here, we present a primer on galactic wind physics and observations, focusing on the currently-visible unsolved problems, directions for new discoveries, and the explorations that could establish a firmer galactic wind phenomenology.

\section{Introduction}
\label{section:introduction}

In the local universe, few  galaxies exhibit the spectacular large-scale outflows of the nearby exemplars M82 and NGC\,253 (see Figure \ref{figure:m82_hst}). These ``starburst" or ``superwind" galaxies display bright extended X-ray emission, extensive multi-kpc scale gas structures moving at hundreds to thousands of km/s, and total mass outflow rates comparable to, or larger than, their star formation rates (SFRs).

While normal galaxies lying on the $z\simeq0$ star-forming main sequence may display some extraplanar gas, dust pillars, radio halos, and occasional super-bubbles, special conditions are evidently required to produce M82-like superwinds today. In contrast, at high-$z$  outflows abound. Surveys find ubiquitous high-velocity blue-shifted absorption and broad emission lines in the rest-frame UV and optical from main sequence galaxies at $z\simeq1-3$. Whatever thresholds are required for launching galactic winds, they are met at cosmic noon. Thus, the rarity of observed winds today notwithstanding, a substantial component of the Universe's $z=0$ stellar inventory was born in a galaxy hosting a galactic outflow.

An important clue to driving galactic winds is that even in M82 and NGC\,253, the outflow originates only from the nuclear region, where the star formation surface density ($\dot{\Sigma}_\star$; see Table \ref{table:symbols} for symbols) is largest. Similarly, otherwise normal local galaxies sometimes exhibit winds driven by rapid circumnuclear star formation, including the Milky Way. Both wind theory and observations imply that the critical physical threshold for winds is in $\dot{\Sigma}_\star$. Whereas an average galactic disk in the local Universe might have ${\rm SFR}\sim1$\,M$_\odot$ yr$^{-1}$ over an area of $\sim50-100$\,kpc$^2$, the cores of M82 and NGC\,253 have $5-10$ times higher SFR over an area 400 times smaller, leading to an $\dot{\Sigma}_\star$ that is $2000-4000$ times higher than typical of the Solar circle. At high-$z$, main sequence galaxies have much higher SFR at fixed stellar mass and are more compact than their low-$z$ counterparts. Typical values are $10-100$\,M$_\odot$ yr$^{-1}$ with sizes of just $\sim1-10$\,kpc$^2$. Such systems also exhibit kpc-scale off-nuclear knots of star formation with properties similar to the M82 starburst. Finally, wind-hosting luminous and ultra-luminous infrared galaxies (ULIRGs) at low- and high-$z$ can be even more stunningly compact. An example is the ULIRG Arp 220, which has ${\rm SFR}\simeq100-200$\,M$_\odot$ yr$^{-1}$ in two merging super-dense galactic nuclei with characteristic sizes of just $\sim50$\,pc each, yielding an $\dot{\Sigma}_\star$ that is $\sim10^6$ times larger than the Galaxy.

Galactic winds are multi-phase and  multi-dynamical, with emission and absorption across the electromagnetic spectrum. Their many components include cold molecular gas, dust grains, neutral HI,  warm ionized atomic gas, hot virial and super-virial gas, relativistic particles evidenced by non-thermal emission, and magnetic fields traced by polarization, Faraday rotation, and Zeeman splitting. Observational tracers include velocity-resolved emission and absorption by the submm and mm rotational lines of molecules, the near-IR  transitions of warm molecular species, the FIR atomic fine structure transitions, 21\,cm HI, and a multitude of optical/UV atomic transitions tracing neutral and ionized gas. While conspicuous X-ray halos and bubbles are prominent in local superwinds, we have little direct dynamical information on the hot and very hot X-ray-emitting phases. Likewise, we lack direct dynamical constraints on the FIR dust continuum or the relativistic particles producing the radio and gamma-ray continua. Current and next-generation X-ray microcalorimeter telescopes may provide much needed constraints on the hot gas dynamics, which at present are sorely lacking. For now, our knowledge of the dynamics of winds is dominated by UV, optical, FIR, and (sub)mm/radio spectroscopy.

\subsection{Winds Writ Large}
\label{section:writ_large}

Although their beauty provides reason enough to study them, galactic winds are also critical to galaxy evolution. Winds directly affect galaxy growth by removing gas and metals. This is ``ejective" feedback: the raw fuel for future star and planet formation is removed from the galaxy and ejected into the circumgalactic and (perhaps) intergalactic medium. But winds are also ``preventive:" they push, stir, and shock the surrounding halo matter by injecting energy and momentum, preventing or delaying the accretion of gas into the central galaxy. Indeed, a motivating fact for galactic wind research is that the stellar masses of galaxies today lie below their primordial budget of hydrogen and helium after the Big Bang (e.g., \citealt{Benson2003,Somerville2015}). Over cosmic time, less massive dark matter halos convert less of their store of primordial gas to the luminous stuff of galaxies. At and below $L_\star$, galactic winds appear to be the answer to the question ``Why?". 

\begin{figure}
\centerline{\includegraphics[width=6.5cm]{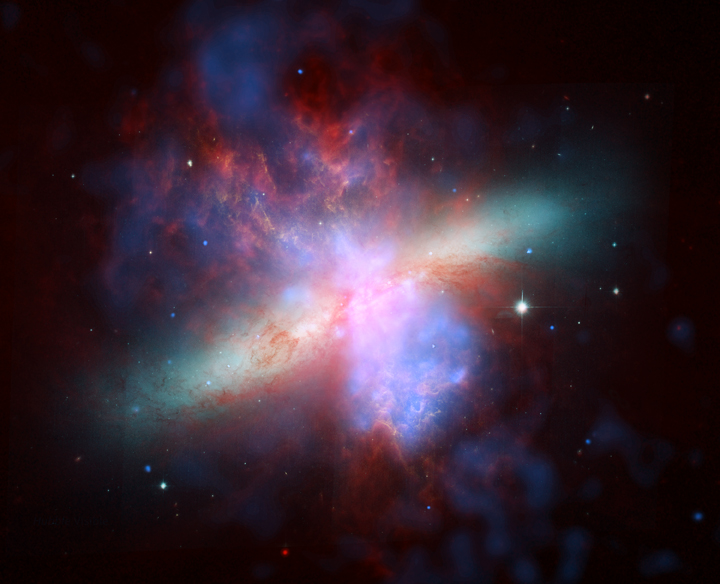}
\hspace*{-6.8cm}
\includegraphics[width=6cm]{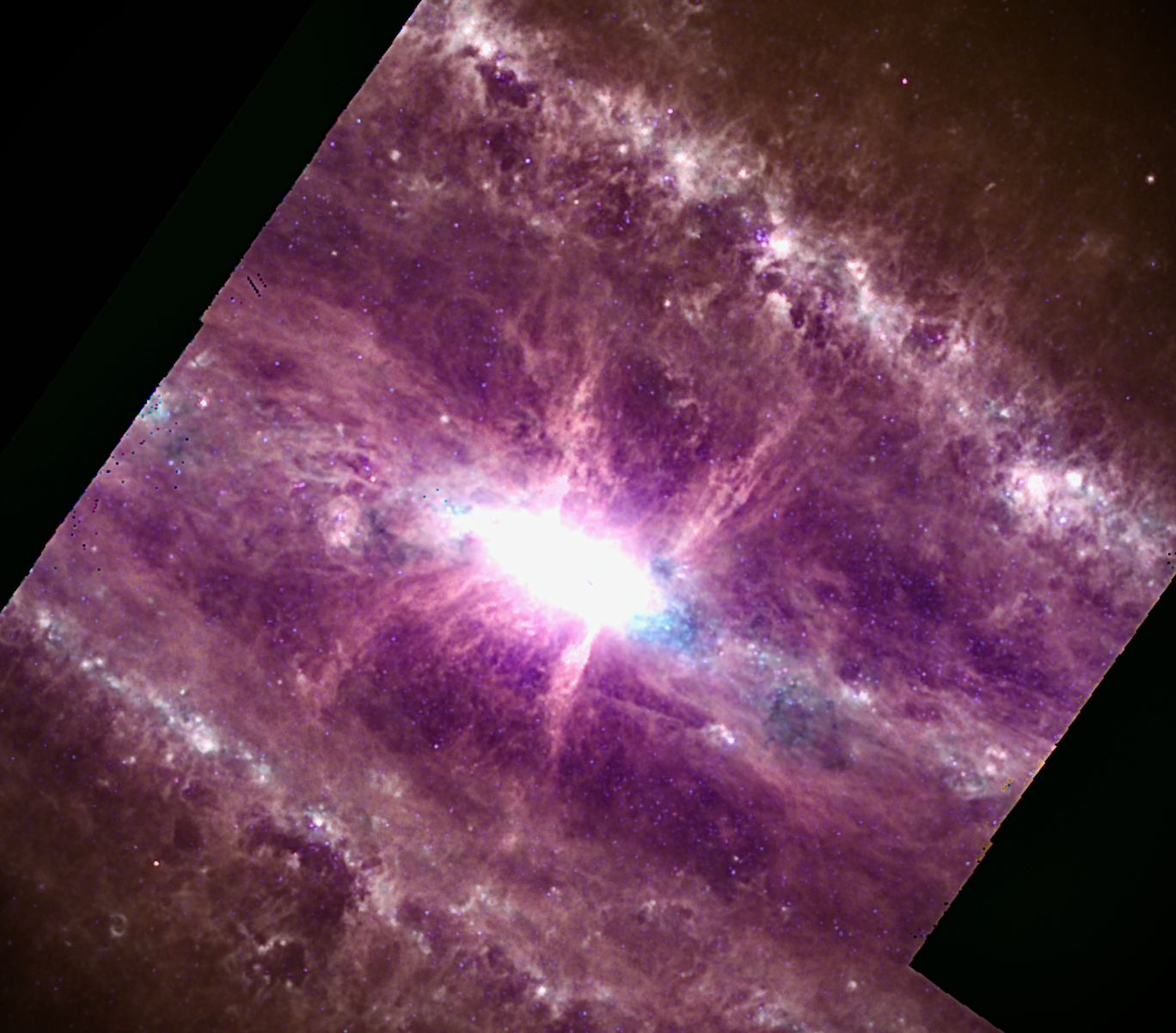}}
\vspace*{.5cm}
\caption{{\it Left:} Starburst superwind exemplar M82, in all its glory. Composite using X-rays (blue), infrared (red), Hydrogen emission (orange), and visible light (yellow-green). Credit: X-ray: NASA/CXC/JHU/D.~Strickland; Optical: NASA/ESA/STScI/AURA/The Hubble Heritage Team; IR: NASA/JPL-Caltech/University of Arizona/C.~Engelbracht. {\it Right:} The core of NGC\,253 showing MIR PAH emission associated with the extended disk and bi-conical outflow (MIRI observations from JWST; 11.3 (red), 7.7 (green), and 5.6\,$\mu$m (blue); part of GO1 project 1701, see also \citealt{Bolatto2024}).}
\label{figure:m82_hst}
\end{figure}

The importance of galactic winds in shaping galaxy growth and enrichment can be understood by considering an equilibrium model that ties the stellar, gaseous, and metal content of galaxies and halos (e.g., \citealt{Dave2012}). Evidence for such a picture comes from galaxy scaling relations and their evolution, including the star-forming main sequence and the mass-metallicity relation, among others. In an equilibrium  picture, the accretion rate into dark matter halos as a function of redshift ($\dot{M}_{\rm in,\,halo}(z)$) is connected to both the SFR and the galactic wind ejection rate ($\dot{M}_{\rm wind}$) such that $\dot{M}_{\rm in,\,halo}={\rm SFR}+\dot{M}_{\rm wind}$. Similarly, the equilibrium metal abundance of any element $Z$ is connected to the elemental yield per star formed ($y$) and the ratio of the inflow abundance to the ISM abundance ($\zeta$). Written in terms of the wind mass-loading $\eta=\dot{M}_{\rm wind}/{\rm SFR}$, its critical importance is manifest:
\beq
{\rm SFR}(z)=
\frac{\dot{M}_{\rm in,\,halo}}{1+\eta}\,\,\,\,{\rm and}\,\,\,\,Z=\frac{y}{1+\eta}\frac{1}{1-\zeta}.
\label{eq:sfr_equilibrium}
\eeq
As written, equation (\ref{eq:sfr_equilibrium}) represents purely ``ejective" feedback from a galaxy: All else constant, higher $\eta$ decreases both the SFR and $Z$, and thus the integrated stellar mass content of halos and their abundances. The possibility of ``preventive" feedback is seen when we imagine the interaction of winds with the CGM, which may prevent or delay gas accretion into the galaxy on long timescales, and with hysterisis, thus breaking the direct mapping between $\dot{M}_{\rm in,\,halo}$, the SFR, and $\dot{M}_{\rm wind}$. With preventive wind feedback, the inflow rate and the abundances of the (re-)accreted gas ($\zeta$) may change in an $\eta$-dependent way, so that any equilibrium depends on the wind's historical momentum ($\dot{p}$) and energy injection rates ($\dot{E}$) (e.g., \citealt{Oppenheimer2008,Oppenheimer2010}). The connection between the ejective properties of galactic winds and their preventive effects on both the star formation and abundances of galaxies remains uncertain, dependent on wind implementation prescription in large-scale simulations, and is an active area of research. Yet, observations clearly suggest that galactic winds are both ejective and preventive, with stunning images of ejection readily beheld (Figs.~\ref{figure:m82_hst}, \ref{figure:m82_HSTzoom}), wind-driven bubbles from star formation observed on $50-100$\,kpc scales around galaxies (e.g., Fig.~\ref{figure:bubble}), and abundant absorption probes of enriched halo gas on the scales of their virial radii \citep{Tumlinson2017}. 

Observed galaxy scaling relations and their interpretation with equations like (\ref{eq:sfr_equilibrium}) frame much of our understanding of galaxy evolution \citep{Somerville2015}. Galactic wind physics is thus deeply connected to galaxy evolution writ large: winds shape our picture of the luminous matter that traces the large-scale structure formation of the Universe, connecting galaxy formation to the cosmological context \citep{Naab2017,Tumlinson2017,Tacconi2020,Faucher-Giguere2023,Crain2023}. 

\subsection{Winds Writ Small}
\label{section:writ_small}

If galactic winds are driven principally by supernovae (SNe), then their physics is linked to the very smallest kilometer scales of neutron stars and the black hole event horizons formed in core collapse. Indeed, the classic argument for the existence of neutron stars on the basis of SN energetics by \cite{Baade1934} might have been advanced ``with all reserve" on the basis of galactic wind phenomenology: a substantial fraction of the kinetic energy released when massive stars explode must be coupled to galactic outflows.

More broadly, winds are connected to the ISM-scale physics of the star formation process itself, including the  gravitational instabilities of star-forming disks, the formation of GMCs, and their disruption. Stars form in clusters and disperse their natal gas  by a combination of feedback mechanisms including ionization heating, protostellar jets, radiation pressure, and stellar  winds. Both deeply-embedded and fully-revealed massive star clusters sit at the base of local superwinds. These clusters drive at first unorganized outflows, entraining mass from the surrounding medium, and then with SN explosions that combine to drive super-bubbles that can blow out of the surrounding disk, again entraining and ram pressure accelerating matter, and venting super-heated gas to the extraplanar near-galactic medium (NGM).\footnote{We prefer``near-galactic medium" to {\it inner CGM}, {\it disk-halo interface}, or {\it galactic corona}.}  In this way, the mass function of star clusters, their star formation efficiency, the physics of their disruption, and the spatial and temporal distribution of SNe and energy injection all matter to the wind physics on larger scales. The collective action of stellar feedback seeds galactic winds directly from the dense star-forming complexes and clusters seen in local starbursts \citep{Westmoquette2007,Westmoquette2009}. Galactic winds are thus closely connected to galaxy evolution writ small: the physics of individual massive star formation \citep{Mckee2007}, massive star and binary star evolution \citep{Eldridge2017}, star cluster formation \citep{Krumholz2019} and GMC disruption (e.g., \citealt{Murray2011}) are the roots at the base of galactic winds.

\subsection{Scope, Framing, \& Wind Archetypes}
\label{section:scope}

The principal question that animates this review is how to accelerate cool neutral and warm ionized gas with sound speed $\lesssim10$\,km/s to the $\sim100-1000$\,km/s velocities seen in emission and absorption with the mass, momentum, and energy budgets observed from star-forming galaxies. The subject of winds driven by AGN is equally rich, but beyond our scope. 

We take a pedagogical approach. In \S\ref{section:theory}, we connect galactic and stellar wind theory, and then in \S\ref{section:thermal} we discuss the Parker wind and provide a more quantitative definition of the problem of cool fast galactic outflows. We then enumerate wind driving mechanisms, including SNe and the very hot phase (\S\ref{section:supernova_driven}), radiation pressure (\S\ref{section:radiation_pressure}), cosmic rays (\S\ref{section:cosmic_rays}), and magneto-thermal winds (\S\ref{section:magnetic}). We focus on analytic results meant to reveal each mechanism's basic physics, summarize numerical results, and highlight physical uncertainties and galaxy regimes where a given mechanism may be more or less effective. In \S\ref{section:observations}, we turn to observations, focusing on the neutral atomic, ionized, and hot phases in emission (\S\ref{section:optical_xray}) and absorption (\S\ref{section:absorption}) at both low- and high-$z$ (\S\ref{section:demographics_cosmic_time}), and in the CGM (\S\ref{section:observations_cgm}). In \S\ref{section:confronting_theory_observations}, we use a particular dataset to sketch arguments that can be used when confronting observations and theory. We highlight observational uncertainties and the need for simulation analysis akin to observations.  In \S\ref{section:discussion_conclusions}, we summarize and conclude. 

Throughout, we reference three wind archetypes. The first and primary archetype is the M82-like winds seen in emission and absorption from local starbursts, as in Figure \ref{figure:m82_hst}, which serve as models for outflows from high-$z$ SFGs. A second archetype is the very dense winds of ULIRGs like Arp 220. This ultra-dense extremum of the starburst phenomenon poses challenges for several wind-driving mechanisms and is thus a useful benchmark. Finally, our third wind archetype may not exist. Works meant to explain the Galactic stellar abundance distributions catalogued by massive spectroscopic surveys (e.g., \citealt{Hayden2015}) require winds with $\eta\sim1$ throughout the past $\sim10$\,Gyr (e.g., \citealt{Johnson2021}). Although this conclusion depends on the (uncertain) absolute yield of iron and alpha elements from core-collapse SNe \citep{Minchev2013,Weinberg2023}, normal $z\sim0$ spirals like the Milky Way may drive massive, but mostly unseen outflows. We keep this possibility in mind.

\section{Theory of Galactic Winds}
\label{section:theory}

Theoretical work on galactic winds began soon after Parker's seminal exchange with Chamberlain on mass loss from the Sun \citep{Parker1958,Parker1960,Chamberlain1960}. Chamberlain's picture was of ``Jeans escape," where the high-velocity tail of the Maxwellian particle distribution escapes the central body's tenuous upper atmosphere and gravitational potential. In contrast, Parker's wind was hydrodynamic: deviation from a purely adiabatic atmosphere in hydrostatic equilibrium produces a net pressure gradient that accelerates gas, starting from subsonic velocities at the wind's base and eventually accelerating through a sonic point (\S\ref{section:thermal}). These works gave rise to thermal wind theory throughout astrophysics, with applications ranging from galaxy clusters to exoplanets. Motivated by early evidence for an outflow from the Galactic center region \citep{Vanwoerden1957,Rougoor1959}, Parker's theory was applied to galaxies, with extensions to winds from stellar systems with spatially-distributed gas and energy injection  (\citealt{Moore1968,Burke1968}; \S\ref{section:supernova_driven}). 

The development of radiation pressure driven galactic wind models is partially mixed in time with that for individual stars. Consideration of radiation pressure on dust goes back at least to  \cite{Poynting1904}'s work on grain dynamics in the Solar System. \cite{Gerasimovic1932,Schoenberg1933,Schalen1939} considered radiation pressure as a mechanism for gas ejection near stars and stellar associations in the ISM with later treatments by \cite{Harwit1962} and \cite{Odell1967} (\S\ref{section:radiation_pressure}). \cite{Wickramasinghe1966} developed the theory for red giant stars. \cite{Chiao1972} discuss radiation pressure driven outflows of dust grains from galaxies and the possibility of significant intergalactic extinction (e.g., \citealt{Aguirre2001,Menard2010,Peek2015}). 

Other classic wind driving mechanisms that are well-developed elsewhere in astrophysics, have seen less development in the galactic wind context. These include the metal ``line-driven" winds of hot stars and quasar disks \citep{Lucy1970}, magnetocentrifugal acceleration \citep{Schatzman1962} and magnetically-driven disk winds \citep{Blandford1982} (see \S\ref{section:magnetic}). In contrast, mechanisms like radiation pressure from Lyman $\alpha$ photon scattering (\citealt{Cox1985}; \S\ref{section:radiation_pressure}) and cosmic rays \citep{Ipavich1975} (see \S\ref{section:cosmic_rays}) have been developed for galaxies, but do not have an obvious parallel for stars and disks in other contexts. Finally, sound wave- and Alfv\'en wave-driven winds and pulsationsal mass loss -- classic topics in stellar wind theory -- may have galactic analogs or provide model problems that could shed light on galaxy simulations, but have not been explored.

Galactic winds have important differences from stellar winds. First, we consider collections of stars that may evolve significantly in time, and whose properties we infer from the IMF, which allows us to translate the light observed to an underlying stellar population, and then to energy/mass injection rates via SNe and other stellar processes (e.g., \citealt{Leitherer1999}). Another important difference is the extended gravitational potential of galaxies, including the old stellar population and dark matter. Differences also manifest in how matter is injected. SNe and stellar winds inject matter directly into the flow as they interact with the ISM, mixing phases. Finally, whereas the atmosphere of a star may be defined by optical depth and radiative equilibrium, the ``atmosphere" of a galaxy is turbulent, and regularly disturbed by SN remnants with characteristic velocities (after the energy-conserving phase) of $\sim200$\,km s$^{-1}$, that can be smaller, of order, or larger than the effective galaxy escape velocity from its ``surface,"  launching outflows that feed the NGM and CGM.

\subsection{Parker Winds \& The Problem of Cool Galactic Outflows}
\label{section:thermal}

We briefly review the thermal Parker wind because of the intuition gained from understanding this model problem (see \citealt{Lamers1999}). Momentarily ignoring the extended galactic gravitational potential, the equations of mass and momentum conservation for a steady-state spherical isothermal flow maintained at constant temperature by heating and cooling processes are
\beq
\frac{d}{dr}\left(r^2\rho v\right)=0\,\,\,\,{\rm and}\,\,\,\,
v\frac{dv}{dr}=-\frac{GM}{r^2}-\frac{1}{\rho}\frac{dP}{dr}
\,\,\Longrightarrow\,\,\frac{r}{v}\frac{dv}{dr}=\left(\frac{2c_T^2-GM/r}{v^2-c_T^2}\right),
\label{parker}
\eeq
where $c_T$ is the isothermal sound speed and $P=\rho c_T^2$. The ``wind equation" for $dv/dr$  shows that if the medium is subsonic $v<c_T$ at the base of the outflow $R$, then positive acceleration requires that $c_T^2$ {\it is less than} $GM/2R$. The flow is thus {\it gravitationally bound} at $R$. The pressure gradient steadily accelerates the matter through the sonic point, where the numerator and denominator of the wind equation vanish simultaneously, such that $v(R_{\rm sonic})=c_T$ at $R_{\rm sonic}=GM/2c_T^2$ and the velocity gradient at $R_{\rm sonic}$ is continuous, positive, and obtained by L'H\^opital's rule. Equation (\ref{parker}) admits a family of outflow and (by $v\rightarrow -v$) accretion solutions, with a single unique profile for the outgoing transonic Parker wind. The ``solution" to the wind problem is the velocity (or density) profile, which yields the mass outflow rate $\dot{M}$, asymptotic velocity, and the wind's kinetic power and force.

Although equation (\ref{parker}) can be solved exactly, an estimate of $\dot{M}$ is more generally applicable to a wider variety of systems. We exploit the fact that $\dot{M}=4\pi r^2\rho v$ is constant everywhere and can thus be evaluated at the sonic point, where we know $v=c_T$ and $r=R_{\rm sonic}=GM/2c_T^2$. The unknown is then the density at the sonic point. A practical expedient is to {\it assume} that the density profile is given by hydrostatic equilibrium (strictly, $v=0$) for radii less than $R_{\rm sonic}$, an approximation that becomes increasingly justified as $r\rightarrow R$. Hydrostatic equilibrium with a point-mass potential demands that
\beq
\frac{1}{\rho}\frac{dP}{dr}=-\frac{GM}{r^2}\,\,\Longrightarrow\,\,\frac{d\ln\rho}{d\ln r}=-\frac{GM}{r\,c_T^2}
\,\,\Longrightarrow\,\,\frac{\rho(r)}{\rho_0}=\exp\left[-\frac{GM}{Rc_T^2}\left(1-\frac{R}{r}\right)\right],
\label{simplehe}
\eeq
where $\rho_0$ is the density at the base of the outflow. While approximate, especially near $R_{\rm sonic}$, equation (\ref{simplehe}) allows us to write down an expression for $\dot{M}$ that captures the physics: 
\beq
\dot{M}\sim4\pi \left(\frac{GM}{2c_T^2}\right)^2 \,c_T \,\rho_0\exp\left[2-\frac{GM}{Rc_T^2}\right].
\eeq
The critical ratio governing $\dot{M}$ is $R_{\rm sonic}/R$, or alternatively, the ratio of the escape velocity $v^2_{\rm esc}=2GM/R$ to $4c_T^2$. As $c_T^2$ approaches $GM/2R$, $R_{\rm sonic}\rightarrow R$ and $\dot{M}\rightarrow4\pi R^2\rho_0 v_{\rm esc}$, which corresponds to dynamical disruption. In the opposite limit $c_T^2\ll GM/R=v^2_{\rm esc}/2$, $R_{\rm sonic}$ is much larger than $R$ and $\dot{M}\propto \exp{(-v_{\rm esc}^2/2c_T^2)}$. The exponential decrease in the density between $\rho_0$ and $\rho(R_{\rm sonic})$ dominates the $\dot{M}$ dependence. Note that a wind will be driven even for {\it arbitrarily small} $v_{\rm esc}/2c_T$, but it will be exponentially weaker as that ratio increases.  While the asymptotic velocity of a truly isothermal wind diverges logarithmically as $r\rightarrow \infty$, real winds are never truly isothermal, and the asymptotic velocity is of order $v_\infty\sim c_T\sim v_{\rm esc}(R_{\rm sonic})$, so that $\dot{E}\sim\dot{M}c_T^2$ and $\dot{p}\sim\dot{M} c_T$. 

The temperature (or sound speed) and density at the base of the outflow are critical, and are generally determined by other pieces of physics. For example, they are often set by ``photospheric" conditions at $R$, e.g., that the  optical depth from the photosphere to infinity is equal to $2/3$ and/or that at sufficiently high densities heating balances cooling or ionizations balance recombinations. Flows heated and cooled by physical processes with these types of inner boundary conditions may achieve near-isothermality over a range of radii, but the assumption is generally not applicable and, in practice, the full problem must be solved with microphysics appropriate to the problem. In doing so, note that the velocity at $R$ ($v_0$) and the base density $\rho_0$ cannot be set simultaneously by hand since this is tantamount to assuming $\dot{M}$. Indeed, in a time-dependent simulation of a wind to a steady-state configuration, one can think about $\dot{M}$ as the ``eigenvalue" of the transonic wind problem. Once $\rho_0$, $M$, $R$, and $c_T$ are specified, the time-dependent system will relax to the transonic solution with the value of $\dot{M}$ required for that critical solution. 

There are many limitations to the isothermal steady-state model, in terms of both its physicality and its application to galactic winds. Yet, it allows for immediate  interesting conclusions. First, we replace $GM/r$ with $GM(r)/r$ for an extended mass distribution. Using a singular isothermal sphere potential of the form $M(r)=2\sigma^2 r/G$ for illustration, where $\sigma$ is the velocity dispersion, the right hand side of the wind equation (\ref{parker}) becomes $2(c_T^2-\sigma^2)/(v^2-c_T^2)$. Because $\sigma$ is approximately constant inside the break radius $R_B$ of the galaxy's inner stellar distribution and/or dark matter halo, if $c_T<\sigma$ at the wind base, then $c_T$ must {\it increase} as a function of distance if there is to be a sonic point very near the host galaxy. This turns out to be the case for the effective CR sound speed in CR-driven winds (see \S\ref{section:cosmic_rays}). Conversely, if $c_T$ is constant from $R$ out to $R_B$, then the sonic point will be set by $R_B$, because it is there that the gravitational term in the numerator of the wind equation (\ref{parker}) starts to decrease and a sonic point can form when $2c_T^2=GM(r)/r$. 

The case of an isothermal gas in an isothermal potential illustrates this physics, highlights the fundamental problem of this review, and is a model for dwarf galaxies during cosmic reionization. Suppose a central concentration of neutral gas sits in a dark matter halo and is exposed to ionizing radiation. An ionization front moves from the outside inwards, heating the gas to $c_T\sim10$\,km/s until either the entire halo is ionized, or recombinations balance ionizations at sufficiently high density. For $c_T>\sigma$ we expect complete disruption, terminating or suppressing star formation (e.g., \citealt{Shapiro2004}). However, for  $c_T<\sigma$, we expect a Parker wind. The outflow base density is set by where recombinations balance ionizations, as in the analogous calculation for irradiated exoplanets (\citealt{Murray-Clay2009}; $\rho_0\simeq[2F\sigma^2/(c_T^2\,R\alpha)]^{1/2}$, where $F$ is the ionizing flux and $\alpha$ is the recombination coefficient). The mass loss rate can be estimated using hydrostatic equilibrium (eq.~\ref{simplehe}), from the outflow base at $R$ to the break radius $R_B$ of the halo.  For an isothermal gas in an isothermal gravitational potential we have $c_T^2 \,d\ln\rho=-2\sigma^2d\ln r$, so that $\rho/\rho_0=(R/r)^{2\sigma^2/c_T^2}$. For $\sigma=10$, 40, or (Milky Way mass) $150$\,km/s and $c_T=10$\,km/s this gives $\dot{M}\sim 4\pi R_B^2 c_T\rho_0\times (R/R_B)^2$, $(R/R_B)^{32}$(!), and $(R/R_B)^{450}$(!!), respectively. Thus, for ``cool" gas in deep potentials ($c_T\ll\sigma$), the density profile is exceptionally steep, leading to vanishingly small  $\dot{M}$ if the cool gas is supported solely by its own pressure.

This brings the problem (\S\ref{section:scope}) of accelerating cool gas to high velocities from galaxies with deep potentials into sharp focus. The question we face is how to accelerate gas with intrinsic sound speed less than $\sim10$\,km/s to the $100-1000$\,km/s wind velocities observed, from galaxies with circular velocities of $\sim30-300$\,km/s, and with $\eta\sim1$ or larger. Although thermal gas pressure is relevant for low-mass dark matter halos with $\sigma\sim10$\,km/s, it simply is not for more massive galaxies. Turned around, gas with $c_T$ of order $\sigma$ or larger ($T\sim m_{\rm p}\,\sigma^2/k_{\rm B}\simeq3\times10^6(\sigma/150\,\rm km/s)^2$\,K) {\it must} form a thermal Parker-type wind. However, this gas is hot for a deep galaxy potential and {\it prima facie} cannot be the cool wind material seen in emission/absorption observations unless there is additional physics. That physics may be mixing or ram pressure from the hot phase, radiation pressure, CRs, turbulence, magnetic acceleration, some other mechanism, or maybe everything, all at once. 

\subsection{Hot Thermal Winds with Mass and Energy Injection}
\label{section:supernova_driven}

The character of the wind problem changes when we consider distributed mass and energy sources like SNe and stellar winds. In classic treatments, matter and energy injected in the host thermalizes, expands outward to set up a large-scale pressure gradient in the galactic potential, and then drives a wind that rapidly accelerates through a sonic point  to produce an outflow \citep{Holzer1970,Johnson1971,Mathews1971}. These additions modify the Parker wind problem schematically in two ways: (1) the continuity equation acquires a source term from mass injection, with a corresponding change in the momentum equation (eqs.~\ref{parker}), and (2) the energy equation explicitly allows for sources/sinks (e.g., photoionization, conduction, SNe, radiative cooling), and energy transfer between thermodynamic phases (e.g., \citealt{Cowie1981,Fielding2022}).

The physics of how mass, energy, and momentum are mixed with the ISM and entrained into an outflow remains a principal topic of research. This is perhaps {\it the} question. Very broadly, we know that molecular and atomic gas in the host galaxy produce (giant) molecular clouds (GMCs) that, in turn, produce new star formation in clusters and stellar associations. These gas agglomerations are disrupted by feedback processes (e.g., \citealt{Krumholz2019}) on the scale of GMCs, a process that also injects momentum and energy into the ISM, and can drive super-bubbles that seed galactic outflows (e.g., \citealt{Murray2011,Kim2017_superbubbles}). Although the exact timing between GMC disruption and the first massive star SNe remains uncertain, very young star clusters become unembedded on Myr timescales. A ZAMS stellar population's most massive stars undergo core collapse about 4\,Myr after birth. Successful SNe explode on timescales of $4-40$\,Myr after cluster formation, inject momentum and energy directly to the ISM, and are considered the dominant contributor to the global energy, momentum, and cosmic ray budgets (\S\ref{section:cosmic_rays}; Sidebar \ref{sidebar:supernovae}). 

Individual SNe go through well-defined stages of evolution, including free expansion, the energy-conserving Sedov-Taylor (ST) phase, and then the momentum-conserving snowplow phase. The latter begins when radiative cooling in the post-shock medium becomes important (``shell formation") and marks the end of the ST phase of evolution. The time, radius, velocity, and momentum of the remnant at shell formation are \citep{Kim2015_supernovae} $t_{\rm SF}\simeq4\times10^4\,{\rm yr}\,E_{\rm SN, 51}^{0.22}n_0^{-0.55}$, $R_{\rm SF}\simeq23\,{\rm pc}\,E_{\rm SN, 51}^{0.29}n_0^{-0.42}$, $v_{\rm SF}\simeq200\,{\rm km/s}\,E_{\rm SN, 51}^{0.07}n_0^{0.13}$, and $p_{\rm SF}=2\times10^{5}\,{\rm M_\odot\,\,km/s}\,E_{\rm SN, 51}^{0.93}n_0^{-0.13}$, where $E_{\rm SN, 51}=E_{\rm SN}/10^{51}$\,ergs is the SN energy and $n_0=n/1$\,cm$^{-3}$ is the density of the surrounding medium. The ST phase is particularly important to models of feedback and turbulence driving in the ISM, but also (potentially) to galactic wind launching, because the total momentum carried by a SN increases as the hot interior does work. In fact, the typical momentum of a SN at the moment of explosion is about 10 times less than $p_{\rm SF}$. After shell formation, $p$ increases by a factor of $\simeq1.6$ in a uniform medium (\citealt{Kim2015_supernovae}; see also \citealt{Walch2015}).

The momentum injection from SN remnants is a key benchmark in understanding SNe as a prime mover for the ISM, for galactic winds, and for comparing with all other proposed wind-driving mechanisms. Assuming a typical SN rate per unit star formation and an energy of $10^{51}$\,ergs per SN (see Sidebar \ref{sidebar:supernovae}), the momentum injection rate 
\beq
\dot{p}_{\rm SN}\simeq20\,(L/c)\,E_{\rm SN,\,51}^{0.93}n_0^{-0.13},
\label{eq:pdot_snr}
\eeq
where $L$ is the bolometric luminosity from continuous star formation (adopting a typical IMF; eq.~\ref{eq:lum_continuous}). Several points bear mention. First, the pre-factor in equation (\ref{eq:pdot_snr}) is significantly larger than for the so-called single-scattering limit ($L/c$) discussed in \S\ref{section:radiation_pressure} for both dust and ionizing radiation. It is comparable to Ly$\alpha$ resonant scattering in dust-less media (eq.~\ref{eq:taueff_lya}), and it is $\sim4$ times larger than the hot gas momentum injection rate discussed below (eq.~\ref{cc85_vinf_pdot}). See Table \ref{table:theory}. The normalization is also dependent on $E_{\rm SN}$ and the rate of successful SNe per unit star formation, highlighting Sidebar \ref{sidebar:supernovae}. The density dependence is weak: even for an ambient density of $10^5$\,cm$^{-3}$, as might be found in the core of a dense ULIRG like Arp 220, $\dot{p}_{\rm SN}$ is reduced by only a factor of $\simeq4.5$. Finally, the turbulence generated by momentum injection is considered critical to models of a self-regulated ISM. In this picture, star formation generates SNe, which inject momentum, driving turbulence on the scale height of the gas, which balances the self-gravity of the disk, maintaining (in a time-averaged sense) vertical hydrostatic equilibrium 
(e.g., \citealt{Thompson2005,Ostriker2011_maximalI,Shetty2012_ostriker_maximalII,Kim2015_ostriker_momentum_feedback}). 

The shell formation time and radius, $t_{\rm SF}$ and $R_{\rm SF}$, are also important because, when combined with the SFR surface density and a measure of the clustering of massive stars, they determine the probability that subsequent SNe will explode within the preceding evacuated SN remnant volume before the momentum and energy is deposited into the ISM at the end of the momentum conserving phase. This and related ``overlap" conditions for SN remnants are important to discussions of the ISM's phase structure \citep{McKee1977}, the thermalization process that seeds hot outflows (\S\ref{section:hot}) \citep{Bregman1978,Dekel1986,Wyse1985}, and the powering of super-bubbles by successive SNe \citep{MacLow1989,Sharma2014,Li2015,Kim2015_supernovae,Gentry2017,Kim2017_superbubbles}. Finally, the numerical value of $v_{\rm SF}$ is important because it is the same scale of galaxy rotation velocities. The critical projected column density required for a SN to reach shell formation is $N_{\rm SF}\simeq R_{\rm SF}n\simeq7\times10^{19}\,{\rm cm^{-2}}E_{\rm SN, 51}^{0.29}n_0^{0.58}$. This relatively small column and the fractional dependence on $n$ imply that most SNe will deposit momentum $p_{\rm SF}$. 

The question of how the collective effects of these feedback processes, and especially the SNe, produce galactic winds, is a major avenue of current theoretical work. Many numerical efforts have explored the process of driving turbulence in the ISM and launching winds into the NGM with SNe. An example is \cite{Fielding2018}, who show that clustered SNe can drive powerful hot outflows after breakout from the gas disk, and highlight the fact that once this breakout occurs, much of the energy is vented into the NGM. Recent simulations like those of \cite{Gatto2017,Kim2020_TIGRESS,Kim2020_SMAUG,Rathjen2021,Rathjen2023} capture the physics of star formation, feedback, and wind launching together.

\begin{textbox}[t]\section{Energy, Momentum, and Element Injection Distribution Functions of Massive Star Supernovae}
In galaxy formation studies, it is often assumed that all massive stars ($\gtrsim10$\,M$_\odot$) produce successful SNe, each with an energy of $10^{51}$\,ergs (but, see \citealt{Gutcke2021}). The SN rate per unit SFR and the SN energy per explosion determine the net ISM energy, momentum, and CR injection rates, but the fraction of collapses yielding explosions, the explosion energy as a function of progenitor mass, and the lower mass limit for explosions are all uncertain from both theory and observation. Theoretical models yield a complicated landscape of (un)successful explosions as a function of progenitor mass, with a range of explosion energies \citep{Ugliano2012,Pejcha2015_landscape}. The more numerous, lowest mass progenitors ($\lesssim10$\,M$_\odot$) generally yield the smallest explosion energies and Fe and $\alpha$-element yields. For example, SN\,1054 (the Crab) has an estimated energy of just $\sim10^{50}$\,ergs \citep{Smith2013}, with a progenitor that may have undergone electron-capture  core collapse (\citealt{Tominaga2013}; see also \citealt{Prieto2008,Thompson2009}). An IMF-average of theoretical models for normal SNe may yield numbers below $10^{51}$\,ergs per collapse \citep{Sukhbold2016,Ertl2020}, potentially with a metallicity dependence from BH formation \citep{Pejcha2015_landscape}.  However, these models do not include rare, very energetic events, including GRB-associated SNe, (pulsational) pair instability SNe, or super-luminous SNe. The energies of observed SNe, the mapping between progenitors and explosions, and the fraction of collapses producing BHs \citep{Kochanek2008,Gerke2015,Adams2017_confirmation}, are topics connected across astrophysics. An important connection between galactic winds, feedback, and nucleosynthesis is that the failure rate of SNe as a function of metallicity is tied to both energy injection, the enrichment history, and the BH mass function  \citep{Griffith2021}. A secure determination of the energy distribution and nucleosynthetic production function of SNe, as a function of metallicity, could have important impact across galaxy formation studies. 
\label{sidebar:supernovae}
\end{textbox}

\subsubsection{Dynamics of the Hot Phase}
\label{section:hot}

Assuming that a very hot phase can develop, many models consider (and simulations produce) something akin to \cite{Chevalier1985} (CC85), a particularly simple model problem for the hot gas that provides a background for understanding mixing with cooler  phases, and the dynamical momentum and energy exchange between them. In the CC85 problem, we take a spherical region $R$ representing the region in the host galaxy where a SN-energized hot phase can be generated. Energy and mass are injected at a constant rate per volume with normalizations
\begin{equation}
    \dot{E}_{\rm hot}=\alpha\dot{E}_{\rm SN}\simeq3\times10^{41}{\rm ergs\,\,s^{-1}}\,\alpha\,({\rm SFR}/{\rm M_\odot/yr})\,\,\,\,\,{\rm and}\,\,\,\,\,\dot{M}_{\rm hot}=\eta\,{\rm SFR},
    \label{eq:alpha_eta}
\end{equation}
where we assume $10^{51}$\,ergs per SN per 100\,M$_\odot$ of new stars formed (see Sidebar \ref{sidebar:supernovae}).  In principle, the thermalization and mass-loading efficiencies ($\alpha$ and $\eta$), which are often taken to be free parameters, should be derived from a comprehensive theory, and may thus be functions of the galaxy gas surface density, $\dot{\Sigma}_\star$, metallicity, and other host properties. 

Assuming constant energy and mass injection, and neglecting gravity, radiative cooling, conduction, self-ionization, or other processes, a self-similar solution inside and the injection region $R$ can be derived.  The sonic point occurs at $R$, and the characteristic temperature, density, and velocity there ($r=R$) are (see Table \ref{table:symbols} for symbols and scalings)
\begin{equation}
    T\simeq1.1\times10^7{\rm K}\,\mu_{0.6}\left(\frac{\alpha}{\eta}\right),\,\,\,\,
    n\simeq1.2\,{\rm cm^{-3}}\frac{\eta^{3/2}}{\alpha^{1/2}}
    \frac{\rm SFR_{10}}{R_{0.3}^2\,\mu_{0.6}},\,\,\,\,
    v\simeq500\,{\rm km/s}\left(\frac{\alpha}{\eta}\right)^{1/2}.
    \label{cc85sonic}
\end{equation}
The gas continues to accelerate outside the sonic point at $R$ and expands adiabatically, with asymptotic velocity $v_\infty=(2\dot{E}_{\rm hot}/\dot{M}_{\rm hot})^{1/2}$
and force $\dot{p}_{\rm hot}=(2\dot{E}_{\rm hot}\dot{M}_{\rm hot})^{1/2}$ of
\beq
v_{\infty}\simeq 1000\,{\rm km/s}\left(\alpha/\eta\right)^{1/2}
\hspace*{0.1cm}{\rm and}\hspace*{0.3cm}
\dot{p}_{\rm hot}\simeq5(\alpha\eta)^{1/2}\left(L/c\right)
\label{cc85_vinf_pdot}
\eeq
This model makes testable predictions. Because there is direct observational evidence from X-ray emission for this super-heated gas in wind-hosting starbursts like  M82  (\S\ref{section:optical_xray}), much work has focused its dynamics and interaction with the surrounding cooler material. If hot gas is the prime mover for the colder phases via ram pressure and/or mixing, its momentum budget is set by equation (\ref{cc85_vinf_pdot}), which will be a point of reference for all other wind theories.

\subsubsection{Ram Pressure Acceleration of Cool Clouds}  
\label{section:ram}
\begin{figure}
\centerline{
\includegraphics[width=\textwidth]{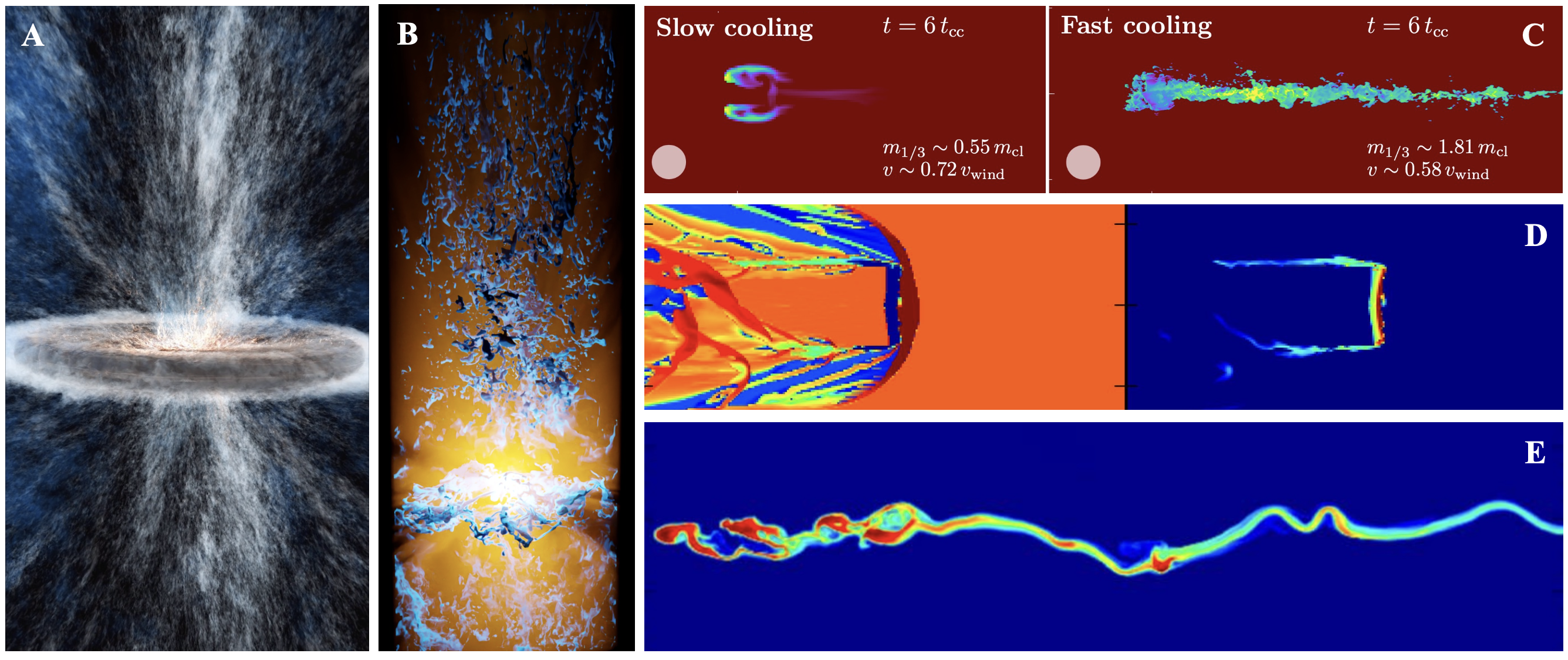}}
\vspace*{.5cm}
\caption{(A) Global star cluster energy-driven multi-phase outflow from \cite{Schneider2020}, exhibiting momentum and energy transfer between phases; (B) Volume rendering of clouds produced by SN-driven outflows from \cite{Tan2024}; (C) Density colormap of cool cloud acceleration and disruption by a hot wind with slow cooling (left), and survival and acceleration with fast cooling (right), from \cite{Gronke2018} (\S\ref{section:entrainment}); (D) Optically-thin radiation pressure accelerated dusty cloud in a background medium from \cite{Zhang2018} (\S\ref{section:radiation_pressure}); (E) Cloud acceleration by CRs from \cite{Bruggen2020} (\S\ref{section:cosmic_rays}).}
\label{figure:theory}
\end{figure}

A principal potential effect of hot winds is in accelerating the leaves and snowflakes of galactic winds, the cool clouds seen in emission and absorption. See Figure \ref{figure:theory} for examples. The classic picture is that the expanding flow ram pressure accelerates cool clouds. Like a person levitated in a wind tunnel on Earth, the critical characteristic of the cloud is its projected mass surface density. For a cloud at rest, with mass $M_{\rm cl}$, projected area $A_{\rm cl}$, and drag coefficient $\xi$ embedded in a wind of density and velocity $\rho_{\rm w}$ and $v_{\rm w}$, 
\beq
M_{\rm cl}\frac{dv_{\rm cl}}{dt}=-gM_{\rm cl}+\frac{1}{2}\xi A_{\rm cl}\rho_{\rm w}v_{\rm w}^2 \,\,\,\,\Longrightarrow\,\,\,\,\frac{dv_{\rm cl}}{dt}=-g+\frac{\xi\rho_{\rm w}v_{\rm w}^2}{2\Sigma_{\rm cl}},
\label{ram}
\eeq
where $\Sigma_{\rm cl}=M_{\rm cl}/A_{\rm cl}$. Setting $dv_{\rm cl}/dt=0$, equation (\ref{ram}) implies a critical ``Eddington" surface density, below which the cloud is accelerated outwards by the wind, and above which it is not: $\Sigma_{\rm cl,\,Edd}\simeq \xi\rho_{\rm w}v_{\rm w}^2/(2g)$. Taking $g=2\sigma^2/r$ for illustration, and using $\dot{M}=\Omega r^2\rho_{\rm w}v_{\rm w}$, the critical ``Eddington" column density is (see Table \ref{table:symbols}; $N_{\rm cl}=\Sigma_{\rm cl}/m$)
\begin{equation}
    N_{\rm cl,\, Edd}\simeq \frac{\xi}{4}n_{\rm w}r\left(\frac{v_{\rm w}}{\sigma}\right)^2\simeq\frac{\xi}{4}\frac{\dot{p}_{\rm hot}}{\Omega \sigma^2 r\,m}\simeq9\times10^{20}\,{\rm cm^{-2}}\,\frac{2\xi(\alpha\eta)^{1/2}{\rm SFR}_{10}}{R_{0.3}\,\Omega_{4\pi}\,\sigma_{150}^2}.
    \label{ram_pressure_edd}
\end{equation}
This expression motivates much of the literature on ram pressure: a hot wind with large $(\alpha \eta)^{1/2}$ can accelerate high-column clouds. Even in galaxies with average gas surface densities above $N_{\rm cl,\,Edd}$, because galaxies are turbulent and present a broad spectrum of cloud columns, low-column clouds with $N_{\rm cl}<N_{\rm cl,\,Edd}$ should be accelerated \citep{Thompson_Krumholz}. Further, since clouds with lower surface density will more rapidly accelerate and reach higher asymptotic velocities, equation (\ref{ram}) suggests a velocity ordering as a function of surface density of the form $v_{\rm cl,\,\infty}\propto\,N_{\rm cl}^{-1/2}$ for low-column (high Eddington ratio) clouds, which translates to a velocity-dependent absorption profile for a given ionic species against the host galaxy's continuum in ``down the barrel" spectroscopy (\S\ref{section:absorption_velocities}).

\subsubsection{Entrainment:} 
\label{section:entrainment}
There has been a long-standing critical problem facing the picture of ram pressure accelerated cool clouds: over a broad range of parameters, previous studies show that cool clouds are shocked, shredded by hydrodynamical instabilities, and rapidly incorporated into the hot background medium (e.g., \citealt{Klein1994,Cooper2009,Scannapieco2015,Bruggen2016,Schneider2015,Schneider2017}). Yet, we see cool gas at high velocity. This ``entrainment problem" is severe. The  timescale to destroy a cloud of radius and density $r_{\rm cl}$ and $\rho_{\rm cl}$ is the ``cloud crushing timescale," $t_{\rm cc}\sim (\rho_{\rm cl}/\rho_{\rm w})^{1/2}(r_{\rm cl}/v_{\rm w})$, while the timescale to accelerate it to $v_{\rm w}$ is $t_{\rm acc}\sim(\rho_{\rm cl}/\rho_{\rm w})(r_{\rm cl}/v_{\rm w})$. Destruction is thus always faster than acceleration by a factor of $t_{\rm cc}/t_{\rm acc}\sim(\rho_{\rm w}/\rho_{\rm cl})^{1/2}$. In short, in the absence of other effects, cool clouds should always be destroyed before they are accelerated \citep{Zhang2017}. This is the ``slow cooling" regime shown in panel (C) in Figure \ref{section:theory}.

The literature addressing the entrainment problem has recently grown, with significant conceptual progress. A flurry of papers show that clouds with particular properties can not only survive, but thrive as they are accelerated. The critical piece of physics that allows this is radiative cooling in the mixing layer between the hot high-velocity wind medium and the shredding cloud in the long cometary tail produced by the wind-cloud interaction. \cite{Gronke2018} argue that earlier simulations of radiative wind-cloud interactions did not run in a computational box large enough to capture cooling in the extended mixing layer, which may be tens to hundreds of times longer than the original cloud size. The criterion distinguishing cloud survival and growth from rapid destruction is the ratio of $t_{\rm cc}$ to the cooling time in the mixing layer ($t_{\rm cool,\,mix}$), which has temperature $T_{\rm mix}\simeq(T_{\rm w}T_{\rm cl})^{1/2}$ and density $n_{\rm mix}\simeq(n_{\rm w}n_{\rm cl})^{1/2}$ \citep{Begelman1990}. Setting these two timescales equal, one finds a critical cloud size, above which clouds should grow \citep{Gronke2018}, which can be written as a critical cloud column density:
\begin{equation}
    N_{\rm cl,\,grow}\simeq n_{\rm cl} r_{\rm cl}\gtrsim \frac{k_{\rm B} T_{\rm mix}}{m_p}\frac{v_{\rm w}}{\Lambda(T_{\rm mix})}\simeq5\times10^{18}\,{\rm cm^{-2}}\,\left(\frac{\alpha}{\eta}\right)^{1/2}T_{\rm mix,\,5.5}\,\,\Lambda_{-21.3},
    \label{ngrow}
\end{equation}
which highlights the role of the mixed phase sound speed, and where $T_{\rm mix,\,5.5}=T_{\rm mix}/10^{5.5}$\,K, $\Lambda_{-21.3}=\Lambda(T_{\rm mix})/10^{-21.3}$\,ergs cm$^3$ s$^{-1}$ is the cooling function evaluated at $T_{\rm mix}$, and where we have taken $v_{\rm w}$ at $r=R$ (eq.~\ref{cc85sonic}). Panel (C) of Figure \ref{figure:theory} shows a cloud in this ``fast cooling" regime, which accelerates and grows. Many recent papers interrogate the physics of mixing between the cool and hot wind phases. Importantly, there is general agreement that above a certain cloud size or column density, clouds can survive, grow, and accelerate \citep{Gronke2020b,Ji2019,Sparre2019,Sparre2020,Tan2021,Fielding2020,Li2020,Kanjilal2021,Abruzzo2022}. The quantitative details depend on the definition of cloud survival, numerical resolution, the parameter space explored, and the physics employed. On the scale of individual clouds, critical theoretical issues include the role of magnetic fields for cloud survival, mixing, and acceleration, and the role of thermal conduction across these interfaces \citep{McCourt2015,Scannapieco2015,Bruggen2016,Armillota2017,Sparre2020,Banda-Baragan2016,Banda-Baragan2018,Banda-Barragan2021,Cottle2020,Sparre2020}. An additional piece of physics in the hot-cool interaction is charge exchange, which is found to contribute significantly to the observed X-ray emission (\S\ref{section:xray_emission}), and which may arise at interfaces in the mixing process.

Critically, $N_{\rm cl\,grow}\ll N_{\rm cl,\,Edd}$ (eqs.~\ref{ram_pressure_edd} and \ref{ngrow}): there is space between the smallest column clouds ($N_{\rm cl}<N_{\rm cl,\,grow}$), which are rapidly destroyed, and the highest-column clouds, which survive, but that cannot be accelerated by ram pressure alone ($N_{\rm cl}>N_{\rm cl,\,Edd}$). Between these two landmarks, we expect a spectrum of cool clouds, with the lowest column material reaching the highest velocity, potentially in accord with observations. Yet, this comparison between $N_{\rm cl,\,grow}$ and $N_{\rm cl,\,Edd}$ is not the whole story. Ram pressure acceleration is only a piece of the puzzle. Cloud-scale \citep{Gronke2020b} and global simulations show that the mixing process itself transmits energy and momentum to the cool phase without ram pressure acceleration per s\'e \citep{Cooper2008,Schneider2020,Tan2024}. See panels (A) and (B) in Figure \ref{figure:theory} for examples. As clouds are accelerated, mixing may increase their surface densities enough that they are no longer super-Eddington. Semi-analytic models of the cool-hot interaction may shed light on how the cool and hot components affect each other by making comparisons with observations of both phases \citep{Fielding2022}. Similarly, models and simulations can be analyzed for absorption and emission diagnostics for direct comparison with imaging and spectroscopy in the UV/optical and X-ray (e.g., \citealt{Krumholz2017,delacruz2021,Yuan2023}). 

These works and others are critically important for comparing the global momentum and energy budgets of observed winds to theory. The missing piece is a complete theory for $\alpha$ and $\eta$ as a function of the star formation surface density and gas surface density, in a self-consistent global calculation with star formation. If $\alpha$ and $\eta$ are large enough ($\dot{E}$ and $\dot{p}$; eqs.~\ref{cc85_vinf_pdot}), and if the dynamical coupling between the hot and cooler phases is efficient enough, scaling relations like those presented in \S\ref{section:confronting_theory_observations} may be understood and explained, together with important additional predictions for the X-ray emission profile, hot gas dynamics, abundance profiles (which can constrain wind-cloud mixing), and the cool gas acceleration profile as a function of cloud column density (e.g., \citealt{Schneider2020}).
 
\subsubsection{Additional Physics of The Hot Phase} 
\label{section:additional_physics_hot_phase}
The physics of the hot wind does not end with equations (\ref{cc85sonic}). Many pieces of physics may be needed to understand its dynamics that are still under investigation and that may prove critical in interpreting observations. 

{\bf Non-spherical Expansion:} X-ray observations of starbursts reveal cylinder- or frustum-like bicone geometries  (\S\ref{section:xray_emission}) \citep{Strickland2002,Lopez2020}, which changes our expectations for the run of density, temperature, and velocity along the wind axis. The collimated nature of the X-ray surface brightness profiles may arise from the interaction of the hot gas with the dense star-forming disk \citep{Strickland2000,Cooper2008}, or from the action of the cocooning cooler wind phases that envelop the hot outflow, magnetic stresses (\S\ref{section:magnetic}), the geometry of energy/mass injection \citep{Nguyen2022}, or all of the above.

{\bf Mass Loading:} 
The dynamics and thermodynamics of the hot phase will be affected as low-column clouds are destroyed and incorporated into the hot flow. The equations governing an initially hot expanding flow with mass-loading from a cooler phase are similar to a free thermal wind, but with important changes to mass, momentum, and energy conservation \citep{Suchkov1996,Nguyen2021}. Along the wind axis $s$,
\beq
\frac{1}{A}\frac{d}{ds}(A\rho v)=\dot{\mu},\hspace{0.3cm}
v\frac{dv}{ds}=-\frac{1}{\rho}\frac{dP}{ds}-\nabla_s\Phi-\frac{\dot{\mu}v}{\rho}, \hspace{0.3cm}
v\frac{d\epsilon}{ds}-\frac{vP}{\rho^2}\frac{d\rho}{ds}=\dot{q}+\frac{\dot{\mu}}{\rho}\left[\frac{v^2}{2}-\epsilon-\frac{P}{\rho}\right],
\label{massloading}
\eeq
where $\epsilon$ is the internal energy, $\nabla_s\Phi$ is the gravitational acceleration, $A(s)$ is the ``areal" function describing the non-spherical geometry of the wind, $\dot{\mu}$ is the mass-loading rate (g/cm$^3$/s) of the cool material into the hot wind, and $\dot{q}$ is the net radiative heating/cooling. Here, the injected gas is assumed to have negligible initial thermal and kinetic energy content for simplicity. In the high Mach number limit ($\mathcal{M} \gg 1$),  the effects of mass-loading and non-spherical areal divergence on the hot phase \citep{Nguyen2021} can be seen: 
\begin{equation}
\frac{d \ln v}{d\ln s} = - \frac{5}{3} \frac{\dot{\mu}s}{v \rho},  \,\,\,\,\, 
\frac{d \ln \rho}{d\ln s} = \frac{8}{3} \frac{\dot{\mu}s}{v \rho} - \dot{a},\,\,\,\,\,
\frac{d\ln T }{d\ln s} =  \frac{5}{9} \mathcal{M}^2 \frac{\dot{\mu}s}{v \rho} - \frac{2\dot{a}}{3},\,\,\,\,\, 
\frac{d\ln K}{d\ln s} = \frac{5k_{\rm B}}{9} \frac{ \dot{\mu}s}{v\rho} \mathcal{M}^2 
\label{eq:mass_loading}
\end{equation}
where $K$ is the entropy and $\dot{a}=d \ln A/d\ln s$. The importance of the mass-loading timescale $\rho/\dot{\mu}$ relative to the advection timescale $s/v$ is evident. The first two equations reflect momentum conservation and continuity: adding mass decelerates the wind and increases $\rho$. The last two expressions reflect the importance of the term in brackets in equation (\ref{massloading}): for $\mathcal{M}>1$, the kinetic energy is partially thermalized through mass-loading, with a heating rate of $\dot{\mu}v^2/(2\rho)$. This process increases the entropy. These expressions show how the density and temperature gradients derived from X-rays can potentially be used to infer $\dot{\mu}$ and to make predictions for the hot gas dynamics, which may be measured by current and upcoming X-ray missions (e.g., \citealt{XRISM2020}).  As an example, using analysis by \cite{Lopez2020}, \cite{Nguyen2021} find that the mass-loading from the cool phase nearly doubles the total mass loss rate from M82, decreasing the asymptotic outflow velocity by a factor of $\simeq1.5$ compared to the case without mass-loading. 

{\bf Radiative Cooling:} If the hot wind is sufficiently mass-loaded, it may become radiative, undergo bulk cooling, and precipitate high-velocity cool gas directly, without ram pressure acceleration or energy/momentum mixing from the hot to the cooler phases. Once a super-heated $\sim10^7-10^8$\,K thermal gas has been produced, the inevitability of expansion and adiabatic cooling, combined with the shape of the radiative cooling function, which becomes more efficient as the temperature decreases below $\sim10^{7.5}$\,K, guarantees that sufficiently mass-loaded winds can become radiative. This process was explored for galactic winds by \cite{Wang1995a,Wang1995b} and by several subsequent works \citep{Efstathiou2000,Silich2003,Silich2004}. For given hot wind parameters, equating the advection and cooling timescales, we derive an expression for the cooling radius (\citealt{Thompson2016}; see Table \ref{table:symbols}),
\begin{equation}
    r_{\rm cool}\simeq1\,{\rm kpc}\,(\alpha\,\mu_{0.6})^{2.13}\eta^{-2.92}R_{0.3}^{1.79}(\Omega_{4\pi}/{\rm SFR}_{10})^{0.789},
    \label{eq:rcool}
\end{equation}
which highlights the strong dependencies on $\alpha$ and $\eta$ (eq.~\ref{eq:alpha_eta}), and the role of the $\dot{\Sigma}_\star={\rm SFR}/(\pi R^2)$: $r_{\rm cool}\propto R^{0.269}{\dot{\Sigma}_\star}^{-0.789}$. Equating $r_{\rm cool}$ with $R$ using the full solution to equations (\ref{cc85sonic}), we derive a critical $\eta$ such that cooling happens on the scale of the host: $\eta_{\rm crit}\simeq1$ for the parameters above. The column density of the cooling material on scale $R$ is $\simeq10^{20}$\,cm$^{-2}$ for the same parameters and its velocity is (\S\ref{section:observations}; Fig.~\ref{figure:xu1}).
\beq
v(r_{\rm cool}=R) \simeq 720\,{\rm km/s} \left[\alpha\xi\,{\rm SFR_{10}}/(\Omega_{4\pi}R_{0.3})\right]^{0.135}.
\label{eq:vcrit}
\eeq
It is tempting to say that bulk cooling from the hot phase is {\it the} answer to the question of how cool material achieves high velocities in galactic winds. But, there are several problems. If the hot medium cools as it expands, it must radiate its thermal energy with $dL_X/d\ln T\simeq(3/2)(k_{\rm B}T/m_p)\dot{M}\simeq2.5\times10^{41}T_{6.5}\eta\,{\rm SFR}_{10}$\,ergs/s \citep{Thompson2016}. This expectation for $L_X$ is generally larger than the band-corrected observed values \citep{Zhang2014}. While the X-ray luminosity at the base of NGC\,253's outflow may meet the criterion for bulk cooling \citep{Lopez2023}, M82 seems not to have high enough $L_X$ to accommodate this hypothesis \citep{Thompson2016}. An additional related issue is that in some systems like the ULIRG Arp 220, we see high-velocity molecular gas \citep{Rangwala2011,Barcos-Munoz2018}. It is unclear if molecular gas can form directly from bulk cooling of an initially super-heated outflow (see \citealt{Richings2018}).

Bulk radiative cooling may also interact with mass loading into the hot phase. Equations (\ref{eq:mass_loading}) are only applicable in the high Mach number and non-radiative limit, corresponding to low $\dot{\mu}$. For large enough $\dot{\mu}$ the character changes: the flow can decelerate sufficiently to pass through a sonic point and become radiative, producing bulk cooling and high-density filaments \citep{Nguyen2024}. This process of cool mass-loading into the hot phase, to produce cooling from the hot phase back into the cool phase -- cool cloud ``transmigration" \citep{Thompson2016} -- is just one part of the phase-dependent interplay at work in simulations \citep{Cooper2008,Schneider2020,Tan2024}.

{\bf Non-Equilibrium Ionization and Self-Ionization:} The estimates above and most observational and theoretical analyses in the literature assume a typical form of the cooling function of hot gas, which generally assumes collisional or photo-ionization equilibrium. However, as the medium rapidly expands from high to low density, ionization equilibrium may break down. Setting the recombination time equal to the advection time for a representative ionic species makes this breakdown evident: lower mass-loading increases the recombination time on a given scale and decreases the advection time. The result of ionization equilibrium breakdown is a rapidly-expanding medium that is ``over-ionized," with abundances for atomic species that one might naively conclude could not exist at the implied temperature. This physics has been explored by \cite{Gray2019a,Gray2019b} and \cite{Sarkar2022}, among others.  \cite{Sarkar2022} presents an analysis including both an accounting of the ionization structure of hot outflows and an assessment of the effect of self-irradiation. As the medium cools, it radiates photons that affect the ionization state of the wind medium itself. A critical outstanding issue is to understand whether these effects make the hypothesis of bulk cooling as a production mechanism for fast cool wind material more or less consistent with observations, and to include these same effects in works on cloud survival and growth via mixing layers at the wind-cloud interface (\S\ref{section:entrainment}).

{\bf Conduction:} While conduction is increasingly discussed in the context of cloud destruction/growth (e.g., \citealt{Armillota2017}), there has been less work on its potential role for the hot phase of galactic winds, where it could shape the thermodynamic gradients and affect the interpretation of X-ray observations. The local volumetric energy deposition rate from conduction is $\dot{\varepsilon}_{\rm cond}=\nabla\cdot[\kappa(T)\nabla T]$ where $\kappa(T)\simeq6\times10^{-7}\,T^{5/2}$\,ergs/cm/s/K is the Spitzer conductivity. Assuming radial magnetic fields in the hot phase and adiabatic expansion, the conduction heating time is $\sim P/\dot{\varepsilon}_{\rm cond}$:
$t_{\rm cond}\simeq9\dot{M}k_{\rm B}/(68\Omega\mu m_p v \kappa(T))\simeq 2\,{\rm Myr}(\eta^4/\alpha^3)({\rm SFR}_{10}/\Omega_{4\pi})$. Setting $t_{\rm cond}$ equal to the advection time at $R$, we obtain a critical mass-loading rate below which conduction dominates: $\eta_{\rm cond}\simeq0.5\,\alpha^{5/7}(\Omega_{4\pi} R_{0.3}/{\rm SFR}_{10})^{2/7}$
Analysis of X-ray observations by \cite{Strickland2009} implies $\eta\simeq0.2-0.5$ and $\alpha\simeq0.5-1$. Comparison with $\eta_{\rm cond}$ shows that conduction may dominate energy transfer on few-hundred pc scales, changing the expected temperature profile. Low $\eta$ will also produce faster advection and a more likely breakdown of ionization equilibrium. The combination of conduction, non-equilibrium ionization, and self-ionization may be required to understand the hot phase well enough to facilitate an apposite comparison with X-ray data. Moreover, comparisons between the electron collisional mean-free path and the temperature gradient, and between the electron-proton equilibration timescale and the advection timescale both indicate that for $\eta\lesssim0.3$, the inherent assumptions underlying a single-temperature collisional and hydrodynamical conducting medium will break down \citep{Thompson2016}. Heat transfer by conduction in low-collisionality plasmas can be suppressed by instabilities (e.g., \citealt{Roberg-Clark2018}), further complicating the physics. More detailed treatments of the hot medium, akin to two-temperature models of the Solar wind, may be required to understand hot winds when $\eta$ is small.

{\bf Dust:} Both large dust grains and PAHs are present throughout an extended halo along the wind axis, both surrounding and perhaps co-spatial with the hot gas in starburst outflows (\citealt{Hoopes2005,Roussel2010,Beirao2015}; Fig.~\ref{figure:m82_hst}). Yet, for the high temperatures of hot thermal winds, dust grains should rapidly sputter on a timescale \citep{Draine1979a,Draine1979b} $t_{\rm sputter}\simeq0.2\,{\rm Myr}(a/{\rm 0.1\,\mu m})({\rm cm^{-3}}/n)$, where $a$ is the grain size and $n$ is the hot gas volume density. On the scale of a few-hundred pc, $t_{\rm sputter}$ is {\it less than} the advection time (eq.~\ref{cc85sonic}) for grains smaller than $\simeq0.4$\,$\mu$m if they are already comoving with the flow. For constant $\dot{M}$  and velocity $n\propto r^{-2}$, and $t_{\rm sputter}$ should increase rapidly with $r$. But grains will be sputtered, charged, and potentially ablated as they are accelerated, affecting all relevant timescales, including the time needed to become comoving, $\simeq (8/3)(a/v_{\rm w})(\rho_{\rm grain}/\rho_{\rm w})$. This picture will be further complicated as cool clouds, which may be dusty, mix material with the hot phase (\S\ref{section:entrainment}). See, e.g., \cite{Kannan2021}.

\subsubsection{Winds from turbulence}
\label{section:turbulence}

Warm/hot winds might be driven without stellar feedback. \cite{Scannapieco2012} identify a critical value of the gas turbulent velocity dispersion of $\simeq35$\,km/s, above which the gas is unstable to runaway heating because of the temperature dependence of the cooling function above $\simeq10^{5.5}$\,K. The work of \cite{Sur2016} explores this physics. In certain regimes, turbulence generated by gravitational instability in marginally Toomre-stable disks produces a turbulent velocity $\gtrsim35$\,km/s, which leads to runaway heating for a fraction of the gas, allowing it to escape the driving region. Over a range of gas surface densities, they find a mass-loading efficiency of $\eta \sim0.1-0.3$.

\subsection{Radiation pressure}
\label{section:radiation_pressure}

Momentum transfer by photons to matter is critical to the dynamics and structure of a broad range of astrophysical systems, from massive stars to quasar disks. The most important radiation forces for galactic outflows may be the scattering and absorption of starlight by dust, the absorption of ionizing photons, and the scattering of Ly$\alpha$ photons.\footnote{Metal and molecular line opacities have not been extensively considered in work on galactic winds. While Thomson scattering is important for many systems, it is likely too small to dominate galactic winds for standard stellar IMFs  (eqs.~\ref{eq:lum_continuous} \& \ref{lm}), although it is close to $\kappa_{\rm R}$ at low $T$ (eq.~\ref{eq:kappa_ross}).}

Much of the physics of radiation pressure driven flows is generic. Consider the case of an optically-thin medium with negligible thermal pressure. The momentum equation for a time-steady spherical flow with opacity $\kappa$, around a point mass $M$ with luminosity $L$, is\footnote{Write eq.~(\ref{radiation1}) as $d/dt(M_{\rm cl} v)$ and add an expression for $dM_{\rm cl}/dr=\Omega_{\rm cl}(r) r^2 \rho(r)dr$ with $dt=dr/v$ to understand how a cloud of solid angle $\Omega_{\rm cl}$ that sweeps up mass affects the dynamics.} 
\beq
v\frac{dv}{dr}=-\frac{GM}{r^2}+\frac{\kappa\,L}{4\pi r^2 c}=\frac{GM}{r^2}\left(\Gamma-1\right)\,\,\,\Longrightarrow\,\,\,({\rm setting}\,\,dv/dr=0) \,\,L_{\rm Edd}=\frac{4\pi G Mc}{\kappa}
\label{radiation1}
\eeq
where $L_{\rm Edd}$ is the Eddington luminosity and $\Gamma=L/L_{\rm Edd}$ is the Eddington ratio. The momentum equation as written is equally correct for an optically-thin cloud of projected area $A_{\rm cl}$, mass $M_{\rm cl}$, and optical depth defined as $\tau_{\rm cl}=\kappa M_{\rm cl}/A_{\rm cl}$. For $\Gamma>1$, an outflow is driven. If the opacity and cloud/shell mass is constant, the momentum equation can be integrated to give the velocity profile and asymptotic velocity \citep{Gerasimovic1932,Schalen1939}
\beq
v(r)=\left[v_0^2+\frac{2GM}{R}\left(1-\frac{R}{r}\right)\left(\Gamma-1\right)\right]^{1/2}\,\,\,\Longrightarrow
\,\,\,v_\infty\simeq v_{\rm esc}\left(\Gamma-1\right)^{1/2},
\label{radp_vel}
\eeq
where $R$ is the launch radius, $v_0=v(R)\ll v_{\rm esc}$, and $v_{\rm esc}=(2GM/R)^{1/2}$ is the escape velocity. The mass loss rate can be calculated from momentum conservation. Multiplying both sides of equation (\ref{radiation1}) by $4\pi r^2\rho$ and employing the continuity equation, 
\beq
\dot{p}_\infty=\dot{M}v_\infty=\frac{\tau\,L}{c}-4\pi G M \int \rho dr =\tau\frac{L}{c}\left(1-\frac{1}{\Gamma}\right)\,\,\,\Longrightarrow\,\,\,\dot{M}\simeq \tau\frac{L}{v_{\rm esc}c}\frac{1}{\Gamma}\left(\Gamma-1\right)^{1/2},
\label{pdot_mdot_radiation}
\eeq
which emphasizes the key role of the wind optical depth $\tau=\int \kappa\rho\,dr$, and shows that  $\dot{p}\rightarrow\tau L/c$ as $\Gamma\rightarrow\infty$. Note that for a given $L$, $\dot{M}$ has a {\it maximum} at $\Gamma=2$ with $v_\infty\simeq v_{\rm esc}$ (eq.~\ref{radp_vel}). Since $\Gamma\propto L$, for all else constant $\dot{M}\propto L^{1/2}$, for large $\Gamma\gg1$.

When the medium is highly optically-thick, many effects can become important, but in its simplest version, the radiation pressure term in equation (\ref{radiation1}) becomes $(1/\rho)\nabla P_{\rm rad}$, where ${\bf F}=-(c/\kappa_{\rm R}\rho)\nabla P_{\rm rad}$ is the flux and $\kappa_{\rm R}$ is the Rosseland-mean opacity. The acceleration is then $\kappa_{\rm R} F/c$, and in analogy with equation (\ref{pdot_mdot_radiation}), the momentum budget is set by $\dot{p}_\infty=\dot{M}v_\infty\simeq\tau_{\rm R} L/c$, where the optical depth $\tau_{\rm R}=\int \kappa_{\rm R}\rho dr$ can in principle be much larger than unity. Photons can give up their momentum many times \citep{Gayley1995}.\footnote{Consider a single photon of initial energy $E$ reflecting between two freely-floating, perfectly reflecting  mirrors of mass $m$, starting at rest with $E\ll mc^2$. Approximately two times the incoming photon momentum is transferred to each mirror {\it per scattering} and the photon is redshifted. Eventually, the photon's initial energy $E=pc$ is shared with each mirror. The final mirror momenta are equal and opposite, save the initial photon momentum, and of order $(mcp)^{1/2}$.  The boost in mirror momentum relative to the initial photon momentum may be very large: $(mc/p)^{1/2}=(mc^2/E)^{1/2}$.} 
However, the momentum content of optically-thick radiation pressure driven flows is bounded by an upper limit on the optical depth set by energy conservation: when an outflow has $\tau\sim c/v$ the work done by the radiation field can be large compared to $L$ (e.g., $L\sim\dot{M} v^2$). This is the ``photon tiring" limit  \citep{Owocki1997,Owocki2017}, analogies of which appear in our discussion of Ly$\alpha$ scattering (\S\ref{section:rp_lyman}) and CR-driven winds (\S\ref{section:cosmic_rays}).

In generalizing these results to galactic winds, the fact that galaxies have extended potentials again affects the dynamics. Assuming an isothermal potential, the gravitational acceleration in equation (\ref{radiation1}) becomes $2\sigma^2/r$. Different from the point-source limit, the ratio between the gravitational and radiation pressure terms grows with distance so that even if the system has $\Gamma > 1$ at $R$, at sufficiently large radius the medium will become sub-Eddington and begin decelerating (see, e.g., \citealt{Murray2005}).  Additionally, the driving luminosity $L$ is related to the stellar population. For continuous star formation, the bolometric luminosity $L$ is proportional to the SFR:
\begin{equation}
    L=\epsilon_{\rm ph}\,{\rm SFR}\, c^2\simeq10^{10}\,{\rm L_\odot}\,({\rm SFR}/{\rm M_\odot\,\,yr^{-1}}), 
    \label{eq:lum_continuous}
\end{equation}
where $\epsilon_{\rm ph}\simeq5-7\times10^{-4}$ for a typical IMF. For an instantaneously-formed ZAMS stellar population, $L$ is initially approximately constant for a timescale set by the lifetime of the most massive stars\footnote{The lifetime of the most massive stars is independent of mass and set by the efficiency of H fusion ($\epsilon\simeq0.007$) and the Thomson Eddington limit: $E/L_{\rm Edd} \sim \epsilon Mc^2/(4\pi GMc/\kappa_{\rm T})\sim\epsilon\kappa_{\rm T} c/4\pi G\sim3$\,Myr with $\kappa_{\rm T}\simeq0.4$\,cm$^2$/g, which is closely related to the Salpeter timescale for black hole growth.} and proportional to the total stellar mass $M$ formed: 
\begin{equation}
    L/M\simeq1000-2000\,\,{\rm L_\odot/M_\odot}\,\,(t\lesssim4\,{\rm Myr})
    \label{lm}
\end{equation}
but then declines with time, roughly as $L\sim t^{-1}$. The normalizations of equations (\ref{lm}) and (\ref{eq:lum_continuous}) are appropriate for populations with a standard low-mass IMF (e.g., Chabrier, Kroupa), and depend on the high-mass cutoff ($100$ or $300$\,M$_\odot$ matters) and whether the shape of the high-mass IMF is fully-populated and precisely Salpeter ($dN/dM\propto M^{-2.35}$; \citealt{Leitherer1999}; for deviations, see \citealt{Schneider2018_IMF}). Stochasticity in IMF sampling can change $L/M$ significantly for ZAMS populations \citep{dasilva2012_slug}.

That radiation pressure may be important for accelerating cool gas in galactic winds is a question of whether or not the system exceeds the Eddington limit. We first discuss the case of dust opacity and then turn to ionizing photons and Ly$\alpha$ scattering. 

\subsubsection{Optically-Thin Limit} 
\label{section:rp_thin}
Because of the large dust opacity, radiation pressure on dusty gas may be an important mechanism in driving cool galactic outflows (e.g., \citealt{Chiao1972}). The radiation pressure-mean opacity per gram of gas is
\beq
\langle \kappa_{\rm rp} \rangle = \frac{f_{\rm dg}\int \pi a^2 (dn/da) \int [Q_{{\rm abs}} (a,\lambda) + (1-g(a,\lambda))Q_{{\rm scatt}}(a,\lambda)]L_{\lambda} \,d\lambda \,da}{\int (dn/da) (4\pi/3) \rho_{\rm grain} \,a^3 da\,\int L_\lambda\,d\lambda},
\label{eq:kapparp}
\eeq
where $a$ is the grain radius, $dn/da$ is the dust grain size distribution, $\rho_{\rm grain}$ is the grain mass density, $f_{\rm dg}$ is the total dust-to-gas mass ratio, $Q_{\rm abs}$ and $Q_{\rm scatt}$ are the dimensionless grain absorption and scattering efficiencies (relative to $\pi a^2$), and $g(a,\lambda)=\cos\theta$ is the size- and wavelength-dependent scattering angle (purely backward(forward) scattering corresponds to $g=-1(+1)$). Young stellar populations are intrinsically bright in the UV and optical, where the scattering is predominantly forward-throwing, decreasing the net momentum transfer per interaction. The steep nature of the dust grain size distributions -- e.g., $dn/da\propto a^{-3.5}$ for ``MRN" \citealt{Mathis1977}) (see also \citealt{Weingartner2001,Hensley2023}) -- means that the smallest dust grains dominate the opacity. Taking both $Q_{\rm abs}$ and $Q_{\rm scatt}$ to be unity (reasonable for $\lambda\lesssim a$), momentarily ignoring anisotropic scattering, and taking $dn/da\propto a^{-3.5}$ from $a_{\rm min}$ to $a_{\rm max}$, an estimate is 
\beq
\langle \kappa_{\rm rp} \rangle\sim\frac{3}{4}\frac{f_{\rm dg}}{\rho_{\rm grain}}\frac{1}{(a_{\rm min}a_{\rm max})^{1/2}}\sim 800\,{\rm cm^2\,\,g^{-1}}\,f_{\rm dg,MW}\,\left(\frac{0.001\,\rm \mu m^2}{a_{\rm min}a_{\rm max}}\right)^{1/2},
\eeq
where $f_{\rm dg,MW}=f_{\rm dg}/(1/100)$ and we have taken $\rho_{\rm grain}\simeq3$\,g\,\,cm$^{-3}$. The full integral (eq.~\ref{eq:kapparp}) with a ZAMS population gives $\simeq500$\,cm$^2$ g$^{-1}$, which we use below. Note the importance of $f_{\rm dg}$ and both extrema of the grain size distribution: larger $a_{\rm min}$ or $a_{\rm max}$ decrease $\langle\kappa_{\rm rp}\rangle$ because of the relative importance of the cross section ($\propto a^2$) and mass ($\propto a^3$). For $\langle\kappa_{\rm rp}\rangle_{500}=\langle \kappa_{\rm rp}\rangle/500$\,cm$^2$ g$^{-1}$, the Eddington luminosity for dusty gas is
\beq
 L_{\rm Edd,\,thin}/M_{\rm tot}=4\pi G c/\langle \kappa_{\rm rp}\rangle\simeq30\,\langle \kappa_{\rm rp}\rangle^{-1}_{500}\,\,L_\odot/M_\odot,
\label{eq:edd_thin}
\eeq
where $M_{\rm tot}$  is the total dynamical mass  enclosed within the medium considered, and where we have implicitly assumed that the dust and gas are well-mixed and fully dynamically coupled. Both assumptions may break down (\S\ref{section:rp_limitations_future}). Note that the Eddington luminosity for the dust alone (uncoupled from the gas) is a factor of $f_{\rm dg}$ {\it lower} than equation (\ref{eq:edd_thin}) (e.g., $\sim0.3$\,L$_\odot$/M$_\odot$), explaining why the radiation force on grains is important around Sun-like stars. The dependence of $\langle\kappa_{\rm rp}\rangle$ on $f_{\rm dg}$ also immediately implies a metallicity dependence for radiation pressure feedback on dusty gas (see \S\ref{section:rp_summary}).

The importance of radiation pressure on dust is evident when we compare equation (\ref{eq:edd_thin}) with $L/M$ for stellar populations.  Optically-thin regions with $L/M>L_{\rm Edd,\,thin}/M_{\rm tot}$ should drive outflows. Comparing equations (\ref{lm}) and (\ref{eq:edd_thin}), $L/M$ for a ZAMS population exceeds $L_{\rm Edd,\,thin}/M_{\rm tot}$ by a factor of $\sim30-50$ for $t\lesssim4$\,Myr and would be super-Eddington for $\simeq30$\,Myr. For continuous star formation, the system is super-Eddington on Gyr timescales \citep{Blackstone2023}. We see optically-thin bright star-forming sub-regions throughout the Universe, and we thus expect these to be super-Eddington. For top-heavy IMFs, as observed by \cite{Schneider2018_IMF} in 30 Doradus, $L/M\simeq4000$\,L$_\odot/M_\odot$ and nominally exceeds the Eddington limit by a factor of over 100. However, $f_{\rm dg}$ is likely lower than $1/100$ in 30 Doradus, emphasizing the role of $f_{\rm dg}$ and metallicity for $\langle \kappa_{\rm rp}\rangle$ (eq.~\ref{eq:kapparp}). If the system is super-Eddington, the characteristic velocity is given by equation (\ref{radp_vel}). For massive star clusters with $v_{\rm esc}\sim10$\,km/s and using $\Gamma\sim10-100$, typical velocities would be $v_\infty\sim30-300$\,km/s, potentially seeding galactic winds \citep{Murray2011}.

For whole galaxies with circular velocities of $\sim100-200$\,km/s and with $L/M\sim50-100$ (local starbursts and high-$z$ SFGs), we would expect optically-thin super-Eddington dusty gas to be accelerated to $\sim200-600$\,km/s. We return to this issue in \S\ref{section:confronting_theory_observations} and \S\ref{section:compare_warm_ionized_theory}, but here note that estimates of the Eddington ratios for high-$z$ SFGs using data from, e.g.,   \cite{Erb2006a,Erb2006b,Erb2006c,Wuyts2011a,Wuyts2011b,Shapley2015} can be in excess of unity \citep{Murray2005}, even with metallicity-corrected $f_{\rm dg}$ (see Fig.~\ref{figure:radp_comp}). This is why radiation pressure on dust remains of interest even though M82 appears to be sub-Eddington on kpc scales along its minor axis \citep{Coker2013}.

Multiplying and dividing by area, we can write equation (\ref{eq:edd_thin}) as an the Eddington flux $F_{\rm Edd,\,thin}$ in terms of the total surface density $\Sigma_{\rm tot, 100}=\Sigma_{\rm tot}/100$\,M$_\odot$ kpc$^{-2}$:
\beq
F_{\rm Edd,\,thin}\simeq3\times10^9\,\Sigma_{\rm tot, 100}\,\langle\kappa_{\rm rp}\rangle_{500}^{-1}\,\,L_\odot/{\rm kpc^2}
\eeq
Using equation (\ref{eq:lum_continuous}) we can rewrite the Eddington flux as a critical $\dot{\Sigma}_\star\simeq0.2$\,M$_\odot$ yr$^{-1}$ kpc$^{-2}$ for the same parameters, which is close to the observed  threshold for galactic wind driving in starbursts (\S\ref{section:observations}). If the system exceeds a flux (or $\dot{\Sigma}_\star$) larger than the expressions given above ($\Gamma>1$), the clouds/shells will be ejected with mass-loading rate $\eta_{\rm rp, thin}=\dot{M}_{\rm rp, thin}/{\rm SFR}$ and force of
\beq
\eta_{\rm rp, thin}
\simeq0.4\langle \kappa_{\rm rp}\rangle_{500} \,\Sigma_{g,5} \,v_{\rm circ,200}^{-1}\,\Gamma^{-1}\left(\Gamma-1\right)^{1/2}\,\,\Longrightarrow\,\,\dot{p}_{\rm rp, thin}= \tau_{\rm rp}(L/c),
\label{eq:massloading_rpthin}
\eeq
where $v_{\rm esc}=\sqrt{2}v_{\rm circ}$, $v_{\rm circ,200}=v_{\rm circ}/200$\,km/s is the circular velocity, and $\Sigma_{g,5}=\Sigma_g/5$\,M$_\odot$/pc$^2$, corresponding to an optically-thin ($\tau_{\rm rp}=\langle\kappa_{\rm rp}\rangle\Sigma_g<1$) projected column density $\simeq5\times10^{20}$\,cm$^{-2}$, assuming a Milky Way-like value of $f_{\rm dg}$. This value of $\eta$ is similar to the estimates for CR-driven winds in \S\ref{section:cosmic_rays}, but it increases linearly with column density (for optically-thin columns). It also increases with decreasing galaxy escape velocity as $v_{\rm circ}^{-1}$ at fixed $\Gamma$, as expected from the ``momentum scaling" of \cite{Murray2005}. (For variable $\Gamma$, the $v_{\rm circ}$ scaling changes since $\Gamma\propto L \langle\kappa_{\rm rp}\rangle/M$.)  The velocity of the flow would be $\simeq v_{\rm esc}(\Gamma-1)^{1/2}$. The force of the outflow is given by equation (\ref{pdot_mdot_radiation}) and can be directed compared with that from hot gas given in equation (\ref{cc85_vinf_pdot}).

Are super-Eddington fluxes common? No. Not galaxy-averaged at $z\simeq0$. Writing equation (\ref{eq:edd_thin}) in terms of the total SFR and stellar mass $M_{\star,10}=M_\star/10^{10}$\,M$_\odot$, the Eddington limit is
\beq
{\rm SFR}_{\rm Edd,\,thin}=4\pi GM_\star/(\epsilon_{\rm ph}\langle\kappa_{\rm rp}\rangle c)\simeq 25\,\,M_{\star,10}\langle\kappa_{\rm rp}\rangle_{500}^{-1}\,\,{\rm M_\odot/yr},
\eeq
which translates to a specific star formation rate of ${\rm SFR}/M_\star\simeq2.5$\,Gyr$^{-1}$ and is about 10 times larger than observed for the galaxy ``main sequence" at $z\simeq0$. Thus, galaxies scattered far above the main sequence with low SED-weighted extinctions may exceed the optically-thin dust Eddington limit. Except in their bright star-forming sub-regions \citep{Blackstone2023}, normal late-type galaxies  have $L/M$ far below $\sim30$\,$L_{\rm \odot}/M_\odot$ (eq.~\ref{eq:edd_thin}). In \S\ref{section:confronting_theory_observations}, we compare data for local starbursts and the optically-thin dust Eddington limit. 

\subsubsection{The Single-Scattering Limit} 
\label{section:rp_ss}
Equation (\ref{eq:edd_thin}) applies when the optical depth to the UV/optical continuum is less than unity, corresponding to a projected surface density of $\Sigma_{\rm gas}\lesssim\langle\kappa_{\rm rp}\rangle^{-1}\simeq10$\,M$_\odot$ pc$^{-2}$ or $E(B-V)\lesssim0.2$. For larger columns, we enter the  ``single-scattering" regime where  the UV/optical continuum is absorbed and re-radiated into the far-infrared (FIR), which is optically-thin. In this limit, the dynamics changes because the cloud or shell absorbs all of the incident momentum, like a sail. This effect is particularly striking when we consider a spherical shell of mass $M_g$ around a central source of luminosity $L$. The momentum equation and Eddington luminosity are
\beq
 \frac{d}{dt}(M_g v)\simeq-\frac{GM_{\rm tot}M_{g}}{r^2}+\frac{L}{c}
\,\,\Longrightarrow\,\, \frac{L_{\rm Edd,\,s}}{M_{\rm tot}}= \frac{GM_g c}{r^2} 
\simeq70\,\frac{L_\odot}{M_\odot}\,\left(\frac{\Sigma_g}{100\,{\rm M_\odot/pc^2}}\right),
\label{eq:ledd_s}
\eeq
where we have taken $\Sigma_g=M_g/\pi r^2$. The value of $L_{\rm Edd,\,s}/M_{\rm tot}$ is about $\sim10-30$ times {\it smaller} than $L/M$ for a ZAMS stellar population (eq.~\ref{lm}), and is thus important in the disruption of star-forming clouds (e.g., \citealt{Lopez2011}). Like the case of cloud accelerated by the ram pressure of a hot wind,  $L_{\rm Edd,\,s}/M_{\rm tot}$ is proportional to the projected column density so that, for a given $L$, the low-$\Sigma_g$ sightlines in a turbulent medium will be preferentially super-Eddington \citep{Thompson_Krumholz}.\footnote{For a cloud of mass $M_{\rm cl}$ and projected area $A_{\rm cl}$, the force in eq.~(\ref{eq:ledd_s}) changes to $(A_{\rm cl}/4\pi r^2) L/c$ and $L_{\rm Edd,\,s}/M_{\rm tot}=4\pi G c (M_{\rm cl}/A_{\rm cl})$, proportional to the cloud's mass surface density, as in \S\ref{section:thermal}.} 

The force $L/c$ in the momentum equation is {\it constant} with radius and independent of dust properties as long as the shell is optically-thick, like a spherical spinnaker. In contrast, the gravitational force may decrease with radius. This leads to a run-away effect where, if the shell is slightly super-Eddington, the acceleration will be positive and {\it increasing} while the shell is optically-thick \citep{Elmegreen1982,Thompson2015}. Once its optical depth drops enough, the dynamics transitions to optically-thin  (\S\ref{section:rp_thin}). The dynamics of the shell (or cloud) may be complicated by the amount of matter swept up and by the shape of the potential as the shell is accelerated. Even so, an estimate for the asymptotic velocity is $v_{\infty, \rm s}\simeq 600\,\,\,{\rm km/s}\,\,L_{11.7}^{1/2}\langle\kappa_{\rm rp}\rangle_{500}^{1/4}\,M_{g,8}^{-1/4}$ (where $L_{11.7}=L/5\times10^{11}$\,L$_\odot$; $M_{g,8}=M_g/10^8$\,M$_\odot$;  \citealt{Thompson2015}). For continuous injection, the mass-loading rate and force are (eq.~\ref{eq:lum_continuous}) \citep{Murray2005}
\beq
\eta_{\rm rp,\,s}=\dot{M}_{\rm rp,\,s}/{\rm SFR}\simeq\epsilon_{\rm ph}\,(c/v_{\infty,\rm s})\simeq0.5\,v_{\infty,500}^{-1} \,\,\,\Longrightarrow\,\,\,\dot{p}_{\rm rp,\,s}\simeq(L/c),
\label{eq:mass_loading_singlescattering}
\eeq
where $v_{\infty,500}=v_\infty/500$\,km/s, which can be compared with equations (\ref{cc85_vinf_pdot}) and (\ref{eq:massloading_rpthin}).

\subsubsection{The optically-thick limit}
\label{section:thick}

For large enough column density, the medium is optically-thick to re-radiated FIR photons. The relevant opacity is the temperature-dependent Rosseland-mean, $\kappa_{\rm R}(T)$. The wavelength dependence of the dust absorption coefficient implies $\kappa_{\rm R}\propto T^2$ so that (e.g., \citealt{Hensley2023}):
\beq
\kappa_{\rm R}(T)\simeq 5\,{\rm cm^2\,\, g^{-1}}\,f_{\rm dg,MW}\left(T/200\,{\rm K}\right)^2
\,\,(T\lesssim300\rm K).
\label{eq:kappa_ross}
\eeq
For higher $T$, $\kappa_{\rm R}$ increases more slowly: e.g., $\kappa_{\rm R}(T=10^3{\rm K})\simeq17\rm\, cm^2/g$. The condition $\tau_{\rm R}\gtrsim1$ requires $\Sigma_g\gtrsim\kappa_{\rm R}(T)^{-1}\simeq10^3$\,M$_\odot/{\rm pc^2} \kappa_{\rm R,5}^{-1}$, where $\kappa_{\rm R,5}=\kappa_{\rm R}/5$\,cm$^2$ g$^{-1}$.  The Eddington limit is \citep{Murray2010}
\beq
L_{\rm Edd,\,R}/M_{\rm tot}=4\pi G c/\kappa_{\rm R}(T)\simeq2600\,\,\kappa_{\rm R,5}^{-1}\,\,L_\odot/M_\odot,
\label{eq:ledd_taur}
\eeq
The numerical value of the Eddington light-to-mass ratio is $\sim1-3$ times larger than that produced by a fully-sampled standard ZAMS IMF (eq.~\ref{lm}). Thus, only the most extreme starbursting systems may reach or exceed the optically-thick dust Eddington limit, either because of a somewhat top-heavy IMF or a larger dust-to-gas ratio $f_{\rm dg}>1/100$.

That star formation itself might be regulated by radiation pressure on dust goes back at least to \cite{Loose1982} and \cite{Firmani1994}. For a disk supported in vertical hydrostatic and thermal equilibrium by radiation pressure with $\kappa_{\rm R}\propto T^2$, \cite{Thompson2005} show the disk picks out a characteristic flux of $\sim10^{13}$\,L$_\odot$/kpc$^2$, corresponding to  $\dot{\Sigma}_\star\sim10^3$\,M$_\odot$/yr/kpc$^2$. For different $\kappa_{\rm R}(T)$ scalings, the Eddington flux can increase with total surface density $\Sigma_{\rm tot}$ because  $F_{\rm Edd,\,R}\simeq2\pi Gc \Sigma_{\rm tot}/\kappa_{\rm R}(T)$. Such conditions may be achieved in the densest  starbursts and in the self-gravitating and potentially star-forming outskirts of quasar-fueling disks \citep{Sirko2003,Thompson2005,Crocker2018}. If a disk is radiating at or near $F_{\rm Edd,\rm R}$, \cite{Zhang_Thompson} show that it may be unstable to wind driving, attaining  $v_\infty\sim1-2 v_{\rm circ}$. If the effective Eddington limit can be exceeded, the mass-loading rate $\eta_{\rm rp,\,thick}=\dot{M}_{\rm rp,\,thick}/{\rm SFR}$ and force are (eq.~\ref{eq:lum_continuous})
\beq
\eta_{\rm rp,\,thick}\simeq\tau_{\rm R}(\epsilon_{\rm ph}c/v_\infty)\simeq 4\, \kappa_{\rm R,5} \,\Sigma_{g,4}\,v_{\infty,500}^{-1}\,\Longrightarrow\, \dot{p}_{\rm rp,\,R}\simeq\tau_{\rm R}(L/c),
\eeq
where 
$\tau_{\rm R}=\kappa_{\rm R}\Sigma_g$ is the optical depth, $\Sigma_{g,4}=\Sigma_g/10^4$\,M$_\odot$ pc$^{-2}$ is the column density, and this expression can be compared with equations (\ref{cc85_vinf_pdot}), (\ref{eq:massloading_rpthin}), and (\ref{eq:mass_loading_singlescattering}).

The key quantitative question is whether or not the outflow observed in extreme systems like Arp 220 and other ULIRGs might plausibly be driven by radiation pressure in the optically-thick limit. Equation (\ref{eq:ledd_taur}) would require a somewhat top-heavy IMF or larger dust-to-gas ratio. For fiducial numbers, most estimates suggest that the Arp 220 nuclei are below, but potentially near, an Eddington ratio of unity \citep{Barcos-Munoz2015,Scoville2015}, particularly for the western nucleus, where an outflow is spatially resolved \citep{Barcos-Munoz2018}. With an inclination-corrected wind velocity of $\simeq840$\,km/s and a mass outflow rate of $\simeq55$\,M$_\odot$/yr (for the northern outflow lobe), the total momentum budget relative to the photon momentum budget $\dot{M} v/(\tau L/c)$ is approximately unity for $\tau=5$ and $L=5\times10^{11}$\,L$_\odot$, corresponding to an SFR of $50$\,M$_\odot$/yr and $\eta\sim1$. For the same assumed $L$ for the western nucleus, an estimate of the Eddington ratio $\Gamma\simeq L \kappa_{\rm R}/(2\pi c r v_{\rm circ}^2)$ gives 0.3 for $\kappa_{\rm R}=5$\,cm$^2$/g and $v_{\rm circ}(50\,\rm pc)\simeq350$\,km/s \citep{Scoville2017}. To reach an outflow velocity as large as observed, the effective Eddington ratio would need to be $\simeq4$, likely requiring a dust-to-gas ratio substantially larger than $f_{\rm dg}=1/100$. If instead the observed outflow was interpreted as a shell, it could reach larger velocity with lower $\Gamma$ if a fraction of the bolometric luminosity was optical/UV radiation.

\subsubsection{Simulation Results}
\label{section:rp_summary}

The radiation pressure force per unit area on a dusty medium can be summarized as 
\beq
\left(1-e^{-\langle\tau_{\rm rp}\rangle}+\tau_{\rm R}\right)(F/c),
\eeq 
which encapsulates the optically-thin, single-scattering, and optically-thick regimes. Simulation results connected to galactic winds exist in each regime, and on different scales.

\cite{Hopkins2012} used approximate prescriptions to treat radiation pressure on dust on the very small scales of  optically-thick dense star clusters, finding that GMC disruption lofted material above the plane, where it was then irradiated by the combined unobscured starlight from the galaxy.  Those prescriptions have been significantly updated on galaxy scales in recent work \citep{Hopkins2020_radiative}. Radiation pressure feedback was also included in galaxy simulations by \cite{Rosdahl2015}. 

In idealized 2D plane-parallel simulations employing flux-limited diffusion (FLD), \cite{Krumholz2012,Krumholz2013} showed that radiation pressure can drive turbulence via the radiation Rayleigh-Taylor instability, and winds for ultra-dense ULIRG-like conditions with $\tau_{\rm R}\gg1$. The simulations produced relatively weak momentum coupling for the outflows because of a strong anti-correlation between the radiative flux and the matter: the radiation escaped through low-density channels. For the same test problem, \cite{Davis2014} found a higher level of radiation-gas momentum coupling using the more accurate variable Eddington tensor (VET) formalism for radiation hydrodynamics, concluding that radiation pressure on dust in the optically-thick limit is a viable mechanism for driving turbulence and winds in extreme star-forming regions. The experiment has since been repeated in several works, with important qualitative differences between different radiation transport methods  \citep{Rosdahl2015,Tsang2015,Zhang_Davis,Kannan2019,Smith2020}. Rather than starting with a monolithic atmosphere, \cite{Zhang2018} performed  numerical experiments with optically-thin/thick clouds (see Figure \ref{figure:theory}d) and compared with the momentum deposition from hot winds (e.g., eqs.~\ref{cc85_vinf_pdot} and \ref{ram_pressure_edd}), finding complicated dynamics for highly optically-thick clouds and an effective momentum coupling no larger than the single-scattering limit, $L/c$. As expected when the radiation can both penetrate and ``go around" a small cloud, one expects the effective ``boost" factor $\tau_{\rm R}$ to be $\lesssim1$. \cite{Huang2020} find that cloud survival and acceleration also depends on the relative mix of incident UV and IR radiation.

On the scale of individual massive star cluster formation, the single-scattering limit has been found to be important for natal cloud disruption at early times in forming star clusters \citep{Skinner2015,Raskutti2016,Raskutti2017,Menon2022,Menon2023,Kim2018_radiative}. \cite{Wibking2018} carried out a model problem similar to \cite{Krumholz2013}, but for column density regimes corresponding to optically-thin and single-scattering conditions, rather than highly optically-thick, and showed that near the Eddington limit the medium becomes unstable to Rayleigh-Taylor-like convection. They concluded that, at least when compared against their average molecular gas surface densities, galaxies are far below the single-scattering Eddington limit, though low-column sightlines might not be. \cite{Tsang2018,Menon2022} concluded that radiation pressure in the highly optically-thick limit on super star cluster scales is either ineffective or contributes modestly to reducing the star formation efficiency, ultimately because of equation (\ref{eq:ledd_taur}) and the temperature-dependence of the Rosseland opacity (eq.~\ref{eq:kappa_ross}; \citealt{Crocker2018_radp_cluster}).

\subsubsection{Future Directions for dust driven winds}
\label{section:rp_limitations_future}

Critical questions facing radiation pressure are whether or not galaxies and their subregions exceed Eddington (see \S\ref{section:confronting_theory_observations}; Fig.~\ref{figure:radp_comp}), and if so, what the dynamics is for clouds of various optical depth in a surrounding medium, and the dust-gas coupling itself. The estimates above assume that the dust is collisionally coupled to the gas: the dust grains absorb/scatter starlight and then collide with the gas, sharing their momentum, like a fully-loaded cargo ship moving in a sea of ping pong balls (mass ratio $\sim10^{10}:1$). In the absence of grain charging and magnetic forces, the lengthscale for coupling is the scale over which the grain sweeps up its own mass $\sim (4/3) a \rho_{\rm grain}/\rho_{\rm ISM}\simeq10\,{\rm pc}(a/0.1\,{\rm \mu m})/(n/{\rm cm^{-3}})$. If dust {\it was not} dynamically coupled to the gas, galaxies would be highly super-Eddington over a broad range of parameters and they would be essentially dustless. Dust-gas dynamical coupling is thus critically important and an active area of current research. Recent work suggests that the two-fluid dust-gas medium is subject to a host of instabilities \citep{Hopkins2018_Squire,Squire2022,Hopkins2022_Dust}. These instabilities may make the dust-gas dynamical coupling dependent on the grain size and structure, the grain charge, the thermodynamic phase of the surrounding medium (hot vs.~cool), the magnetic field strength, and the character of the ambient radiation field. 

\subsubsection{Lyman continuum and Lyman $\alpha$ radiation pressure}
\label{section:rp_lyman}

A ZAMS stellar population radiates about $30-50$\% of its bolometric luminosity in ionizing radiation, corresponding to  $\simeq4-7\times10^{46}$\,ionizing photons s$^{-1}$ M$_\odot^{-1}$ of new stars formed (e.g., \citealt{Murray_Rahman2010}). Lyman continuum radiation photoionizes and heats the medium, which is an important feedback process in the disruption of star-forming clouds. These photons also deposit momentum  through photoionization or dust absorption. The cross section for photoionization is $\simeq6.3\times10^{-18}$\,cm$^{2}$ at the Lyman edge, implying large optical depths for star-forming regions and galaxies.\footnote{The escape of ionizing photons from star-forming regions and galaxies is directly connected to the reionization history of the universe. See, e.g., \cite{Robertson2022} for a recent review.}  Because the photons deposit their momentum $\sim$once, the force is\footnote{Lyman-Werner photons are efficiently absorbed in H$_2$ and heat the gas, but their contribution to the radiation pressure momentum budget is much smaller than eq.~\ref{eq:pdot_ion}.}   
\beq
\dot{p}_{\rm rp,\,ion}\simeq f_{\rm abs}\,L_{\rm ion}/c, 
\label{eq:pdot_ion}
\eeq
where $f_{\rm abs} = (1-f_{\rm esc})$ and $f_{\rm esc}$ is the escape fraction. $\dot{p}_{\rm rp,\,ion}$ is similar to the single-scattering limit for dust and about 6 times less than for a hot wind with $\alpha=\eta=1$ (eq.~\ref{cc85_vinf_pdot}). However, these photons can be absorbed even in dust-free gas (i.e., at very low metallicity).

Subsequent recombinations in the ionized gas produce Ly$\alpha$ radiation.\footnote{Ly$\alpha$ radiation can also be produced after collisional excitation of HI, which is connected to discussions of the cooling of halo gas and the observed Ly$\alpha$ emitter populations.}   In ionization equilibrium, the production rate of Ly$\alpha$ photons is related to the ionizing photon absorption rate: $\dot{N}_{\rm Ly\alpha}\simeq0.68\dot{N}_{\rm ion,\,abs}$, assuming Case B recombination and $T_4=T/10^4$\,K \citep{Dijkstra2014}. The emitted Ly$\alpha$ photons undergo strong absorption and (nearly instantaneous) re-emission by neutral Hydrogen atoms (``resonant scattering"). The corresponding cross section is $\simeq5.9\times10^{-14}$\,cm$^2\,T_4^{-1/2}$, implying enormous optical depths at line center ($\tau_0$) for typical galaxy HI column densities: $\tau_0=\sigma_0 N_{\rm HI}\simeq6\times10^7 N_{\rm HI,\,21}T_4^{-1/2}$, where $N_{\rm HI,21}=N_{\rm HI}/10^{21}$\,cm$^{-2}$. One might naively expect a correspondingly large radiation pressure force, but the momentum coupling to the gas is muted by multiple effects. In dust-free gas, the {\it effective} optical depth of the medium is dramatically reduced relative to $\tau_0$ by Doppler broadening, which leads to frequency redistribution during scattering. Photons diffuse in energy space to the wings of the absorption line until they can escape, leading to characteristic double-horned emission line profiles (e.g.,  \citealt{Dijkstra2014}). 

Because of its importance across astrophysics, the effective optical depth (``force multiplier" for momentum transfer) for Ly$\alpha$ has been explored by many authors (e.g., \citealt{Adams1972,Adams1975,Bonilha1979,Smith2017,Kimm2018,Lao2020,Tomaselli2021,McClellan2022}). Analytic results are limited to idealized geometries and source distributions. For dust-less and static spherical gas configurations with a central point source, the effective optical depth in the limit of  large $\tau_0$ is 
\beq
\tau_{\rm eff}\simeq3.5(\tau_0 a)^{1/3}\simeq100\,N_{\rm HI,\,21}^{1/3}\,T_4^{-1/3}
\,\,\,\,(\tau_0\gtrsim10^6),
\label{eq:taueff_lya}
\eeq
where $a\simeq4.7\times10^{-4}T_4^{-1/2}$ and the pre-factor depends on the geometry (for a slab, $3.5\rightarrow2.2$), its density profile, the source distribution, and boundary conditions. For lower line-center optical depth $\tau_0\lesssim10^6$, $\tau_{\rm eff}$ varies even more slowly with $\tau_0$ so that the approximation above underpredicts $\tau_{\rm eff}$: e.g., $\tau_{\rm eff}\propto\tau_0^{0.2}$ for $10\lesssim\tau_0\lesssim10^6$  \citep{Bonilha1979}. 

The overall force on the medium is then set by the Ly$\alpha$ luminosity and the effective optical depth: $\dot{p}_{\rm rp,\,Ly\alpha}\simeq\tau_{\rm eff}L_{\rm Ly\alpha}/c$. Approximating $L_{\rm Ly\alpha}\simeq 0.68\,L_{\rm ion}f_{\rm abs}/2$, taking $L_{\rm ion}\simeq0.1\,L$ for continuous star formation on Gyr timescales \citep{Leitherer1999}, and writing $\xi_{0.1}=L_{\rm ion}/(0.1L)$ the force and bolometric Eddington luminosity ($4\pi G M_{\rm tot}c m_p N_{\rm HI}/\tau_{\rm eff}$)
\begin{equation}
\dot{p}_{\rm rp,\,Ly\alpha}\simeq4\left(\frac{L}{c}\right)\,
\frac{f_{\rm abs}\,\xi_{0.1}\,N_{\rm HI,21}^{1/3}}{T_4^{1/3}}\,\,\,{\rm and}\,\,\,
\frac{L_{\rm Edd,\, Ly\alpha}}{M_{\rm tot}}\simeq 6\frac{L_\odot}{M_\odot}\frac{N_{\rm HI,21}^{2/3}T_4^{1/3}}{f_{\rm abs}\,\xi_{0.1}}\,\,\,(\tau_0\gtrsim10^6).
\label{eq:pdot_lymanalpha}
\end{equation} 
Note that for a ZAMS population $\xi_{0.1}\simeq3-5$, which increases the force and decreases the Eddington luminosity. Additionally, using $T=10^2$ or $10^3$\,K for the cool or unstable neutral medium increases the radiation pressure force by factors of 4.64 and 2.15, respectively. For dust-less media, comparing equations (\ref{eq:pdot_lymanalpha}) and (\ref{cc85_vinf_pdot}) implies that the Ly$\alpha$ radiation pressure force could be as large or larger than the ram pressure force from the hot wind, depending on $N_{\rm HI}$, $T$, and the thermalization ($\alpha$) and mass-loading efficiency ($\eta$) in the hot phase. 

These considerations motivate the idea that Ly$\alpha$ radiation pressure might regulate star formation and drive galactic winds in \cite{Cox1985} (see also \citealt{Haehnelt1995,Oh2002,Smith2017,Kimm2018}). In particular, \cite{Dijkstra2008,Dijkstra2009} consider galactic supershells driven by Ly$\alpha$ scattering, finding that dustless outflows can be accelerated to hundreds of km/s. Recent discussion and calculations are provided by \cite{Smith2017}, who find that Ly$\alpha$ can enhance supershell velocities, and by \cite{Kimm2018}, who find that Ly$\alpha$ feedback suppresses star formation in metal-poor galaxies by regulating the pre-SN dynamics of star-forming clouds. Counter-intuitively, \cite{Kimm2018} find {\it weaker} galactic winds with Ly$\alpha$ radiation pressure because the star formation suppression leads to less-strong SN-driven outflows.

Several additional physical effects limit $\tau_{\rm eff}$. The above expressions apply for static and dust-less gas configurations. Yet, while Ly$\alpha$ photons are scattering in the medium, they can be absorbed by dust grains, which then re-radiate in the FIR. Random walk arguments combined with typical dust opacities limit $\tau_{\rm eff}$ to \citep{Kimm2018,Tomaselli2021} $\tau_{\rm eff,\,max}\simeq35\,f_{\rm dg,\,MW}^{-1/4}T_4^{-1/4}$, so that for a static medium, $\tau_{\rm eff}$ should be the minimum of this expression and equation (\ref{eq:taueff_lya}), and  equations (\ref{eq:pdot_lymanalpha}) become $(\tau_{\rm eff}>\tau_{\rm eff,\,max})$
\begin{equation}
\dot{p}_{\rm rp,\,Ly\alpha}\simeq1\left(\frac{L}{c}\right)\frac{f_{\rm abs}\,\xi_{0.1}}{f_{\rm dg,MW}^{1/4}\,T_4^{1/4}}
\,\,\,{\rm and}\,\,\,
\frac{L_{\rm Edd, Ly\alpha}}{M_{\rm tot}}\simeq20\frac{L_\odot}{M_\odot}
\frac{N_{\rm HI,21}f_{\rm dg,MW}^{1/4}\,T_4^{1/4}}{f_{\rm abs}\,\xi_{0.1}}.
\label{eq:pdot_lymanalpha_dust}
\end{equation} 
We note that the estimate  of the force here is substantially larger than that by  \cite{Henney1998} (their eq.~12), which is equivalent to zero spatial diffusion, so that all Ly$\alpha$ photons are absorbed in a single dust mean-free-path light-crossing-time. This gives $\dot{p}_{\rm rp, Ly\alpha}\sim (L_{\rm Ly\alpha}/c)(1/\sigma_{\rm dust} N_{\rm H})\sim0.02(L/c) (f_{\rm abs} \xi_{0.1}/N_{\rm H,21}f_{\rm dg,\,MW})$,  50 times less than in equation (\ref{eq:pdot_lymanalpha_dust}), using $\sigma_{\rm dust}=1.6\times10^{-21}$\,cm$^2$ at Ly$\alpha$, and where $N_{\rm H}$ is the total H column density. In addition to dust absorption, both velocity gradients across the medium and the intrinsic velocities of the medium as it is accelerated can hasten Ly$\alpha$ escape \citep{Seon2020}, decreasing the force multiplier in equation (\ref{eq:taueff_lya}) as the velocity of the shell increases \citep{Dijkstra2008,Kimm2018,Tomaselli2021}. Indeed, although equation (\ref{eq:taueff_lya}) predicts $\tau_{\rm eff}\sim100$, 50, and 20 for $N_{\rm HI}=10^{21}$, 10$^{20}$, and $10^{19}$\,cm$^{-2}$ and $T=10^4$\,K, respectively, using Monte Carlo calculations, \cite{Dijkstra2008} find that $\tau_{\rm eff}$ decreases to values $\lesssim1$ for $v\gtrsim1000$, 400, and 150\,km/s, respectively. 

An important caveat for the importance of Ly$\alpha$ radiation pressure and a direction for future research is to understand $\tau_{\rm eff}$ for geometrically small HI clouds embedded in a tenuous and (potentially) ionized medium. High resolution simulations of multi-phase flows with Ly$\alpha$ radiation transport are required to assess this question, but like the case of $\tau_{\rm R}\gg1$ for dusty clouds \citep{Zhang2018}, one imagines that the effective momentum coupling will be closer to the expectation from the single-scattering limit ($\tau_{\rm eff}\sim1$) when small clouds have nominally high $\tau_{\rm eff}\gg1$. Even so, the importance of Ly$\alpha$ radiation pressure for natal cloud disruption and for super-shell acceleration is evident from equations (\ref{eq:pdot_lymanalpha}) and (\ref{eq:pdot_lymanalpha_dust}), especially at low metallicity and low $f_{\rm dg}$.

\subsection{Cosmic Rays (CRs)}
\label{section:cosmic_rays}

CR-driven winds were first discussed by \cite{Ipavich1975}, motivated by measurements of the local CR pressure $P_{\rm cr}$ that show it is comparable to that required for vertical hydrostatic equilibrium of the Galactic disk (\citealt{Boulares1990}),
\beq
P_{\rm HSE}\simeq\pi G\Sigma_{\rm tot}\Sigma_g \simeq9\times10^{-12}\,\Sigma_{\rm tot,100}\,\Sigma_{g,10}\,{\rm ergs/cm^3},
\label{eq:phse}
\eeq
where $\Sigma_{\rm tot,100}=\Sigma_{\rm tot}/100$\,M$_\odot$ pc$^{-2}$ and $\Sigma_{g,10}=\Sigma_g/10$\,M$_\odot$ pc$^{-2}$ are the total and gas surface densities, respectively (see Table \ref{table:symbols}).

The picture is as follows. CR protons (and nuclei) and ``primary" CR $e^-$ are injected into the ISM via Fermi acceleration in SN shockwaves and other stellar processes. Assuming one SN per 100\,M$_\odot$ of new stars formed (see Sidebar \ref{sidebar:supernovae}), and that 10\% of the $10^{51}$\,ergs is injected into CR protons, the CR energy injection rate is\footnote{The CR energy injected per SN is uncertain, but is constrained by the FIR-radio and FIR-$\gamma$-ray correlations \citep{Lacki2010}, and by studies of SN remnants \citep{Ackermann2013}. The primary CR $e^-$ injection rate is $\sim10$ times smaller than for protons and nuclei.}
\beq
\dot{E}_{\rm cr}=\epsilon_{\rm cr}\,{\rm SFR}\, c^2\simeq3\times10^{40}\,{\rm ergs\,\,s^{-1}}\,({\rm SFR}/{\rm M_\odot\,\,yr^{-1}}),
\label{eq:edotcr}
\eeq
where $\epsilon_{\rm cr}\simeq5\times10^{-7}$. Once injected, the CRs propagate within and then out of the host galaxy by scattering off of magnetic field fluctuations, with an effective mean-free path of $\sim0.1$\,pc \citep{Longair1994}. These fluctuations may be generated extrinsically by turbulence (``diffusion;" \citealt{Yan2002}) or intrinsically by the CRs via self-excited Alfv\'en wave scattering (``streaming" or ``self-confinement;" \citealt{Kulsrud1969}). CR transport may be dominated by either or both, depending on the physical conditions and scale. Transport establishes a large-scale $P_{\rm cr}$ gradient that can in principle accelerate gas from galaxies without heating it. Indeed, \cite{Ipavich1975} constructed CR-driven winds in the zero-temperature limit, and \cite{Socrates2008} argued for the importance of CRs solely on the basis of their macroscopic force. The complications of CR-driven wind models come in how $P_{\rm cr}$ is set at the base of the outflow, and how the transport of CRs is mediated through both the neutral and ionized phases of the magnetized and turbulent medium. Recent works show that neither the diffusion nor the streaming pictures alone can be reconciled with the Galaxy's CR spectrum \citep{Kempski2022,Hopkins2021}.

\begin{figure}
\centerline{\includegraphics[width=6.36cm]{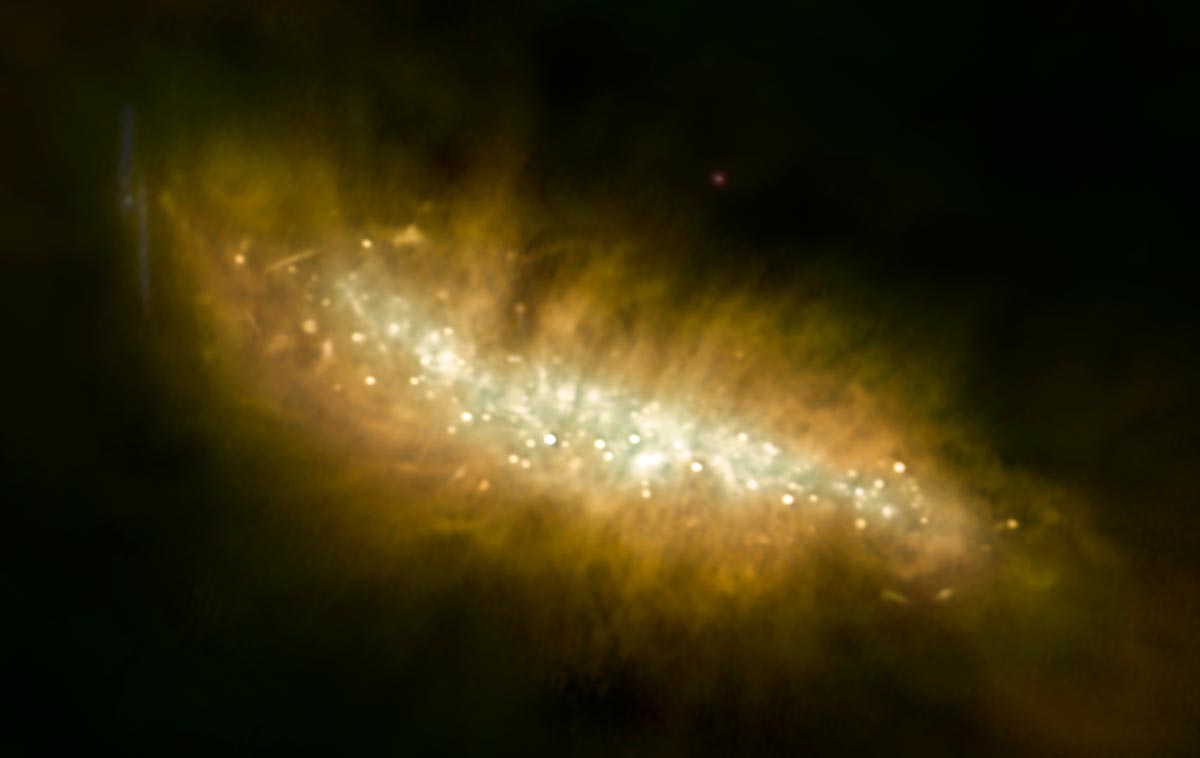}\hspace*{-6.75cm}\includegraphics[width=5.8cm]{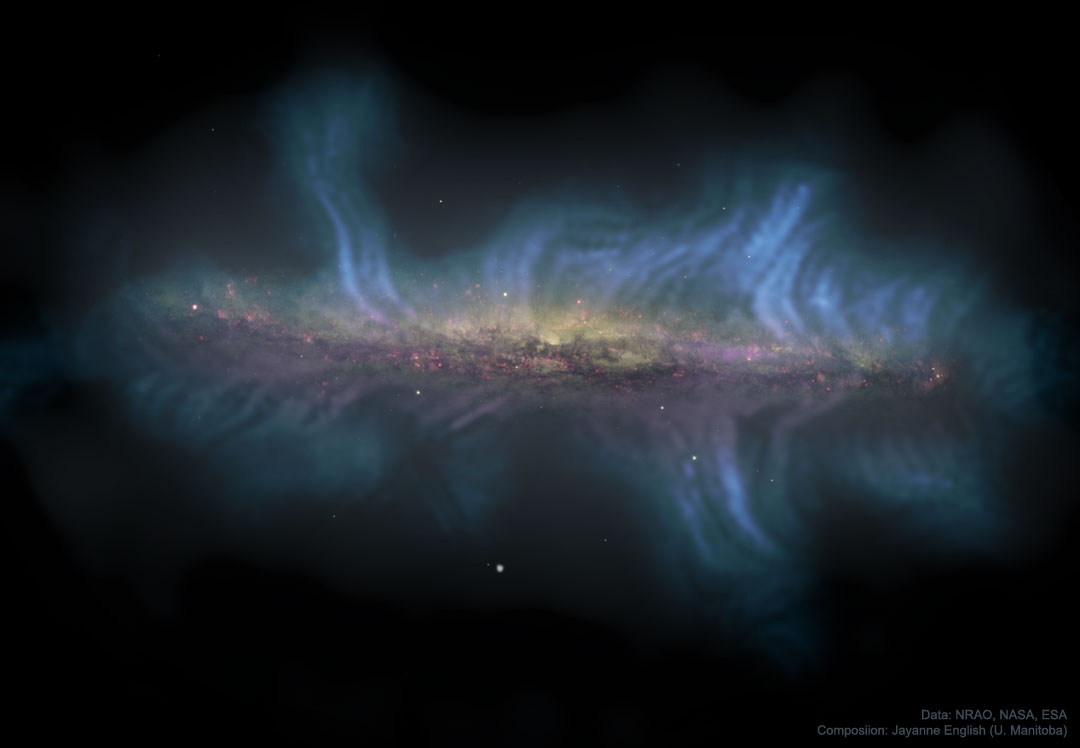}} 
\vspace*{.5cm}
\caption{{\it Left:} Radio continuum image of the M82 starburst with the JVLA showing compact HII regions, SN remnants, and diffuse emission sculpted by the global outflow and magnetic fields. Credit: J.~Marvil (NM Tech/NRAO), B.~Saxton (NRAO/AUI/NSF), NASA. {\it Right:} Composite image of NGC 5775 showing extended synchrotron emission and magnetic field structure. Credit: Jayanne English (U.\ Manitoba) with Y.~Stein, (CDS), A.~Miskolczi (Ruhr-University Bochum), and J.~Irwin (Queens U.)/CHANG-ES/NRAO/HST.}
\label{figure:marvil}
\end{figure}

A limitation in assessing CRs as a wind mechanism is that, outside of the Milky Way, we have few probes of the CR proton population, which dominates the energetics. The most direct probe is the $\gamma$-ray emission from $\pi^0$ decay produced by inelastic CR collisions with the host's ISM. These collisions also produce $\pi^\pm$, which subsequently decay to ``secondary" CR $e^\pm$ and neutrinos. The Fermi satellite and atmospheric Cerenkov telescopes like HESS, VERITAS, and others allow us to observe diffuse $\sim0.1-100$\,GeV and $\sim1-10$\,TeV $\gamma$-ray emission in the Milky Way, the Magellanic Clouds, M31, and some nearby starbursts and Seyferts, including M82, NGC\,253, NGC\,1068, and the ULIRG Arp 220, among others (e.g., \citealt{abdo2010_m82_ngc253,veritas2009_m82,hess2009_ngc253,Griffin2016,Ajello2020}). However, $\gamma$-ray emission is not spatially resolved beyond the Local Group. Thus, while the integrated  $\gamma$-ray emission constrains CR-driven wind models (\S\ref{section:cr_limitations}), the continuum radio synchrotron emission from CR leptons is our principal (albeit indirect) probe of the spatial distribution of CRs (see Fig.~\ref{figure:marvil}; \citealt{Adebahr2013,Adebahr2017}).

Treatments of CR transport start from the Boltzmann equation for the time evolution of the CR phase space distribution function (e.g., \citealt{Zweibel2017}). As in radiation hydrodynamics, a hierarchy of moments to the Boltzmann equation can be derived, yielding a coupled set of partial differential equations for their evolution. See \cite{Pakmor2013,Pakmor2016,Pfrommer2017,Jiang2018,Ruszkowski2023} and references therein for the motivating equations, approximations, and closures. For our purposes, the CR energy equation and the ``equilibrium" time-steady CR flux including both streaming and diffusion can be written as 
\beq
\frac{\p U_{\rm cr}}{\p t} +\nabla\cdot {\bf F}_{\rm cr}=\left({\bf v}+{\bf v_{\rm s}}\right)\cdot \nabla P_{\rm cr} \,\,\,{\rm and} \,\,\,{\bf F}_{\rm cr}= 4P_{\rm cr} \left({\bf v}+{\bf v_{\rm str}}\right)-3 D \,{\bf n}\left({\bf n}\cdot\nabla P_{\rm cr}\right),
\label{eq:cr_energy}
\eeq
where ${\bf v}_{\rm str}=-{\bf v}_{\rm A}\,|{\bf\nabla} P_{\rm cr}|/{\bf \nabla} P_{\rm cr}$,  is the streaming velocity, ${\bf v}_{\rm A}={\bf B}/(4\pi \rho)^{1/2}$ is the Alfv\'en speed, ${\bf n}={\bf v}_{\rm A}/|{\bf v}_{\rm A}|$, $U_{\rm cr}=3P_{\rm cr}$ is the CR energy density, and  $D[=]\,{\rm cm^2/s}$ is the (energy-dependent) diffusion coefficient ($c\lambda/3$). Ignoring the time-dependence of ${\bf F}_{\rm cr}$  is an approximation that breaks down in certain regimes. The full equations can be solved via two-moment and reduced speed of light methods (see, e.g., \citealt{Jiang2018}). Solutions yield predictions for the $P_{\rm cr}$, $F_{\rm cr}$, and the non-thermal emission that can be compared with observations (e.g., \citealt{Strong2007,Werhahn_2021_voyager}).

Although CR transport is complicated and ISM phase-dependent, we can estimate $P_{\rm cr}$ inside the host galaxy from the CR escape time $t_{\rm esc}$ using $P_{\rm cr,\,0}\sim(1/3)(\dot{E}_{\rm cr}\,t_{\rm esc}/{\rm Vol})$, where ${\rm Vol}=(2h)\pi R^2$ is the confinement volume, $R$ is the galaxy radius, and $h$ is the CR scale height. For example, if diffusion dominates transport, $P_{\rm cr,\,diff,0}\sim(1/3)(\dot{E}_{\rm cr}\,t_{\rm diff}/{\rm Vol})$, and
\beq
P_{\rm cr,\,diff,0} 
\sim\epsilon_{\rm cr}\dot{\Sigma}_\star c^2/(6D/h)
\sim2\times10^{-12}\,\dot{\Sigma}_{\star,0.01}\,(h_{\rm kpc}/D_{28})\,\,{\rm ergs/cm^3}
\label{eq:pcr_tdiff}
\eeq
where $h_{\rm kpc}=h/{\rm kpc}$ and $\dot{\Sigma}_{\star,0.01}=\dot{\Sigma}_{\star}/0.01$\,M$_\odot$/yr/kpc$^2$, similar to the Milky Way, and $t_{\rm diff}=h^2/D$ ($\simeq30$\,Myr for the parameters above). See Table \ref{table:symbols}. In the Galaxy, $t_{\rm diff}$ and/or $h$ and $D$ (and their energy dependence) can be constrained from the abundance of spallation products, the equilibrium CR spectrum, and the observed CR angular anisotopy at Earth \citep{Longair1994}.\footnote{In models that reproduce Galactic constraints, values of $D$ in the range $10^{27}-10^{29}$\,cm$^2$/s are inferred at GeV energies. As implied by eq.~(\ref{eq:pcr_tdiff}), $D$ is degenerate with $h$ (e.g., \citealt{Linden2010}).} For comparison with diffusion, if streaming dominates transport, the CR escape time is $t_{\rm str}\sim h/v_{\rm A}$, $P_{\rm cr,\,str,0}\sim(1/3)(\dot{E}_{\rm cr}\,t_{\rm str}/{\rm Vol})$, and 
\beq
P_{\rm cr,\,str,0}
\sim\epsilon_{\rm cr} \dot{\Sigma}_\star c^2/(6\,v_{\rm A})
\sim2\times10^{-12}\,\dot{\Sigma}_{\star. 0.01}/v_{\rm A, 30} \,\,{\rm ergs/cm^3}.
\label{eq:pcr_tstream}
\eeq
where $v_{\rm A}\simeq35$\,km/s\,$(B/5{\rm \mu G})(0.1\,{\rm cm^{-3}}/n)^{1/2}$. For typical parameters,  $P_{\rm cr,\,0}$ is of order $P_{\rm HSE}$ (eq.~\ref{eq:phse}), and thus CRs might be dynamically important. This comparison motivates CR wind theories. Note that for sufficiently high ISM densities the pionic loss timescale can become shorter than $t_{\rm diff}$ or $t_{\rm str}$, and our expectations for $P_{\rm cr,0}$ change (see \S\ref{section:cr_limitations}).

\subsubsection{Diffusion} The potential of CRs to drive galactic winds in the diffusive limit can be readily seen. By combining equations (\ref{eq:edotcr}) and (\ref{eq:lum_continuous}), and taking the diffusion optical depth $\tau=h/\lambda$, the momentum budget is \citep{Socrates2008}
\beq
\dot{p}_{\rm cr,\,diff}\simeq\tau\frac{\dot{E}_{\rm cr}}{c}\simeq\left(\frac{h}{\lambda}\right)\left(\frac{\epsilon_{\rm cr}}{\epsilon_{\rm ph}}\right)\frac{L}{c}\simeq7\left(\frac{h/\lambda}{10^4}\right)\frac{L}{c}.
\label{eq:pdotcr_diff_tau}
\eeq
Thus, the CR force is potentially larger than radiation pressure (eq.~\ref{eq:mass_loading_singlescattering}) and as large as the force from the very hot gas (eq.~\ref{cc85_vinf_pdot}). 

As in the discussion of the photon tiring limit, consideration of the CR energy equation is needed when  $\tau\gg1$ and the wind material is accelerated. To write down a wind model, we note that the continuity and momentum equations are the same as for the Parker wind (eq.~\ref{parker}), but with the addition of $\nabla P_{\rm cr}/\rho$. Assuming pure diffusion with ${\bf v}_{\rm str}=0$, the steady-state CR energy equation is (eq.~\ref{eq:cr_energy}) 
\beq
v\frac{dP_{\rm cr}}{dr}  + \frac{4}{3}\frac{P_{\rm cr}}{r^2} \frac{d (r^2 v)}{dr} - \frac{1}{r^2}\frac{d}{dr}\left(r^2 D \frac{d P_{\rm cr}}{dr}\right)=0.
\label{eq:cr_energy_diffusion}
\eeq
To make progress, we assume that $v$ is small near the base of the outflow so that the diffusion term dominates the other two terms. This requires $D/r\gg v(r)$, ``fast" diffusion.  Using this approximation, we can make estimates that illustrate the physics and that correspond well with numerical simulations \citep{Quataert2022a}. If diffusion is fast and the flow is planar, then $3D\,dP_{\rm cr}/dz = -F_{\rm cr}$. Using $d\ln v=-d\ln\rho$ from continuity, and assuming the gas is isothermal with sound speed $c_T$, the momentum equation can be recast as a ``wind" equation:
\begin{equation}
 v\,\frac{dv}{dz} = -c_T^2 \frac{d\ln\rho}{dz} + \frac{F_{\rm cr}}{3D\rho} - g \Longrightarrow    \frac{d\ln v}{dz} = \frac{(F_{\rm cr}/(3D\rho) - g)}{(v^2 - c_T^2)}.
\end{equation}
If $v\ll c_T$ at the outflow's base, for positive acceleration $F_{\rm cr}/(3D\rho)<g$. Because $v$ increases with distance, $\rho$ must decrease, increasing $F_{\rm cr}/(3D\rho)$ until it reaches $g$ at the sonic point (near the outflow base in a planar approximation). From the denominator we have that $v = c_T = \dot{m}/\rho$ at the sonic point, where $\dot{m}=\rho v$ is the mass-loss rate per area, while from the numerator we have that $\dot{m} = c_T F_{\rm cr}/(3g D)$. Substituting $g=2\sigma^2/R$ and multiplying by $\pi R^2$, we have an estimate for the mass loss rate and loading efficiency:
\beq
    \dot{M}_{\rm cr,\,diff}\simeq\left(\frac{c_T \,R}{3D}\right)\left(\frac{\dot{E}_{\rm cr}}{2\sigma^2}\right)\propto P_{\rm cr,\,diff,0}
    \,\,\,\Longrightarrow\,\,\,\eta_{\rm cr,\,diff} = \frac{\dot{M}_{\rm cr,\,diff}}{\rm SFR}\simeq1\,\frac{c_{T,10}R_5}{\sigma_{100}^{2}\,D_{28}},
 \label{eq:mdot_crdiffusion}
 \eeq
where $R_5=R/5$\,kpc, $c_{T,10}=c_T/10$\,km/s, $D_{28}=D/10^{28}$\,cm$^2$/s, and $\sigma_{100}=\sigma/100$\,km/s.  The first equality illustrates the dependence on $D$ and $\dot{E}_{\rm cr}$ (tied to the SFR; eq.~\ref{eq:edotcr}). The proportionality illustrates the dependence on the base CR pressure (eq.~\ref{eq:pcr_tdiff}): faster diffusion implies lower $P_{\rm cr,\,diff,0}$ for the same $\dot{E}_{\rm cr}$, which is fundamentally why $D$ enters the denominator in these expressions. The result for $\eta_{\rm cr,\,diff}$ is important. Significant mass-loading is possible for massive galaxies, $\eta$ is independent of the SFR, and it exhibits an $\propto \sigma^{-2}$ ``energy"-like scaling that enhances mass loss in low-mass galaxies. The assumption of fast diffusion can be checked post hoc: $D/R\simeq30\,{\rm km/s}$ (for the parameters above) is larger, but not much larger, than $v$ at the sonic point $v=c_T=10$\,km/s.\footnote{For hotter ISM phases, fast diffusion ($D/R>c_T$) breaks down. Models with $D/R<c_T$ produce low-velocity breeze-like outflows that can have large mass loading \citep{Quataert2022a}.} 

To estimate the kinematics of the wind, we substitute the full CR energy equation into the momentum equation to derive the Bernoulli integral:
\beq
\dot{M}_{\rm cr,\,diff}\left[(v^2/2)+c_T^2\ln(\rho)+(4P_{\rm cr}/\rho)-2\sigma^2\ln(r)\right]+\dot{E}_{\rm cr}={\rm constant}.
\label{eq:diffusion_bernoulli}
\eeq
At the outflow's base, the term in brackets is small and $\dot{E}_{\rm cr}$ dominates. At large scales, where the assumption of an isothermal potential breaks down, the asymptotic kinetic energy $v_\infty^2/2$ dominates. Thus,  $\dot{E}_{\rm cr}\sim\dot{M}_{\rm cr,\,diff}v_\infty^2/2$ and $\dot{p}_{\rm cr,\,diff}=(2\dot{E}_{\rm cr}\dot{M}_{\rm cr,diff})^{1/2}$ (using eq.~\ref{eq:mdot_crdiffusion}), 
\beq
v_\infty\sim2\sigma\left(\frac{3D}{Rc_T}\right)^{1/2} \hspace*{-0.2cm}\sim 300\,\frac{\rm km}{\rm s} \frac{\sigma_{100}D_{28}^{1/2}}{R_5^{1/2}c_{T,10}^{1/2}}\,\,\,{\rm and}\,\,\,
\dot{p}_{\rm cr}\sim\left(\frac{\dot{E}_{\rm cr}}{\sigma}\right)\left(\frac{c_T R}{3D}\right)^{1/2}
\hspace*{-.3cm}\sim2\frac{L}{c}
\frac{R_5^{1/2}c_{T,10}^{1/2}}{\sigma_{100}D_{28}^{1/2}}
\label{vinfty_diffusion}
\eeq
Like $\eta$ (eq.~\ref{eq:mdot_crdiffusion}), $v_\infty$ has no explicit SFR dependence and it exhibits a $v_\infty\propto \sigma$ scaling that is similar to observational determinations (\S\S\ref{section:observations} \& \ref{section:confronting_theory_observations}, \citealt{Martin2005}). The wind force is similar to that carried by the radiation field $L/c$ (compare with eqs.~\ref{cc85_vinf_pdot}, \ref{eq:massloading_rpthin}, and \ref{eq:mass_loading_singlescattering}).

For large enough $c_T/D$, one might expect $\eta$ in equation (\ref{eq:mdot_crdiffusion}) to diverge. However, the mass loss rate is bounded by equation (\ref{eq:diffusion_bernoulli}). When $\sigma$ is large enough, we expect $\dot{E}_{\rm cr}-\dot{M}_{\rm cr,\,diff}\,2\sigma^2\ln(r/R)\sim0$: in essence, all of the injected CR power goes into lifting the material out of the gravitational potential, like the ``photon tiring limit" (\S\ref{section:radiation_pressure}), as discussed for CRs by \cite{Socrates2008}. Taking $\ln(r/R)\sim1$, the maximum possible $\eta$ is
\beq
\eta_{\rm cr,\,max,\,diff}=\dot{M}_{\rm cr,\,diff,\,max}/\rm SFR\sim(\epsilon_{\rm cr}/2)(c/\sigma)^2\sim2\,\sigma_{100}^{-2}.
\label{eq:eta_cr_diff_max}
\eeq
These estimates and the numerical work of \cite{Pakmor2016,Girichidis2016,Wiener2017,Chan2019} and others indicate that diffusive CR-driven winds may be efficient in removing cool gas from normal main sequence Milky Way-type galaxies, as required by some calculations of nucleosynthetic enrichment (\S\ref{section:scope}). To our knowledge, no other wind-driving mechanism is as promising for normal $z\sim0$ spirals. The main limitation is the applicability of diffusion itself. In contrast, for dense starbursts, pion losses may decrease $P_{\rm cr,0}$ (eq.~\ref{eq:mdot_crdiffusion}), limiting the role of CRs (\S\ref{section:cr_limitations}).

\subsubsection{Streaming} The original works on CR driven winds were done in the ``streaming" (or ``self-confinement") limit, where CRs self-excite and scatter off of Alfv\'en waves generated by the gyroresonant interaction that then damp  \citep{Ipavich1975,Breitschwerdt1991,Everett2008}. The physical mechanisms governing damping  -- e.g., ion-neutral, nonlinear Landau, turbulent damping, etc. -- remain active topics of investigation and are ISM phase-dependent (see, e.g., \citealt{Zweibel2017}). Here, we provide a sketch. 

CR-driven winds with streaming  are qualitatively different from diffusion models for two reasons: (1) the local Alfv\'en speed controls the CR flux and (2) there is a microphysical gas heating term arising from wave damping that is proportional to $v_{\rm A}\nabla P_{\rm cr}$. The contrast between winds in the diffusive and streaming limits has been highlighted in simulations (e.g., \citealt{Wiener2017}).  Although the full time-dependent problem can be simulated \citep{Chan2019,Hopkins2020_CRs,Hopkins2022_CRs,Thomas2023}, in order to understand the underlying physics, we consider a spherical time-steady model problem. The magnetic field has a split monopole geometry and equations~(\ref{eq:cr_energy}) with ($D=0$) combine to give
\beq
\frac{dP_{\rm cr}}{dr}=\frac{4}{3}\frac{P_{\rm cr}}{\rho}\left(\frac{v_{\rm A}/2+v}{v_{\rm A}+v}\right)\frac{d\rho}{dr}=c_{\rm eff}^2\frac{d\rho}{dr},
\label{eq:cr_energy_streaming}
\eeq
where the last equality defines the effective CR sound speed. In the limit $v\gg v_{\rm A}$, $c^2_{\rm eff}=(4/3)P_{\rm cr}/\rho$ ($P_{\rm cr}\propto\rho^{4/3}$), the equation of state of an adiabatic relativistic gas. Because deviations from adiabaticity drive winds, we expect weak outflows if $v\gg v_{\rm A}$. However, in the opposite limit ($v\ll v_{\rm A}$), which we expect near the wind launching radius, $c^2_{\rm eff}=(2/3)P_{\rm cr}/\rho$ and $P_{\rm cr}\propto\rho^{2/3}$. This remarkable equation of state is the basis for the theory of \cite{Ipavich1975} and  subsequent works, fundamentally because the CR sound speed $(P_{\rm cr}/\rho)^{1/2}\propto\rho^{-1/6}$ {\it increases} as the density {\it decreases}. Thus, in contrast to an isothermal gas (\S\ref{section:thermal}), $c_{\rm eff}$ can increase as the wind expands and accelerates, exactly the properties needed to meet the extended and deep gravitational potentials of galaxies.

Again assuming the gas is isothermal, combining equation (\ref{eq:cr_energy_streaming}) with the momentum and continuity equations, we derive a wind equation analogous to equation (\ref{parker}):
\beq
(r/v)(dv/dr)=2(c_T^2+c_{\rm eff}^2-\sigma^2)/(v^2-c_T^2-c_{\rm eff}^2).
\label{eq:cr_wind_streaming}
\eeq
Taking the $v\ll v_{\rm A}$ limit in the subsonic portion of the flow, and assuming hydrostatic equilibrium below the sonic point (\S\ref{section:thermal}) the mass loss rate is \citep{Quataert2022b}
\beq
\dot{M}_{\rm cr,\,str,\,classic}\simeq  \, 4\pi R^2\,\rho_0\,\sigma \, e\,(c_{\rm eff,0}/\sigma)^6\propto P_{\rm cr,\,str,0}^3,
\label{mdot_stream_classic}
\eeq
where $P_{\rm cr,\,str,0}$ is the pressure (eq.~\ref{eq:pcr_tstream}) and $c^2_{\rm eff,0}=(2/3)P_{\rm cr,\,str,0}/\rho_0$ is the CR sound speed at the outflow's base. This expression predicts a strong dependence on the $c_{\rm eff,0}$ and $\sigma$. 

However, the steady-state analysis of \cite{Ipavich1975} and subsequent works turns out to be incomplete. The solutions are unstable to time-dependent perturbations whose non-linear development changes the outflow structure \citep{Huang2022a,Huang2022b,Quataert2022b,Tsung2022,Modak2023}. To see this, note that the steady-state flux requires $F_{\rm cr}\sim 4 v_{\rm A}P_{\rm cr}$ and that CRs must stream down the density gradient. However, if there is a small local perturbation in density, then $v_{\rm A}$ will decrease. To maintain a constant CR flux,  $P_{\rm cr}$ must then increase. This situation violates the requirement that $\nabla P_{\rm cr}<0$ and leads to a ``bottleneck" effect that causes shocks with a staircase-like structure: the bulk of the volume has $dP_{\rm cr}/dr\simeq0$ and shocks cause sudden drops in $P_{\rm cr}$. These solutions are qualitatively and quantitatively different from the classic treatments.

The role of this instability in shaping winds is still being studied. The shocks can become radiative, producing a thermally-unstable multi-phase medium \citep{Huang2022a,Huang2022b,Modak2023}. CR heating further invalidates an isothermal gas equation of state in some regimes. An effective theory can be written for the case where the instability develops and the volume is permeated by a shocktrain \citep{Quataert2022b}. With $dP_{\rm cr}/dr\simeq0$, we expect $r^2 F_{\rm cr}\sim  r^2 v_{\rm A}P_{\rm cr}\sim r^2 B(r)\rho^{-1/2} P_{\rm cr}={\rm constant}$. Because $B(r)\propto r^{-2}$ for a split-monopole field, $P_{\rm cr}$ must then be $\propto \rho^{1/2}$. This is different from equation (\ref{eq:cr_energy_streaming}), but still has the property that $c_{\rm eff}$ increases as $\rho$ decreases. Again, a wind equation can be derived, the density profile can be estimated from the assumption of HSE below the sonic point  (\S\ref{section:thermal}), and  
\beq
\dot{M}_{\rm cr,\,str}\simeq  \, 4\pi R^2\,\rho_0\,c_{\rm eff,0} \, e^2\,\left(c_{\rm eff,0}/\sigma\right)^3
\propto P_{\rm cr,0}^2
\label{mdot_streaming}
\eeq
which can be compared with equation (\ref{mdot_stream_classic}) and the diffusive limit (eq.~\ref{eq:mdot_crdiffusion}). Using equation (\ref{eq:pcr_tstream}), the mass loading rate $\dot{M}_{\rm cr,\,str}/{\rm SFR}$ can be written as
\beq
\eta_{\rm cr,\,str}
\sim(e^2/36)(\epsilon_{\rm cr}\dot{\Sigma}_\star c^2/\rho_0\sigma^3)(\epsilon_{\rm cr}c^2/v_{\rm A, 0}^2)
\sim1 \,\dot{\Sigma}_{\star, 0.01} \,B_{0, 5\rm \mu G}^{-2} \sigma_{100}^{-3}.
\label{cr_massload_stream}
\eeq
CR-driven winds in the streaming limit can have significant mass loading, but unlike the case for diffusion, here we find a strong dependence on  $\dot{\Sigma}_\star$ and $B_0$, and a $\sigma^{-3}$ scaling that is steeper than the $\sigma^{-2}$ typically invoked from energy conservation. The term $\dot{\Sigma}_\star/B_0^2$ is particularly intriguing because it is proportional to the ratio of photon energy density ($U_{\rm ph}$) to the magnetic energy density ($U_B$), and thus measures the ratio of the synchrotron and inverse Compton cooling timescales for CR $e^\pm$. Because $\dot{\Sigma}_\star\propto U_{\rm ph}$ varies over more than $10^6$ across the Schmidt law from Milky Way-like galaxies to the most intense ULIRGs, and yet the FIR-radio correlation persists across that sequence, an approximate scaling of a form similar to $\dot{\Sigma}_\star\propto B_0^2$ must exist  \citep{Thompson2006,Lacki2010}, likely as a result of dynamo action. For this reason, $\dot{\Sigma}_\star/B_0^2$ may be approximately constant.

The Bernoulli integral for the streaming case can be used to estimate $v_\infty$, the force, and the maximum possible mass loss rate. These yield
\beq
v_\infty \simeq (2 \sigma/e)(v_{\rm A,0} \,\sigma/c_{\rm eff,\,0}^2)^{1/2}   \simeq 230 {\rm \,km\,\,s^{-1}}\, \sigma_{100}^{3/2} \,v_{\rm A, 10}^{1/2} \,c_{\rm eff, 10}^{-1},
\label{eq:cr_vinf_stream}
\eeq
with a unique $\sigma^{3/2}$ scaling. When combined with equation (\ref{eq:pcr_tstream}), $v_\infty\propto \sigma^{3/2} (B^2/\dot{\Sigma}_\star)^{1/2}$, which is again proportional to the ratio of the inverse Compton and synchrotron cooling times for CR leptons. The force of the outflow is similar to $L/c$ for constant SFR (product of eqs.~\ref{eq:cr_vinf_stream} \& \ref{cr_massload_stream}; Table \ref{table:theory}). Like the case for diffusion, energy conservation through the Bernoulli integral limits the maximum possible mass-loading rate to
\beq
\eta_{\rm cr,\,str,\,max}=\dot{M}_{\rm cr,\,str,\,max}/{\rm SFR}\simeq2\zeta\,\sigma_{100}^{-2}
\label{eq:max_mass_loading_streaming}
\eeq
where $\zeta<1$ is the amount of energy transferred from the CRs to the gas in the shock trains \citep{Quataert2022b}. Recent work by \cite{Modak2023} implies that when the isothermal approximation to the gas thermodynamics is relaxed a relatively large fraction of the CR energy is lost to radiation and to work against gravity. Their results indicate a wind power that is ${\rm few}-10$ times lower than the total CR energy budget given in equation (\ref{eq:edotcr}), which is in tension with the observations presented in \S\ref{section:confronting_theory_observations}.

More work is required to understand CR streaming winds and the role of instabilities. The above analysis employs overly idealized geometry and thermodynamics. Indeed, the spherical calculations show that the Alfv\'en speed dominates all other characteristic speeds, indicating that $B$ may be dynamically dominant in the galaxy's NGM, requiring a multi-dimensional treatment \citep{Thomas2023}. Additionally, both CR heating and radiative cooling in the high-density shocks necessitates a more complete treatment of the gas thermodynamics and very high spatial resolution \citep{Huang2022b,Modak2023}. 

\subsubsection{CR-driven Winds in Dense Starbursts}
\label{section:cr_limitations}
The scalings above imply that CR-driven winds may be important in Milky Way-like and low-density dwarf galaxies. However, an important potential limitation to CR wind models in dense starbursts like M82, NGC\,253, or the normal star-forming galaxies we see at $z\sim1-3$ is that pion production from CR proton collisions with the ISM of the host galaxy can decrease the CR pressure at the base of the outflow relative to expectations without pion losses. Whether driven by diffusion or streaming, a decrease in the base pressure would decrease the mass-loss rate (eqs.~\ref{eq:mdot_crdiffusion} \& \ref{mdot_streaming}). 

The pion loss timescale is \citep{Gaisser1990} $t_\pi \simeq5\times10^7\,{\rm yr}/n_{\rm eff}$, where $n_{\rm eff}$ is the gas number density {\it seen} by CRs in cm$^{-3}$, which may in principle be larger or smaller than the average density of the host galaxy depending on the phase-dependent transport of CRs. {\it Prima facie}, high-density galaxies could have $t_\pi$ less than other loss timescales. If so, neglecting CR transport before pion losses, the CR pressure is (compare with eqs.~\ref{eq:pcr_tdiff} \& \ref{eq:pcr_tstream})
\beq
P_{\rm cr,\,\pi,0}\sim (\dot{E}_{\rm CR}t_\pi/6\pi R^2h)
\sim 6\times10^{-12}\,\dot{\Sigma}_{\star, 0.01}\,\Sigma_{g,\,\rm eff, 10}^{-1} \,\,{\rm ergs \,\,cm^{-3}}\,
\label{eq:pcr_tpi}
\eeq
where $\Sigma_{g,\,\rm eff}=2h n_{\rm eff} m$ is the gas surface density ``seen" by CRs. If we compare $P_{\rm cr,\,\pi,0}$ to the pressure needed for vertical hydrostatic equilibrium (eq.~\ref{eq:phse}), we find
\beq
P_{\rm cr,\,\pi,0}/P_{\rm HSE} \simeq0.7\,(f_g/0.1) \dot{\Sigma}_{\star, 0.01} \,\Sigma_{g,\,\rm eff, 10}^{-1} \,\Sigma_{g,\, 10}^{-2},
\label{ratio}
\eeq
where $f_g=\Sigma_g/\Sigma_{\rm tot}$. Note the strong dependence on $\Sigma_g$. Given the form of the observed Schmidt Law ($\dot{\Sigma}_\star\propto \Sigma_g^N$), with $N$ ranging from values as low as 1, to $N\simeq1.4$  \citep{Kennicutt1998}, to steeper determinations ($N\simeq2$), it is clear that the ratio $P_{\rm cr,\,\pi,0}/P_{\rm HSE}$ strongly {\it decreasing} as the galaxy gas surface density increases {\it if} CRs interact with gas at approximately the mean density, i.e., if $\Sigma_{g,\,\rm eff}\simeq\Sigma_g$.  For $N=1.4$ (or $2.0$), $P_{\rm CR,\,0}/P_{\rm HSE}\propto f_g\Sigma_g^{-1.6}$ (or $\propto f_g\Sigma_g^{-1.0}$). That is, if CRs interact with the mean density of the ISM, and if pion losses dominate other losses, then the central CR pressure becomes dynamically weak with respect to gravity at high $\Sigma_g$, precisely where we see dramatic starburst superwinds. 

Consider M82. The mean gas density in the starburst is of order $n\simeq250$\,cm$^{-3}$ \citep{Weiss2001}, implying $t_\pi\sim2\times10^5$\,yr if CRs interact with gas at mean density, smaller than the diffusion or streaming timescales. Using equation (\ref{eq:pcr_tpi}), \cite{Lacki2011} find that $P_{\rm cr,\,\pi,0}/P_{\rm HSE}\simeq0.02$ \citep{Persic2008,Yoast-Hull2013_M82,Buckman2020}. The wind mass loss rate would then be $50$ (diffusion) or $50^2$ (streaming) {\it smaller} than equations (\ref{eq:mdot_crdiffusion}) and (\ref{cr_massload_stream}). The nuclei of Arp 220 are even more extreme, with an average density of $\gtrsim10^4$\,cm$^{-3}$ \citep{Scoville2017}, implying $t_\pi\lesssim5000$\,yr. If CRs interact with the mean density \citep{Torres2004,Yoast-Hull2015_Arp220,Yoast-Hull2019_Arp220}, $P_{\rm cr,\,\pi,0}/P_{\rm HSE}$ is very small, and there is a correspondingly tiny predicted outflow rate. Yet, Arp 220 exhibits a strong wind \citep{Rangwala2011,Barcos-Munoz2018}. In models of the observed FIR-radio correlation and gamma-ray correlations ranging from dense starbursts to normal SFGs, this decrease in $P_{\rm cr,\,\pi,0}/P_{\rm HSE}$ as a function of density is seen \citep{Lacki2010,Lacki2011,Crocker2021_I,Crocker2021_IIthreshold,Werhahn2021_gamma,Werhahn2021_FRC}.
A potential counterargument could be that the CRs simply interact with a medium of much lower than average density (i.e., 50 times lower in M82), thus escaping pionic losses. However, the wind advection time is sufficiently short -- $\sim0.5$\,Myr in M82 -- that one would then require much larger $\dot{E}_{\rm cr}$ (eq.~\ref{eq:edotcr}) to maintain the observed gamma-ray luminosity. This increase would far exceed the energy budget since M82 is observed to be close to the ``calorimetric limit" \citep{Lacki2011}. 

When do CRs become weak(er) in galaxies? Setting $t_{\rm diff}=h^2/D$ or $t_{\rm str}=h/v_A$ equal to $t_\pi$, we derive a value of $n_{\rm eff}$ above which pion losses dominate: $n_{\rm eff,\,diff}\simeq1.5\,{\rm cm^{-3}}D_{28}/h_{\rm kpc}^2$ and $n_{\rm eff,\,str}\simeq1.5\,{\rm cm^{-3}}\,v_{\rm A, 30}/h_{\rm kpc}$, respectively. Both are about 10 times larger than the density seen by Galactic CRs \citep{Longair1994}. Taking the Schmidt Law from \cite{Kennicutt1998}, equation (\ref{ratio}) implies that $P_{\rm cr,0}/P_{\rm HSE}$ should become increasingly small above a gas surface density of $\Sigma_{g}\gtrsim50$\,M$_\odot$/pc$^2$ \citep{Lacki2010,Crocker2021_I,Crocker2021_IIthreshold}. Finally, note that although equation (\ref{eq:pcr_tpi}) is a good estimate for the {\it central} CR pressure when $t_\pi\ll t_{\rm diff}$ and $t_{\rm str}$, \cite{Socrates2008} show that because of CR scattering before pion losses, the reduction in pressure is not as strong at the galaxy's {\it surface}, where the wind is driven.

\subsubsection{Summary of CR-Driven Winds:} CR-driven winds in both the diffusive and streaming limits are promising for explaining large-scale outflows from SFGs along the galaxy main sequence, especially relatively low gas density systems like the Milky Way, other late-type spiral galaxies in the local universe, and star-forming systems like the Magellanic Clouds. Simple estimates give mass loading, energy, and momentum loss rates comparable to those seen in observed winds. While pion losses may prevent large mass-loading in dense starbursts, more work is required to understand how both CR nuclei and leptons interact with the ISM. Remaining pressing issues include the energy- and ISM phase-dependent transport of CRs. The question of diffusion or streaming is not either/or. They may operate together. Setting $t_{\rm diff}=t_{\rm stream}=h^2/D = h/v_A$ implies a critical scale 
\begin{equation}
h_{\rm \,diff\,=\,stream}\sim D/v_A\sim1\,{\rm kpc}
\,\left(D/10^{28}\,{\rm cm^2\,s^{-1}}\right)\left(30\,{\rm km\,\,s^{-1}}/v_A\right), 
\end{equation}
below which diffusion dominates, and above which streaming dominates. More realistically, CR transport may vary by phase and may only in some regions be well-characterized as ``diffusion" or ``streaming" (e.g., \citealt{Thomas2023}). Cool cloud acceleration experiments, like those done for hot winds, are needed to more fully understand CR coupling (e.g., \citealt{Bruggen2020,Ruszkowski2023}; Panel E in Fig.~\ref{figure:theory}).

\subsection{Magnetically-Driven Winds}
\label{section:magnetic}

The idea that magnetic fields are critically important for accelerating stellar winds and for affecting their angular momentum evolution goes back at least to \cite{Schatzman1962}. The theory of magnetic rotators was extended by \cite{Weber1967}. \cite{Blandford1982} developed the magneto-thermal wind theory for disks (see \citealt{Pudritz1986,Pelletier1992,Kudoh1997,Konigl2000}). While magneto-thermal winds are thought to play a critical role in the mass loss, accretion, and evolution of proto-planetary and AGN disks, there has been relatively little application to galaxies (but, see \citealt{Pino1999,Steinwandel2020}).

The reader may find it odd that we review the physics of a mechanism that is undeveloped in the current literature, and for which galaxy simulations provide scant (or negative) evidence. Galaxy MHD simulations to date do not drive the strong magneto-thermal winds found in the proto-planetary disk literature. For example, \cite{Pakmor2013} find relatively weak fountain-like flows originating from highly-magnetized bubbles that fall back to the disk, \cite{Thomas2023} find strong CR-driven outflows, but show quantitatively that magnetic fields provide little (if any) wind acceleration, and \cite{Wibking2023} find no difference in the SN-driven wind outflow rates between their purely hydrodynamic and fully MHD galaxy simulations (when SFR-normalized). Thus, there is evidence from theory against galactic magneto-thermal winds. Even so, we find four reasons to proceed: (1) the prevalence and importance of magneto-thermal winds throughout astrophysics; (2) the simulation results so far for galaxies may be sensitive to the initial {\it poloidal} magnetic flux, to the dynamo physics, to the galactic potential and its dynamical state (e.g., bars, spiral waves, circumnuclear rings), to the feedback physics employed, and to their coupled interaction; (3) for the most extreme starbursts, magneto-thermal winds do not suffer from the same problems as hot winds (radiative cooling, molecule formation) or CR-driven winds (pion losses), although they may suffer from others; (4) most importantly, observations may demand magneto-thermal winds for some systems, theory notwithstanding. Specifically, observations by \cite{Heesen2011} of NGC\,253's core and outflow indicate strong extraplanar toroidal magnetic fields that are correlated with the H$\alpha$-emitting filaments, that sheath the X-ray emitting gas, and that may be strong enough to channel the wind  -- properties similar to expectations for a magneto-thermal wind. For these reasons, we sketch the theory here as a guide for future works.

The controlling parameters for magneto-thermal disk winds are the cylindrical launch radius ($R$) and height ($z_0$), the base {\it ordered poloidal} field strength $B_{p0}$, the corresponding Alfv\'en speed $v_{Ap0}=B_{p0}/(4\pi\rho_0)^{1/2}$, the gas sound speed $c_s$, and the azimuthal velocity of the field line footpoints ($v_{\phi0}$), which may be approximated by the orbital velocity. In steady-state models, the field geometry is typically set by hand since a self-consistent configuration would necessitate solving the Grad-Shafranov equation for cross-field force balance. A magneto-thermal wind is said to be ``cold" if $c_s\ll v_{\phi0}$, which is the case of interest for massive galaxies  driving cool outflows. In this regime, the wind is principally magnetically-driven, even though finite $c_s$ is important for setting the subsonic density profile and the mass loss rate. The magnetization, $\mu$, appears throughout magnetic rotator theory: 
\begin{equation}
    \mu
    =\frac{v_{\phi0}\,v_{\rm p0}}{v_{Ap0}^2}
    \sim\frac{\dot{M}_{\rm wind} v_{\phi0}}{B_{p0}^2 R^2}\sim
    8\left(\frac{\dot{M}_{\rm wind}}{\rm M_\odot/yr}\right)\left(\frac{v_{\phi 0}}{200\,\rm km/s}\right)
    \left(\frac{\rm \mu G}{B_{p0}}\right)^2
    \left(\frac{4 \, \rm kpc}{R}\right)^2,
    \label{massloadingmu}
\end{equation}
where $v_{\rm p0}$ is the base poloidal outflow velocity, and where we have scaled to values appropriate for a Milky Way-like disk with $\dot{M}_{\rm wind}$ comparable to the SFR. Motivated by observations, the normalization for $B_{\rm p0}$ is deliberately chosen to be significantly below the average {\it total} magnetic field strength of the disk, which is dominated by the toroidal and turbulent components. Like the Milky Way, for parameters appropriate to an M82-like starburst or high-$z$ star-forming clump ($B_{p0}\sim10-100$\,$\mu$G, $R_0\sim0.25$\,kpc, $\dot{M}_{\rm wind}\sim5-10$\,M$_\odot$/yr), or for a nucleus of an Arp 220-like ULIRG ($B_{p0}\sim0.1-1$\,mG, $R_0\sim0.05$\,kpc, $\dot{M}_{\rm wind}\sim100$\,M$_\odot$/yr), equation (\ref{massloadingmu}) shows that we expect $\mu\gtrsim1$. In this regime, disk winds are driven by the vertical magnetic pressure gradient, the Alfv\'en radius $R_A$ is near $R_0$, the $B_\phi$ is much larger than $B_{p0}$ at the wind's base ($|B_{\phi0}/B_{p0}|\sim \mu$),  $v_\infty\sim v_{\phi0}\mu^{-1/3}$ for cold winds \citep{Spruit2010}, and the wind dynamics is {\it not} dominated by the magneto-centrifugal ``beads-on-a-wire" slinging effect, as in strongly-magnetized winds with $\mu\ll 1$ and $R_A\gg R_0$. Note that the gas density at the wind's base $\rho_0=\rho(z_0)$ may be much less than the disk midplane density at $R$, that $\rho(z)$ decreases much more rapidly than $B(z)$ so that $v_{\rm A}$ rapidly increases along the minor axis, and that additional physics determines the base density $\rho_0$ (\S\ref{section:theory}).

Because the wind carries away angular momentum, the outflow is accompanied by accretion (e.g., \citealt{Pelletier1992,Konigl2000}):
\begin{equation}
    \dot{M}_{\rm acc}=\frac{2R_0}{\Omega}|B_{p0}B_{\phi0}|\sin\theta\sim2{\rm M_\odot/yr}\left(\frac{R_0}{4\,{\rm kpc}}\right)^2\left(\frac{200\,{\rm km/s}}{v_{\phi0}}\right)\left(\frac{B_{p0}B_{\phi0}}{10^{-11}\,{\rm \mu G}^2}\right),
    \label{eq:magnetic_accretion}
\end{equation}
where $\theta$ is the angle between the disk normal and the poloidal field line ($B_{z0}=B_{p0}\sin\theta$),  $\Omega$ is the orbital frequency at $R_0$, and where the scaling above assumes $B_{\rm p0}\sim$\,$\mu$G, $|B_{\phi0}/B_{p0}|\sim10$, and $\sin\theta\sim1$. Because $|B_{\phi0}/B_{p0}|\sim\mu$ for $\mu\gg1$, $\dot{M}_{\rm acc}\sim \dot{M}_{\rm wind}$. If such a relation holds for galaxies, this would have important implications for their metallicity evolution by directly shaping the connection between inflow and outflow, and thus the enrichment and SFR histories, and the metallicity gradients (in stars and gas). As an example, we use the code from \cite{Bai2016} to generate a wind model in a  point-mass potential $v_{\phi0}$ $=$ $200$\,km/s, $R_0$ $=$ $8$\,kpc, $z_0$ $=$ $100$\,pc, an ordered poloidal field of $B_{p0}$ $\simeq$ $0.6$ $\mu$G, $n(z_0)$ $=$ $0.1$ cm$^{-3}$ ($v_{Ap0}\simeq=4$\,km/s), $\theta$ $=$ $45^o$, and $c_s$ $=$ $10$\,km/s. We find that $\mu$ $\simeq$ $7$ (eq.~\ref{massloadingmu}), $v_\infty$ $\simeq$ $200$\,km/s, $B_{\phi,0}$ $\simeq$ $5\,\mu$G, $\dot{M}_{\rm wind}\simeq0.9$\,M$_\odot$ yr$^{-1}$, and an accretion rate of $\simeq2.1$\,M$_\odot$ yr$^{-1}$ (eq.~\ref{eq:magnetic_accretion}). Because $\mu>1$, the poloidal profile of $v_\phi$ of the outflowing gas closely follows angular momentum conservation and not the spinning porcupine beads-on-a-wire $v_\phi\sim R$ (for $R$ $\lesssim$ $ R_A$) expected from strong magneto-centrifugal acceleration.

The magneto-thermal wind theory should be more fully generalized to the galaxy context. We currently do not have an analytic theory for $\dot{M}_{\rm wind}$ and are thus unable to provide estimates and scaling relations for the force and energy content as we have for other potential wind-driving mechanisms. Global MHD galaxy simulations with a range of initial poloidal field strengths may be needed to understand whether or not magneto-thermal winds can affect galaxies. We are aware of no limit in the literature where, below a certain $B_{p0}$, no wind is driven. There may be such a limit that shows why galactic magneto-thermal winds are impossible, given their small primordial halo fields and their non-point-mass potentials. If so, it should be understood and articulated.

As mentioned above, \cite{Heesen2011} find a strong $\sim20$\,$\mu$G helical magnetic fields in outflow cone of NGC\,253 (Fig.~\ref{figure:m82_hst}). While these observations may point to the importance of a magneto-thermal wind in this specific case, the inferred field geometry in NGC\,253 is apparently not generic for either normal SFGs or starbursts. The CHANG-ES survey provides constraints on extraplanar magnetic fields in local edge-on spirals via radio continuum observations and Faraday rotation \citep{Irwin2012,Krause2020}. While strong magnetic fields appear ubiquitous, simple field topologies do not. Instead, large-scale field reversals and complicated geometries are generic (Fig.~\ref{figure:marvil}). In contrast to NGC\,253, the wind in M\,82 exhibits nearly vertical magnetic field orientation as inferred from dust polarization observations with SOPHIA \citep{Jones2019_SophiaM82_NGC253,Lopez-Rodriguez2021_SophiaM82}. This purely vertical geometry is similar to that found in simulations by \cite{Thomas2023}, where $B$ plays little dynamical role. Thus, it may be that in some systems the hot wind or CR-driven wind dominates, producing a purely vertical field, while in other systems (e.g., NGC\,253) the field is able to exert forces that contribute to cool wind acceleration. For a review of galactic magnetic fields, see, e.g., \cite{Beck2015}.

\section{Observed Properties of Galactic Winds}
\label{section:observations}

Rather than provide a comprehensive review of galactic wind observations, we focus on several specific aspects. First, we focus on starburst-driven winds in the local Universe. Although winds are rare at $z\simeq0$, nearby galaxies have the most pan-chromatic information about different wind phases, can be studied in greater detail, and the information extracted can be used to understand high-$z$ systems where winds are prevalent. Second, we focus on the warm and hot phases, motivated by the fact that the former provides the majority of information on outflows at higher $z$ and the latter might be the cool gas wind driver. Finally, we emphasize how physical properties of winds are derived from data, where the key uncertainties are, and what constraints the data provide on wind models. 

We first summarize results from emission and absorption line observations (\S\S\ref{section:optical_xray} \& \ref{section:absorption}). We briefly discuss the impact of winds on the CGM (\S\ref{section:observations_cgm}) and their demographics over cosmic time (\S\ref{section:demographics_cosmic_time}). \S\ref{section:confronting_theory_observations} gives a preliminary comparison between observations and theory.

\begin{figure}[t]
\centering\includegraphics[width=\textwidth]{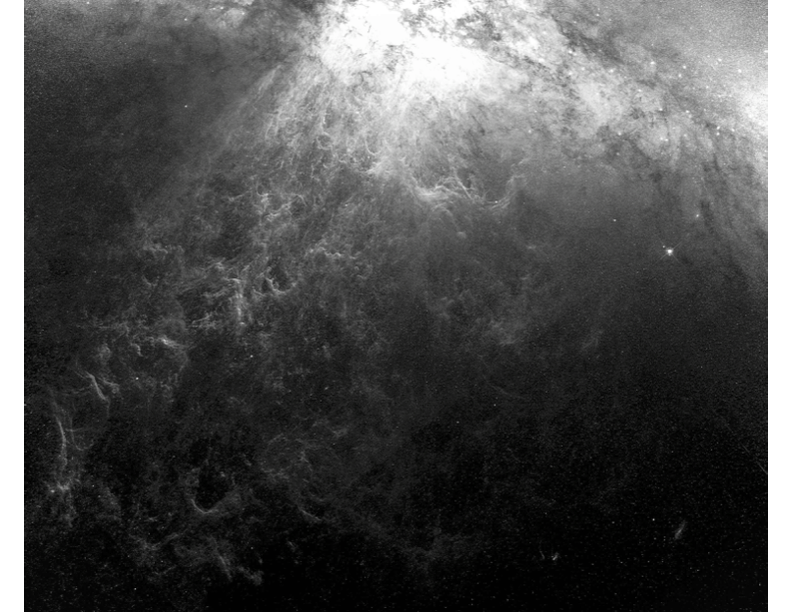}
\caption{Detail of an HST H$\alpha$ image in the southern outflow of M82 \citep{Mutchler2007}. The image has been stretched and sharpened to highlight the cloud structure. The starburst is discernible in the upper middle. The abundant characteristic arc-like structures are only marginally-resolved ($\sim5$\,pc) in the direction along the minor axis. They do not have cometary morphologies, as might be expected if they were the contrail-like structures created in simulations of the wind-cloud interaction (e.g., \citealt{Scannapieco2015}).}
\label{figure:m82_HSTzoom}
\end{figure}

\subsection{Emission}
\label{section:optical_xray}

\subsubsection{Optical/UV Lines}
\label{section:optical_uv_emission}
Extended extraplanar emission line regions are a defining characteristic of starburst winds. Evidence for M\,82's outflow dates back 60 years to the discovery of a system of filamentary emission-line gas extending kpc scales along its minor axis  (\citealt{Lynds1963}; Figs.~\ref{figure:m82_hst} \& \ref{figure:m82_HSTzoom}). The intricate small-scale structure of this material is evident in the H$\alpha$ emission shown in Figure \ref{figure:m82_HSTzoom}. Similar spectacular morphology is seen in many other starbursts, including NGC\,6240 (Fig.~\ref{figure:bubble}). This gas can also be observed through nebular line and continuum emission in the vacuum UV, and through mid- and far-IR fine-structure line emission \citep{Beirao2015, Hoopes2005,Contursi2013}. Diagnostic diagrams involving pairs of emission-line ratios \citep{Veilleux1987} imply that the gas is predominantly photoionized near the starburst, but that at larger distances there is a significant and increasing contribution from collisional ionization, possibly due to wind-driven shocks \citep{Armus1989,Shopbell1998,Lehnert1996,Veilleux1995}. Nebular diagnostics can also be used to measure the spatial variation in the electron density (e.g., [SII] 6717/6731), with typical values dropping from $n_e \sim 10^3$\,cm$^{-3}$ in the starburst to $\sim 10^2$\,cm$^{-3}$ at radii of $\sim1-3$\,kpc \citep{Heckman1990,Lehnert1996}, with a corresponding pressure decrease for $10^4$\,K gas. The extinction-corrected H$\alpha$ surface brightness is proportional to the path integral of the $n_e^2\,f$, where $f$ is the volume filling factor. Knowing $n_e$ from line diagnostics, and adopting a path-length through the outflow, one can solve for $f$. In M82, \citet{Xu2023_M82} find that $f$ is $< 10^{-3}$, at least in the inner 2\,kpc region where $n_e$ can be reliably measured (see also \citealt{Heckman1990,Westmoquette2007,Westmoquette2009}).
 
\begin{figure}
\centerline{\includegraphics[width=12cm]
{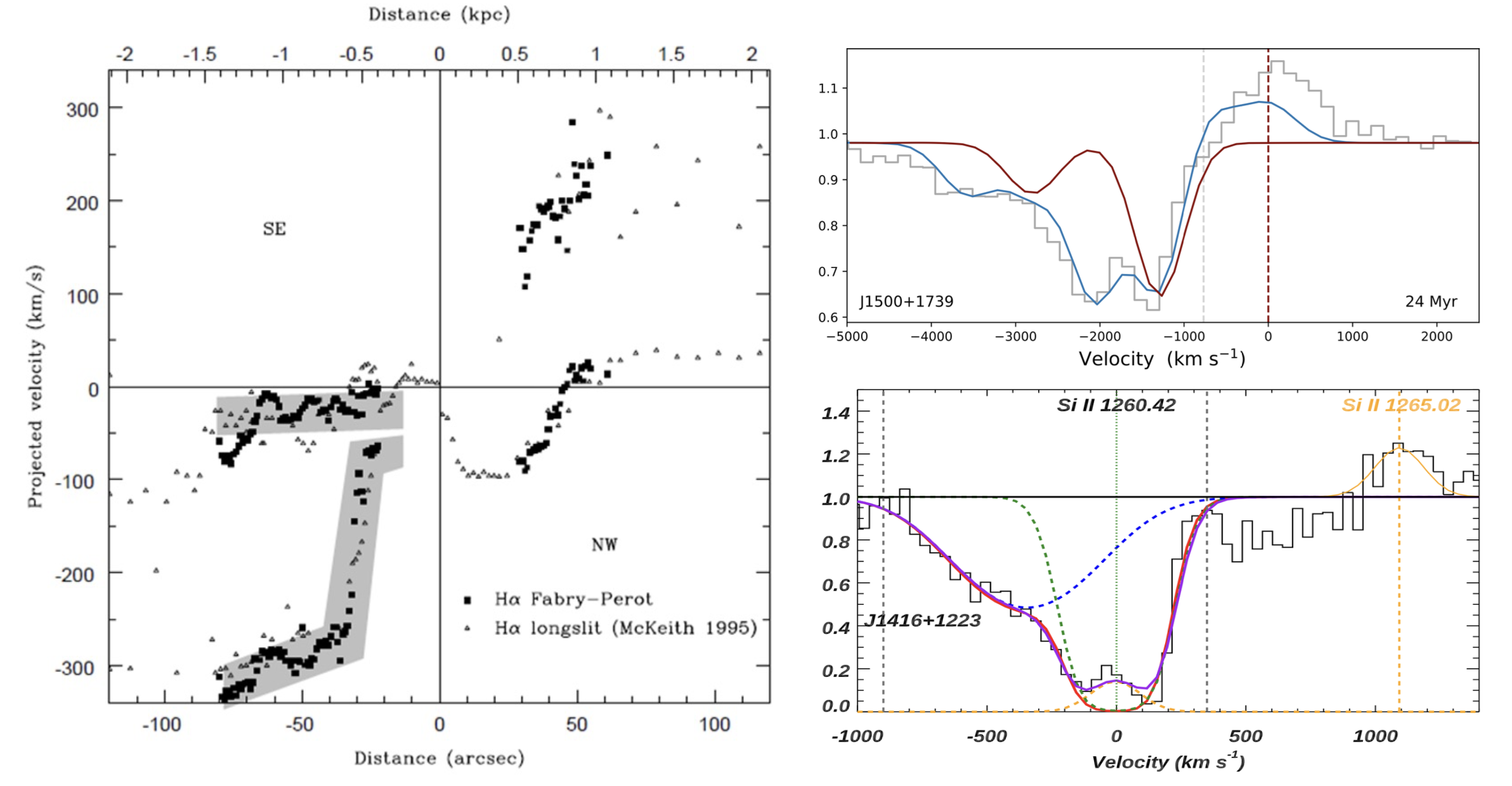}}
\hspace*{0.3cm}
\caption{{\it Left:} Kinematics of the optical emission-line gas in the M\,82 outflow \citep{Shopbell1998}. Double-peaked emission-line profiles are observed along the minor axis, and can be modeled as an outflow along a hollow biconical structure with a opening angle of $\simeq70$ degrees and a de-projected velocity of $\sim600$\,km s$^{-1}$. {\it Right:} Blue-shifted absorption line fitting of Mg\,II (top; \citealt{Davis2023}) and Si\,II (bottom; \citealt{Xu2022}).} 
\label{figure:line_montage}
\end{figure}

In M82, the kinematics and morphology of the line emitting gas (left panel of Fig.~\ref{figure:line_montage}) implies that it lies largely along the surface of a bi-cone that originates near the starburst \citep{Shopbell1998}, with a deprojected outflow speed of $\sim600$\,km s$^{-1}$, larger than M\,82's escape velocity \citep{Greco2012}. The velocity field displays rapid acceleration at a distance of $\sim500$\,pc and beyond $\sim700$\,pc the flow speed is roughly constant out to 3\,kpc \citep{Yoshida2019}. As in M\,82, in other systems like NGC\,253, the structure and kinematics of the warm ionized gas, as traced by optical emission lines, imply that it lies along the surface of an expanding hollow structure. Large-scale radial filaments extend out to scales of $\sim1-10$\,kpc or larger. In some cases, the emission-line gas is seen to trace a pair of bubble-like structures extending into the CGM (see Fig.~\ref{figure:bubble}). For both bi-cones and bubbles, optical spectra reveal double-peaked emission-line profiles from the near and far sides of the hollow structure. The observed line-peak separations ($\sim100-1000$\,km s$^{-1}$) are similar to M\,82 in most cases \citep{Heckman1990}, but are significantly smaller in dwarf starbursts ($\sim10{\rm s}-100$\,km s$^{-1}$). In the majority of these cases, the expanding structures seem to be kpc-scale bubbles that may be embedded within the dwarf host galaxy's ISM \citep{Marlowe1995,Martin1998,Martin1999}. 

\subsubsection{Molecular gas, Atomic gas, \& Dust}
\label{section:molecular_atomic_dust_emission}
Multi-wavelength observations of M82 reveal a plethora of outflowing phases. Cold and warm molecular gas have been mapped through mm-wave and mid/near IR spectroscopy, respectively \citep{Veilleux2009,Beirao2015,Leroy2015}. Atomic gas is detected in the HI 21\,cm line \citep{Leroy2015,Martini2018} and with Far-IR spectroscopy of the [OI] and [CII] fine-structure lines \citep{Contursi2013}.  The double-peaked line profiles seen in CO and HI in the outflow suggest that -- like the gas traced by H$\alpha$ -- the molecular and atomic gas lies along the surface of a bi-cone \citep{Leroy2015}. However, the observed outflow speeds for the molecular and atomic gas are much smaller than those measured for the warm-ionized phase ($\sim10^2$\,km/s), implying a fountain flow rather than an unbound wind that may again trace the interaction of the hot fast wind with denser, higher column density ambient gas (\S\ref{section:ram}).

In local starbursts, spectroscopy in the MIR also establishes the presence of PAHs throughout the outflow (Fig.~\ref{figure:m82_hst}; \citealt{Beirao2015,Bolatto2024}). Dust can be traced by emission in the IR \citep{Contursi2013,Leroy2015}, and by scattering of far-UV light from the starburst into our line of sight \citep{Hoopes2005}. Scattering also produces polarized emission in the outflow \citep{Scarrott1991}. \cite{Yoshida2019} use spectropolarimetry to disentangle the scattered line emission from that produced {\it in situ} and estimate an dust outflow speed of $300-450$\,km/s. 

\subsubsection{X-rays}
\label{section:xray_emission}
\begin{figure}
    \centerline{
    \includegraphics[width=12.5cm]{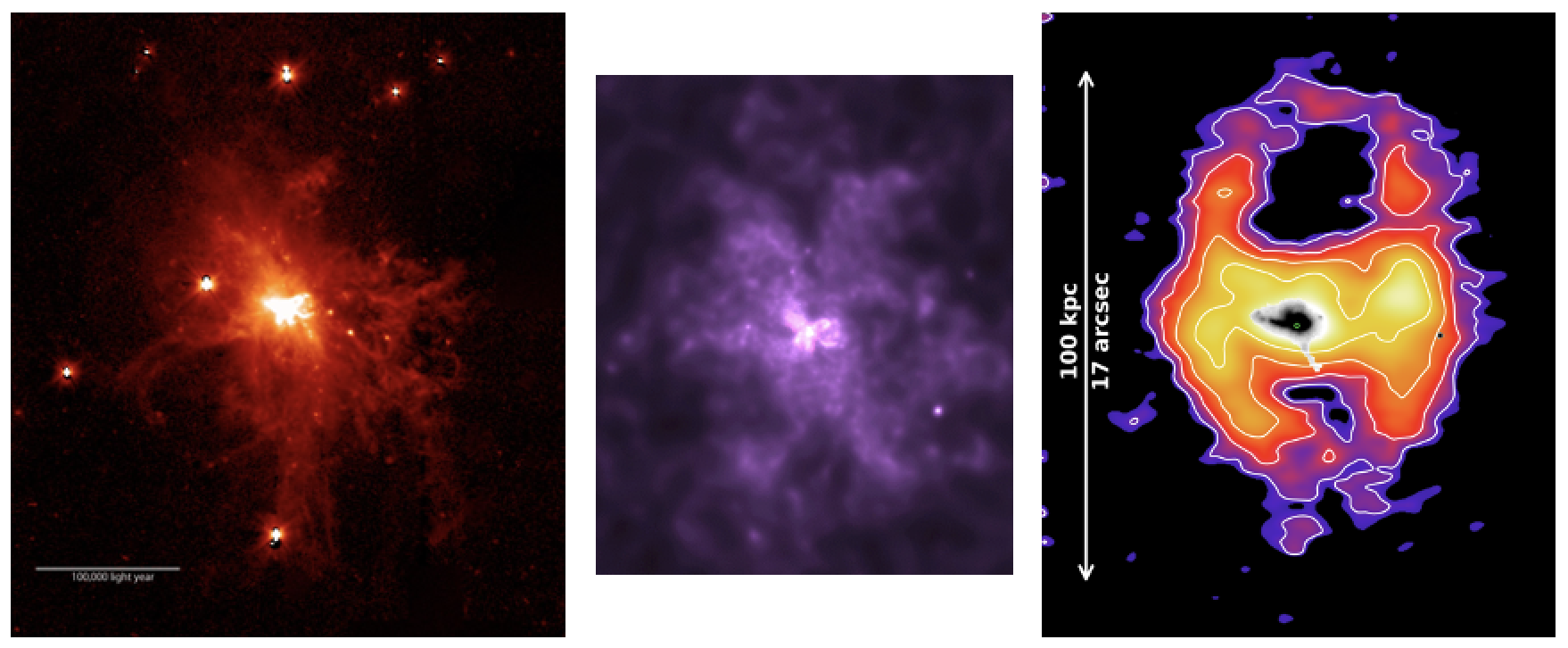}}
    \vspace*{0.5cm}
    \caption{{\it Left:} $\sim100$\,kpc H$\alpha$ nebula of NGC\,6240 by \cite{Yoshida2016} (Credit: Hiroshima University and NAOJ). {\it Middle:} Corresponding soft X-ray emission in NGC\,6240 (Credit: NASA/CXC/SAO/\citealt{Nardini2013}). {\it Right:} [OII] 3727$\rm \AA$ image of the double-bubble outflow in Makani \citep{Rupke2019}, with a diameter of nearly 100\,kpc.}
    \label{figure:bubble}
\end{figure}

Diffuse X-ray emission is a hallmark of starburst superwinds, with extended X-ray emission seen in many systems (Figs.~\ref{figure:m82_hst} \& \ref{figure:bubble}). When they can be measured, the properties of the diffuse X-ray emission are broadly consistent with M82, where the data are most complete, and where multiple emission components are seen.

The very hot phase produces both X-ray continuum emission via bremsstrahlung and line emission from He- and H-like Fe at 6.7 and 6.9\,keV, respectively. In collisional ionization equilibrium, He-like Fe is the dominant ion for $\log_{10}[T/{\rm K}] = 7.9-8.1$ and H-like Fe dominates at higher $T$  \citep{Sutherland1993}. Although the hard X-ray continuum from starbursts could be used to establish $T$, it may have contributions from high-mass X-ray binaries and/or inverse Compton or synchrotron emission from CR leptons \citep{Strickland2009,Lacki2013,Iwasawa2009,Iwasawa2023}. Thus, line emission is the cleanest signature of the very hot phase in starbursts. Analysis of the hard X-ray spectrum of M82's core shows that dominant ionic stage of Fe is He-like. Including information from the H- and He-like S, Ar, and Ca lines \cite{Strickland2009} find a best fit $T\simeq5-8 \times 10^7$\,K. The very hot gas properties are consistent with the \cite{Chevalier1985} thermal wind model (\S\ref{section:thermal}) with a thermalization efficiency of $\alpha\sim0.3-1.0$ and a mass-loading factor of $\eta\sim0.2-0.6$ (eq.~\ref{eq:alpha_eta}). The implied asymptotic velocity for this ``very hot" gas is $v_\infty\simeq1400-2200$\,km s$^{-1}$ (eq.~\ref{cc85_vinf_pdot}) if it does not entrain more gas as it emerges from the nucleus (see \S\ref{section:additional_physics_hot_phase}). The hard X-ray emission from starbursts is summarized in Figure \ref{figure:heckman_various} (see \S\ref{section:confronting_theory_observations}).

The extended soft X-ray emission ($<2$\,keV)  (see Fig.~\ref{figure:m82_hst} \& \ref{figure:bubble}) traces $T\sim5-10\times 10^6$\,K gas (e.g., \citealt{Strickland2004,Ranalli2008,Lopez2020}). This is the ``hot" phase (not to be confused the ``very hot" phase). In general, the temperature and composition of the medium are derived by comparing model spectra folded through the known X-ray system response matrix. Two-temperature models are typically required to obtain good fits to the spectra, and yield components with $T \sim 3 \times 10^6$\,K and $10^7$\,K, respectively \citep{Strickland2004}. In cases with the best data (like M\,82) a three-component fit is needed that includes charge exchange (e.g., \citealt{Liu2012,Zhang2014,Lopez2020,Lopez2023}). These models may be overly idealized, and a broad range of temperatures could be present. Additionally, non-equilibrium ionization and self-irradiation may affect the X-ray emission (\S\ref{section:additional_physics_hot_phase}), requiring new types of analysis.

Although the low spectral resolution of the imaging X-ray instruments on Chandra and XMM-Newton make it challenging, current X-ray data also constrain the chemical composition of the hot gas. In the dwarf starburst NGC\,1569, \citet{Martin2002} find $\alpha$ element abundances that are about five times higher in the hot gas than in the cool/warm ISM. \citet{Lopez2020,Lopez2023} mapped chemical abundances in the central starbursts and outflows of M\,82 and NGC\,253, finding elevated central abundances. In these systems, the hot gas has abundance ratios of $\alpha$ elements (O, Ne, Mg, Si, S) to iron that are $2-4$ times solar (see also \citealt{Iwasawa2023}), implying that the hot gas is enriched by the ejecta of core-collapse SNe. Abundance gradients can be used to assess the mixing of more pristine ISM material into the X-ray emitting gas as it expands (\S\ref{section:additional_physics_hot_phase}; \citealt{Nguyen2021}).   

The density of the hot gas can be constrained from the X-ray luminosity, which is proportional to the volume integral of the density squared: $L_{\rm X} \propto f \,n^2\times {\rm Vol}$, where ${\rm Vol}$ is the observed volume, and $f$ is the volume filling fraction. The number density in the region is thus $n \propto (L_{\rm X}/{\rm Vol} f)^{1/2}$. This is the most direct measure we have for $n$, but it provides only a lower limit by assuming $f=1$. The corresponding {\it minimum} gas pressure is $P = n k_{\rm B} T$. For $f < 1$, $P$ increases as $f^{-1/2}$.

The soft X-ray emission morphology corresponds closely with the warm ionized phase \citep{Cecil2002,Strickland2004,Grimes2005}. See Figures~\ref{figure:m82_hst} and \ref{figure:bubble}. However, on a local level in M82 and other systems, the soft X-ray emission is systematically located interior to the warm ionized gas (Fig.~\ref{figure:m82_hst}), perhaps indicating that the soft X-rays trace the interface between the {\it very} hot and tenuous wind fluid and the cooler denser gas. This X-ray-inside-warm phase morphology includes the ``cap," a filamentary structure about 9\,kpc above M82 that is oriented roughly perpendicular to the outflow \citep{Lehnert1999}.

Importantly, while the soft X-ray luminosity is correlated with the SFR, it is typically only about 1\% of the kinetic energy injection rate supplied by the starburst (eq.~\ref{eq:alpha_eta}; \citealt{Grimes2005}). Thus, radiative losses via X-ray emission must be negligible in influencing the dynamics of the hot gas in most systems. In fact, the lack of strong X-ray emission puts strong constraints on the mass-loading and thermalization efficiencies \citep{Zhang2014} and argues against strong bulk radiative cooling as the production mechanism for the cool outflowing gas (\S\ref{section:additional_physics_hot_phase}). As counterpoint, \cite{Thompson2016} and \cite{Lopez2023} present evidence that the base of the hot outflow in NGC\,253 might be radiative.

\subsubsection{Radio Continuum \& $\gamma$-rays}
\label{section:radio_gamma_emission}
In M82, radio continuum observations reveal an extended synchrotron halo produced by a magnetized relativistic phase (\citealt{Seaquist1991,Adebahr2013,Adebahr2017}; Fig.~\ref{figure:marvil}). The synchrotron-emitting primary CR $e^-$ and secondary $e^\pm$ are advected out into the halo by the wind, during which time they suffer both adiabatic and radiative losses via synchrotron and inverse Compton scattering. Comparison of the integrated $\gamma$-ray luminosity of M\,82 with the spatially- and spectrally-resolved radio continuum images indicates strong pionic losses in the core (see \S\ref{section:cr_limitations}), large values for the Alfv\'en velocity in the halo, and potentially dynamically important CR pressure gradients \citep{Buckman2020}. See Fig.~\ref{figure:marvil} and \S\ref{section:cosmic_rays} and \S\ref{section:magnetic}, where we briefly discuss radio halos.

\subsubsection{Outflow Rates}
\label{section:emission_outflow_rates}

Mass outflow rates can be estimated from optical emission-line measurements using  $\dot{M}=\Omega r^2 \rho v\sim \Omega (r^3\rho)(v/r)$, where the wind gas mass is $M_{\rm gas}\sim \Omega r^3\rho$ and the outflow time is ($r/v$), where $r$ is the radius and $v$ is the outflow velocity. Like the X-ray emitting phase, the  luminosity of the warm ionized gas is proportional to the volume integral of  density squared ($L\propto {\rm Vol}\,n^2\,f\propto M_{\rm gas}\,n\,f$) and hence to the product of $M_{\rm gas} n$. For the warm ionized phase, the [SII]\,6717,6731 line ratio provides an independent measurement of $n$ near the starburst. For M\,82, $\dot{M}\sim5$\,M$_{\odot}$ yr$^{-1}$ using $v=600$\,km/s at $\sim 1$\,kpc from the starburst core \citep{Xu2023_M82}, corresponding to $\eta=\dot{M}/{\rm SFR}\sim0.6$. The implied momentum and kinetic energy injection rates are $\sim 2 \times 10^{34}$\,dynes and $6 \times 10^{41}$\,erg s$^{-1}$, respectively, representing $\sim50-100$\% of $L/c$ and $\sim10-20$\% of $\dot{E}_{\rm SN}$ supplied by the starburst (eq.~\ref{cc85_vinf_pdot}). More systems are summarized in Figure \ref{figure:heckman_various} and \S\ref{section:confronting_theory_observations}. This technique has also been applied to SFGs at both low- and high-$z$ \citep{Forster2019,Davies2019,Perna2020,Fluetsch2021}, but a single characteristic value of $n_e$ is adopted, so the outflow rates of ionized gas are uncertain.

The outflow rates for the hot gas traced by soft X-ray emission are even more uncertain because we lack independent constraints on the gas density, and we have no information on the hot gas dynamics.  Because the physical volume occupied by the hot gas is $\propto {\rm Vol} \,f$, using $n$ derived from the X-ray emission yields $M_{\rm gas}\propto (L_{\rm X} {\rm Vol} \,f)^{1/2}$. Thus, $f = 1$ yields an upper limit on $M_{\rm gas}$ and the total thermal energy content. Assuming that the hot gas outflow velocity is the same as the warm ionized gas (motivated by the morphological correspondence), the maximum mass outflow rates are comparable to the SFR and the energy outflow rates are $\sim30$\% of the starburst injection rate (eq.~\ref{eq:alpha_eta}; \S\ref{section:confronting_theory_observations}; \citealt{Armus1995,Wang1997,Heckman1999,Grimes2006,Grimes2007,Strickland2002}). The most important observational piece missing in our understanding of the hot phase is its kinematics: a measurement of the hot gas velocity via X-ray spectroscopy will transform galactic wind studies by definitively constraining its energy and momentum content.

\subsection{Absorption-Line Probes of Galactic Winds at Low and High Redshift}
\label{section:absorption}

The majority of the available data on winds across cosmic time consists of spectra of blue-shifted absorption lines of both neutral and ionized gas, as in the right two panels of Figure \ref{figure:line_montage} (e.g., \citealt{Heckman2000,Shapley2003,Rupke2005,Martin2005,Grimes2009,Sato2009,Weiner2009,Rubin2010,Steidel2010,Chen2010,Erb2012,Kornei2012,Martin2012,Bordoloi2014,Rubin2014,Heckman2015,Zhu2015,Chisholm2015,Chisholm2017,Sugahara2019,Wang2022,Xu2022,Xu2023,Davis2023}. While we have such data for relatively large samples of low- and high-$z$ galaxies, the information provided directly from the observations is limited because of uncertainties in the physical size of the absorbing medium along the line-of-sight and in converting from the ions probed by the data to the total gas mass. 

\subsubsection{Physical Conditions in the Absorbing Outflow}
\label{section:absorption_physical_conditions}

The gas seen in UV absorption lines can be readily fit by a model where it is photo-ionized by radiation leaking out of the host galaxy and/or local EUV radiation from fast shocks driven by the wind. Models of collisionally-ionized gas yield lines ratios of low- to high-ionization column densities that are larger than observed \citep{ Chisholm2016,Chisholm2017}. For photo-ionized gas, the best-fit  models \citep{Xu2022} yield ionization parameters of $U \sim 10^{-2.5}$, implying that in most cases the gas is almost fully ionized with $N_{\rm HI}/N_{\rm H} < 0.1$. The outflows measured with these UV absorption lines are therefore different from those in dusty starbursts traced with Na\,I D in the rest-frame optical, even though the kinematics are broadly similar \citep{Heckman2000,Rupke2005,Martin2005}. The low Na\,I ionization potential of only 5.1\,eV means that it will be photoionized along lines-of-sight that are UV-translucent (as in galaxies with UV absorption spectra), and thus that the outflows it traces are likely of predominantly cooler atomic or possibly even molecular gas \citep{Veilleux2020}. 

More direct density measurements in absorption-line outflows use the ratio of the column density 
in the ${\rm Si\,\,II}^*$ $1265\,{\rm \AA}$ excited transition to the ${\rm Si\,\,II}$ $1260\,{\rm \AA}$ resonance transition \citep{Xu2023}, yielding a median value of $n_e\simeq23$\,cm$^{-3}$ (with a range of $4-60$\,cm$^{-3}$). By combining $n_e$ and $N_{\rm H}$, cloud radii and masses can be estimated using $R_{\rm cl} = N_{\rm H}/(n_{\rm H} \,C)$, where $C$ is the areal covering factor (see \S\ref{section:absorption_outflow_rates}). The median values are $\simeq5$\,pc and $\simeq200$\,M$_{\odot}$, respectively. \cite{Xu2023} further discuss three independent estimates for the distance from the host to the absorbing medium, each of which yields $\sim 1-5$ times the host radius (i.e., the absorbing gas is in the inner outflow region). The median value for the volume filling-factor is $f \simeq 4 \times 10^{-3}$, similar to  values derived for the optical emission-line gas in M82 (\S\ref{section:optical_xray}). The physical picture is then of a large population of pc-scale, dense, clumpy, low filling factor clouds being accelerated by some prime mover.

\subsubsection{Constraints on the Global Structure of the Absorbing Outflow}
\label{section:absorption_global_structure}

Perhaps the greatest impediment to unlocking the full diagnostic power of absorption-line measurements is the uncertainty in the size and solid angle of the region in which the absorption-lines arise. 

In standard ``down the barrel" observations, the absorption lines are produced only by gas directly along the line-of-sight to the continuum from the starburst. They correspond to transitions up from the ground state of a particular ionic species. Absorption is quickly followed by radiative decay to the ground-state (``resonance scattering") or (in the case of some ionic species) to a fine-structure level located just above the ground-state (``fluorescence"). Unlike the absorption lines, these emission line photons will be produced from throughout the portion of the outflow enclosed within the spectroscopic aperture (e.g. \citealt{Zhu2015,Prochaska2011,Scarlata2015,Carr2018}). Importantly, in the absence of dust, photons are conserved: for every photon absorbed, a photon must be emitted (via resonance or fluorescence). It is this crucial difference between the locations of the absorption and emission that can be used to constrain the structure of the outflow.

IFU observations of Mg\,II emission surrounding galaxies at intermediate- and high-$z$ provide direct measurements of the outflow structure. \citet{Dutta2023} used MUSE to map Mg\,II emission in stacked spectra constructed from a sample of 600 SFGs at $z =0.7-1.5$ with a median stellar mass of $M_\star\simeq2 \times 10^9$\,M$_{\odot}$. They found a median half-light diameter of $\simeq30$\,kpc (3 to 4 arcsec), and an intrinsic surface brightness that increases with both $z$ and $M_\star$. Mg\,II emission has also been mapped out to radii of about 20\,kpc in two starbursts at $z = 0.7$ \citep{Zabl2021} and $z = 1.7$ \citep{Shaban2022}, and by \citet{Rupke2019} out to a radius of about 12\,kpc in a powerful starburst at $z = 0.46$. A detailed analysis of Mg\,II was carried out by \citet{Burchett2021}, who detected emission out to a radius of $\simeq18$\,kpc and used radiative transfer models to show both that the outflow decelerates as it expands and that the observations favor a relatively shallow $n \propto r^{-1}$ density profile.

Similarly, \citet{Erb2022} use KCWI to map Ly$\alpha$ emission from a sample 12 low-mass (median $M_\star\sim 10^{9.2}$\,M$_{\odot}$) LBGs at $z\sim2.3$. They find that the emission can be detected out to radii of $\sim20$\,kpc, with half-light diameters of $12-20$\,kpc. Similar results were obtained by \cite{Guo2023a}. An outflow model in which clouds are accelerated by the momentum supplied by massive stars provides a good fit to the data, and is consistent with a roughly constant velocity profile and a cloud density that drops as $r^{-2}$.  \citet{Runnholm2023} use HST to image Ly$\alpha$ in seven starburst galaxies at $z\sim0.2-0.3$ and find that the surface-brightness scale-lengths are about six times larger than the FUV continuum (starlight) on-average.  

Outflows mapped in optical emission-lines and X-rays in nearby starbursts typically reveal bi-conical or bubble-like structures (\S\ref{section:optical_xray}). What constraints can be placed on the opening angles of the outflows traced in absorption? At low-$z$, \citet{Chen2010} use SDSS data to show that outflows traced by Na\,I D are only seen in disk galaxies viewed at angles $<60^{o}$ with respect to the galaxy minor axis, implying an opening angle $\sim 2\pi$ (see also \citealt{Borsani2019}). Qualitatively similar results were found using Mg\,II absorption-lines at $z \sim 1$ for SFGs by \citet{Kornei2012}, \citet{Bouche2012}, \citet{Bordoloi2014} and \citet{Rubin2014}. \citet{Lan2018}  use background QSOs proximate to a large sample of emission-line galaxies at $z \sim1$ and find excess Mg\,II absorption along the minor axis relative to the major axis out to impact parameters of nearly 100\,kpc, which they attribute to a bi-conical outflows with a full opening angle of about 90$^o$ (see also \citet{Bordoloi2011}). \cite{Guo2023b} use stacked images of Mg II in emission for edge-on and face-on galaxies at $z \sim 1$ to show that the outflowing gas is aligned with the galaxy minor axis out to distances of more than $10$\,kpc. All these results are consistent with the optical and X-ray emission of low-$z$ winds driven by compact central starbursts located within larger disk galaxies (see \S\ref{section:optical_xray}).  

In contrast, at $z > 2$, blueshifted absorption lines are seen in the spectra of $\sim90$\% of LBGs \citep{Shapley2003,Steidel2010}, indicating nearly isotropic outflows, consistent with the spatially-resolved structure of Ly$\alpha$ emission \citep{Erb2022}. The same high incidence rate is seen in the most extreme local starbursts, which are the best analogs to  $z > 2$ galaxies \citep{Heckman2015,Xu2022}. This near-isotropy of winds in these complex, clumpy galaxies may arise from the lack of organized large-scale gas disks, which may act to collimate outflows \citep{Overzier2008,Overzier2010,Goncalves2010,Wu2019,Law2009,Price2016}. 

\subsubsection{Outflow Velocities}
\label{section:absorption_velocities}

Deriving the physical parameter most directly connected to the data, the outflow velocity, is not straightforward. The  absorption-line profiles are wide, ranging from just redward of the galaxy's systemic velocity, to blue-shifted velocities of $100-1000$\,km s$^{-1}$. There is no {\it single} velocity. See the right two panels of Figure \ref{figure:line_montage} for examples. Estimates of outflow velocities are typically either the line centroid or the ``maximum" velocity. Neither is ideal. The former is affected by absorption produced by the quiescent ISM of the host galaxy and by infilling o f the absorption profile by wind emission via resonance scattering from outflowing gas not directly along the line-of-sight (\S\ref{section:absorption_global_structure}). The ``maximum velocity" is insensitive to these contaminating sources, but depends on both the S/N of the data and the modelling of the continuum. It is also clearly not representative of the ``average" or ``typical" wind velocity, though it may provide clues to the underlying acceleration physics. A reasonable approach is to fit line profiles with a double-Gaussian. One component represents the host galaxy ISM, with its centroid defined by the systemic redshift. The second blue-shifted component represents the outflow, whose centroid defines a characteristic outflow velocity and whose width is a measure of the maximum velocity and column-dependent acceleration profile \citep{Weiner2009,Erb2012,Martin2012,Bordoloi2014,Xu2022, Wang2022}. Scaling relations and comparisons with theory are provided below and in \S\ref{section:confronting_theory_observations}.

\subsubsection{Outflow Rates}
\label{section:absorption_outflow_rates}

Using absorption-line data to estimate outflow rates is challenging, with important limitations. As in emission-line estimates, $\dot{M} \simeq\Omega r^2\rho v \sim \Omega(r^3\rho)(v/r)\sim M_{\rm gas} (v/r)$. However, the strength of the observed absorption-lines depends on the column density of the outflowing gas projected along the line of sight. Idealized examples help to understand the geometric uncertainties. First, consider an expanding shell with column density $N_{\rm H}$ that subtends a solid angle $\Omega$ with respect to the host galaxy. In this case, $\dot{M} = \Omega r N_{\rm H} v m$  where $m$ is the mean particle mass per H. If the shell geometry is correct, then $r$ should be larger than the starburst radius $R$, but the product $\Omega r$ is not further constrained, except by total mass arguments, or by noting that smaller $r$ (larger $N_{\rm H}$) is favored by absorption line observations. Next, consider two cases of continuous mass-conserving outflows. A plausible example has a constant $v$ and $\Omega$ so that the density profile $n(r) = n_0 (R/r)^{2}$, where $N_{\rm H}=\int n(r) dr=n_0R$. In this case, $\dot{M} = \Omega R N_{\rm H} v m$. Contrast this with a shallower density profile $n(r) = n_0 (R/r)\rightarrow N_{\rm H}=n_0 R\ln(r_{\rm max}/R)$ (see \S\ref{section:absorption_global_structure}), where $r_{\rm max}$ is the maximum radius of the outflow. For constant $\Omega$, mass conservation requires $v(r) = v_0 (R/r)$, which yields $\dot{M} = \Omega R N_{\rm H} v_0 \langle m\rangle/\ln(r_{\rm max}/R)$. In all these cases, the values of $\dot{M}$ will be the same to within a factor of two for plausible $r_{\rm max}/R$, but the overall normalization $\Omega r$ (shell) or $\Omega R$ (continuous) are uncertain. 

Determining the total column density $N_{\rm H}$ is also uncertain. First, the strongest, most easily measured absorption lines are usually optically-thick, so that the line depth is only a weak function of the column density. Second, the directly measured values of a given ion's column density $N_{\rm ion}$ pertain only to that specific ion. Converting $N_{\rm ion}$ to $N_{\rm H}$ thus requires uncertain ionization, metallicity, and dust depletion corrections (e.g., \citealt{Murray2007}). 

A more rigorous analysis of the data to derive column densities is possible \citep{Chisholm2015,Chisholm2016,Xu2022}. The first step is to fit a model where only part of the UV continuum source is covered by the absorbing outflow at a given $v$ (the ``covering fraction" $C(v)$). For ions like Si\,II and Si\,IV where two or more lines with different optical depths are present, a simultaneous fit to these lines yields each ion's $C(v)$ and column density distribution $dN/dv$. The ratio of the Si\,II and Si\,IV column densities together with photoionization models yields a total Si column density. Assuming that the outflow metallicity is the same as that measured from optical emission-lines, the total H column $dN_{\rm H}/dv$ can be derived. Integrating over $v$ with a choice for the outflow radius and $\Omega$ yields outflow rates of mass, momentum, and kinetic energy. See \S\ref{section:absorption_scaling_relations} and \S\ref{section:confronting_theory_observations} for discussion.

\begin{figure}
\centerline{\includegraphics[width=6.8cm]{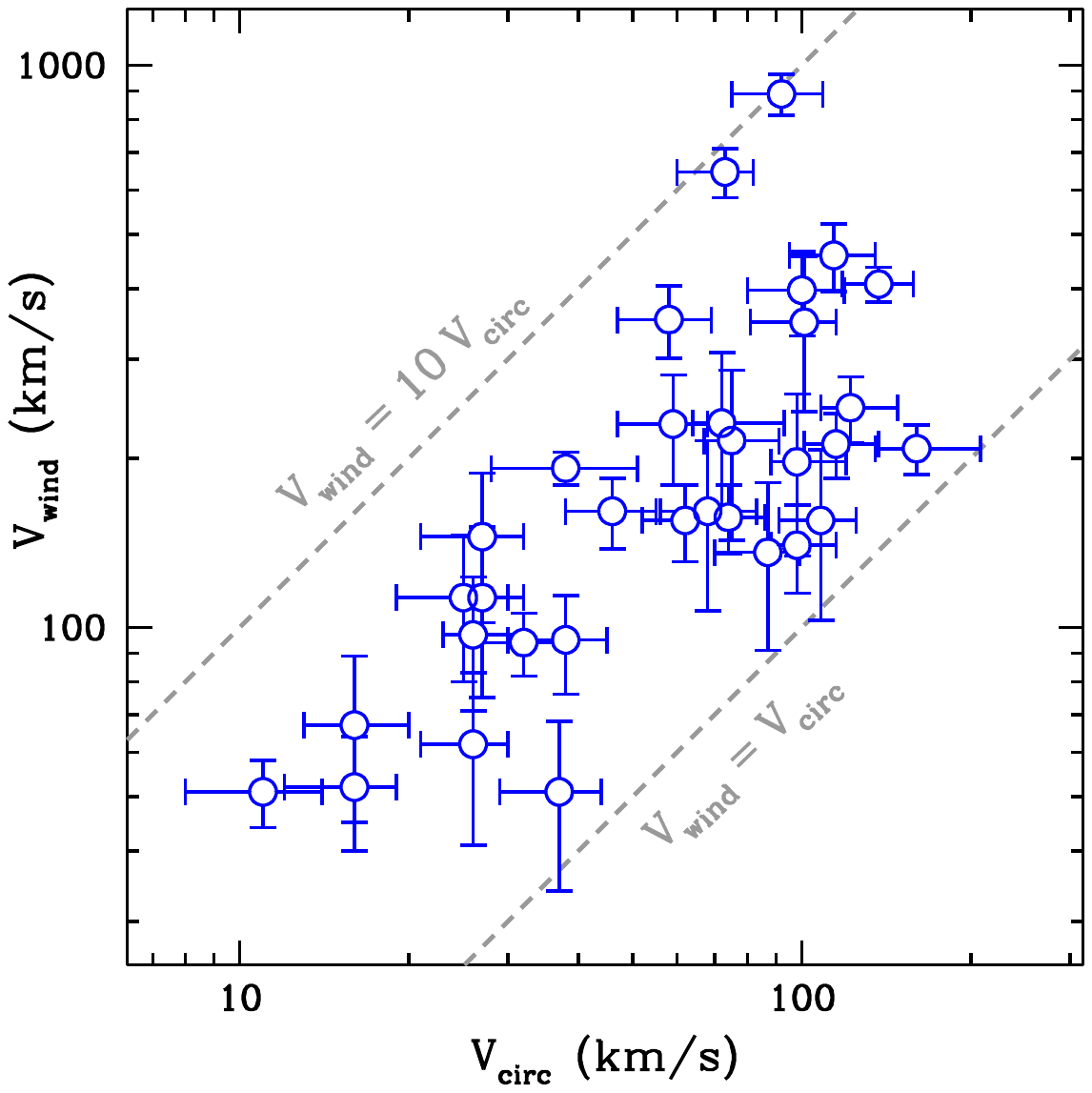}
\hspace*{-6.8cm}
\includegraphics[width=6.8cm]{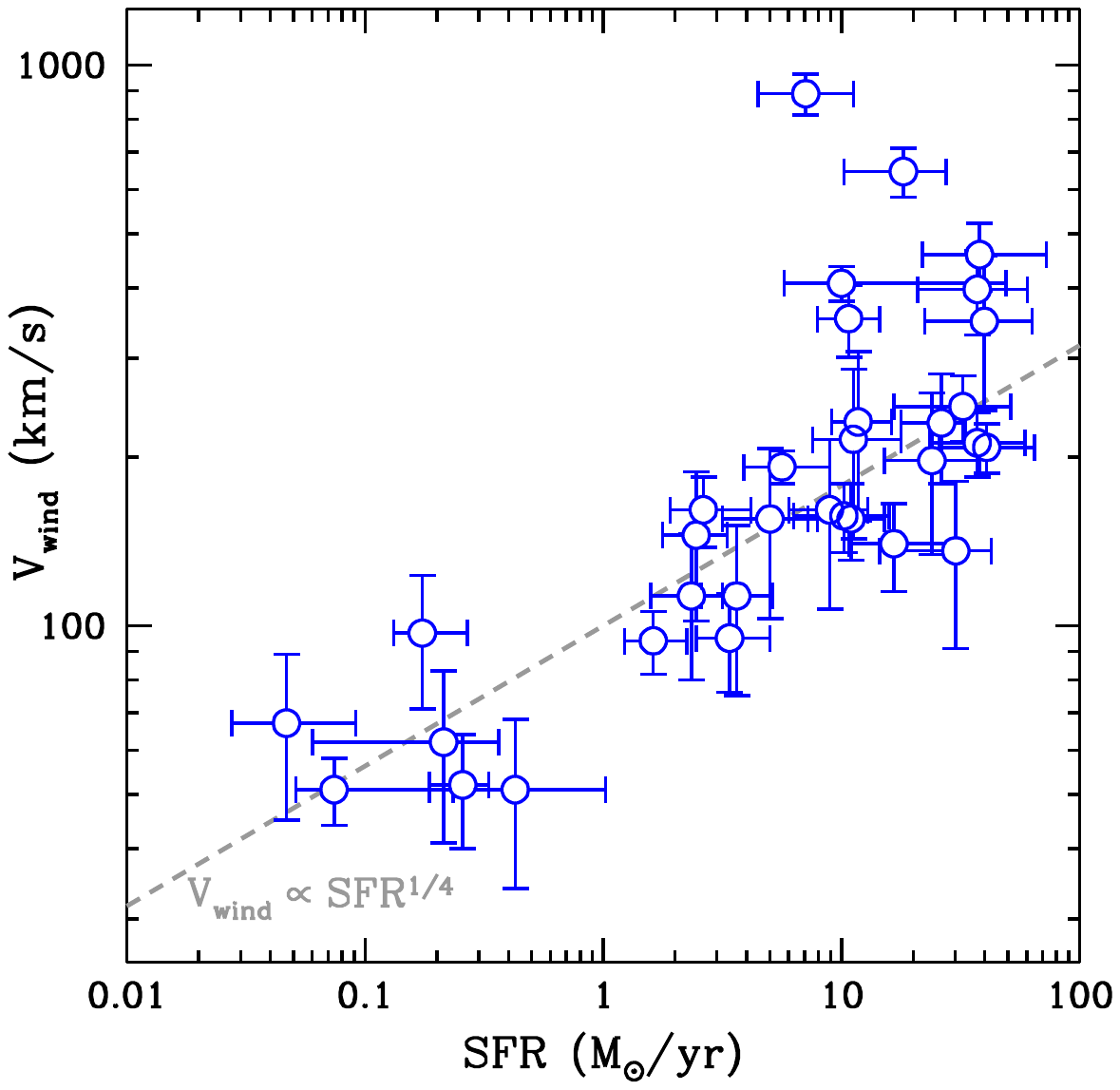}}
\vspace*{-1.0cm}
\caption{The mean blueshift of UV absorption-lines versus the galaxy circular velocity (left) and SFR (right) for outflows in a sample of 52 low-$z$ starbursts \citep{Xu2022}.} 
\label{figure:xu1}
\end{figure}

\subsubsection{Inferred Outflow Properties and Scaling Relations}
\label{section:absorption_scaling_relations}

The influence of galactic winds on the evolution of galaxies is determined by how their bulk properties scale with the properties of their host galaxies (\S\ref{section:introduction}). The ultimate goal is to derive empirical relations between outflow and galaxy properties as a function of redshift, that will allow us to understand their basic physics and their impacts. This work is on-going at low-$z$ and explorations of these scaling relations at higher $z$ has just begun \citep{Sugahara2019,Wang2022}. 

We caution that the physical interpretation of how scaling relations evolve with $z$ is challenging because the easiest-to-measure wind parameter (outflow velocity $V_{\rm out}$) may depend on more than just the easiest-to-measure galaxy properties (SFR \& $M_\star$). For example, at fixed SFR, physically smaller starbursts drive faster outflows, and at fixed $M_\star$,  galaxies with higher SFR also have higher $V_{\rm out}$ (e.g., \citealt{Heckman2015}). Thus, there is no {\it single} scaling relation between $V_{\rm out}$ and {\rm SFR} or $M_\star$. This complicates the interpretation of how $V_{\rm out}$ evolves, since the relationships between {\rm SFR}, $M_\star$, and size all evolve with cosmic time. It is also important to emphasize that the scaling relations at $z \sim0$ are based on rare starbursts, while at higher $z$ they are based on normal SFGs.

Originally, studies of outflow scaling relations at low-$z$ were undertaken using Na\,I absorption \citep{Heckman2000,Rupke2005,Martin2005}, which because of its fragile nature is only abundant in gas that is heavily shielded from UV light by dust \citep{Chen2010}. The analyses performed were therefore mostly of dusty LIRGs and ULIRGs. The fragility of Na\,I and its refractory nature makes it difficult to make the ionization and dust-depletion corrections needed to determine $N_{\rm H}$. As a result, Na\,I proved more robust for measuring outflow kinematics than mass loss rates. Soon after, ground-based surveys began to use Mg\,II (and Fe\,II) to study outflows from SFGs at intermediate $z$ \citep{Weiner2009,Rubin2010,Kornei2012,Erb2012,Martin2012,Wang2022}. These data suffer some of the same limitations as Na\,I D, in that the Mg\,II lines are typically optically-thick, and there are uncertain ionization and dust depletion corrections.

With the launch of FUSE and the deployment of COS on HST, the far-UV window in the low-$z$ universe was opened, making it possible to measure outflow properties using species spanning a wide range of ionization states (e.g., O\,I through O\,VI). This diagnostic-rich window has also been used to study outflows from $z>2$ LBGs \citep{Shapley2003,Steidel2010,Steidel2018,Sugahara2017,Sugahara2019}. Important studies were undertaken at low-$z$ using this far-UV data to revisit the scaling relations deduced earlier from Na\,I D \citep{Grimes2009,Heckman2015,Chisholm2015,Heckman2016,Sugahara2017,Chisholm2016,Chisholm2018}. 

Because it is impossible to discuss all of the disparate observations across tracers, hosts, and $z$, we opt to summarize work by \citet{Xu2022}, who use a sample of 52 low-$z$ starbursts drawn from the union of the COS samples in CLASSY (\citealt{Berg2022}). They use the ``double-Guassian" procedure described in \S\ref{section:absorption_velocities}, and found a consistent value of the mean outflow velocity for each galaxy across both the ionization states and optical depths of the different absorption lines. However, the more optically-thick lines were broader, implying lower covering fractions and lower column density for material far from line center. \cite{Xu2022} adopted a minimum outflow radius of twice the starburst UV half-light radius (\S\ref{section:absorption_global_structure}), consistent with observations of the outflow in M\,82.

Figure \ref{figure:xu1} shows the relationship between the kinematics of the outflow as measured by $V_{\rm out}$, which is defined as the centroid of the blueshifted absorption-line as a function (\S\ref{section:absorption_velocities}) of the host circular velocity $V_{\rm circ}$ and the SFR. The measured relations are similar to those determined in previous studies based on far-UV data (e.g., \citealt{Chisholm2015,Chishlom2016,Heckman2016}), but the CLASSY sample provides broader coverage in the low-mass, low-SFR regime. Although there is significant scatter, $V_{\rm out}$ is broadly correlated with both $V_{\rm circ}$ (see \citealt{Martin2005}) and SFR, providing clues to the underlying physics (\S\ref{section:confronting_theory_observations}).

\begin{figure}
\centerline{\includegraphics[width=6.5cm]{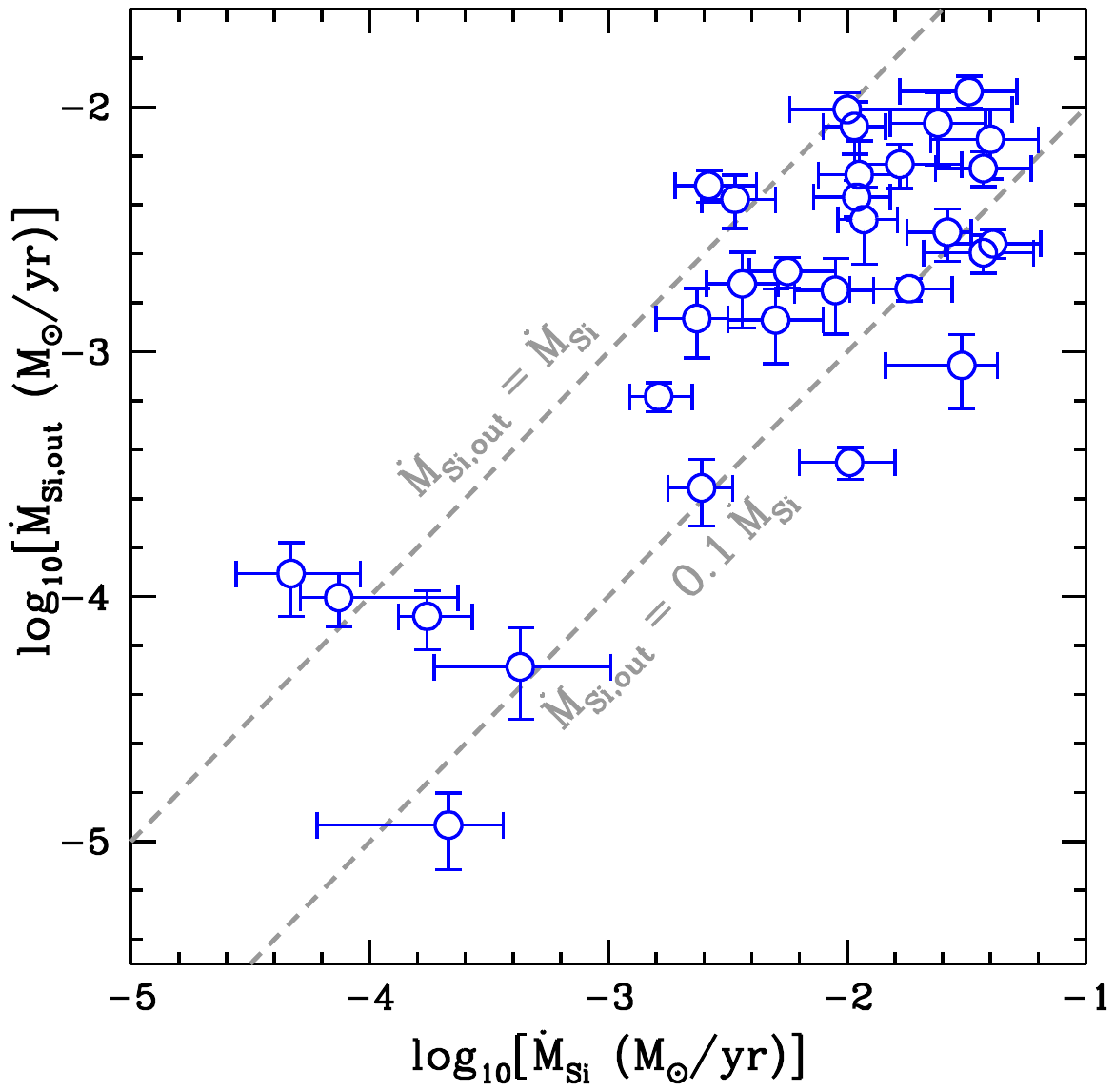}
\hspace*{-7cm}
\includegraphics[width=6.5cm]{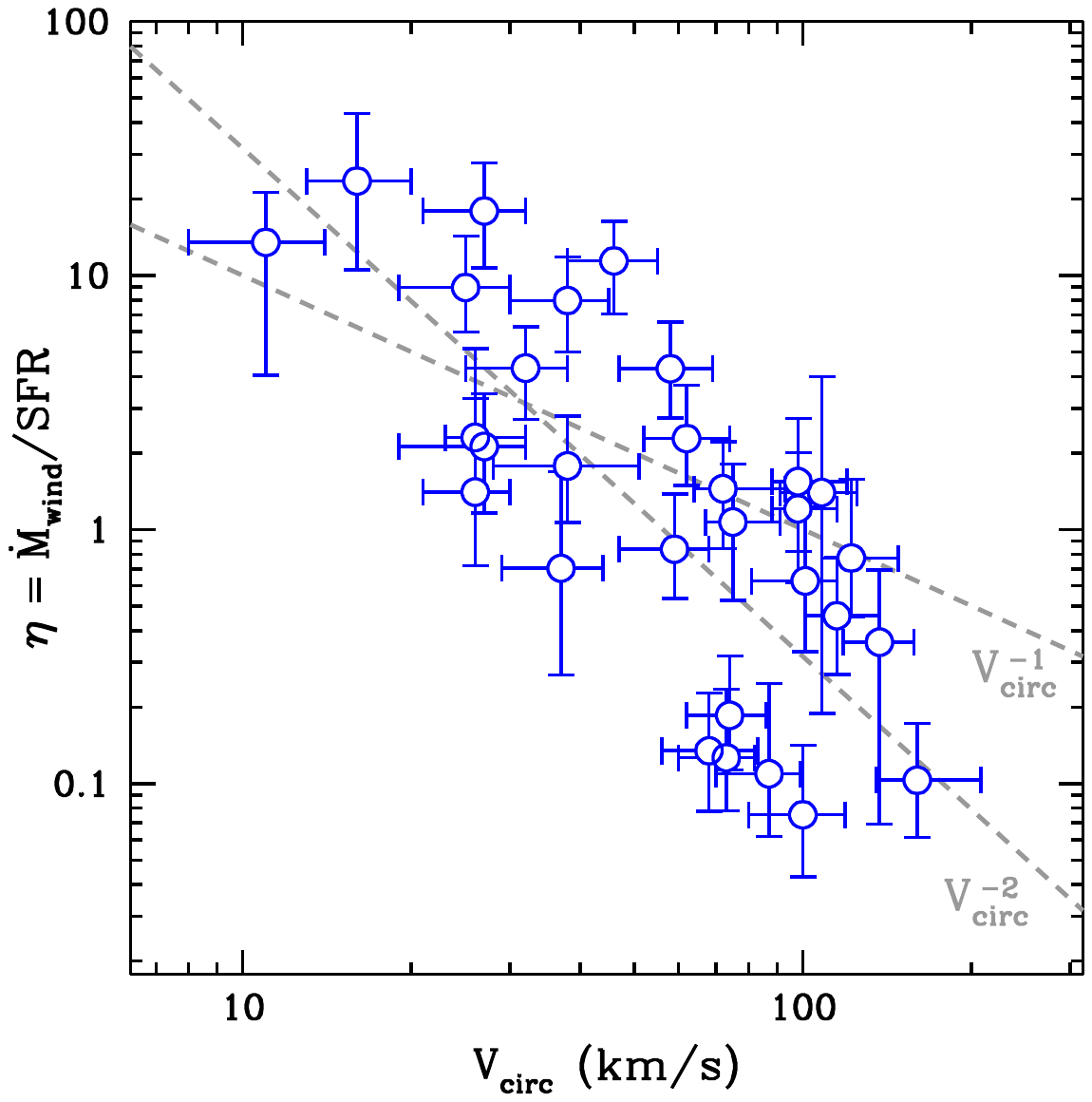}}
\vspace*{-1cm}
\caption{
Outflow rates of warm ionized gas in absorption for low-$z$ starbursts (\S\ref{section:absorption}; \citealt{Xu2022}). {\it Left:} Outflow rate of Si vs.\ the production rate of Si by SNe, based on the SFR. For many systems, much of the Si must reside in an unobserved phase. {\it Right:} $\eta$ vs.\ galaxy circular velocity. Energy-like ($\eta \propto V_{\rm circ}^{-2}$) and momentum-like ($\eta\propto V_{\rm circ}^{-1}$) scalings are added for reference. See discussion \S\ref{section:absorption_scaling_relations}.}
\label{figure:xu2}
\end{figure}

Figure \ref{figure:xu2} shows key relationships involving mass outflow rates. The left panel shows that the mass loss rate of Si in the warm outflow $\dot{M}_{\rm Si,\,out}$ is in many cases less than the rate Si is being created/injected by SNe (by a median factor of $\sim 5$). Since some of the observed Si in the outflow will be from ISM clouds entrained by the wind \citep{Chisholm2018}, the great majority of the newly-created Si from SNe must reside in a different phase of the outflow (plausibly the hot and very hot phases), and so is invisible in the Si\,II, III, IV ions probed in the UV. The right panel shows that $\eta=\dot{M}_{\rm wind}/{\rm SFR}$ has a steep inverse $V_{\rm circ}^{-1.6}$ dependence, intermediate between  the so-called ``momentum-driven" and ``energy-driven" models. In the lowest mass galaxies, $\eta\sim 10-30$, underscoring the powerful impact of winds in galaxy evolution (see \S\ref{section:introduction}). In \S\ref{section:confronting_theory_observations}, we combine Figures \ref{figure:xu1} and \ref{figure:xu2} to discuss the momentum and kinetic energy content of these outflows.

These relations show scatter. While there may be physical secondary correlations, absorption-line data only probe a single line-of-sight, through which the outflow properties may differ from those measured in a global angle-averaged sense \citep{Gazagnes2023}. 

In the spirit of the caveats highlighted at the start of this subsection, note that this sample consists entirely of strong starbursts, with a limited range in $\dot{\Sigma}_\star$ and SFR/M$_\star$ that is $\sim2$ dex higher than normal $z\sim0$ systems. They therefore resemble typical SFGs at $z \sim2.5$ \citep{Berg2022}. Previous results for samples with greater dynamical range in these ratios show that $V_{\rm out}$ correlates more strongly with SFR than with $M_\star$ and that $V_{\rm out}$ correlates with $\dot{\Sigma}_\star$. The importance of $\dot{\Sigma}_\star$ for $V_{\rm out}$ has been confirmed for a large sample of low- and intermediate-$z$ starbursts spanning a range of about six dex in $\dot{\Sigma}_\star$ \citep{Davis2023}. Even more extreme outflows are observed in a rare population of massive galaxies at intermediate $z$, characterized by exceptionally high outflow velocities ($\gtrsim 10^3$ km s$^{-1}$) originating in massive (few $\times 10^{10}$ M$_{\odot}$) and ultra-compact (radii of a few $10^2$ pc) post-starbursts \cite{Tremonti2007,Diamond-Stanic2021,Perrotta2023,Davis2023}. See Figure \ref{figure:line_montage}. The outflows appear consistent with an extension of the trend in $V_{\rm out}-\dot{\Sigma}_\star$ \citep{Heckman2016}, but strain theoretical interpretation (\S\ref{section:confronting_theory_observations}).

\subsection{Observations of the Impact of Winds on the Circumgalactic Medium}
\label{section:observations_cgm}

Over cosmic time, the fraction of accreted baryons turned into stars and molecular gas has been roughly constant at 4\% \citep{Madau2014,Behroozi2019,Peroux2020}. Galactic winds may strongly affect this conversion efficiency through ``preventive" feedback, by delaying or prolonging accretion into the central galaxy and enriching the accreted material (\S\ref{section:writ_large}; \citealt{Tumlinson2017,Faucher-Giguere2023}). It is physically plausible that the coupling of feedback from winds is more effective in the CGM than in the disk ISM. The latter has high volume densities, which implies rapid radiative cooling, high column densities, which are difficult to accelerate, and a geometry where winds can blow out along the minor axis without coupling to the bulk of the ISM. In contrast, the CGM has much lower densities and a more spherical geometry.

Evidence that winds impact the CGM in strong SFGs at all $z$ comes from the UV absorption-lines that the outflowing gas imprints on the spectra of background QSOs or galaxies. The largest body of such data is for galaxies at $z \sim 1$. \citet{Lan2018} created stacked spectra of QSOs located behind the CGM of 70,000 star-forming Emission-Line Galaxies (ELGs) at a median $z$ of 0.85 from the SDSS eBOSS survey and 360,000 quiescent Luminous Red Galaxies (LRGs) at a median $z$ of 0.55 from the SDSS BOSS survey. They found that the Mg\,II absorption lines were significantly stronger in the ELGs out to impact parameters of $\sim80$\,kpc, and that the lines were stronger along the minor axis than along the major axis. They interpreted this as reflecting the impact of galactic winds on the inner CGM. At $z \sim 1.3$, \citet{Schroetter2019} used VLT MUSE data to study Mg\,II absorption in the CGM of SFGs. They found excess absorption along both the major and minor axes, and modeled this as inflow and outflow, respectively. The average outflow mass-loading factor were of order unity for $M_\star$ galaxies ($M_{\rm halo} \sim 10^{12}$\,M$_{\odot}$).

At $z \sim 2-3$, \cite{Steidel2010} stacked spectra of background galaxies located behind the foreground CGM of LBGs to trace absorption-lines of C\,IV, C\,II, Si\,IV, and Si\,II. The line strengths decline smoothly with increasing impact parameter with a sharp cut-off at $\sim100$\,kpc, which may represent the outer edge of the ``zone-of-influence" of the galactic winds in the CGM. At low-$z$, \citet{Heckman2017} compared HST COS absorption-line spectra of QSOs located behind the CGM of starbursts to the CGM of a control sample of typical SFGs. The column densities of the metal absorption-lines (Si\,III and C\,IV) and the equivalent widths of the Ly$\alpha$ lines were significantly larger in the outer CGM ($\sim R_{\rm Vir}$) in the starbursts, and the velocity dispersions were roughly twice as large as in the control sample and roughly twice the expectation for virialized gas. They argued that an expanding bubble driven into the CGM by the starburst could account for these properties (see  \S\ref{section:confronting_theory_observations}). 

\subsection{The Demographics of Winds Over Cosmic Time}
\label{section:demographics_cosmic_time}

In the contemporary universe, evidence for winds is clear only in starbursts. For example, \citet{Chen2010} used the SDSS\,I survey to study a complete sample of $\sim140,000$ SFGs at $z \leq 0.2$. They find that outflows traced by Na\,I D in absorption are strongest in galaxies viewed at angles $<60^{o}$ with respect to the minor axis, and that the incidence of detected outflows increases from near zero for galaxies with $\dot{\Sigma}_\star\simeq10^{-1.5}$\,M$_{\odot}$\,yr$^{-1}$\,kpc$^{-2}$ to $\sim60$\% for galaxies with  $\dot{\Sigma}_\star\simeq10^{-0.5}$\,M$_{\odot}$\,yr$^{-1}$\,kpc$^{-2}$. The latter represents just the top few\% of systems in the sample and is similar to the lower limit for starbursts \citep{Kennicutt2012}. Similar incidence and thresholds for wind driving,  are found in studies of optical emission-line gas \citep{Lehnert1996}, soft X-ray emission \citep{Strickland2004}, and UV absorption lines (\citealt{Heckman2015}). Today, only a few\% of SFGs -- those with the highest $\dot{\Sigma}_\star$ and/or ${\rm SFR/M_\star}$ -- have strong outflows.

At higher redshift, the situation changes. At $z \sim1$, several works probe outflows using the near-UV Mg\,II doublet and/or Fe\,II multiplets in absorption (see \S\ref{section:absorption_global_structure}). These samples do not all cover the same ranges in $z$, SFR, and $M_\star$, but the incidence of outflows found range from 20\% to near 90\%, much higher than at $z \sim 0$. In particular, \cite{Weiner2009} find ubiquitous outflows in $z\sim1$ galaxies with typical columns of $N_{\rm H}\simeq10^{20}$\,cm$^{-2}$, velocities from $V_{\rm out}\sim300-500$\,km/s up to $1000$\,km/s, and with $V_{\rm out}\propto{\rm SFR}^{0.3}$, similar to Figures \ref{figure:xu1} \& \ref{figure:xu2}. At higher redshifts ($z > 2$) outflows traced using far-UV absorption-lines like the Si\,II 1260 and C\,II 1334 line and/or Ly$\alpha$ emission-lines are seen in over 90\% of LBGs \citep{Shapley2003,Steidel2010,Shapley2011}.

The evolution from rare outflows at $z\sim0$ to ubiquitous outflows at high-$z$ may have a simple physical basis. As discussed in \S\ref{section:theory}, the critical condition for gas expulsion can be written as the ratio between the outward force of the prime mover to the inward force of gravity. For stellar feedback, the force $\dot{p}\propto{\rm SFR}\propto L$, as in the case ram-pressure accelerated clouds in a hot wind (eq.~\ref{ram_pressure_edd}) or by radiation pressure (eq.~\ref{eq:ledd_s}) (see Table \ref{table:theory}). For a cloud of mass $M_{\rm cl}$ and area $A_{\rm cl}$ 
\beq
F_{\rm out}/F_{\rm grav} \simeq \dot{p}\,(A_{\rm cl}/\Omega r^2)/(GM_{\rm tot}M_{\rm cl}/r^2)\propto ({\rm SFR}/M_\star)N_{\rm H}.
\label{eq:fout_fin}
\eeq
Between $z=0$ and $z=3$, the main sequence shifts to higher SFR/$M_\star \propto (1 + z)^{2}$ \citep{Forster2020}. We therefore expect $F_{\rm out}/F_{\rm grav}\propto (1+z)^2 N_{\rm H}$ to increase by an order-of-magnitude or more between now and cosmic noon at fixed $M_\star$ and $N_{\rm H}$ for typical main sequence galaxies. 

\section{Confronting Theory and Observations}
\label{section:confronting_theory_observations}

The wind theories surveyed in \S\ref{section:theory} make predictions for their mass, energy, and momentum content, their acceleration profiles, and for various ancillary supporting observations. However, different mechanisms are at different stages of theoretical development and observational scrutiny. For some theories we can ask sharp questions, while for others we cannot. Here, we thus provide a necessarily incomplete discussion, that first focuses on assessing the hot wind as the prime mover (\S\ref{section:observations_constraints_windfluid}), and then evidence for and against other mechanisms in light of the wind scaling relations (\S\ref{section:compare_warm_ionized_theory}).

\subsection{The Very Hot Phase as Prime Mover}
\label{section:observations_constraints_windfluid}

The thermalized ejecta of massive stars -- the ``very hot" phase (see \S\ref{section:xray_emission}) -- may be the prime mover of the  cooler phases: high $\dot{\Sigma}_\star$ may allow for the efficient conversion of the kinetic energy supplied by SNe and massive star winds into the thermal energy of a very hot fluid, with the subsequent conversion of this thermal energy into the bulk kinetic energy of a very hot outflow, whose energy and momentum is shared with the cooler phases via mixing and ram pressure acceleration (\S\ref{section:thermal}).  

\begin{figure}
\captionsetup[subfigure]{labelformat=empty}
	\begin{subfigure}[t]{0.33\linewidth}
		\centering
		\includegraphics[width=\textwidth]{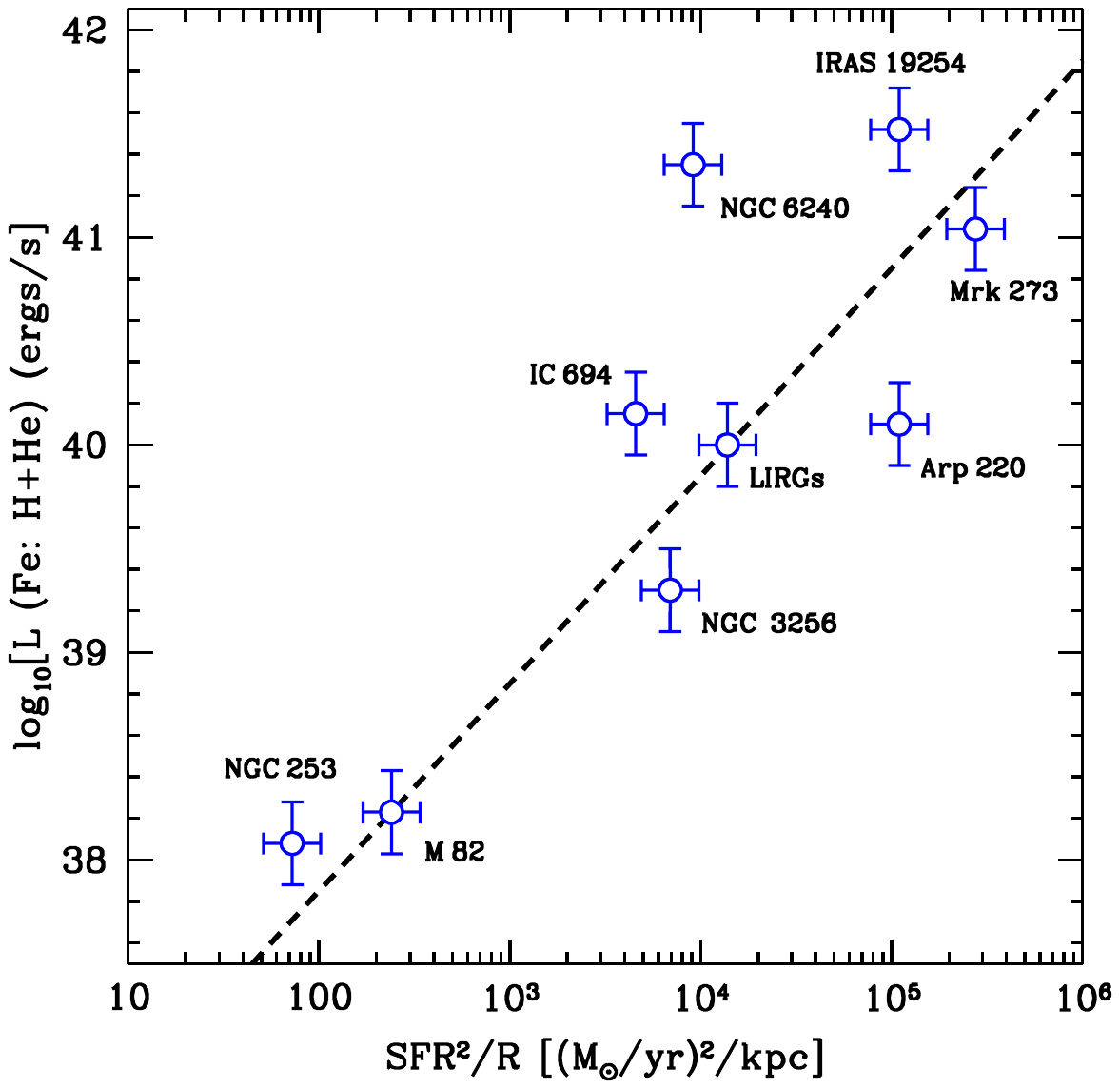}
		\caption{}
	\end{subfigure}\hfill
	\begin{subfigure}[t]{0.33\linewidth}
		\centering
		\includegraphics[width=\textwidth]{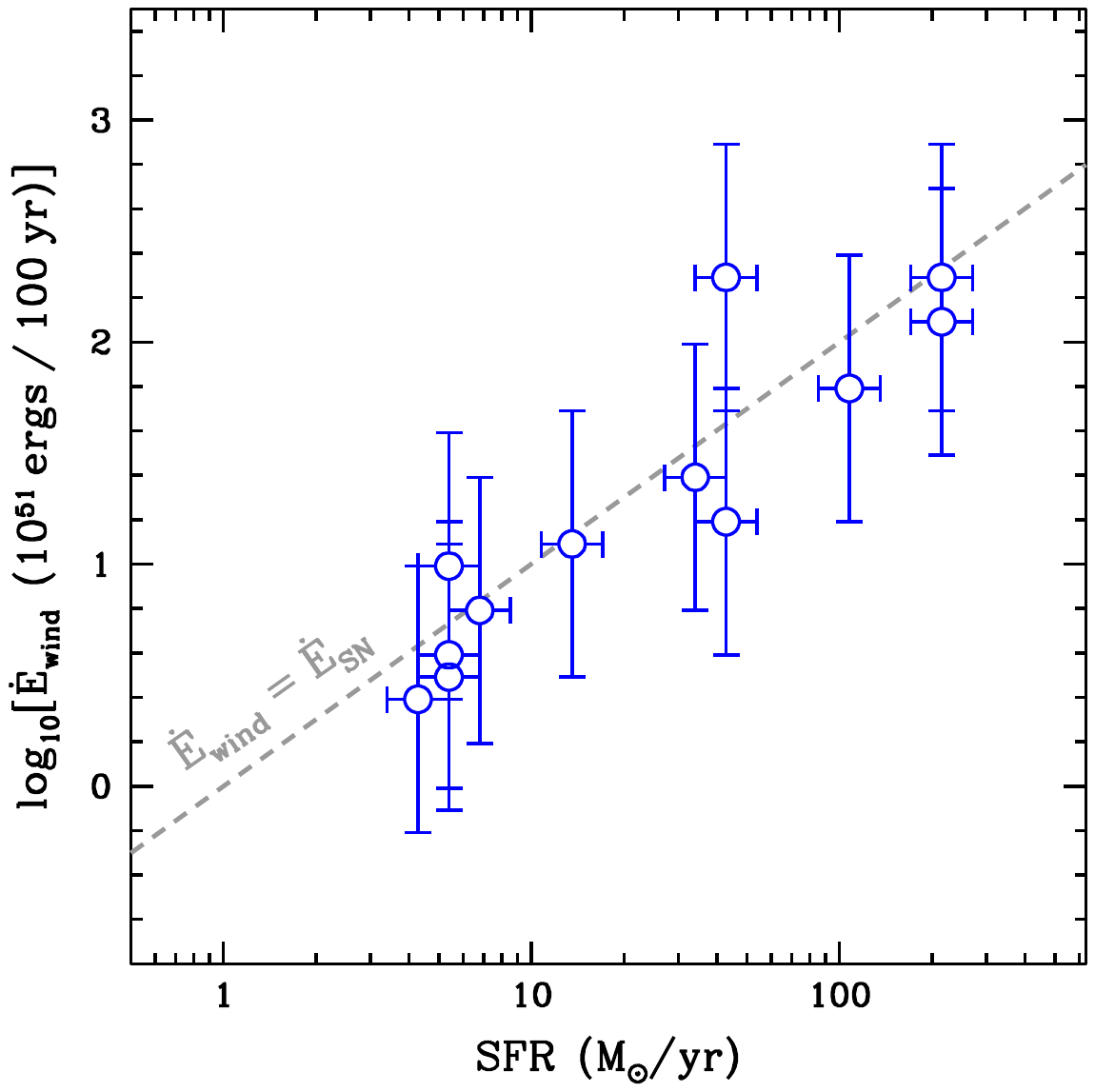}
		\caption{}
	\end{subfigure}
		\begin{subfigure}[t]{0.33\linewidth}
		\centering
		\includegraphics[width=\textwidth]{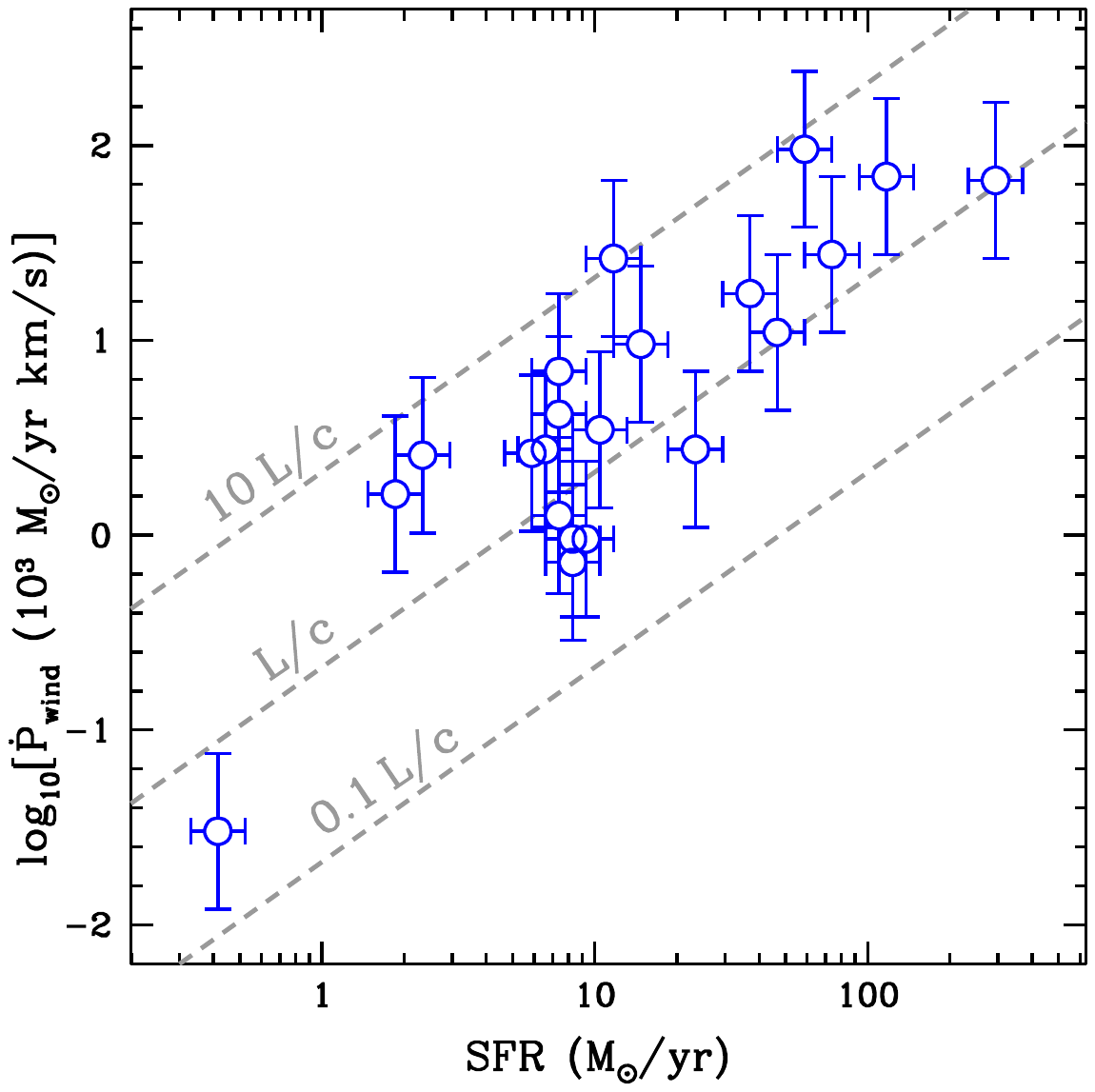}
		\caption{}
	\end{subfigure}
	\vspace*{-1.5cm} 
\caption{{\it Left:} Sum of He-like and H-like Fe\,K$\alpha$ emission-line luminosities versus ${\rm SFR}^2/R$ for 8 starbursts \citep{Balestra2005,Iwasawa2005,Strickland2007,Jia2012,Mitsuishi2011,Pereira-Santaella2011,Wang2014} and a stack of 28 LIRGs \citep{Iwasawa2009}. The line shown has a slope of unity, with the normalization set by M\,82 (see \S\ref{section:observations_constraints_windfluid} \& eqs.~\ref{cc85sonic}). {\it Center:} The volume integral of the radial pressure profiles measured in the warm ionized gas divided by the outflow time $(r/v)$ ($\dot{E}_{\rm wind}$; see Table 6 in \citealt{Heckman1990}) versus SFR. The line shows $\dot{E}_{\rm wind}=\dot{E}_{\rm SN}$ (eq.~\ref{eq:alpha_eta}). {\it Right:} Wind force $\dot{p}_{\rm wind}$ assuming that the thermal pressure in the emission-line gas measures the hot wind ram pressure for an outflow into $2\pi$\,str (based on data in \citealt{Lehnert1996}). The lines shown compare the derived force in the wind to $L/c$. See discussion \S\ref{section:observations_constraints_windfluid} and compare with Fig.~\ref{figure:xu3}, which shows $\dot{p}$ for absorption line measurements.}
\label{figure:heckman_various}
\end{figure}

Is this the answer? We note 3 checks on this picture. First, we have direct observational evidence for the very hot phase in M82 and other starbursts from X-ray Fe line emission, which constrains its total thermal energy content and thus its ability to do work on cool gas. Since the Fe K$\alpha$ line is collisionally excited, its luminosity is $L_{\rm Fe}\propto n^2 R^3$, where $n$ is the density and $R$ is the starburst radius. Mass conservation in a thermal wind model dictates that $n\propto{\rm SFR}/R^2$ in the starburst core (eq.~\ref{cc85sonic}), implying that $L_{\rm Fe}\propto{\rm SFR}^2/R$. In the left panel of Figure \ref{figure:heckman_various} we compare this scaling relation with the extant starburst X-ray Fe line data. The dashed line shows the expectation with normalization set by M82 with $\alpha=1$ and $\eta =0.2$ (see \ref{section:xray_emission}), and overall good correspondence.  A second check  is provided by the radial pressure profiles measured from the warm ionized gas in emission from $\sim1-{\rm few}$\,kpc using direct measurements of the electron density and temperature from line emission  (see \S\ref{section:optical_uv_emission}). Assuming that the warm ionized gas is pressure-confined by a volume-filling very hot wind fluid (e.g., via ram pressure or thermal pressure), its energy content will then be the volume integral of the warm ionized gas pressure. Dividing by the dynamical expansion time $(r/v)$ of the warm ionized gas gives the kinetic power $\dot{E}_{\rm wind}$ of the putative driving wind. The middle panel of Figure \ref{figure:heckman_various} shows that $\dot{E}_{\rm wind}\sim\dot{E}_{\rm SN}$, the power provided by star formation    (eq.~\ref{eq:alpha_eta}; \citealt{Heckman1990}), consistent with $\alpha\sim1$ derived from the M\,82 X-ray data, and implying that bulk radiative losses in the hot wind fluid are not dynamically important (\S\ref{section:additional_physics_hot_phase}). As a final check, assuming that the pressure in the warm outflowing gas is set by the hot wind ram pressure ($P_{\rm ram}\sim\rho_{\rm wind} v_{\rm wind}^2$), we can estimate the force of the wind at $r$ via $\dot{p} \sim P_{\rm ram} (\Omega r^2) $, where $\Omega$ is the solid angle occupied by the hot wind. The right panel of Figure \ref{figure:heckman_various} shows $\dot{p}$ versus the SFR for $\Omega \sim \pi$ \citep{Heckman1990,Lehnert1996}. The force of different prime movers relative to $L/c$ (gray lines) is summarized in Table \ref{table:theory}. Turning this argument around, the agreement between the measured radial pressure profiles and ram pressure profile created by the momentum injection argues that the clouds are in fact confined by the wind ram pressure, rather than its thermal pressure, as expected for a rapidly expanding supersonic flow. 

With these checks passed, the remaining missing pieces are (1) the very hot gas dynamics, which may eventually be probed by X-ray spectroscopy, (2) a more complete theory for the mixing of energy and momentum between the fast wind and the cooler observed phases (\S\ref{section:entrainment}, Fig.~\ref{figure:theory}), (3) a consistent explanation for the soft X-ray emission, and (4) models that can produce the cloud structures observed in emission and absorption. The primary challenges to the picture of the very hot phase as prime mover is that the absorption line observations favor high $(\alpha \eta)$ compared to inferences from the hard X-ray (\S\ref{section:compare_warm_ionized_theory}), and that the  cloud morphology seen in Figure~\ref{figure:m82_HSTzoom} notably lacks the cometary features seen in some thermally-driven simulations (e.g., \citealt{Cooper2008}).

\subsection{Scaling Relations for the Warm Ionized Gas and Theory Comparisons}
\label{section:compare_warm_ionized_theory}

\begin{figure}
\centerline{\includegraphics[width=6.cm]{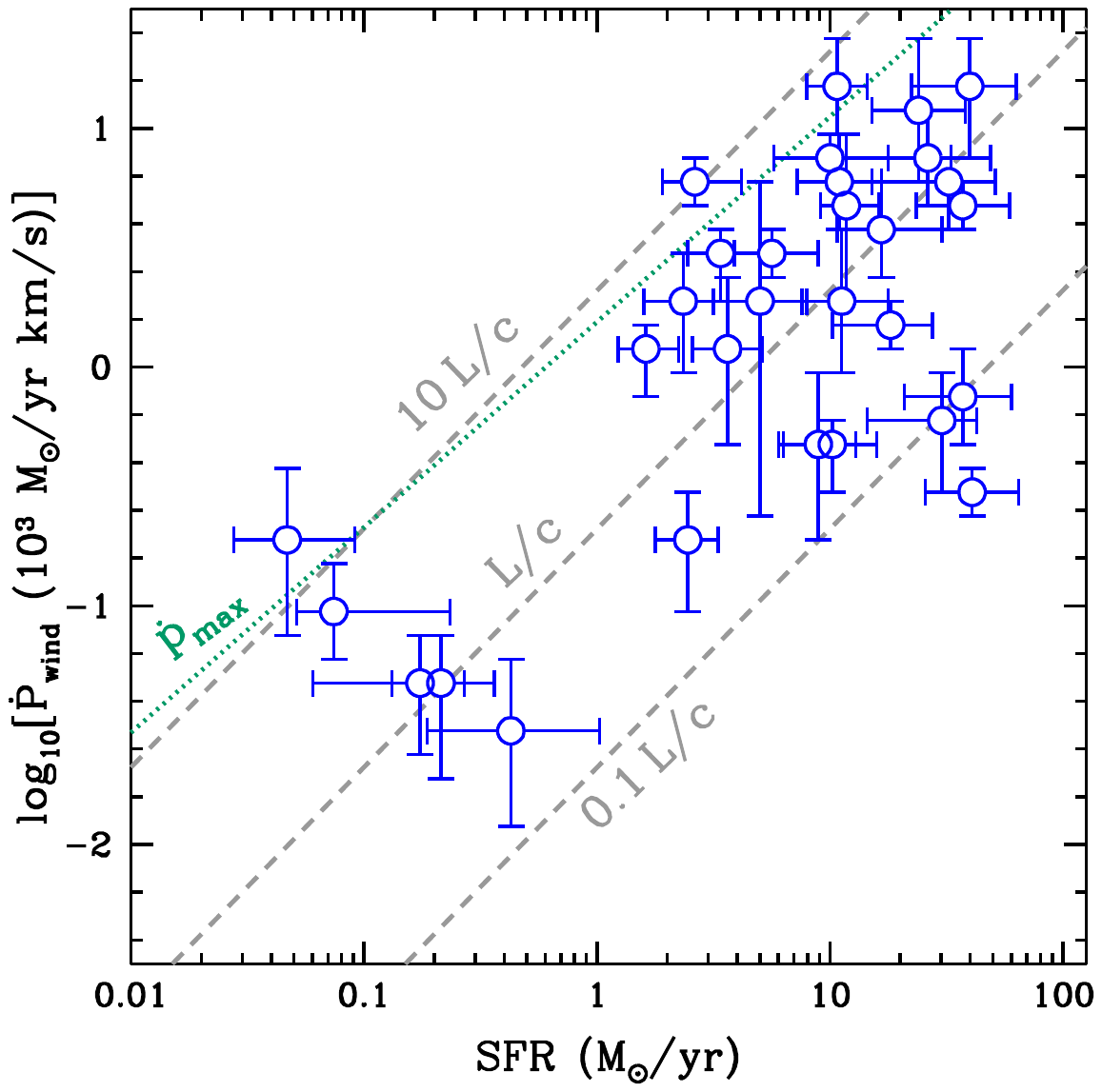}
\hspace*{-7cm}
\includegraphics[width=6.cm]{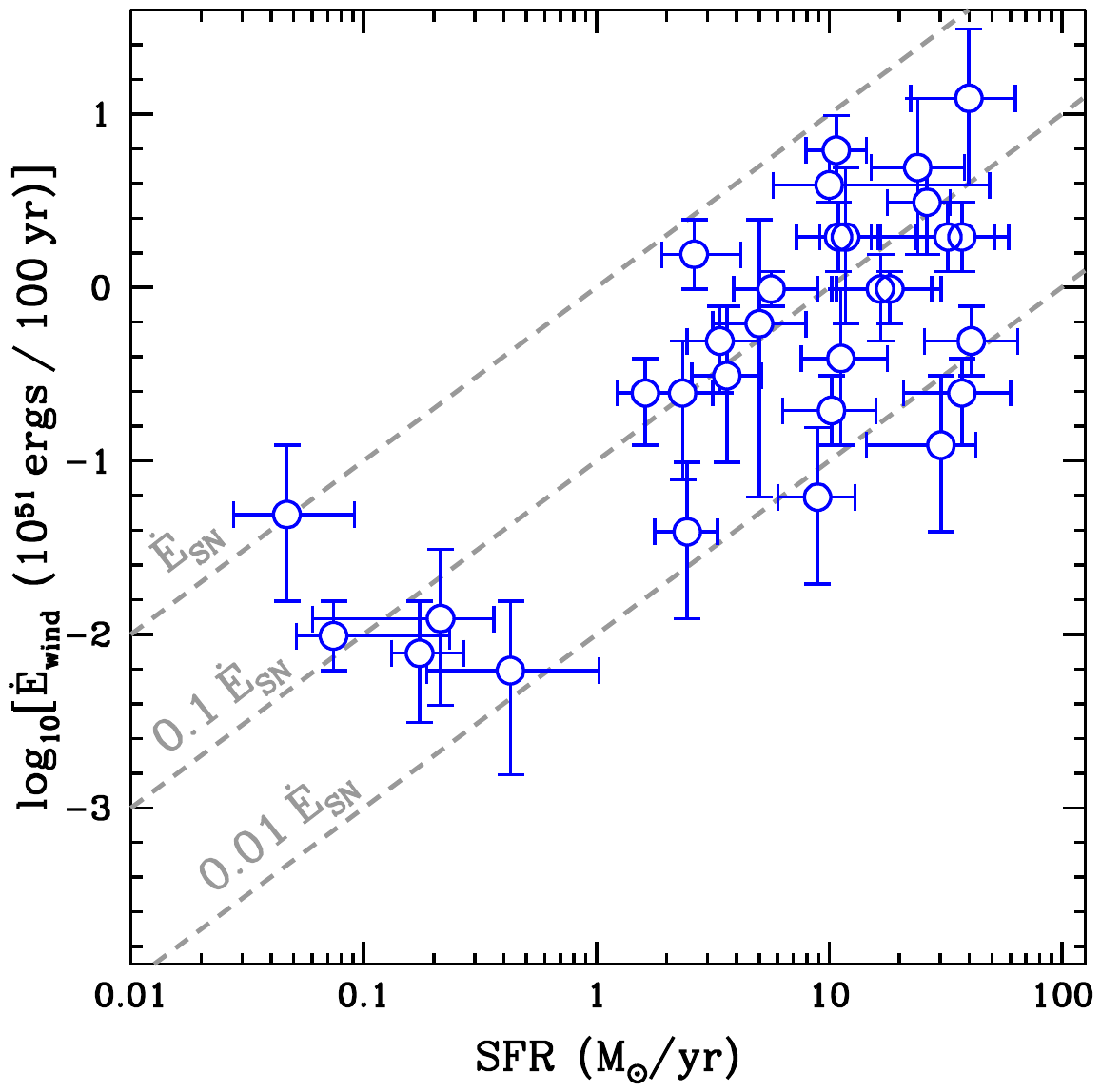}}
\vspace*{-0.8cm}
\caption{The momentum rate (left) and kinetic power (right) carried by the warm outflow as a function of the SFR \citep{Xu2022}. Although there is significant scatter, the warm gas carries a significant share of the available momentum, as measured by $L/c$ (see Table \ref{table:theory}), and a smaller share of the available kinetic energy (eq.~\ref{eq:alpha_eta}). In the left panel, $\dot{p}_{\rm max}$ for the hot gas is also shown (eq.~\ref{eq:hot_pmax} assuming $\alpha=1$ and $R=300$\,pc; green dotted). In the right panel, note that the maximum power available in cosmic rays is of order $\sim0.1\dot{E}_{\rm SN}$ (eq.~\ref{eq:edotcr}).}
\label{figure:xu3}
\end{figure}

Figure \ref{figure:xu3} shows the force and kinetic power carried by the warm ionized outflow, as measured with absorption lines in the CLASSY data set (Fig.~\ref{figure:line_montage}), as described in \S\ref{section:absorption} \citep{Xu2022}. For a hot wind, this flux is given by equation (\ref{cc85_vinf_pdot}) (see Table \ref{table:theory}). For a constant SFR on $0.1-1$\,Gyr timescales, the corresponding momentum flux associated with radiation pressure is $L/c$ (eq.~\ref{eq:lum_continuous}) in the single-scattering limit. In the left panel of Figure \ref{figure:xu3}, we see that the measured momentum fluxes in the warm ionized gas range from $\sim0.1-10(L/c)$ and are generally similar to the predicted momentum fluxes in the hot wind, radiation pressure, and cosmic rays (Table \ref{table:theory}). Thus, to zeroth order, the observed outflows carry an amount of momentum consistent with the amount supplied. This implies efficient transfer of momentum to the outflowing absorbing gas, especially as evidenced by the systems reaching $\sim10(L/c)$. Interpreted as entirely driven by the ram pressure of the very hot flow, the implication from equation (\ref{cc85_vinf_pdot}) would be that $(\alpha\eta)^{1/2}\sim2$ for such systems. In contrast, the X-ray analysis of M\,82 in \S\ref{section:xray_emission} indicates $(\alpha\eta)^{1/2}\sim 0.3-0.8$ ($\alpha\sim0.5-1$ and $\eta\sim0.2-0.6$). Such a large force from the hot flow could be accommodated by mass-loading into the hot phase as the flow expands \citep{Suchkov1996,Nguyen2021}. An additional point of reference is useful for the judging the momentum content of the very hot phase. Following \cite{Wunsch2008}, \cite{Lochhaas2021} showed that for large enough $\eta$, the interior of the very hot wind driving region becomes radiative, and this limits the total force of the hot outflow at the host galaxy's radius $R$ to a maximum of (their eq.~19)
\begin{equation}
\dot{p}_{\rm max} \simeq 7 \times 10^{34} R_{0.3}^{0.14} (\alpha\, {\rm SFR_{10}})^{0.86}\,\,{\rm dynes}\simeq 6\,(L/c) R_{0.3}^{0.14} (\alpha\, {\rm SFR_{10}})^{0.86},
\label{eq:hot_pmax}
\end{equation}
which is shown in the left panel of Figure \ref{figure:xu3} for $\alpha=1$. It patrols the upper envelope of the observations and is thus consistent with being a maximum for the very hot phase.

\begin{figure}
\centerline{\includegraphics[width=4.5cm]{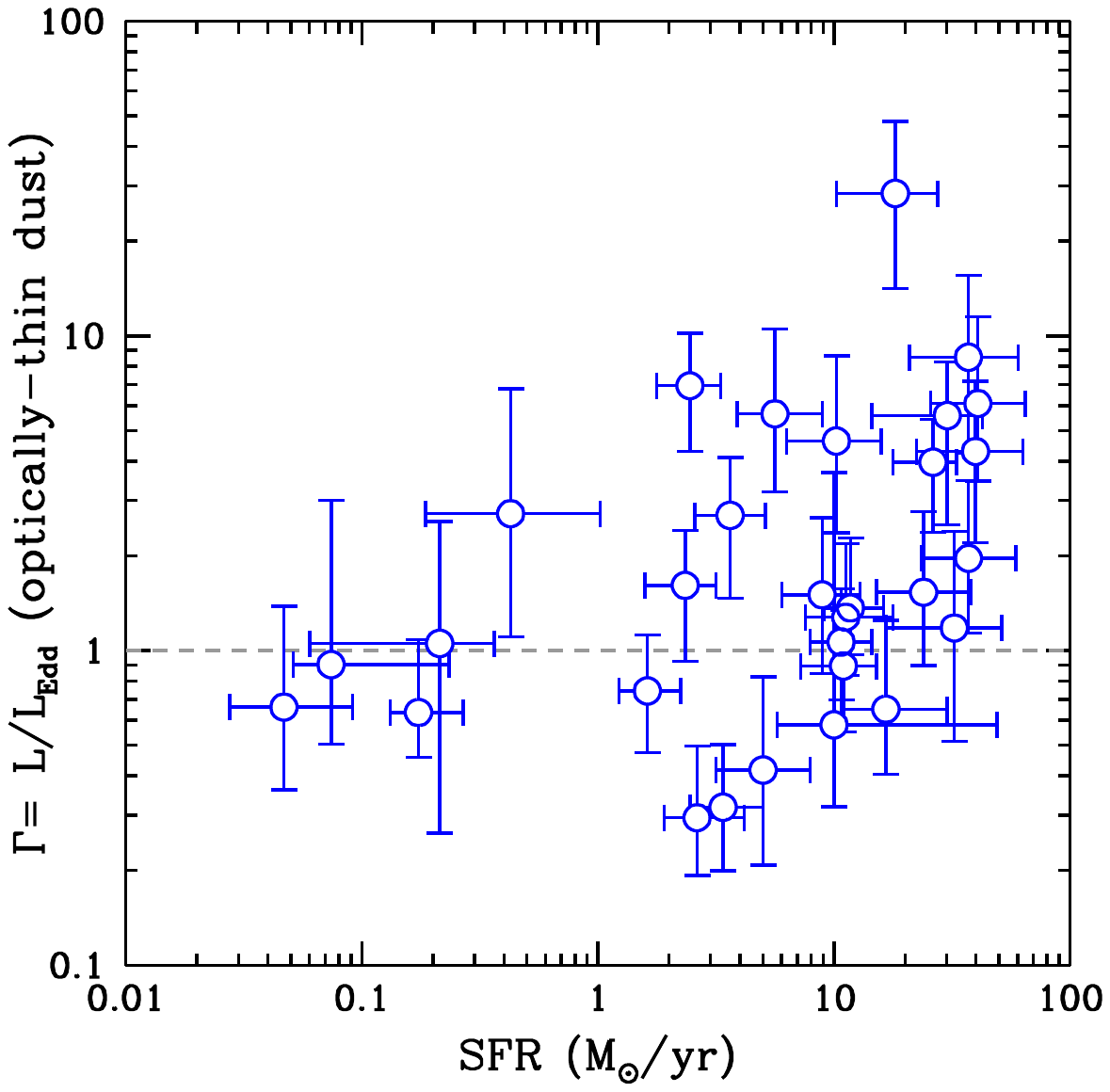}
\hspace*{-9cm}
\includegraphics[width=4.5cm]{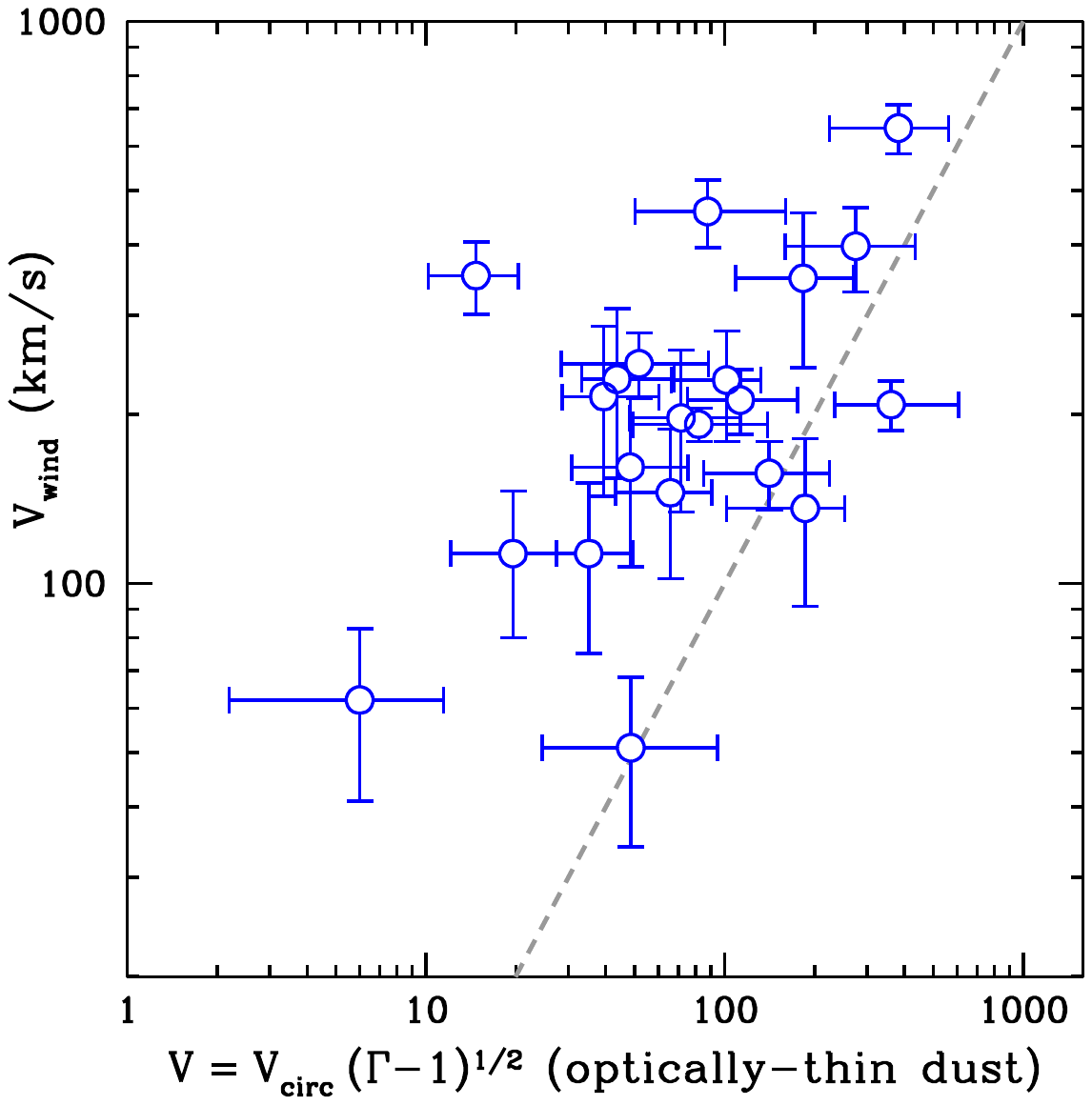}
\hspace*{-9cm}
\includegraphics[width=4.5cm]{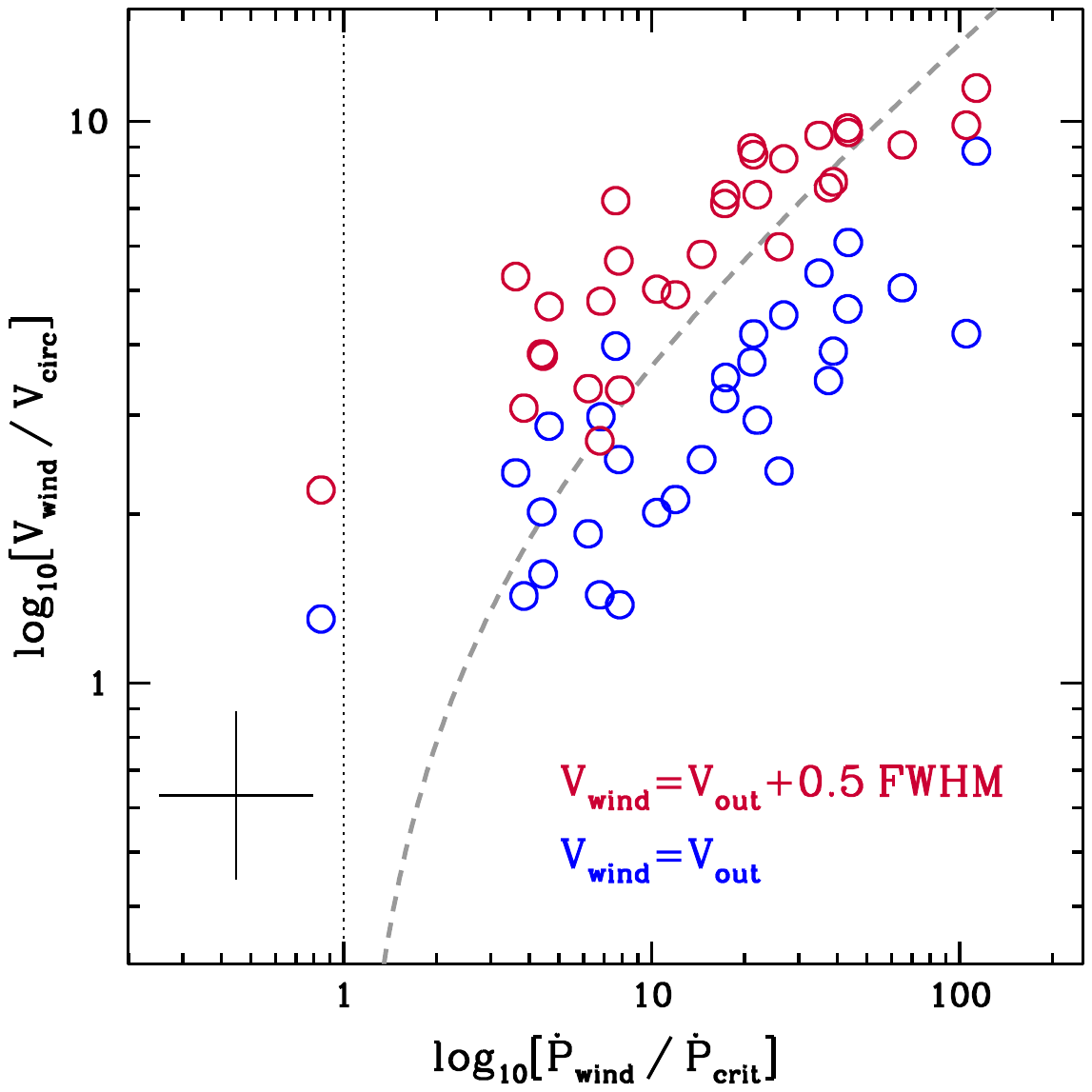}}
\vspace*{-0.5cm}
\caption{The Eddington ratio ($\Gamma$) versus SFR (left) and the measured outflow velocity $V_{\rm wind}$ versus the predicted outflow velocity $V_{\rm circ}(\Gamma-1)^{1/2}$ (middle; gray dashed line is one-to-one) for systems with $\Gamma >1$ under the assumption of optically-thin dust (see \S\ref{section:compare_warm_ionized_theory}).  The predicted velocities are somewhat lower than $V_{\rm wind}$. Right: Normalized outflow velocity computed in two ways versus the observed wind momentum relative to the critical momentum input rate for acceleration (eqs.~\ref{ram_pressure_edd}, \S\ref{section:confronting_theory_observations}).}
\label{figure:radp_comp}
\end{figure}

In the case of radiation pressure on dust, we can go further. Estimates of the UV dust opacity in the warm ionized outflows imply that they are optically-thin: the median optical depth for far-UV photons is about $\simeq0.3$ \citep{Xu2022}. Thus, the observed momentum flux for many systems in Figure \ref{figure:xu3} is significantly larger than $\langle\tau_{\rm rp}\rangle(L/c)$, the expectation for optically-thin radiation pressure on dust (\S\ref{section:rp_thin}). We can estimate the Eddington ratio $\Gamma$ for the outflows in these galaxies using equation (\ref{radiation1}). To do so, we take $M_{\rm tot}=V_{\rm circ}^2 (2R_{1/2})/G$ where $R_{1/2}$ is the half-light radius, and we scale $f_{\rm dg}$ in $\langle\kappa_{\rm rp}\rangle$ (eq.~\ref{eq:kapparp}) with the measured values of the metallicity using $12+\log[\rm O/H]=8.69$ as a reference Solar value. Figure \ref{figure:radp_comp} shows  $\Gamma$ for each galaxy (left panel) and the observed and predicted velocities (middle panel) for those with $\Gamma>1$, assuming that radiation pressure on dust is the only driver. Radiation pressure underpredicts the momentum and velocity of the winds for most systems. Still, this comparison shows that it may play an important role in accelerating optically-thin gas for some systems. It is notable that multiplying the derived values of $\Gamma$ by $\sim5$ makes essentially all observed systems super-Eddington and brings the predicted and observed velocities into much closer agreement. This again provides support for a picture of momentum accelerated clouds.

Another way to assess the momentum budget of the observed outflows in light of theoretical expectations is to compare the critical momentum input rate needed to eject the gas  for a given observed column density $N_{\rm H}$, $\dot{p}_{\rm crit} = 4\pi R N_{\rm H} m V^{2}_{\rm circ}$ (see eqs.~\ref{ram_pressure_edd} and \ref{eq:fout_fin}). For a cloud with {\it constant} radius and mass, the maximum velocity attained by the cloud will be $V_{\rm max}/V_{\rm circ} = [2(\Gamma - 1) -2\ln(\Gamma)]^{1/2}$, where $\Gamma = \dot{p}_{\rm wind}/\dot{p}_{\rm crit}$, analogous to the Eddington ratio. In Figure \ref{figure:radp_comp}, we compare two measures of the normalized outflow velocity -- $V_{\rm wind}/V_{\rm circ}$ and $[V_{\rm wind}+(1/2){\rm FWHM}]/V_{\rm circ}$ (where the FWHM measures the observed wind line-width and not just the centroid; see Fig.~\ref{figure:line_montage} and \S\ref{section:absorption_velocities}) versus the normalized outflow momentum flux delivered to the clouds: $\dot{p}_{\rm wind}/\dot{p}_{\rm crit}$, where we take $R$, $v_{\rm circ}$, and $N_{\rm H}$ directly from the reported data. The dashed line shows the expectation from these two equations. There is a good match between the expectation and the observations, but with uncertainty in how the outflow velocity is defined. As implied by Figure \ref{figure:xu3}, the momentum flux needed to match $p_{\rm crit}$ can be accounted for by the very hot wind fluid with  $(\alpha\eta)^{1/2}\sim0.5-3$ (eq.~\ref{cc85_vinf_pdot}). Note that the total projected observed column densities are above the minimum individual cloud column densities required for cloud survival in a hot supersonic flow (eq.~\ref{ngrow}) and below the maximum cloud column required for acceleration via ram pressure (eq.~\ref{ram_pressure_edd}).

We can do a similar comparison between the kinetic energy carried by the observed outflow and the energy injection rate provided by the direct thermalization of SNe and stellar ejecta into the very hot wind ($\dot{E}_{\rm SN}$; eq.~\ref{eq:alpha_eta}) and (more indirectly) by the conversion of some of the injected energy into cosmic rays ($\dot{E}_{\rm cr}$; eq.~\ref{eq:edotcr}). Note that $\dot{E}_{\rm cr}\simeq0.1\dot{E}_{\rm SN}$ and that massive star winds contribute to the total energy injection rate at the 25\% level. In the right panel of Figure \ref{figure:xu3}, we plot the measured value of $\dot{E}_{\rm wind}$ versus the SFR. We see that the outflows carry a median value of $\dot{E}_{\rm wind}\sim0.1\dot{E}_{\rm SN}$, or about $\dot{E}_{\rm cr}$, and $5-10$ times below the thermalization efficiency $\alpha\simeq0.5-1.0$ inferred from the hard X-ray observations in the core of M82 (\S\ref{section:xray_emission}). If the outflowing absorption line gas is driven by a CC85-like outflow (\S\ref{section:supernova_driven}), the low value for the kinetic energy content of the wind is consistent with recent models and  simulations that find that the bulk of the kinetic energy is carried by the hot wind fluid \citep{Schneider2020,Kim2020_TIGRESS,Fielding2022}. 

Are CRs important? Since $\dot{E}_{\rm cr}\simeq0.1\dot{E}_{\rm SN}$, the right panel of Figure \ref{figure:xu3} shows that there needs to be very efficient energetic coupling between the CRs and the absorption line gas.  In both the ``diffusive" and ``streaming" CR-driven winds discussed in \S\ref{section:cosmic_rays}, an outflow kinetic power of $\dot{E}_{\rm cr}\simeq0.1\dot{E}_{\rm SN}$ is  expected (eqs.~\ref{eq:eta_cr_diff_max} \& \ref{eq:max_mass_loading_streaming}) and with outflow velocities of $\sim V_{\rm circ}$ (Fig.~\ref{figure:xu1}) if there are no strong radiative losses and if the winds are maximal. However, this analysis did not treat cloud acceleration, but instead treated the wind as a monolithic medium at fixed sound speed. Additionally, recent work by \cite{Modak2023} shows that in the case of streaming transport, a substantial fraction of the available CR energy is radiated away and used to do work against gravity as the wind is accelerating, perhaps challenging this mechanism in the face of Figure \ref{figure:xu3}. Finally, it would be difficult to explain systems with $\dot{E}_{\rm wind}\simeq10\dot{E}_{\rm cr}$ and \cite{Quataert2022a} argue that gamma-ray observations are not compatible with strong CR-driven winds in the diffusive limit (\S\ref{section:cr_limitations}). Even so, equations (\ref{eq:pdotcr_diff_tau}) and (\ref{vinfty_diffusion}) provide a promising start to explaining the observed correlations.

To contrast with these more ``continuous" ways of thinking about the outflow, we briefly consider a wind-blown bubble driven by the momentum flux provided by the very hot wind or one of the other potential prime movers (Table \ref{table:theory}). Energy-driven bubbles could also be considered. We are motivated in part by the spatially-resolved structure and kinematics of the outflowing warm ionized gas seen in optical emission-lines (see \S\ref{section:optical_xray} and Figs.~\ref{figure:m82_hst} and \ref{figure:bubble}). In this picture, the gas we see in absorption is swept-up ambient ISM and NGM gas at the surface of the expanding bubble. For simplicity, we use a spherical model with uniform ambient density \citep{Dyson1989}. Neglecting gravity, the bubble radius and velocity are $R_{\rm b} \simeq 2.2\,{\rm kpc}\,\,\dot{p}_{\rm 35}^{1/4} n_0^{-1/4} t_7^{1/2} \simeq 1.8\,{\rm kpc}\,\,{\rm SFR}_{10}^{1/4} n_0^{-1/4} t_7^{1/2}$ and $V_{\rm b} \simeq 110\,{\rm km/s}\,\,\dot{p}_{35}^{1/4} n_0^{-1/4} t_7^{-1/2} \simeq 90\,{\rm km/s}\,\,{\rm SFR}_{10}^{1/4} n_0^{-1/4} t_7^{-1/2}$,
where $\dot{p}_{35}=\dot{p}/10^{35}$\,dynes, ${\rm SFR}_{10}={\rm SFR}/10$\,M$_{\odot}$ yr$^{-1}$, $n_0$ is the gas density in cm$^{-3}$ in the region into which the bubble expands, and $t_7=t/10^7$\,yr is the time since the expansion began, similar to the ages derived from fits to the CLASSY data (P.~Senchyna \& J.~Chisholm, private communication). In the right panel of Figure \ref{figure:xu1}, we show the $V_b\propto{\rm SFR}^{1/4}$ with $n_0\simeq0.07$\,cm$^{-3}$ treated as a free parameter (dashed line). Such low densities implies material swept up from  outside the starburst. Excepting the very high velocity systems shown, this expression provides a reasonable fit to the data. However, any such  model cannot be complete. First, additional physics would be needed to establish an observed correlation between the $V_{\rm wind}$ and $V_{\rm circ}$. Second, for an expanding bubble, the absorption lines will only come from the part of the bubble located directly along the line-of-sight to the starburst. We would thus expect a relatively narrow blue-shifted absorption-line with ${\rm FWHM} < V_{\rm out}$, which is inconsistent with the finding that ${\rm FWHM} \sim 2 V_{\rm out}$ (Figs.~\ref{figure:line_montage} \& \ref{figure:xu3}; \citealt{Xu2022}). 

An alternative model for generating the warm ionized outflow is via bulk cooling directly from the hot phase (\S\ref{section:additional_physics_hot_phase}). For high enough mass-loading parameter $\eta\simeq1$, the cooling radius (eq.~\ref{eq:rcool}) becomes equal to the host galaxy radius $R$, producing an outflow velocity given by equation (\ref{eq:vcrit}). Comparing with Figures \ref{figure:xu1} and \ref{figure:xu2},  the normalization is too high and the scaling with the SFR is too weak to match the absorption line data, at least for fixed thermalization efficiency $\alpha$. Additionally, it is not immediately obvious how an expression like equation (\ref{eq:vcrit}) produces $V_{\rm wind}\propto V_{\rm circ}$ (e.g., Fig.~\ref{figure:xu1}), although $\alpha$, $\eta$, and the $\dot{\Sigma}_\star$ may conspire to yield such a relation in a complete theory. 

\section{Discussion \& Conclusions}
\label{section:discussion_conclusions}

Galactic winds are critical to galaxy evolution (\S\ref{section:introduction}). We have presented a discussion of proposed physical driving mechanisms meant to explain their basic physics, their energy and momentum budgets, and their potential limitations (\S\ref{section:theory}). We have also presented a discussion of galactic wind observations focused the hot and warm ionized phases of winds driven from low-$z$ starburst galaxies, as a model for their higher-$z$ counterparts (\S\ref{section:observations}). 

These observations are in general consistent with the hot thermal wind models in which the ejecta of massive stars (winds and supernovae) are thermalized inside the starburst, producing high (super-virial) temperatures and a strong pressure gradient (\S\ref{section:thermal}). This hot wind fluid is directly probed in local starburst cores using hard X-ray observations of the H-like and He-like Fe K$\alpha$ emission lines (\S\ref{section:xray_emission}). The observed luminosities of these lines scale with starburst properties as expected (Fig.~\ref{figure:heckman_various}). This hot fluid then expands and cools adiabatically and potentially radiatively, forming a tenuous and high-velocity wind fluid. As it does, it interacts with ambient gas in the ISM, NGM, and CGM to produce direct signatures of outflowing gas in phases spanning a wide range in temperature and column density. While the conventional picture is one where the ram pressure of the hot fluid accelerates the cooler material, a more complicated mixing process may transfer energy and momentum between the phases (\S\ref{section:supernova_driven}; Fig.~\ref{figure:theory}). Still, as emphasized in \S\ref{section:confronting_theory_observations}, an ``Eddington-like" critical momentum injection rate seems to be implied for the cool gas acceleration, and the correlation of the outflow velocity with the circular velocity provides additional evidence for this picture. The primary problems with the idea that the hot gas drives the cool gas dynamics is that we lack a complete theory for the thermalization efficiency and mass-loading of the very hot phase, that the required values of these parameters needed to match absorption line data (Fig.~\ref{figure:xu3}; e.g., $(\alpha\eta)^{1/2}\sim{\rm few}$ with $\alpha\sim0.1$) may not match the existing X-ray data ($\alpha\sim0.5-1$ and $\eta\sim0.2-0.6$ for M82; \S\ref{section:xray_emission}), and that we have yet to observe or directly constrain the hot gas kinematics from X-ray spectroscopy.

These issues motivate consideration of other potential contributors to wind driving. A simple calculation of the dust Eddington ratio for the local starbursts indicates that some are above unity and should drive outflows along low column density sightlines in optically-thin gas (Fig.~\ref{figure:radp_comp}). If radiation pressure in the optically-thick limit is ever important it may be for driving the exceptionally high column dusty molecular components of the highly optically-thick star-forming nuclei of ULIRGs like Arp 220 (\S\ref{section:radiation_pressure}) where radiative cooling may prohibit the formation of a dynamically important hot wind-driving phase and where pion losses may prevent high cosmic ray pressures (\S\ref{section:cr_limitations}).  Although transport microphysics remains a critical uncertainty in assessing cosmic rays as a wind driver, models indicate that there is sufficient momentum and energy (in most cases) to match observations (Fig.~\ref{figure:xu3}; Table \ref{table:theory}; \S\ref{section:cosmic_rays}) if CRs strongly couple their energy to the observed absorption line gas. When pion losses are weak, the current theory indicates that outflows driven in both the CR streaming and diffusion limits may produce highly mass loaded outflows. Current streaming models face challenges in matching the observed kinetic power of winds, which can be higher than $\dot{E}_{\rm cr}\simeq0.1\dot{E}_{\rm SN}$ (Fig.~\ref{figure:xu3}). Cosmic ray driven winds may be important in typical low-$z$ main sequence Milky Way-like galaxies, as perhaps evidenced by their extended radio halos (Fig.~\ref{figure:marvil}). Because the total momentum injection rate from ionizing photons and from Ly$\alpha$ scattering can be comparable to or larger than the ``single scattering" $L/c$ benchmark from dust, we cannot yet eliminate them as potentially important, especially at low metallicity. Magneto-thermal winds akin to proto-planetary disk outflows may be important in some cases (\S\ref{section:magnetic}). Additionally, SN remnants that complete the Sedov-Taylor phase inject significant momentum into the ISM and the NGM, nominally more than other sources, and are no doubt critical in driving ISM turbulence, which sets the stage for star formation and the galactic ``atmospheres" from which winds are driven (\S\ref{section:theory}).

A simple model where we observe a population of outflowing clouds accelerated by a combination of the hot wind fluid and other momentum sources, combined with the action of gravity appears to be able to explain the observed scaling relations for warm outflows. The ``mass-loading factor" ($\eta=$\,mass outflow rate/SFR) has a steep inverse correlation with both the galaxy circular velocity and outflow velocity that appears in between the so-called ``momentum" and ``energy" limits. The total momentum of the outflows ranges from $1-10(L/c)$ and the kinetic power is of order $0.1\dot{E}_{\rm SN}$, albeit with large scatter (Figures \ref{figure:xu1}, \ref{figure:xu2}, \& \ref{figure:xu3}). Outflows capable of escaping a galaxy are favored in galaxies with high $\dot{\Sigma}_\star$ and small circular velocities. These conditions are rare in the current universe, but become increasingly common with increasing redshift, consistent with the rarity of outflows today, and the near-ubiquity of galactic winds at $z \gtrsim 2$. As discussed in \S\ref{section:demographics_cosmic_time} (eq.~\ref{eq:fout_fin}), at fixed wind column density, the increase in outflow prevalence may be directly related to the increasing star formation rate per unit stellar mass (SFR/$M_\star$) of the star-forming galaxy main sequence with redshift.

\begin{summary}[SUMMARY POINTS]
\begin{enumerate}
\item The dominant driving mechanism for the acceleration of gas in galactic winds may vary across the diversity of galaxies, as a function of gas surface density and star formation rate surface density, but also as a function of thermodynamic phase.
\item Mechanisms that can contribute to cool gas acceleration or production include the dynamical coupling of the cool gas with a hot thermal wind phase, cosmic rays, radiation pressure, and magnetic fields, among others. 
\item Comparing theoretical mechanisms to absorption line kinematics from local starbursts, a picture emerges where cool clouds are accelerated by momentum input out of the galaxy potential with momentum loss rate $\sim L/c$ and with kinetic power $\sim0.1\dot{E}_{\rm SN}$, albeit with significant observed scatter.
\item The very hot gas seen in X-ray emission in the cores of local starburst exemplars like M82 may be the dominant wind driver, but more work is needed to understand the energy and momentum exchange between the hot and cooler phases, and the emergence of the emission and absorption line kinematics.
\item The strong observed correlations between the mass-loading rate, momentum rate, and kinetic power with the circular velocity of the host galaxy and the star formation rate constrain any theory for wind driving.  
\end{enumerate}
\end{summary}

\begin{issues}[FUTURE ISSUES \& DIRECTIONS]
\begin{enumerate}
\item The thermodynamics of the hot phase, including mixing with the cooler phases, conduction, dust, non-equilibrium ionization, radiative cooling, and other processes needs to be understood for a more complete comparison of the hot gas properties with both current and next-generation X-ray data.
\item The relative role of supernovae in driving turbulence in the ISM, depositing momentum in the near-galactic medium, and venting their thermal energy to drive the hot phase needs to continue to be investigated at high fidelity, and especially in extreme starburst environments.  
\item High-resolution wind tunnel-type numerical experiments of cloud acceleration by radiation pressure (dust, ionizing radiation, and Lyman $\alpha$) and cosmic rays (for a variety of physical transport models) should be pursued to bring cloud acceleration by other potential mechanisms to the same level as is being achieved for the hot gas-cool cloud interaction.
\item Velocity measurements or constraints for the hot X-ray emitting gas are critically needed to understand the dynamical importance of the hot phase and its energy and momentum content. Upcoming missions like XRISM and Athena may provide these measurements (or constraints) for the first time.
\item The most abundant and easily-obtained data on outflows from low- to high-redshift are UV absorption-lines. The interpretation of these data is primarily hampered by a lack of knowledge as to the structure and size of the outflowing gas they measure. More detailed comparisons of the 3-D data cubes (MUSE/VLT, KCWI/Keck) that map the resonantly scattered emission from these outflows to the predictions of radiative transfer models are needed to make additional progress.   
\end{enumerate}
\end{issues}

\begin{table}[h]
\tabcolsep7.5pt
\caption{Some symbols and scalings used in this review}
\label{table:symbols}
\begin{center}
\begin{tabular}{@{}l|l|l@{}}
\hline
Symbol & Typical Value/Units & Meaning\\
\hline
$R$ & kpc & Radius of host galaxy or wind launching radius. $R_x=R/x$\,kpc.\\
$\Omega$ & str & Solid angle subtended by wind. $\Omega_x=\Omega/x$\,str.\\
$L$ & $L_\odot$ & Bolometric luminosity of stellar population. \\
$n=\rho/m$ & cm$^{-3}$ & Gas number density, mass density ($\rho$), average particle mass ($m$).\\ 
$\alpha$ & $0-1$ & Thermalization efficiency in hot winds (eq.~\ref{eq:alpha_eta}). \\
$\eta$ & $0-1$ & Mass-loading efficiency, normalized by SFR   (eq.\ref{eq:alpha_eta}). \\
${E}_{\rm SN}$ & $10^{51}$\,ergs/SN & SN energy. See Sidebar \ref{sidebar:supernovae}. $\dot{E}_{\rm SN, 51}=E_{\rm SN}/10^{51}$\,ergs.\\
$N_{\rm H}$ & cm$^{-2}$ & Or $N$, $N_{\rm HI}$. Gas column density. E.g., $N_{\rm H,\, x}=N_{\rm H}/10^x$\,cm$^{-2}$. \\
$\Sigma_g$, $\Sigma_{\rm tot}$  & $1-10^3$\,M$_\odot$/pc$^2$ & Gas and total surface density. E.g., $\Sigma_{g, x}=\Sigma_g/x$\,M$_\odot$/pc$^2$. \\
$D$ & $10^{27}-10^{29}$\,cm$^2$/s & Cosmic ray diffusion constant (\S\ref{section:cosmic_rays}). $D_x=D/10^x$\,cm$^2$/s. \\
$\sigma$ & $10-300$\,km/s & Velocity dispersion for singular isothermal sphere. $\sigma_x=\sigma/x$\,km/s. \\
$V_{\rm circ}$ & $10-300$\,km/s & Galaxy circular velocity (e.g., Fig.~\ref{figure:xu1}); $V_{\rm circ}\simeq\sqrt{2}\sigma$. \\
SFR& $0.1-10^3$\,M$_\odot$/yr & Star formation rate. ${\rm SFR}_x={\rm SFR}/x$\,M$_\odot$/yr. \\
$\dot{\Sigma}_\star$ & M$_\odot$/yr/kpc$^2$ & Star formation rate surface density: $\dot{\Sigma}_{\star,x}=\dot{\Sigma}_\star/x$\,M$_\odot$/yr/kpc$^2$.\\
$\dot{M}$, $\dot{M}_{\rm wind}$ & $0.1-1000$\,M$_\odot$/yr & Wind mass loss rate. $\dot{M}=\eta {\rm SFR}$.\\
$\dot{p}_{\rm wind}$ & $L/c$ or dynes  & Wind momentum injection rate or force; $\dot{p}\simeq \dot{M} v_\infty$. \\
$\dot{E}_{\rm wind}$ & $\dot{E}_{\rm SN}$ or ergs/s & Wind energy injection/loss rate; $\dot{E}\simeq (1/2)\dot{M} v_\infty^2$. \\
$v_\infty$ & $100-1000$\,km/s & Asymptotic wind velocity. \\
$\epsilon_{\rm ph}$ & $4-7\times10^{-4}$ & Photon production efficiency: $L=\epsilon_{\rm ph}{\rm SFR} c^2$ (eq.~\ref{eq:lum_continuous}). IMF-dependent.\\
$f_{\rm dg}$ & $1/100$ & Total dust-to-gas mass ratio. $f_{\rm dg,\,MW}=f_{\rm dg}/(1/100)$.\\
$B$ & $0.1-10^3$\,$\mu$G & Magnetic field strength in host galaxy or outflow. \\ 
$v_{\rm A}$ & $1-100$\,km/s & Alfv\'en speed, $B/(4\pi\rho)^{1/2}$. $v_{\rm A, x}=v_{\rm A}/x$\,km/s. \S\ref{section:cosmic_rays} \& \ref{section:magnetic}. \\
$c_T$ & $0.1-1000$\,km/s & Isothermal gas sound speed. $c_{T, x}=c_T/x$\,km/s \\
$\epsilon_{\rm cr}$ & $10^{-6.3}$ & Efficiency of CR acceleration: $\dot{E}_{\rm cr}=\epsilon_{\rm cr}{\rm SFR} c^2$ (eq.~\ref{eq:edotcr}).\\
$\dot{E}_{\rm cr}$, $F_{\rm cr}$ & ergs/s, ergs/s/cm$^2$ & CR energy injection rate (eq.~\ref{eq:edotcr}) and flux ($F_{\rm cr}=\dot{E}_{\rm cr}/{\rm Area}$). \\
$P_{\rm cr}$, $U_{\rm cr}$ & ergs/cm$^3$ & CR pressure and energy density. $U_{\rm cr}=3P_{\rm cr}$.\\
\hline
\end{tabular}
\end{center}
\begin{tabnote}
\end{tabnote}
\end{table}

\begin{table}[h]
\tabcolsep7.5pt
\caption{Summary of wind prime movers}
\label{table:theory}
\begin{center}
\begin{tabular}{@{}l|l|l}
\hline
\hline
Name & Force $\dot{p}$ ($L/c$; eq.~\ref{eq:lum_continuous}) & Equation/Section \\
\hline
\\
Supernova Remnants$^{\rm a}$ & $20\,E_{\rm SN,\,51}^{0.93}n_0^{-0.13}$ & eq.~\ref{eq:pdot_snr}; \S\ref{section:thermal} \\
\\
Hot Wind Fluid$^{\rm b}$ & $5(\alpha\eta)^{1/2}$ & eq.~\ref{cc85_vinf_pdot}; \S\ref{section:thermal}  \\
\\
Radiation Pressure (on dust)$^{\rm c}$ & $\left(1-e^{-\langle\tau_{\rm rp}\rangle}+\tau_{\rm R}\right)$ & eq.~\ref{pdot_mdot_radiation};  \S\ref{section:radiation_pressure}  \\
\\
Radiation Pressure (ionizing photons)$^{\rm d}$  & $(0.1-0.5)(1-f_{\rm esc})$  & eq.~\ref{eq:pdot_ion}; \S\ref{section:radiation_pressure} \\
\\
Radiation Pressure (Ly$\alpha$)$^{\rm e}$  & $(4-20) N_{\rm HI,\,21}^{1/3}\,T_4^{-1/3}$ & eq.~\ref{eq:taueff_lya}; \S\ref{section:radiation_pressure} \\
\\
Cosmic Rays (Diffusion)$^{\rm f}$  & $1R^{1/2}_{\rm kpc}D_{28}^{-1/2}\sigma_{100}^{-1}$
& eq.~\ref{vinfty_diffusion}; \S\ref{section:cosmic_rays} \\
\\
Cosmic Rays (Streaming)$^{\rm g}$ & 
$1\dot{\Sigma}_{\star,\,0.01}^{1/2} B_{5\mu\rm G}^{-1}\sigma_{100}^{-3/2}$
& eqs.~\ref{cr_massload_stream} \& \ref{eq:cr_vinf_stream}; \S\ref{section:cosmic_rays}   \\
\\
Magneto-thermal$^{\rm h}$ & $\dot{M}_{\rm acc}v_{\rm circ}/(L/c)$ & \S\ref{section:magnetic} \\
\\
\hline
\hline
\end{tabular}
\end{center}
\begin{tabnote}
$^{\rm a}$This momentum input is thought to drive turbulence in the ISM. How it couples to galactic winds is a critical open question. See Sidebar \ref{sidebar:supernovae}.\\
$^{\rm b}$The pre-factor assumes continuous star formation and supernova energy injection (eqs.~\ref{eq:alpha_eta} \& \ref{cc85_vinf_pdot}). $\alpha$ and $\eta$ are the hot wind thermalization and mass-loading parameters. Cool clouds with column density below a critical value are destroyed (eq.~\ref{ngrow}). High column density clouds above a critical Eddington-like value are not accelerated by ram pressure (eq.~\ref{ram}). \\
$^{\rm c}$The system must exceed Eddington limit for wind driving. There is a strong implied metallicity dependence for optically-thin clouds/shells through the dust-to-gas mass ratio in the radiation pressure optical depth and opacity (eq.~\ref{eq:kapparp}). For geometrically small clouds with Rosseland optical depth $\tau_{\rm R}\gg1$ (eq.~\ref{eq:kappa_ross}) in a more tenuous medium, the momentum boost has a maximum at $\tau_{\rm R}\simeq1$. 
$^{\rm d}$The pre-factor depends on the fraction of the bolometric luminosity $L$ that goes into ionizing radiation,  which depends predominantly on the stellar population age, IMF, metallicity. $f_{\rm esc}$ is the global escape fraction of ionizing photons from the host galaxy or region. \\
$^{\rm e}$The pre-factor depends on the ionizing-to-bolometric luminosity ratio for the stellar population. This estimate assumes an idealized static dust-less and spherical geometry, not clouds. $N_{\rm HI,\,21}=N_{\rm HI}/10^{21}\,{\rm cm^{-2}}$ and $T_4=T/10^4$\,K for the warm ionized medium. Dust absorption of Ly$\alpha$ photons is an important limitation to Ly$\alpha$ radiation pressure (see eq.~\ref{eq:pdot_lymanalpha_dust}).\\ 
$^{\rm f}$Here, $R_{\rm kpc}=R/{\rm kpc}$, $D_{28}=D/10^{28}$\,cm$^2$/s, and $\sigma_{100}=\sigma/100$\,km/s. The maximum value of the momentum input rate is set by energy conservation and given by $(\epsilon_{\rm cr}/\epsilon_{\rm ph})(c/v_\infty)$ (eqs.~\ref{eq:edotcr} \& \ref{eq:eta_cr_diff_max}). \\
$^{\rm g}$Here, $\dot{\Sigma}_{\star,\,0.01}=\dot{\Sigma}_\star/0.01$\,M$_\odot$/yr/kpc$^2$, $B_{5\mu\rm G}=B/5\,\mu$G. As in the case for CR diffusion, the maximum momentum input rate is set by energy conservation: $(\epsilon_{\rm cr}/\epsilon_{\rm ph})(c/v_\infty)$ (eqs.~\ref{eq:edotcr} \& \ref{eq:max_mass_loading_streaming}).\\
$^{\rm h}$The mass loss rate from a magneto-thermal wind is of order the accretion rate in the disk $\dot{M}_{\rm acc}$. The asymptotic velocity is approximately the circular velocity from which the wind is launched for $\mu\gtrsim 1$ in equation (\ref{massloadingmu}). We do not currently have an analytic scaling for the mass outflow rate. See \S\ref{section:magnetic}. 
\end{tabnote}
\end{table}

\section*{DISCLOSURE STATEMENT}
The authors are not aware of any affiliations, memberships, funding, or financial holdings that might be perceived as affecting the objectivity of this review. 

\section*{ACKNOWLEDGMENTS}
TAT thanks E.~Quataert, N.~Murray, and M.~Krumholz for collaboration on topics related to this review, and E.~Schneider, A.~Leroy, P.~Martini, L.~Lopez, D.~Weinberg, E.~Ostriker, J.~Ostriker, R.~Crocker, P.~Hopkins, P.~Chang, B.~Draine, B.~Robertson, B.~Lacki, D.~Zhang, C.~Lochhaas, B.~Wibking, D.~Nguyen, O.~Pejcha, C.~Martin, I.~Blackstone, S.~Lopez, A.~Smith, A.~Burrows, N.~Scoville, and C.~Pfrommer for discussions of wind physics. TAT thanks A.~Shapley, J.~Davis, and D.~Rupke for galaxy wind data, and X.~Bai for the use of his code (\S\ref{section:magnetic}).  TAT thanks K.~Byram for encouragement and support. TAT acknowledges financial support from NSF grant \#1516967 and NASA grants \#80NSSC18K0526, \#80NSSC20K0531, and \#80NSSC23K1480. TMH acknowledges the contributions of former postdocs and graduate students over many years of research together on galactic winds, including L.~Armus, S.~Borthakur, Y.-M.~Chen, M.~Dahlem, J.~Grimes, M.~Lehnert, A.~Marlowe, D.~Strickland, C.~Tremonti, B.~Wang, J.~Wang, W.~Wang, K.~Weaver, and X.~Xu. TAT and TMH thank the Simons Foundation and the organizers of the workshop {\it Galactic Winds: Beyond Phenomenology} (J.\ Kollmeier \& A.\ Benson).


%

\bibliographystyle{ar-style2}
\bibliography{bibliography.bib}

\end{document}


%% file: review.bbl
\begin{thebibliography}{}
\expandafter\ifx\csname natexlab\endcsname\relax\def\natexlab#1{#1}\fi

\bibitem[{{Abdo} et~al.(2010){Abdo}, {Ackermann}, {Ajello}, {Atwood},
  {Axelsson} et~al.}]{abdo2010_m82_ngc253}
{Abdo} AA, {Ackermann} M, {Ajello} M, {Atwood} WB, {Axelsson} M, et~al. 2010.
\textit{\apjl} 709:L152--L157

\bibitem[{{Abruzzo} et~al.(2022){Abruzzo}, {Bryan} \& {Fielding}}]{Abruzzo2022}
{Abruzzo} MW, {Bryan} GL, {Fielding} DB. 2022.
\textit{\apj} 925:199

\bibitem[{{Acero} et~al.(2009){Acero}, {Aharonian}, {Akhperjanian}, {Anton},
  {Barres de Almeida} et~al.}]{hess2009_ngc253}
{Acero} F, {Aharonian} F, {Akhperjanian} AG, {Anton} G, {Barres de Almeida} U,
  et~al. 2009.
\textit{Science} 326:1080

\bibitem[{{Ackermann} et~al.(2013){Ackermann}, {Ajello}, {Allafort}, {Baldini},
  {Ballet} et~al.}]{Ackermann2013}
{Ackermann} M, {Ajello} M, {Allafort} A, {Baldini} L, {Ballet} J, et~al. 2013.
\textit{Science} 339:807--811

\bibitem[{{Adams} et~al.(2017){Adams}, {Kochanek}, {Gerke}, {Stanek} \&
  {Dai}}]{Adams2017_confirmation}
{Adams} SM, {Kochanek} CS, {Gerke} JR, {Stanek} KZ, {Dai} X. 2017.
\textit{\mnras} 468:4968--4981

\bibitem[{{Adams}(1972)}]{Adams1972}
{Adams} TF. 1972.
\textit{\apj} 174:439

\bibitem[{{Adams}(1975)}]{Adams1975}
{Adams} TF. 1975.
\textit{\apj} 201:350--351

\bibitem[{{Adebahr} et~al.(2017){Adebahr}, {Krause}, {Klein}, {Heald} \&
  {Dettmar}}]{Adebahr2017}
{Adebahr} B, {Krause} M, {Klein} U, {Heald} G, {Dettmar} RJ. 2017.
\textit{\aap} 608:A29

\bibitem[{{Adebahr} et~al.(2013){Adebahr}, {Krause}, {Klein}, {We{\.z}gowiec},
  {Bomans} \& {Dettmar}}]{Adebahr2013}
{Adebahr} B, {Krause} M, {Klein} U, {We{\.z}gowiec} M, {Bomans} DJ, {Dettmar}
  RJ. 2013.
\textit{\aap} 555:A23

\bibitem[{{Aguirre} et~al.(2001){Aguirre}, {Hernquist}, {Schaye}, {Katz},
  {Weinberg} \& {Gardner}}]{Aguirre2001}
{Aguirre} A, {Hernquist} L, {Schaye} J, {Katz} N, {Weinberg} DH, {Gardner} J.
  2001.
\textit{\apj} 561:521--549

\bibitem[{{Ajello} et~al.(2020){Ajello}, {Di Mauro}, {Paliya} \&
  {Garrappa}}]{Ajello2020}
{Ajello} M, {Di Mauro} M, {Paliya} VS, {Garrappa} S. 2020.
\textit{\apj} 894:88

\bibitem[{{Armillotta} et~al.(2017){Armillotta}, {Fraternali}, {Werk},
  {Prochaska} \& {Marinacci}}]{Armillota2017}
{Armillotta} L, {Fraternali} F, {Werk} JK, {Prochaska} JX, {Marinacci} F. 2017.
\textit{\mnras} 470:114--125

\bibitem[{{Armus} et~al.(1989){Armus}, {Heckman} \& {Miley}}]{Armus1989}
{Armus} L, {Heckman} TM, {Miley} GK. 1989.
\textit{\apj} 347:727

\bibitem[{{Armus} et~al.(1995){Armus}, {Heckman}, {Weaver} \&
  {Lehnert}}]{Armus1995}
{Armus} L, {Heckman} TM, {Weaver} KA, {Lehnert} MD. 1995.
\textit{\apj} 445:666

\bibitem[{{Baade} \& {Zwicky}(1934)}]{Baade1934}
{Baade} W, {Zwicky} F. 1934.
\textit{Proceedings of the National Academy of Science} 20:259--263

\bibitem[{{Bai} et~al.(2016){Bai}, {Ye}, {Goodman} \& {Yuan}}]{Bai2016}
{Bai} XN, {Ye} J, {Goodman} J, {Yuan} F. 2016.
\textit{\apj} 818:152

\bibitem[{{Balestra} et~al.(2005){Balestra}, {Boller}, {Gallo}, {Lutz} \&
  {Hess}}]{Balestra2005}
{Balestra} I, {Boller} T, {Gallo} L, {Lutz} D, {Hess} S. 2005.
\textit{\aap} 442:469--478

\bibitem[{{Banda-Barrag{\'a}n} et~al.(2021){Banda-Barrag{\'a}n}, {Br{\"u}ggen},
  {Heesen}, {Scannapieco}, {Cottle} et~al.}]{Banda-Barragan2021}
{Banda-Barrag{\'a}n} WE, {Br{\"u}ggen} M, {Heesen} V, {Scannapieco} E, {Cottle}
  J, et~al. 2021.
\textit{\mnras} 506:5658--5680

\bibitem[{{Banda-Barrag{\'a}n} et~al.(2018){Banda-Barrag{\'a}n}, {Federrath},
  {Crocker} \& {Bicknell}}]{Banda-Baragan2018}
{Banda-Barrag{\'a}n} WE, {Federrath} C, {Crocker} RM, {Bicknell} GV. 2018.
\textit{\mnras} 473:3454--3489

\bibitem[{{Banda-Barrag{\'a}n} et~al.(2016){Banda-Barrag{\'a}n}, {Parkin},
  {Federrath}, {Crocker} \& {Bicknell}}]{Banda-Baragan2016}
{Banda-Barrag{\'a}n} WE, {Parkin} ER, {Federrath} C, {Crocker} RM, {Bicknell}
  GV. 2016.
\textit{\mnras} 455:1309--1333

\bibitem[{{Barcos-Mu{\~n}oz} et~al.(2018){Barcos-Mu{\~n}oz}, {Aalto},
  {Thompson}, {Sakamoto}, {Mart{\'{\i}}n} et~al.}]{Barcos-Munoz2018}
{Barcos-Mu{\~n}oz} L, {Aalto} S, {Thompson} TA, {Sakamoto} K, {Mart{\'{\i}}n}
  S, et~al. 2018.
\textit{\apjl} 853:L28

\bibitem[{{Barcos-Mu{\~n}oz} et~al.(2015){Barcos-Mu{\~n}oz}, {Leroy}, {Evans},
  {Privon}, {Armus} et~al.}]{Barcos-Munoz2015}
{Barcos-Mu{\~n}oz} L, {Leroy} AK, {Evans} AS, {Privon} GC, {Armus} L, et~al.
  2015.
\textit{\apj} 799:10

\bibitem[{{Beck}(2015)}]{Beck2015}
{Beck} R. 2015.
\textit{\aapr} 24:4

\bibitem[{{Begelman} \& {Fabian}(1990)}]{Begelman1990}
{Begelman} MC, {Fabian} AC. 1990.
\textit{\mnras} 244:26P--29

\bibitem[{{Behroozi} et~al.(2019){Behroozi}, {Wechsler}, {Hearin} \&
  {Conroy}}]{Behroozi2019}
{Behroozi} P, {Wechsler} RH, {Hearin} AP, {Conroy} C. 2019.
\textit{\mnras} 488:3143--3194

\bibitem[{{Beir{\~a}o} et~al.(2015){Beir{\~a}o}, {Armus}, {Lehnert},
  {Guillard}, {Heckman} et~al.}]{Beirao2015}
{Beir{\~a}o} P, {Armus} L, {Lehnert} MD, {Guillard} P, {Heckman} T, et~al.
  2015.
\textit{\mnras} 451:2640--2655

\bibitem[{{Benson} et~al.(2003){Benson}, {Bower}, {Frenk}, {Lacey}, {Baugh} \&
  {Cole}}]{Benson2003}
{Benson} AJ, {Bower} RG, {Frenk} CS, {Lacey} CG, {Baugh} CM, {Cole} S. 2003.
\textit{\apj} 599:38--49

\bibitem[{{Berg} et~al.(2022){Berg}, {James}, {King}, {McDonald}, {Chen}
  et~al.}]{Berg2022}
{Berg} DA, {James} BL, {King} T, {McDonald} M, {Chen} Z, et~al. 2022.
\textit{\apjs} 261:31

\bibitem[{{Blackstone} \& {Thompson}(2023)}]{Blackstone2023}
{Blackstone} I, {Thompson} TA. 2023.
\textit{\mnras} 523:4309--4325

\bibitem[{{Blandford} \& {Payne}(1982)}]{Blandford1982}
{Blandford} RD, {Payne} DG. 1982.
\textit{\mnras} 199:883--903

\bibitem[{{Bolatto} et~al.(2024){Bolatto}, {Levy}, {Tarantino}, {Boyer},
  {Fisher} et~al.}]{Bolatto2024}
{Bolatto} AD, {Levy} RC, {Tarantino} E, {Boyer} ML, {Fisher} DB, et~al. 2024.
\textit{arXiv e-prints} :arXiv:2401.16648

\bibitem[{{Bonilha} et~al.(1979){Bonilha}, {Ferch}, {Salpeter}, {Slater} \&
  {Noerdlinger}}]{Bonilha1979}
{Bonilha} JRM, {Ferch} R, {Salpeter} EE, {Slater} G, {Noerdlinger} PD. 1979.
\textit{\apj} 233:649--660

\bibitem[{{Bordoloi} et~al.(2014){Bordoloi}, {Lilly}, {Hardmeier}, {Contini},
  {Kneib} et~al.}]{Bordoloi2014}
{Bordoloi} R, {Lilly} SJ, {Hardmeier} E, {Contini} T, {Kneib} JP, et~al. 2014.
\textit{\apj} 794:130

\bibitem[{{Bordoloi} et~al.(2011){Bordoloi}, {Lilly}, {Knobel}, {Bolzonella},
  {Kampczyk} et~al.}]{Bordoloi2011}
{Bordoloi} R, {Lilly} SJ, {Knobel} C, {Bolzonella} M, {Kampczyk} P, et~al.
  2011.
\textit{\apj} 743:10

\bibitem[{{Bouch{\'e}} et~al.(2012){Bouch{\'e}}, {Hohensee}, {Vargas},
  {Kacprzak}, {Martin} et~al.}]{Bouche2012}
{Bouch{\'e}} N, {Hohensee} W, {Vargas} R, {Kacprzak} GG, {Martin} CL, et~al.
  2012.
\textit{\mnras} 426:801--815

\bibitem[{{Boulares} \& {Cox}(1990)}]{Boulares1990}
{Boulares} A, {Cox} DP. 1990.
\textit{\apj} 365:544

\bibitem[{{Bregman}(1978)}]{Bregman1978}
{Bregman} JN. 1978.
\textit{\apj} 224:768--781

\bibitem[{{Breitschwerdt} et~al.(1991){Breitschwerdt}, {McKenzie} \&
  {Voelk}}]{Breitschwerdt1991}
{Breitschwerdt} D, {McKenzie} JF, {Voelk} HJ. 1991.
\textit{\aap} 245:79--98

\bibitem[{{Br{\"u}ggen} \& {Scannapieco}(2016)}]{Bruggen2016}
{Br{\"u}ggen} M, {Scannapieco} E. 2016.
\textit{\apj} 822:31

\bibitem[{{Br{\"u}ggen} \& {Scannapieco}(2020)}]{Bruggen2020}
{Br{\"u}ggen} M, {Scannapieco} E. 2020.
\textit{\apj} 905:19

\bibitem[{{Buckman} et~al.(2020){Buckman}, {Linden} \&
  {Thompson}}]{Buckman2020}
{Buckman} BJ, {Linden} T, {Thompson} TA. 2020.
\textit{\mnras} 494:2679--2705

\bibitem[{{Burchett} et~al.(2021){Burchett}, {Rubin}, {Prochaska}, {Coil},
  {Vaught} \& {Hennawi}}]{Burchett2021}
{Burchett} JN, {Rubin} KHR, {Prochaska} JX, {Coil} AL, {Vaught} RR, {Hennawi}
  JF. 2021.
\textit{\apj} 909:151

\bibitem[{{Burke}(1968)}]{Burke1968}
{Burke} JA. 1968.
\textit{\mnras} 140:241

\bibitem[{{Carr} et~al.(2018){Carr}, {Scarlata}, {Panagia} \&
  {Henry}}]{Carr2018}
{Carr} C, {Scarlata} C, {Panagia} N, {Henry} A. 2018.
\textit{\apj} 860:143

\bibitem[{{Cecil} et~al.(2002){Cecil}, {Bland-Hawthorn} \&
  {Veilleux}}]{Cecil2002}
{Cecil} G, {Bland-Hawthorn} J, {Veilleux} S. 2002.
\textit{\apj} 576:745--752

\bibitem[{{Chamberlain}(1960)}]{Chamberlain1960}
{Chamberlain} JW. 1960.
\textit{\apj} 131:47

\bibitem[{{Chan} et~al.(2019){Chan}, {Kere{\v{s}}}, {Hopkins}, {Quataert}, {Su}
  et~al.}]{Chan2019}
{Chan} TK, {Kere{\v{s}}} D, {Hopkins} PF, {Quataert} E, {Su} KY, et~al. 2019.
\textit{\mnras} 488:3716--3744

\bibitem[{{Chen} et~al.(2010){Chen}, {Tremonti}, {Heckman}, {Kauffmann},
  {Weiner} et~al.}]{Chen2010}
{Chen} YM, {Tremonti} CA, {Heckman} TM, {Kauffmann} G, {Weiner} BJ, et~al.
  2010.
\textit{\aj} 140:445--461

\bibitem[{{Chevalier} \& {Clegg}(1985)}]{Chevalier1985}
{Chevalier} RA, {Clegg} AW. 1985.
\textit{\nat} 317:44--45

\bibitem[{{Chiao} \& {Wickramasinghe}(1972)}]{Chiao1972}
{Chiao} RY, {Wickramasinghe} NC. 1972.
\textit{\mnras} 159:361

\bibitem[{{Chisholm} et~al.(2018){Chisholm}, {Tremonti} \&
  {Leitherer}}]{Chisholm2018}
{Chisholm} J, {Tremonti} C, {Leitherer} C. 2018.
\textit{\mnras} 481:1690--1706

\bibitem[{{Chisholm} et~al.(2017){Chisholm}, {Tremonti}, {Leitherer} \&
  {Chen}}]{Chisholm2017}
{Chisholm} J, {Tremonti} CA, {Leitherer} C, {Chen} Y. 2017.
\textit{\mnras} 469:4831--4849

\bibitem[{{Chisholm} et~al.(2016{\natexlab{a}}){Chisholm}, {Tremonti},
  {Leitherer}, {Chen} \& {Wofford}}]{Chisholm2016}
{Chisholm} J, {Tremonti} CA, {Leitherer} C, {Chen} Y, {Wofford} A.
  2016{\natexlab{a}}.
\textit{\mnras} 457:3133--3161

\bibitem[{{Chisholm} et~al.(2016{\natexlab{b}}){Chisholm}, {Tremonti},
  {Leitherer}, {Chen} \& {Wofford}}]{Chishlom2016}
{Chisholm} J, {Tremonti} CA, {Leitherer} C, {Chen} Y, {Wofford} A.
  2016{\natexlab{b}}.
\textit{\mnras} 457:3133--3161

\bibitem[{{Chisholm} et~al.(2015){Chisholm}, {Tremonti}, {Leitherer}, {Chen},
  {Wofford} \& {Lundgren}}]{Chisholm2015}
{Chisholm} J, {Tremonti} CA, {Leitherer} C, {Chen} Y, {Wofford} A, {Lundgren}
  B. 2015.
\textit{\apj} 811:149

\bibitem[{{Coker} et~al.(2013){Coker}, {Thompson} \& {Martini}}]{Coker2013}
{Coker} CT, {Thompson} TA, {Martini} P. 2013.
\textit{\apj} 778:79

\bibitem[{{Contursi} et~al.(2013){Contursi}, {Poglitsch}, {Graci{\'a} Carpio},
  {Veilleux}, {Sturm} et~al.}]{Contursi2013}
{Contursi} A, {Poglitsch} A, {Graci{\'a} Carpio} J, {Veilleux} S, {Sturm} E,
  et~al. 2013.
\textit{\aap} 549:A118

\bibitem[{{Cooper} et~al.(2008){Cooper}, {Bicknell}, {Sutherland} \&
  {Bland-Hawthorn}}]{Cooper2008}
{Cooper} JL, {Bicknell} GV, {Sutherland} RS, {Bland-Hawthorn} J. 2008.
\textit{\apj} 674:157--171

\bibitem[{{Cooper} et~al.(2009){Cooper}, {Bicknell}, {Sutherland} \&
  {Bland-Hawthorn}}]{Cooper2009}
{Cooper} JL, {Bicknell} GV, {Sutherland} RS, {Bland-Hawthorn} J. 2009.
\textit{\apj} 703:330--347

\bibitem[{{Cottle} et~al.(2020){Cottle}, {Scannapieco}, {Br{\"u}ggen},
  {Banda-Barrag{\'a}n} \& {Federrath}}]{Cottle2020}
{Cottle} J, {Scannapieco} E, {Br{\"u}ggen} M, {Banda-Barrag{\'a}n} W,
  {Federrath} C. 2020.
\textit{\apj} 892:59

\bibitem[{{Cowie} et~al.(1981){Cowie}, {McKee} \& {Ostriker}}]{Cowie1981}
{Cowie} LL, {McKee} CF, {Ostriker} JP. 1981.
\textit{\apj} 247:908--924

\bibitem[{{Cox}(1985)}]{Cox1985}
{Cox} DP. 1985.
\textit{\apj} 288:465--480

\bibitem[{{Crain} \& {van de Voort}(2023)}]{Crain2023}
{Crain} RA, {van de Voort} F. 2023.
\textit{\araa} 61:473--515

\bibitem[{{Crocker} et~al.(2021{\natexlab{a}}){Crocker}, {Krumholz} \&
  {Thompson}}]{Crocker2021_I}
{Crocker} RM, {Krumholz} MR, {Thompson} TA. 2021{\natexlab{a}}.
\textit{\mnras} 502:1312--1333

\bibitem[{{Crocker} et~al.(2021{\natexlab{b}}){Crocker}, {Krumholz} \&
  {Thompson}}]{Crocker2021_IIthreshold}
{Crocker} RM, {Krumholz} MR, {Thompson} TA. 2021{\natexlab{b}}.
\textit{\mnras} 503:2651--2664

\bibitem[{{Crocker} et~al.(2018{\natexlab{a}}){Crocker}, {Krumholz},
  {Thompson}, {Baumgardt} \& {Mackey}}]{Crocker2018_radp_cluster}
{Crocker} RM, {Krumholz} MR, {Thompson} TA, {Baumgardt} H, {Mackey} D.
  2018{\natexlab{a}}.
\textit{\mnras} 481:4895--4906

\bibitem[{{Crocker} et~al.(2018{\natexlab{b}}){Crocker}, {Krumholz}, {Thompson}
  \& {Clutterbuck}}]{Crocker2018}
{Crocker} RM, {Krumholz} MR, {Thompson} TA, {Clutterbuck} J.
  2018{\natexlab{b}}.
\textit{\mnras} 478:81--94

\bibitem[{{da Silva} et~al.(2012){da Silva}, {Fumagalli} \&
  {Krumholz}}]{dasilva2012_slug}
{da Silva} RL, {Fumagalli} M, {Krumholz} M. 2012.
\textit{\apj} 745:145

\bibitem[{{Dav{\'e}} et~al.(2012){Dav{\'e}}, {Finlator} \&
  {Oppenheimer}}]{Dave2012}
{Dav{\'e}} R, {Finlator} K, {Oppenheimer} BD. 2012.
\textit{\mnras} 421:98--107

\bibitem[{{Davies} et~al.(2019){Davies}, {F{\"o}rster Schreiber}, {{\"U}bler},
  {Genzel}, {Lutz} et~al.}]{Davies2019}
{Davies} RL, {F{\"o}rster Schreiber} NM, {{\"U}bler} H, {Genzel} R, {Lutz} D,
  et~al. 2019.
\textit{\apj} 873:122

\bibitem[{{Davis} et~al.(2023){Davis}, {Tremonti}, {Swiggum}, {Moustakas},
  {Diamond-Stanic} et~al.}]{Davis2023}
{Davis} JD, {Tremonti} CA, {Swiggum} CN, {Moustakas} J, {Diamond-Stanic} AM,
  et~al. 2023.
\textit{\apj} 951:105

\bibitem[{{Davis} et~al.(2014){Davis}, {Jiang}, {Stone} \&
  {Murray}}]{Davis2014}
{Davis} SW, {Jiang} YF, {Stone} JM, {Murray} N. 2014.
\textit{\apj} 796:107

\bibitem[{{de Gouveia Dal Pino} \& {Tanco}(1999)}]{Pino1999}
{de Gouveia Dal Pino} EM, {Tanco} GAM. 1999.
\textit{\apj} 518:129--137

\bibitem[{{de la Cruz} et~al.(2021){de la Cruz}, {Schneider} \&
  {Ostriker}}]{delacruz2021}
{de la Cruz} LM, {Schneider} EE, {Ostriker} EC. 2021.
\textit{\apj} 919:112

\bibitem[{{Dekel} \& {Silk}(1986)}]{Dekel1986}
{Dekel} A, {Silk} J. 1986.
\textit{\apj} 303:39--55

\bibitem[{{Diamond-Stanic} et~al.(2021){Diamond-Stanic}, {Moustakas}, {Sell},
  {Tremonti}, {Coil} et~al.}]{Diamond-Stanic2021}
{Diamond-Stanic} AM, {Moustakas} J, {Sell} PH, {Tremonti} CA, {Coil} AL, et~al.
  2021.
\textit{\apj} 912:11

\bibitem[{{Dijkstra}(2014)}]{Dijkstra2014}
{Dijkstra} M. 2014.
\textit{\pasa} 31:e040

\bibitem[{{Dijkstra} \& {Loeb}(2008)}]{Dijkstra2008}
{Dijkstra} M, {Loeb} A. 2008.
\textit{\mnras} 391:457--466

\bibitem[{{Dijkstra} \& {Loeb}(2009)}]{Dijkstra2009}
{Dijkstra} M, {Loeb} A. 2009.
\textit{\mnras} 396:377--384

\bibitem[{{Draine} \& {Salpeter}(1979{\natexlab{a}})}]{Draine1979b}
{Draine} BT, {Salpeter} EE. 1979{\natexlab{a}}.
\textit{\apj} 231:438--455

\bibitem[{{Draine} \& {Salpeter}(1979{\natexlab{b}})}]{Draine1979a}
{Draine} BT, {Salpeter} EE. 1979{\natexlab{b}}.
\textit{\apj} 231:77--94

\bibitem[{{Dutta} et~al.(2023){Dutta}, {Fossati}, {Fumagalli}, {Revalski},
  {Lofthouse} et~al.}]{Dutta2023}
{Dutta} R, {Fossati} M, {Fumagalli} M, {Revalski} M, {Lofthouse} EK, et~al.
  2023.
\textit{\mnras} 522:535--558

\bibitem[{{Dyson}(1989)}]{Dyson1989}
{Dyson} JE. 1989.
\textit{{Interstellar Wind-Blown Bubbles}}. In \textit{IAU Colloq. 120:
  Structure and Dynamics of the Interstellar Medium}, eds. G~{Tenorio-Tagle},
  M~{Moles}, J~{Melnick}, vol. 350.  137

\bibitem[{{Efstathiou}(2000)}]{Efstathiou2000}
{Efstathiou} G. 2000.
\textit{\mnras} 317:697--719

\bibitem[{{Eldridge} et~al.(2017){Eldridge}, {Stanway}, {Xiao}, {McClelland},
  {Taylor} et~al.}]{Eldridge2017}
{Eldridge} JJ, {Stanway} ER, {Xiao} L, {McClelland} LAS, {Taylor} G, et~al.
  2017.
\textit{\pasa} 34:e058

\bibitem[{{Elmegreen} \& {Chiang}(1982)}]{Elmegreen1982}
{Elmegreen} BG, {Chiang} WH. 1982.
\textit{\apj} 253:666--678

\bibitem[{{Erb} et~al.(2023){Erb}, {Li}, {Steidel}, {Chen}, {Gronke}
  et~al.}]{Erb2022}
{Erb} DK, {Li} Z, {Steidel} CC, {Chen} Y, {Gronke} M, et~al. 2023.
\textit{\apj} 953:118

\bibitem[{{Erb} et~al.(2012){Erb}, {Quider}, {Henry} \& {Martin}}]{Erb2012}
{Erb} DK, {Quider} AM, {Henry} AL, {Martin} CL. 2012.
\textit{\apj} 759:26

\bibitem[{{Erb} et~al.(2006{\natexlab{a}}){Erb}, {Shapley}, {Pettini},
  {Steidel}, {Reddy} \& {Adelberger}}]{Erb2006a}
{Erb} DK, {Shapley} AE, {Pettini} M, {Steidel} CC, {Reddy} NA, {Adelberger} KL.
  2006{\natexlab{a}}.
\textit{\apj} 644:813--828

\bibitem[{{Erb} et~al.(2006{\natexlab{b}}){Erb}, {Steidel}, {Shapley},
  {Pettini}, {Reddy} \& {Adelberger}}]{Erb2006c}
{Erb} DK, {Steidel} CC, {Shapley} AE, {Pettini} M, {Reddy} NA, {Adelberger} KL.
  2006{\natexlab{b}}.
\textit{\apj} 647:128--139

\bibitem[{{Erb} et~al.(2006{\natexlab{c}}){Erb}, {Steidel}, {Shapley},
  {Pettini}, {Reddy} \& {Adelberger}}]{Erb2006b}
{Erb} DK, {Steidel} CC, {Shapley} AE, {Pettini} M, {Reddy} NA, {Adelberger} KL.
  2006{\natexlab{c}}.
\textit{\apj} 646:107--132

\bibitem[{{Ertl} et~al.(2020){Ertl}, {Woosley}, {Sukhbold} \&
  {Janka}}]{Ertl2020}
{Ertl} T, {Woosley} SE, {Sukhbold} T, {Janka} HT. 2020.
\textit{\apj} 890:51

\bibitem[{{Everett} et~al.(2008){Everett}, {Zweibel}, {Benjamin}, {McCammon},
  {Rocks} \& {Gallagher}}]{Everett2008}
{Everett} JE, {Zweibel} EG, {Benjamin} RA, {McCammon} D, {Rocks} L, {Gallagher}
  III JS. 2008.
\textit{\apj} 674:258--270

\bibitem[{{Faucher-Gigu{\`e}re} \& {Oh}(2023)}]{Faucher-Giguere2023}
{Faucher-Gigu{\`e}re} CA, {Oh} SP. 2023.
\textit{\araa} 61:131--195

\bibitem[{{Fielding} et~al.(2018){Fielding}, {Quataert} \&
  {Martizzi}}]{Fielding2018}
{Fielding} D, {Quataert} E, {Martizzi} D. 2018.
\textit{\mnras} 481:3325--3347

\bibitem[{{Fielding} \& {Bryan}(2022)}]{Fielding2022}
{Fielding} DB, {Bryan} GL. 2022.
\textit{\apj} 924:82

\bibitem[{{Fielding} et~al.(2020){Fielding}, {Ostriker}, {Bryan} \&
  {Jermyn}}]{Fielding2020}
{Fielding} DB, {Ostriker} EC, {Bryan} GL, {Jermyn} AS. 2020.
\textit{\apjl} 894:L24

\bibitem[{{Firmani} \& {Tutukov}(1994)}]{Firmani1994}
{Firmani} C, {Tutukov} AV. 1994.
\textit{\aap} 288:713--730

\bibitem[{{Fluetsch} et~al.(2021){Fluetsch}, {Maiolino}, {Carniani}, {Arribas},
  {Belfiore} et~al.}]{Fluetsch2021}
{Fluetsch} A, {Maiolino} R, {Carniani} S, {Arribas} S, {Belfiore} F, et~al.
  2021.
\textit{\mnras} 505:5753--5783

\bibitem[{{F{\"o}rster Schreiber} et~al.(2019){F{\"o}rster Schreiber},
  {{\"U}bler}, {Davies}, {Genzel}, {Wisnioski} et~al.}]{Forster2019}
{F{\"o}rster Schreiber} NM, {{\"U}bler} H, {Davies} RL, {Genzel} R, {Wisnioski}
  E, et~al. 2019.
\textit{\apj} 875:21

\bibitem[{{F{\"o}rster Schreiber} \& {Wuyts}(2020)}]{Forster2020}
{F{\"o}rster Schreiber} NM, {Wuyts} S. 2020.
\textit{\araa} 58:661--725

\bibitem[{{Gaisser}(1990)}]{Gaisser1990}
{Gaisser} TK. 1990.
\textit{{Cosmic rays and particle physics.}}

\bibitem[{{Gatto} et~al.(2017){Gatto}, {Walch}, {Naab}, {Girichidis},
  {W{\"u}nsch} et~al.}]{Gatto2017}
{Gatto} A, {Walch} S, {Naab} T, {Girichidis} P, {W{\"u}nsch} R, et~al. 2017.
\textit{\mnras} 466:1903--1924

\bibitem[{{Gayley} et~al.(1995){Gayley}, {Owocki} \& {Cranmer}}]{Gayley1995}
{Gayley} KG, {Owocki} SP, {Cranmer} SR. 1995.
\textit{\apj} 442:296

\bibitem[{{Gazagnes} et~al.(2023){Gazagnes}, {Mauerhofer}, {Berg}, {Blaizot},
  {Verhamme} et~al.}]{Gazagnes2023}
{Gazagnes} S, {Mauerhofer} V, {Berg} DA, {Blaizot} J, {Verhamme} A, et~al.
  2023.
\textit{\apj} 952:164

\bibitem[{{Gentry} et~al.(2017){Gentry}, {Krumholz}, {Dekel} \&
  {Madau}}]{Gentry2017}
{Gentry} ES, {Krumholz} MR, {Dekel} A, {Madau} P. 2017.
\textit{\mnras} 465:2471--2488

\bibitem[{{Gerasimovi{\v{c}}}(1932)}]{Gerasimovic1932}
{Gerasimovi{\v{c}}} BP. 1932.
\textit{\zap} 4:265

\bibitem[{{Gerke} et~al.(2015){Gerke}, {Kochanek} \& {Stanek}}]{Gerke2015}
{Gerke} JR, {Kochanek} CS, {Stanek} KZ. 2015.
\textit{\mnras} 450:3289--3305

\bibitem[{{Girichidis} et~al.(2016){Girichidis}, {Naab}, {Walch}, {Hanasz},
  {Mac Low} et~al.}]{Girichidis2016}
{Girichidis} P, {Naab} T, {Walch} S, {Hanasz} M, {Mac Low} MM, et~al. 2016.
\textit{\apjl} 816:L19

\bibitem[{{Gon{\c{c}}alves} et~al.(2010){Gon{\c{c}}alves}, {Basu-Zych},
  {Overzier}, {Martin}, {Law} et~al.}]{Goncalves2010}
{Gon{\c{c}}alves} TS, {Basu-Zych} A, {Overzier} R, {Martin} DC, {Law} DR,
  et~al. 2010.
\textit{\apj} 724:1373--1388

\bibitem[{{Gray} et~al.(2019{\natexlab{a}}){Gray}, {Oey}, {Silich} \&
  {Scannapieco}}]{Gray2019b}
{Gray} WJ, {Oey} MS, {Silich} S, {Scannapieco} E. 2019{\natexlab{a}}.
\textit{\apj} 887:161

\bibitem[{{Gray} et~al.(2019{\natexlab{b}}){Gray}, {Scannapieco} \&
  {Lehnert}}]{Gray2019a}
{Gray} WJ, {Scannapieco} E, {Lehnert} MD. 2019{\natexlab{b}}.
\textit{\apj} 875:110

\bibitem[{{Greco} et~al.(2012){Greco}, {Martini} \& {Thompson}}]{Greco2012}
{Greco} JP, {Martini} P, {Thompson} TA. 2012.
\textit{\apj} 757:24

\bibitem[{{Griffin} et~al.(2016){Griffin}, {Dai} \& {Thompson}}]{Griffin2016}
{Griffin} RD, {Dai} X, {Thompson} TA. 2016.
\textit{\apjl} 823:L17

\bibitem[{{Griffith} et~al.(2021){Griffith}, {Sukhbold}, {Weinberg}, {Johnson},
  {Johnson} \& {Vincenzo}}]{Griffith2021}
{Griffith} EJ, {Sukhbold} T, {Weinberg} DH, {Johnson} JA, {Johnson} JW,
  {Vincenzo} F. 2021.
\textit{\apj} 921:73

\bibitem[{{Grimes} et~al.(2009){Grimes}, {Heckman}, {Aloisi}, {Calzetti},
  {Leitherer} et~al.}]{Grimes2009}
{Grimes} JP, {Heckman} T, {Aloisi} A, {Calzetti} D, {Leitherer} C, et~al. 2009.
\textit{\apjs} 181:272--320

\bibitem[{{Grimes} et~al.(2006){Grimes}, {Heckman}, {Hoopes}, {Strickland},
  {Aloisi} et~al.}]{Grimes2006}
{Grimes} JP, {Heckman} T, {Hoopes} C, {Strickland} D, {Aloisi} A, et~al. 2006.
\textit{\apj} 648:310--322

\bibitem[{{Grimes} et~al.(2007){Grimes}, {Heckman}, {Strickland}, {Dixon},
  {Sembach} et~al.}]{Grimes2007}
{Grimes} JP, {Heckman} T, {Strickland} D, {Dixon} WV, {Sembach} K, et~al. 2007.
\textit{\apj} 668:891--905

\bibitem[{{Grimes} et~al.(2005){Grimes}, {Heckman}, {Strickland} \&
  {Ptak}}]{Grimes2005}
{Grimes} JP, {Heckman} T, {Strickland} D, {Ptak} A. 2005.
\textit{\apj} 628:187--204

\bibitem[{{Gronke} \& {Oh}(2018)}]{Gronke2018}
{Gronke} M, {Oh} SP. 2018.
\textit{\mnras} 480:L111--L115

\bibitem[{{Gronke} \& {Oh}(2020)}]{Gronke2020b}
{Gronke} M, {Oh} SP. 2020.
\textit{\mnras} 494:L27--L31

\bibitem[{{Guo} et~al.(2023{\natexlab{a}}){Guo}, {Bacon}, {Bouch{\'e}},
  {Wisotzki}, {Schaye} et~al.}]{Guo2023b}
{Guo} Y, {Bacon} R, {Bouch{\'e}} NF, {Wisotzki} L, {Schaye} J, et~al.
  2023{\natexlab{a}}.
\textit{\nat} 624:53--56

\bibitem[{{Guo} et~al.(2023{\natexlab{b}}){Guo}, {Bacon}, {Wisotzki}, {Garel},
  {Blaizot} et~al.}]{Guo2023a}
{Guo} Y, {Bacon} R, {Wisotzki} L, {Garel} T, {Blaizot} J, et~al.
  2023{\natexlab{b}}.
\textit{arXiv e-prints} :arXiv:2309.05513

\bibitem[{{Gutcke} et~al.(2021){Gutcke}, {Pakmor}, {Naab} \&
  {Springel}}]{Gutcke2021}
{Gutcke} TA, {Pakmor} R, {Naab} T, {Springel} V. 2021.
\textit{\mnras} 501:5597--5615

\bibitem[{{Haehnelt}(1995)}]{Haehnelt1995}
{Haehnelt} MG. 1995.
\textit{\mnras} 273:249--256

\bibitem[{{Harwit}(1962)}]{Harwit1962}
{Harwit} M. 1962.
\textit{\apj} 136:832

\bibitem[{{Hayden} et~al.(2015){Hayden}, {Bovy}, {Holtzman}, {Nidever}, {Bird}
  et~al.}]{Hayden2015}
{Hayden} MR, {Bovy} J, {Holtzman} JA, {Nidever} DL, {Bird} JC, et~al. 2015.
\textit{\apj} 808:132

\bibitem[{{Heckman} et~al.(2017){Heckman}, {Borthakur}, {Wild}, {Schiminovich}
  \& {Bordoloi}}]{Heckman2017}
{Heckman} T, {Borthakur} S, {Wild} V, {Schiminovich} D, {Bordoloi} R. 2017.
\textit{\apj} 846:151

\bibitem[{{Heckman} et~al.(2015){Heckman}, {Alexandroff}, {Borthakur},
  {Overzier} \& {Leitherer}}]{Heckman2015}
{Heckman} TM, {Alexandroff} RM, {Borthakur} S, {Overzier} R, {Leitherer} C.
  2015.
\textit{\apj} 809:147

\bibitem[{{Heckman} et~al.(1990){Heckman}, {Armus} \& {Miley}}]{Heckman1990}
{Heckman} TM, {Armus} L, {Miley} GK. 1990.
\textit{\apjs} 74:833

\bibitem[{{Heckman} et~al.(1999){Heckman}, {Armus}, {Weaver} \&
  {Wang}}]{Heckman1999}
{Heckman} TM, {Armus} L, {Weaver} KA, {Wang} J. 1999.
\textit{\apj} 517:130--147

\bibitem[{{Heckman} \& {Borthakur}(2016)}]{Heckman2016}
{Heckman} TM, {Borthakur} S. 2016.
\textit{\apj} 822:9

\bibitem[{{Heckman} et~al.(2000){Heckman}, {Lehnert}, {Strickland} \&
  {Armus}}]{Heckman2000}
{Heckman} TM, {Lehnert} MD, {Strickland} DK, {Armus} L. 2000.
\textit{\apjs} 129:493--516

\bibitem[{{Heesen} et~al.(2011){Heesen}, {Beck}, {Krause} \&
  {Dettmar}}]{Heesen2011}
{Heesen} V, {Beck} R, {Krause} M, {Dettmar} RJ. 2011.
\textit{\aap} 535:A79

\bibitem[{{Henney} \& {Arthur}(1998)}]{Henney1998}
{Henney} WJ, {Arthur} SJ. 1998.
\textit{\aj} 116:322--335

\bibitem[{{Hensley} \& {Draine}(2023)}]{Hensley2023}
{Hensley} BS, {Draine} BT. 2023.
\textit{\apj} 948:55

\bibitem[{{Holzer} \& {Axford}(1970)}]{Holzer1970}
{Holzer} TE, {Axford} WI. 1970.
\textit{\araa} 8:31

\bibitem[{{Hoopes} et~al.(2005){Hoopes}, {Heckman}, {Strickland}, {Seibert},
  {Madore} et~al.}]{Hoopes2005}
{Hoopes} CG, {Heckman} TM, {Strickland} DK, {Seibert} M, {Madore} BF, et~al.
  2005.
\textit{\apjl} 619:L99--L102

\bibitem[{{Hopkins} et~al.(2022{\natexlab{a}}){Hopkins}, {Butsky},
  {Panopoulou}, {Ji}, {Quataert} et~al.}]{Hopkins2022_CRs}
{Hopkins} PF, {Butsky} IS, {Panopoulou} GV, {Ji} S, {Quataert} E, et~al.
  2022{\natexlab{a}}.
\textit{\mnras} 516:3470--3514

\bibitem[{{Hopkins} et~al.(2020{\natexlab{a}}){Hopkins}, {Chan},
  {Garrison-Kimmel}, {Ji}, {Su} et~al.}]{Hopkins2020_CRs}
{Hopkins} PF, {Chan} TK, {Garrison-Kimmel} S, {Ji} S, {Su} KY, et~al.
  2020{\natexlab{a}}.
\textit{\mnras} 492:3465--3498

\bibitem[{{Hopkins} et~al.(2020{\natexlab{b}}){Hopkins}, {Grudi{\'c}},
  {Wetzel}, {Kere{\v{s}}}, {Faucher-Gigu{\`e}re}
  et~al.}]{Hopkins2020_radiative}
{Hopkins} PF, {Grudi{\'c}} MY, {Wetzel} A, {Kere{\v{s}}} D,
  {Faucher-Gigu{\`e}re} CA, et~al. 2020{\natexlab{b}}.
\textit{\mnras} 491:3702--3729

\bibitem[{{Hopkins} et~al.(2012){Hopkins}, {Quataert} \&
  {Murray}}]{Hopkins2012}
{Hopkins} PF, {Quataert} E, {Murray} N. 2012.
\textit{\mnras} 421:3522--3537

\bibitem[{{Hopkins} et~al.(2022{\natexlab{b}}){Hopkins}, {Rosen}, {Squire},
  {Panopoulou}, {Soliman} et~al.}]{Hopkins2022_Dust}
{Hopkins} PF, {Rosen} AL, {Squire} J, {Panopoulou} GV, {Soliman} NH, et~al.
  2022{\natexlab{b}}.
\textit{\mnras} 517:1491--1517

\bibitem[{{Hopkins} \& {Squire}(2018)}]{Hopkins2018_Squire}
{Hopkins} PF, {Squire} J. 2018.
\textit{\mnras} 479:4681--4719

\bibitem[{{Hopkins} et~al.(2022{\natexlab{c}}){Hopkins}, {Squire}, {Butsky} \&
  {Ji}}]{Hopkins2021}
{Hopkins} PF, {Squire} J, {Butsky} IS, {Ji} S. 2022{\natexlab{c}}.
\textit{\mnras} 517:5413--5448

\bibitem[{{Huang} \& {Davis}(2022)}]{Huang2022a}
{Huang} X, {Davis} SW. 2022.
\textit{\mnras} 511:5125--5141

\bibitem[{{Huang} et~al.(2020){Huang}, {Davis} \& {Zhang}}]{Huang2020}
{Huang} X, {Davis} SW, {Zhang} D. 2020.
\textit{\apj} 893:50

\bibitem[{{Huang} et~al.(2022){Huang}, {Jiang} \& {Davis}}]{Huang2022b}
{Huang} X, {Jiang} Yf, {Davis} SW. 2022.
\textit{\apj} 931:140

\bibitem[{{Ipavich}(1975)}]{Ipavich1975}
{Ipavich} FM. 1975.
\textit{\apj} 196:107--120

\bibitem[{{Irwin} et~al.(2012){Irwin}, {Beck}, {Benjamin}, {Dettmar}, {English}
  et~al.}]{Irwin2012}
{Irwin} J, {Beck} R, {Benjamin} RA, {Dettmar} RJ, {English} J, et~al. 2012.
\textit{\aj} 144:43

\bibitem[{{Iwasawa} et~al.(2023){Iwasawa}, {Norman}, {Gilli}, {Gandhi} \&
  {Per{\'e}z-Torres}}]{Iwasawa2023}
{Iwasawa} K, {Norman} C, {Gilli} R, {Gandhi} P, {Per{\'e}z-Torres} MA. 2023.
\textit{\aap} 674:A77

\bibitem[{{Iwasawa} et~al.(2009){Iwasawa}, {Sanders}, {Evans}, {Mazzarella},
  {Armus} \& {Surace}}]{Iwasawa2009}
{Iwasawa} K, {Sanders} DB, {Evans} AS, {Mazzarella} JM, {Armus} L, {Surace} JA.
  2009.
\textit{\apjl} 695:L103--L106

\bibitem[{{Iwasawa} et~al.(2005){Iwasawa}, {Sanders}, {Evans}, {Trentham},
  {Miniutti} \& {Spoon}}]{Iwasawa2005}
{Iwasawa} K, {Sanders} DB, {Evans} AS, {Trentham} N, {Miniutti} G, {Spoon} HWW.
  2005.
\textit{\mnras} 357:565--571

\bibitem[{{Ji} et~al.(2019){Ji}, {Oh} \& {Masterson}}]{Ji2019}
{Ji} S, {Oh} SP, {Masterson} P. 2019.
\textit{\mnras} 487:737--754

\bibitem[{{Jia} et~al.(2012){Jia}, {Ptak}, {Heckman}, {Braito} \&
  {Reeves}}]{Jia2012}
{Jia} J, {Ptak} A, {Heckman} TM, {Braito} V, {Reeves} J. 2012.
\textit{\apj} 759:41

\bibitem[{{Jiang} \& {Oh}(2018)}]{Jiang2018}
{Jiang} YF, {Oh} SP. 2018.
\textit{\apj} 854:5

\bibitem[{{Johnson} \& {Axford}(1971)}]{Johnson1971}
{Johnson} HE, {Axford} WI. 1971.
\textit{\apj} 165:381

\bibitem[{{Johnson} et~al.(2021){Johnson}, {Weinberg}, {Vincenzo}, {Bird},
  {Loebman} et~al.}]{Johnson2021}
{Johnson} JW, {Weinberg} DH, {Vincenzo} F, {Bird} JC, {Loebman} SR, et~al.
  2021.
\textit{\mnras} 508:4484--4511

\bibitem[{{Jones} et~al.(2019){Jones}, {Dowell}, {Lopez Rodriguez}, {Zweibel},
  {Berthoud} et~al.}]{Jones2019_SophiaM82_NGC253}
{Jones} TJ, {Dowell} CD, {Lopez Rodriguez} E, {Zweibel} EG, {Berthoud} M,
  et~al. 2019.
\textit{\apjl} 870:L9

\bibitem[{{Kanjilal} et~al.(2021){Kanjilal}, {Dutta} \&
  {Sharma}}]{Kanjilal2021}
{Kanjilal} V, {Dutta} A, {Sharma} P. 2021.
\textit{\mnras} 501:1143--1159

\bibitem[{{Kannan} et~al.(2019){Kannan}, {Vogelsberger}, {Marinacci},
  {McKinnon}, {Pakmor} \& {Springel}}]{Kannan2019}
{Kannan} R, {Vogelsberger} M, {Marinacci} F, {McKinnon} R, {Pakmor} R,
  {Springel} V. 2019.
\textit{\mnras} 485:117--149

\bibitem[{{Kannan} et~al.(2021){Kannan}, {Vogelsberger}, {Marinacci}, {Sales},
  {Torrey} \& {Hernquist}}]{Kannan2021}
{Kannan} R, {Vogelsberger} M, {Marinacci} F, {Sales} LV, {Torrey} P,
  {Hernquist} L. 2021.
\textit{\mnras} 503:336--343

\bibitem[{{Kempski} \& {Quataert}(2022)}]{Kempski2022}
{Kempski} P, {Quataert} E. 2022.
\textit{\mnras} 514:657--674

\bibitem[{{Kennicutt}(1998)}]{Kennicutt1998}
{Kennicutt} Robert~C. J. 1998.
\textit{\apj} 498:541--552

\bibitem[{{Kennicutt} \& {Evans}(2012)}]{Kennicutt2012}
{Kennicutt} RC, {Evans} NJ. 2012.
\textit{\araa} 50:531--608

\bibitem[{{Kim} \& {Ostriker}(2015{\natexlab{a}})}]{Kim2015_supernovae}
{Kim} CG, {Ostriker} EC. 2015{\natexlab{a}}.
\textit{\apj} 802:99

\bibitem[{{Kim} \&
  {Ostriker}(2015{\natexlab{b}})}]{Kim2015_ostriker_momentum_feedback}
{Kim} CG, {Ostriker} EC. 2015{\natexlab{b}}.
\textit{\apj} 815:67

\bibitem[{{Kim} et~al.(2020{\natexlab{a}}){Kim}, {Ostriker}, {Fielding},
  {Smith}, {Bryan} et~al.}]{Kim2020_TIGRESS}
{Kim} CG, {Ostriker} EC, {Fielding} DB, {Smith} MC, {Bryan} GL, et~al.
  2020{\natexlab{a}}.
\textit{\apjl} 903:L34

\bibitem[{{Kim} et~al.(2017){Kim}, {Ostriker} \&
  {Raileanu}}]{Kim2017_superbubbles}
{Kim} CG, {Ostriker} EC, {Raileanu} R. 2017.
\textit{\apj} 834:25

\bibitem[{{Kim} et~al.(2020{\natexlab{b}}){Kim}, {Ostriker}, {Somerville},
  {Bryan}, {Fielding} et~al.}]{Kim2020_SMAUG}
{Kim} CG, {Ostriker} EC, {Somerville} RS, {Bryan} GL, {Fielding} DB, et~al.
  2020{\natexlab{b}}.
\textit{\apj} 900:61

\bibitem[{{Kim} et~al.(2018){Kim}, {Kim} \& {Ostriker}}]{Kim2018_radiative}
{Kim} JG, {Kim} WT, {Ostriker} EC. 2018.
\textit{\apj} 859:68

\bibitem[{{Kimm} et~al.(2018){Kimm}, {Haehnelt}, {Blaizot}, {Katz},
  {Michel-Dansac} et~al.}]{Kimm2018}
{Kimm} T, {Haehnelt} M, {Blaizot} J, {Katz} H, {Michel-Dansac} L, et~al. 2018.
\textit{\mnras} 475:4617--4635

\bibitem[{{Klein} et~al.(1994){Klein}, {McKee} \& {Colella}}]{Klein1994}
{Klein} RI, {McKee} CF, {Colella} P. 1994.
\textit{\apj} 420:213

\bibitem[{{Kochanek} et~al.(2008){Kochanek}, {Beacom}, {Kistler}, {Prieto},
  {Stanek} et~al.}]{Kochanek2008}
{Kochanek} CS, {Beacom} JF, {Kistler} MD, {Prieto} JL, {Stanek} KZ, et~al.
  2008.
\textit{\apj} 684:1336--1342

\bibitem[{{Konigl} \& {Pudritz}(2000)}]{Konigl2000}
{Konigl} A, {Pudritz} RE. 2000.
\textit{{Disk Winds and the Accretion-Outflow Connection}}. In
  \textit{Protostars and Planets IV}, eds. V~{Mannings}, AP~{Boss},
  SS~{Russell}

\bibitem[{{Kornei} et~al.(2012){Kornei}, {Shapley}, {Martin}, {Coil}, {Lotz}
  et~al.}]{Kornei2012}
{Kornei} KA, {Shapley} AE, {Martin} CL, {Coil} AL, {Lotz} JM, et~al. 2012.
\textit{\apj} 758:135

\bibitem[{{Krause} et~al.(2020){Krause}, {Irwin}, {Schmidt}, {Stein},
  {Miskolczi} et~al.}]{Krause2020}
{Krause} M, {Irwin} J, {Schmidt} P, {Stein} Y, {Miskolczi} A, et~al. 2020.
\textit{\aap} 639:A112

\bibitem[{{Krumholz} et~al.(2019){Krumholz}, {McKee} \&
  {Bland-Hawthorn}}]{Krumholz2019}
{Krumholz} MR, {McKee} CF, {Bland-Hawthorn} J. 2019.
\textit{\araa} 57:227--303

\bibitem[{{Krumholz} \& {Thompson}(2012)}]{Krumholz2012}
{Krumholz} MR, {Thompson} TA. 2012.
\textit{\apj} 760:155

\bibitem[{{Krumholz} \& {Thompson}(2013)}]{Krumholz2013}
{Krumholz} MR, {Thompson} TA. 2013.
\textit{\mnras} 434:2329--2346

\bibitem[{{Krumholz} et~al.(2017){Krumholz}, {Thompson}, {Ostriker} \&
  {Martin}}]{Krumholz2017}
{Krumholz} MR, {Thompson} TA, {Ostriker} EC, {Martin} CL. 2017.
\textit{\mnras} 471:4061--4086

\bibitem[{{Kudoh} \& {Shibata}(1997)}]{Kudoh1997}
{Kudoh} T, {Shibata} K. 1997.
\textit{\apj} 474:362--377

\bibitem[{{Kulsrud} \& {Pearce}(1969)}]{Kulsrud1969}
{Kulsrud} R, {Pearce} WP. 1969.
\textit{\apj} 156:445

\bibitem[{{Lacki} \& {Thompson}(2013)}]{Lacki2013}
{Lacki} BC, {Thompson} TA. 2013.
\textit{\apj} 762:29

\bibitem[{{Lacki} et~al.(2010){Lacki}, {Thompson} \& {Quataert}}]{Lacki2010}
{Lacki} BC, {Thompson} TA, {Quataert} E. 2010.
\textit{\apj} 717:1--28

\bibitem[{{Lacki} et~al.(2011){Lacki}, {Thompson}, {Quataert}, {Loeb} \&
  {Waxman}}]{Lacki2011}
{Lacki} BC, {Thompson} TA, {Quataert} E, {Loeb} A, {Waxman} E. 2011.
\textit{\apj} 734:107

\bibitem[{{Lamers} \& {Cassinelli}(1999)}]{Lamers1999}
{Lamers} HJGLM, {Cassinelli} JP. 1999.
\textit{{Introduction to Stellar Winds}}

\bibitem[{{Lan} \& {Mo}(2018)}]{Lan2018}
{Lan} TW, {Mo} H. 2018.
\textit{\apj} 866:36

\bibitem[{{Lao} \& {Smith}(2020)}]{Lao2020}
{Lao} BX, {Smith} A. 2020.
\textit{\mnras} 497:3925--3942

\bibitem[{{Law} et~al.(2009){Law}, {Steidel}, {Erb}, {Larkin}, {Pettini}
  et~al.}]{Law2009}
{Law} DR, {Steidel} CC, {Erb} DK, {Larkin} JE, {Pettini} M, et~al. 2009.
\textit{\apj} 697:2057--2082

\bibitem[{{Lehnert} \& {Heckman}(1996)}]{Lehnert1996}
{Lehnert} MD, {Heckman} TM. 1996.
\textit{\apj} 462:651

\bibitem[{{Lehnert} et~al.(1999){Lehnert}, {Heckman} \& {Weaver}}]{Lehnert1999}
{Lehnert} MD, {Heckman} TM, {Weaver} KA. 1999.
\textit{\apj} 523:575--584

\bibitem[{{Leitherer} et~al.(1999){Leitherer}, {Schaerer}, {Goldader},
  {Delgado}, {Robert} et~al.}]{Leitherer1999}
{Leitherer} C, {Schaerer} D, {Goldader} JD, {Delgado} RMG, {Robert} C, et~al.
  1999.
\textit{\apjs} 123:3--40

\bibitem[{{Leroy} et~al.(2015){Leroy}, {Walter}, {Martini}, {Roussel},
  {Sandstrom} et~al.}]{Leroy2015}
{Leroy} AK, {Walter} F, {Martini} P, {Roussel} H, {Sandstrom} K, et~al. 2015.
\textit{\apj} 814:83

\bibitem[{{Li} et~al.(2015){Li}, {Ostriker}, {Cen}, {Bryan} \& {Naab}}]{Li2015}
{Li} M, {Ostriker} JP, {Cen} R, {Bryan} GL, {Naab} T. 2015.
\textit{\apj} 814:4

\bibitem[{{Li} et~al.(2020){Li}, {Hopkins}, {Squire} \& {Hummels}}]{Li2020}
{Li} Z, {Hopkins} PF, {Squire} J, {Hummels} C. 2020.
\textit{\mnras} 492:1841--1854

\bibitem[{{Linden} et~al.(2010){Linden}, {Profumo} \& {Anderson}}]{Linden2010}
{Linden} T, {Profumo} S, {Anderson} B. 2010.
\textit{\prd} 82:063529

\bibitem[{{Liu} et~al.(2012){Liu}, {Wang} \& {Mao}}]{Liu2012}
{Liu} J, {Wang} QD, {Mao} S. 2012.
\textit{\mnras} 420:3389--3395

\bibitem[{{Lochhaas} et~al.(2021){Lochhaas}, {Thompson} \&
  {Schneider}}]{Lochhaas2021}
{Lochhaas} C, {Thompson} TA, {Schneider} EE. 2021.
\textit{\mnras} 504:3412--3423

\bibitem[{{Longair}(1994)}]{Longair1994}
{Longair} MS. 1994.
\textit{{High energy astrophysics}}.
vol.~2

\bibitem[{{Loose} et~al.(1982){Loose}, {Kruegel} \& {Tutukov}}]{Loose1982}
{Loose} HH, {Kruegel} E, {Tutukov} A. 1982.
\textit{\aap} 105:342--350

\bibitem[{{Lopez} et~al.(2011){Lopez}, {Krumholz}, {Bolatto}, {Prochaska} \&
  {Ramirez-Ruiz}}]{Lopez2011}
{Lopez} LA, {Krumholz} MR, {Bolatto} AD, {Prochaska} JX, {Ramirez-Ruiz} E.
  2011.
\textit{\apj} 731:91

\bibitem[{{Lopez} et~al.(2020){Lopez}, {Mathur}, {Nguyen}, {Thompson} \&
  {Olivier}}]{Lopez2020}
{Lopez} LA, {Mathur} S, {Nguyen} DD, {Thompson} TA, {Olivier} GM. 2020.
\textit{\apj} 904:152

\bibitem[{{Lopez} et~al.(2023){Lopez}, {Lopez}, {Nguyen}, {Thompson}, {Mathur}
  et~al.}]{Lopez2023}
{Lopez} S, {Lopez} LA, {Nguyen} DD, {Thompson} TA, {Mathur} S, et~al. 2023.
\textit{\apj} 942:108

\bibitem[{{Lopez-Rodriguez} et~al.(2021){Lopez-Rodriguez}, {Guerra},
  {Asgari-Targhi} \& {Schmelz}}]{Lopez-Rodriguez2021_SophiaM82}
{Lopez-Rodriguez} E, {Guerra} JA, {Asgari-Targhi} M, {Schmelz} JT. 2021.
\textit{\apj} 914:24

\bibitem[{{Lucy} \& {Solomon}(1970)}]{Lucy1970}
{Lucy} LB, {Solomon} PM. 1970.
\textit{\apj} 159:879

\bibitem[{{Lynds} \& {Sandage}(1963)}]{Lynds1963}
{Lynds} CR, {Sandage} AR. 1963.
\textit{\apj} 137:1005

\bibitem[{{Mac Low} et~al.(1989){Mac Low}, {McCray} \& {Norman}}]{MacLow1989}
{Mac Low} MM, {McCray} R, {Norman} ML. 1989.
\textit{\apj} 337:141

\bibitem[{{Madau} \& {Dickinson}(2014)}]{Madau2014}
{Madau} P, {Dickinson} M. 2014.
\textit{\araa} 52:415--486

\bibitem[{{Marlowe} et~al.(1995){Marlowe}, {Heckman}, {Wyse} \&
  {Schommer}}]{Marlowe1995}
{Marlowe} AT, {Heckman} TM, {Wyse} RFG, {Schommer} R. 1995.
\textit{\apj} 438:563

\bibitem[{{Martin}(1998)}]{Martin1998}
{Martin} CL. 1998.
\textit{\apj} 506:222--252

\bibitem[{{Martin}(1999)}]{Martin1999}
{Martin} CL. 1999.
\textit{\apj} 513:156--160

\bibitem[{{Martin}(2005)}]{Martin2005}
{Martin} CL. 2005.
\textit{\apj} 621:227--245

\bibitem[{{Martin} et~al.(2002){Martin}, {Kobulnicky} \&
  {Heckman}}]{Martin2002}
{Martin} CL, {Kobulnicky} HA, {Heckman} TM. 2002.
\textit{\apj} 574:663--692

\bibitem[{{Martin} et~al.(2012){Martin}, {Shapley}, {Coil}, {Kornei}, {Bundy}
  et~al.}]{Martin2012}
{Martin} CL, {Shapley} AE, {Coil} AL, {Kornei} KA, {Bundy} K, et~al. 2012.
\textit{\apj} 760:127

\bibitem[{{Martini} et~al.(2018){Martini}, {Leroy}, {Mangum}, {Bolatto},
  {Keating} et~al.}]{Martini2018}
{Martini} P, {Leroy} AK, {Mangum} JG, {Bolatto} A, {Keating} KM, et~al. 2018.
\textit{\apj} 856:61

\bibitem[{{Mathews} \& {Baker}(1971)}]{Mathews1971}
{Mathews} WG, {Baker} JC. 1971.
\textit{\apj} 170:241

\bibitem[{{Mathis} et~al.(1977){Mathis}, {Rumpl} \& {Nordsieck}}]{Mathis1977}
{Mathis} JS, {Rumpl} W, {Nordsieck} KH. 1977.
\textit{\apj} 217:425--433

\bibitem[{{McClellan} et~al.(2022){McClellan}, {Davis} \&
  {Arras}}]{McClellan2022}
{McClellan} BC, {Davis} SW, {Arras} P. 2022.
\textit{\apj} 934:37

\bibitem[{{McCourt} et~al.(2015){McCourt}, {O'Leary}, {Madigan} \&
  {Quataert}}]{McCourt2015}
{McCourt} M, {O'Leary} RM, {Madigan} AM, {Quataert} E. 2015.
\textit{\mnras} 449:2--7

\bibitem[{{McKee} \& {Ostriker}(2007)}]{Mckee2007}
{McKee} CF, {Ostriker} EC. 2007.
\textit{\araa} 45:565--687

\bibitem[{{McKee} \& {Ostriker}(1977)}]{McKee1977}
{McKee} CF, {Ostriker} JP. 1977.
\textit{\apj} 218:148--169

\bibitem[{{M{\'e}nard} et~al.(2010){M{\'e}nard}, {Scranton}, {Fukugita} \&
  {Richards}}]{Menard2010}
{M{\'e}nard} B, {Scranton} R, {Fukugita} M, {Richards} G. 2010.
\textit{\mnras} 405:1025--1039

\bibitem[{{Menon} et~al.(2022){Menon}, {Federrath} \& {Krumholz}}]{Menon2022}
{Menon} SH, {Federrath} C, {Krumholz} MR. 2022.
\textit{\mnras} 517:1313--1338

\bibitem[{{Menon} et~al.(2023){Menon}, {Federrath} \& {Krumholz}}]{Menon2023}
{Menon} SH, {Federrath} C, {Krumholz} MR. 2023.
\textit{\mnras} 521:5160--5176

\bibitem[{{Minchev} et~al.(2013){Minchev}, {Chiappini} \&
  {Martig}}]{Minchev2013}
{Minchev} I, {Chiappini} C, {Martig} M. 2013.
\textit{\aap} 558:A9

\bibitem[{{Mitsuishi} et~al.(2011){Mitsuishi}, {Yamasaki} \&
  {Takei}}]{Mitsuishi2011}
{Mitsuishi} I, {Yamasaki} NY, {Takei} Y. 2011.
\textit{\apjl} 742:L31

\bibitem[{{Modak} et~al.(2023){Modak}, {Quataert}, {Jiang} \&
  {Thompson}}]{Modak2023}
{Modak} S, {Quataert} E, {Jiang} YF, {Thompson} TA. 2023.
\textit{\mnras} 524:6374--6391

\bibitem[{{Moore} \& {Spiegel}(1968)}]{Moore1968}
{Moore} DW, {Spiegel} EA. 1968.
\textit{\apj} 154:863

\bibitem[{{Murray} et~al.(2007){Murray}, {Martin}, {Quataert} \&
  {Thompson}}]{Murray2007}
{Murray} N, {Martin} CL, {Quataert} E, {Thompson} TA. 2007.
\textit{\apj} 660:211--220

\bibitem[{{Murray} et~al.(2011){Murray}, {M{\'e}nard} \&
  {Thompson}}]{Murray2011}
{Murray} N, {M{\'e}nard} B, {Thompson} TA. 2011.
\textit{\apj} 735:66

\bibitem[{{Murray} et~al.(2005){Murray}, {Quataert} \& {Thompson}}]{Murray2005}
{Murray} N, {Quataert} E, {Thompson} TA. 2005.
\textit{\apj} 618:569--585

\bibitem[{{Murray} et~al.(2010){Murray}, {Quataert} \& {Thompson}}]{Murray2010}
{Murray} N, {Quataert} E, {Thompson} TA. 2010.
\textit{\apj} 709:191--209

\bibitem[{{Murray} \& {Rahman}(2010)}]{Murray_Rahman2010}
{Murray} N, {Rahman} M. 2010.
\textit{\apj} 709:424--435

\bibitem[{{Murray-Clay} et~al.(2009){Murray-Clay}, {Chiang} \&
  {Murray}}]{Murray-Clay2009}
{Murray-Clay} RA, {Chiang} EI, {Murray} N. 2009.
\textit{\apj} 693:23--42

\bibitem[{{Mutchler} et~al.(2007){Mutchler}, {Bond}, {Christian}, {Frattare},
  {Hamilton} et~al.}]{Mutchler2007}
{Mutchler} M, {Bond} HE, {Christian} CA, {Frattare} LM, {Hamilton} F, et~al.
  2007.
\textit{\pasp} 119:1--6

\bibitem[{{Naab} \& {Ostriker}(2017)}]{Naab2017}
{Naab} T, {Ostriker} JP. 2017.
\textit{\araa} 55:59--109

\bibitem[{{Nardini} et~al.(2013){Nardini}, {Wang}, {Fabbiano}, {Elvis},
  {Pellegrini} et~al.}]{Nardini2013}
{Nardini} E, {Wang} J, {Fabbiano} G, {Elvis} M, {Pellegrini} S, et~al. 2013.
\textit{\apj} 765:141

\bibitem[{{Nguyen} \& {Thompson}(2021)}]{Nguyen2021}
{Nguyen} DD, {Thompson} TA. 2021.
\textit{\mnras} 508:5310--5325

\bibitem[{{Nguyen} \& {Thompson}(2022)}]{Nguyen2022}
{Nguyen} DD, {Thompson} TA. 2022.
\textit{\apjl} 935:L24

\bibitem[{{Nguyen} et~al.(2024){Nguyen}, {Thompson}, {Schneider} \&
  {Tarrant}}]{Nguyen2024}
{Nguyen} DD, {Thompson} TA, {Schneider} EE, {Tarrant} AP. 2024.
\textit{\mnras}

\bibitem[{{O'dell} et~al.(1967){O'dell}, {York} \& {Henize}}]{Odell1967}
{O'dell} CR, {York} DG, {Henize} KG. 1967.
\textit{\apj} 150:835

\bibitem[{{Oh} \& {Haiman}(2002)}]{Oh2002}
{Oh} SP, {Haiman} Z. 2002.
\textit{\apj} 569:558--572

\bibitem[{{Oppenheimer} \& {Dav{\'e}}(2008)}]{Oppenheimer2008}
{Oppenheimer} BD, {Dav{\'e}} R. 2008.
\textit{\mnras} 387:577--600

\bibitem[{{Oppenheimer} et~al.(2010){Oppenheimer}, {Dav{\'e}}, {Kere{\v s}},
  {Fardal}, {Katz} et~al.}]{Oppenheimer2010}
{Oppenheimer} BD, {Dav{\'e}} R, {Kere{\v s}} D, {Fardal} M, {Katz} N, et~al.
  2010.
\textit{\mnras} 406:2325--2338

\bibitem[{{Ostriker} \& {Shetty}(2011)}]{Ostriker2011_maximalI}
{Ostriker} EC, {Shetty} R. 2011.
\textit{\apj} 731:41

\bibitem[{{Overzier} et~al.(2008){Overzier}, {Heckman}, {Kauffmann}, {Seibert},
  {Rich} et~al.}]{Overzier2008}
{Overzier} RA, {Heckman} TM, {Kauffmann} G, {Seibert} M, {Rich} RM, et~al.
  2008.
\textit{\apj} 677:37--62

\bibitem[{{Overzier} et~al.(2010){Overzier}, {Heckman}, {Schiminovich},
  {Basu-Zych}, {Gon{\c{c}}alves} et~al.}]{Overzier2010}
{Overzier} RA, {Heckman} TM, {Schiminovich} D, {Basu-Zych} A, {Gon{\c{c}}alves}
  T, et~al. 2010.
\textit{\apj} 710:979--991

\bibitem[{{Owocki} \& {Gayley}(1997)}]{Owocki1997}
{Owocki} SP, {Gayley} KG. 1997.
\textit{{ThePhysics of Stellar Winds Near the Eddington Limit}}. In
  \textit{Luminous Blue Variables: Massive Stars in Transition}, eds. A~{Nota},
  H~{Lamers}, vol. 120 of \textit{Astronomical Society of the Pacific
  Conference Series}

\bibitem[{{Owocki} et~al.(2017){Owocki}, {Townsend} \& {Quataert}}]{Owocki2017}
{Owocki} SP, {Townsend} RHD, {Quataert} E. 2017.
\textit{\mnras} 472:3749--3760

\bibitem[{{Pakmor} et~al.(2016){Pakmor}, {Pfrommer}, {Simpson} \&
  {Springel}}]{Pakmor2016}
{Pakmor} R, {Pfrommer} C, {Simpson} CM, {Springel} V. 2016.
\textit{\apjl} 824:L30

\bibitem[{{Pakmor} \& {Springel}(2013)}]{Pakmor2013}
{Pakmor} R, {Springel} V. 2013.
\textit{\mnras} 432:176--193

\bibitem[{{Parker}(1958)}]{Parker1958}
{Parker} EN. 1958.
\textit{\apj} 128:664

\bibitem[{{Parker}(1960)}]{Parker1960}
{Parker} EN. 1960.
\textit{\apj} 132:175

\bibitem[{{Peek} et~al.(2015){Peek}, {M{\'e}nard} \& {Corrales}}]{Peek2015}
{Peek} JEG, {M{\'e}nard} B, {Corrales} L. 2015.
\textit{\apj} 813:7

\bibitem[{{Pejcha} \& {Thompson}(2015)}]{Pejcha2015_landscape}
{Pejcha} O, {Thompson} TA. 2015.
\textit{\apj} 801:90

\bibitem[{{Pelletier} \& {Pudritz}(1992)}]{Pelletier1992}
{Pelletier} G, {Pudritz} RE. 1992.
\textit{\apj} 394:117

\bibitem[{{Pereira-Santaella} et~al.(2011){Pereira-Santaella},
  {Alonso-Herrero}, {Santos-Lleo}, {Colina}, {Jim{\'e}nez-Bail{\'o}n}
  et~al.}]{Pereira-Santaella2011}
{Pereira-Santaella} M, {Alonso-Herrero} A, {Santos-Lleo} M, {Colina} L,
  {Jim{\'e}nez-Bail{\'o}n} E, et~al. 2011.
\textit{\aap} 535:A93

\bibitem[{{Perna} et~al.(2020){Perna}, {Arribas}, {Catal{\'a}n-Torrecilla},
  {Colina}, {Bellocchi} et~al.}]{Perna2020}
{Perna} M, {Arribas} S, {Catal{\'a}n-Torrecilla} C, {Colina} L, {Bellocchi} E,
  et~al. 2020.
\textit{\aap} 643:A139

\bibitem[{{P{\'e}roux} \& {Howk}(2020)}]{Peroux2020}
{P{\'e}roux} C, {Howk} JC. 2020.
\textit{\araa} 58:363--406

\bibitem[{{Perrotta} et~al.(2023){Perrotta}, {Coil}, {Rupke}, {Tremonti},
  {Davis} et~al.}]{Perrotta2023}
{Perrotta} S, {Coil} AL, {Rupke} DSN, {Tremonti} CA, {Davis} JD, et~al. 2023.
\textit{\apj} 949:9

\bibitem[{{Persic} et~al.(2008){Persic}, {Rephaeli} \& {Arieli}}]{Persic2008}
{Persic} M, {Rephaeli} Y, {Arieli} Y. 2008.
\textit{\aap} 486:143--149

\bibitem[{{Pfrommer} et~al.(2017){Pfrommer}, {Pakmor}, {Simpson} \&
  {Springel}}]{Pfrommer2017}
{Pfrommer} C, {Pakmor} R, {Simpson} CM, {Springel} V. 2017.
\textit{\apjl} 847:L13

\bibitem[{{Poynting}(1904)}]{Poynting1904}
{Poynting} JH. 1904.
\textit{Philosophical Transactions of the Royal Society of London Series A}
  202:525--552

\bibitem[{{Price} et~al.(2016){Price}, {Kriek}, {Shapley}, {Reddy}, {Freeman}
  et~al.}]{Price2016}
{Price} SH, {Kriek} M, {Shapley} AE, {Reddy} NA, {Freeman} WR, et~al. 2016.
\textit{\apj} 819:80

\bibitem[{{Prieto} et~al.(2008){Prieto}, {Kistler}, {Thompson}, {Y{\"u}ksel},
  {Kochanek} et~al.}]{Prieto2008}
{Prieto} JL, {Kistler} MD, {Thompson} TA, {Y{\"u}ksel} H, {Kochanek} CS, et~al.
  2008.
\textit{\apjl} 681:L9

\bibitem[{{Prochaska} et~al.(2011){Prochaska}, {Kasen} \&
  {Rubin}}]{Prochaska2011}
{Prochaska} JX, {Kasen} D, {Rubin} K. 2011.
\textit{\apj} 734:24

\bibitem[{{Pudritz} \& {Norman}(1986)}]{Pudritz1986}
{Pudritz} RE, {Norman} CA. 1986.
\textit{\apj} 301:571

\bibitem[{{Quataert} et~al.(2022{\natexlab{a}}){Quataert}, {Jiang} \&
  {Thompson}}]{Quataert2022b}
{Quataert} E, {Jiang} YF, {Thompson} TA. 2022{\natexlab{a}}.
\textit{\mnras} 510:920--945

\bibitem[{{Quataert} et~al.(2022{\natexlab{b}}){Quataert}, {Thompson} \&
  {Jiang}}]{Quataert2022a}
{Quataert} E, {Thompson} TA, {Jiang} YF. 2022{\natexlab{b}}.
\textit{\mnras} 510:1184--1203

\bibitem[{{Ranalli} et~al.(2008){Ranalli}, {Comastri}, {Origlia} \&
  {Maiolino}}]{Ranalli2008}
{Ranalli} P, {Comastri} A, {Origlia} L, {Maiolino} R. 2008.
\textit{\mnras} 386:1464--1480

\bibitem[{{Rangwala} et~al.(2011){Rangwala}, {Maloney}, {Glenn}, {Wilson},
  {Rykala} et~al.}]{Rangwala2011}
{Rangwala} N, {Maloney} PR, {Glenn} J, {Wilson} CD, {Rykala} A, et~al. 2011.
\textit{\apj} 743:94

\bibitem[{{Raskutti} et~al.(2016){Raskutti}, {Ostriker} \&
  {Skinner}}]{Raskutti2016}
{Raskutti} S, {Ostriker} EC, {Skinner} MA. 2016.
\textit{\apj} 829:130

\bibitem[{{Raskutti} et~al.(2017){Raskutti}, {Ostriker} \&
  {Skinner}}]{Raskutti2017}
{Raskutti} S, {Ostriker} EC, {Skinner} MA. 2017.
\textit{\apj} 850:112

\bibitem[{{Rathjen} et~al.(2021){Rathjen}, {Naab}, {Girichidis}, {Walch},
  {W{\"u}nsch} et~al.}]{Rathjen2021}
{Rathjen} TE, {Naab} T, {Girichidis} P, {Walch} S, {W{\"u}nsch} R, et~al. 2021.
\textit{\mnras} 504:1039--1061

\bibitem[{{Rathjen} et~al.(2023){Rathjen}, {Naab}, {Walch}, {Seifried},
  {Girichidis} \& {W{\"u}nsch}}]{Rathjen2023}
{Rathjen} TE, {Naab} T, {Walch} S, {Seifried} D, {Girichidis} P, {W{\"u}nsch}
  R. 2023.
\textit{\mnras} 522:1843--1862

\bibitem[{{Richings} \& {Faucher-Gigu{\`e}re}(2018)}]{Richings2018}
{Richings} AJ, {Faucher-Gigu{\`e}re} CA. 2018.
\textit{\mnras} 474:3673--3699

\bibitem[{{Roberg-Clark} et~al.(2018){Roberg-Clark}, {Drake}, {Reynolds} \&
  {Swisdak}}]{Roberg-Clark2018}
{Roberg-Clark} GT, {Drake} JF, {Reynolds} CS, {Swisdak} M. 2018.
\textit{\prl} 120:035101

\bibitem[{{Roberts-Borsani} \& {Saintonge}(2019)}]{Borsani2019}
{Roberts-Borsani} GW, {Saintonge} A. 2019.
\textit{\mnras} 482:4111--4145

\bibitem[{{Robertson}(2022)}]{Robertson2022}
{Robertson} BE. 2022.
\textit{\araa} 60:121--158

\bibitem[{{Rosdahl} \& {Teyssier}(2015)}]{Rosdahl2015}
{Rosdahl} J, {Teyssier} R. 2015.
\textit{\mnras} 449:4380--4403

\bibitem[{{Rougoor} \& {Oort}(1959)}]{Rougoor1959}
{Rougoor} GW, {Oort} JH. 1959.
\textit{{Neutral hydrogen in the central part of the galactic system}}. In
  \textit{URSI Symp. 1: Paris Symposium on Radio Astronomy}, ed.
  RN~{Bracewell}, vol.~9 of \textit{IAU Symposium}

\bibitem[{{Roussel} et~al.(2010){Roussel}, {Wilson}, {Vigroux}, {Isaak},
  {Sauvage} et~al.}]{Roussel2010}
{Roussel} H, {Wilson} CD, {Vigroux} L, {Isaak} KG, {Sauvage} M, et~al. 2010.
\textit{\aap} 518:L66

\bibitem[{{Rubin} et~al.(2014){Rubin}, {Prochaska}, {Koo}, {Phillips}, {Martin}
  \& {Winstrom}}]{Rubin2014}
{Rubin} KHR, {Prochaska} JX, {Koo} DC, {Phillips} AC, {Martin} CL, {Winstrom}
  LO. 2014.
\textit{\apj} 794:156

\bibitem[{{Rubin} et~al.(2010){Rubin}, {Weiner}, {Koo}, {Martin}, {Prochaska}
  et~al.}]{Rubin2010}
{Rubin} KHR, {Weiner} BJ, {Koo} DC, {Martin} CL, {Prochaska} JX, et~al. 2010.
\textit{\apj} 719:1503--1525

\bibitem[{{Runnholm} et~al.(2023){Runnholm}, {Hayes}, {Lin}, {Melinder},
  {Scarlata} et~al.}]{Runnholm2023}
{Runnholm} A, {Hayes} MJ, {Lin} YH, {Melinder} J, {Scarlata} C, et~al. 2023.
\textit{\mnras} 522:4275--4293

\bibitem[{{Rupke} et~al.(2005){Rupke}, {Veilleux} \& {Sanders}}]{Rupke2005}
{Rupke} DS, {Veilleux} S, {Sanders} DB. 2005.
\textit{\apjs} 160:115--148

\bibitem[{{Rupke} et~al.(2019){Rupke}, {Coil}, {Geach}, {Tremonti},
  {Diamond-Stanic} et~al.}]{Rupke2019}
{Rupke} DSN, {Coil} A, {Geach} JE, {Tremonti} C, {Diamond-Stanic} AM, et~al.
  2019.
\textit{\nat} 574:643--646

\bibitem[{{Ruszkowski} \& {Pfrommer}(2023)}]{Ruszkowski2023}
{Ruszkowski} M, {Pfrommer} C. 2023.
\textit{\aapr} 31:4

\bibitem[{{Sarkar} et~al.(2022){Sarkar}, {Sternberg} \& {Gnat}}]{Sarkar2022}
{Sarkar} KC, {Sternberg} A, {Gnat} O. 2022.
\textit{\apj} 940:44

\bibitem[{{Sato} et~al.(2009){Sato}, {Martin}, {Noeske}, {Koo} \&
  {Lotz}}]{Sato2009}
{Sato} T, {Martin} CL, {Noeske} KG, {Koo} DC, {Lotz} JM. 2009.
\textit{\apj} 696:214--232

\bibitem[{{Scannapieco} \& {Br{\"u}ggen}(2015)}]{Scannapieco2015}
{Scannapieco} E, {Br{\"u}ggen} M. 2015.
\textit{\apj} 805:158

\bibitem[{{Scannapieco} et~al.(2012){Scannapieco}, {Gray} \&
  {Pan}}]{Scannapieco2012}
{Scannapieco} E, {Gray} WJ, {Pan} L. 2012.
\textit{\apj} 746:57

\bibitem[{{Scarlata} \& {Panagia}(2015)}]{Scarlata2015}
{Scarlata} C, {Panagia} N. 2015.
\textit{\apj} 801:43

\bibitem[{{Scarrott} et~al.(1991){Scarrott}, {Eaton} \& {Axon}}]{Scarrott1991}
{Scarrott} SM, {Eaton} N, {Axon} DJ. 1991.
\textit{\mnras} 252:12P

\bibitem[{{Schal{\'e}n}(1939)}]{Schalen1939}
{Schal{\'e}n} C. 1939.
\textit{\zap} 17:260

\bibitem[{{Schatzman}(1962)}]{Schatzman1962}
{Schatzman} E. 1962.
\textit{Annales d'Astrophysique} 25:18

\bibitem[{{Schneider} et~al.(2020){Schneider}, {Ostriker}, {Robertson} \&
  {Thompson}}]{Schneider2020}
{Schneider} EE, {Ostriker} EC, {Robertson} BE, {Thompson} TA. 2020.
\textit{\apj} 895:43

\bibitem[{{Schneider} \& {Robertson}(2015)}]{Schneider2015}
{Schneider} EE, {Robertson} BE. 2015.
\textit{\apjs} 217:24

\bibitem[{{Schneider} \& {Robertson}(2017)}]{Schneider2017}
{Schneider} EE, {Robertson} BE. 2017.
\textit{\apj} 834:144

\bibitem[{{Schneider} et~al.(2018){Schneider}, {Sana}, {Evans}, {Bestenlehner},
  {Castro} et~al.}]{Schneider2018_IMF}
{Schneider} FRN, {Sana} H, {Evans} CJ, {Bestenlehner} JM, {Castro} N, et~al.
  2018.
\textit{Science} 359:69--71

\bibitem[{{Schoenberg} \& {Jung}(1933)}]{Schoenberg1933}
{Schoenberg} E, {Jung} B. 1933.
\textit{Astronomische Nachrichten} 247:413

\bibitem[{{Schroetter} et~al.(2019){Schroetter}, {Bouch{\'e}}, {Zabl},
  {Contini}, {Wendt} et~al.}]{Schroetter2019}
{Schroetter} I, {Bouch{\'e}} NF, {Zabl} J, {Contini} T, {Wendt} M, et~al. 2019.
\textit{\mnras} 490:4368--4381

\bibitem[{{Scoville} et~al.(2017){Scoville}, {Murchikova}, {Walter},
  {Vlahakis}, {Koda} et~al.}]{Scoville2017}
{Scoville} N, {Murchikova} L, {Walter} F, {Vlahakis} C, {Koda} J, et~al. 2017.
\textit{\apj} 836:66

\bibitem[{{Scoville} et~al.(2015){Scoville}, {Sheth}, {Walter}, {Manohar},
  {Zschaechner} et~al.}]{Scoville2015}
{Scoville} N, {Sheth} K, {Walter} F, {Manohar} S, {Zschaechner} L, et~al. 2015.
\textit{\apj} 800:70

\bibitem[{{Seaquist} \& {Odegard}(1991)}]{Seaquist1991}
{Seaquist} ER, {Odegard} N. 1991.
\textit{\apj} 369:320

\bibitem[{{Seon} \& {Kim}(2020)}]{Seon2020}
{Seon} Ki, {Kim} CG. 2020.
\textit{\apjs} 250:9

\bibitem[{{Shaban} et~al.(2022){Shaban}, {Bordoloi}, {Chisholm}, {Sharma},
  {Sharon} et~al.}]{Shaban2022}
{Shaban} A, {Bordoloi} R, {Chisholm} J, {Sharma} S, {Sharon} K, et~al. 2022.
\textit{\apj} 936:77

\bibitem[{{Shapiro} et~al.(2004){Shapiro}, {Iliev} \& {Raga}}]{Shapiro2004}
{Shapiro} PR, {Iliev} IT, {Raga} AC. 2004.
\textit{\mnras} 348:753--782

\bibitem[{{Shapley}(2011)}]{Shapley2011}
{Shapley} AE. 2011.
\textit{\araa} 49:525--580

\bibitem[{{Shapley} et~al.(2015){Shapley}, {Reddy}, {Kriek}, {Freeman},
  {Sanders} et~al.}]{Shapley2015}
{Shapley} AE, {Reddy} NA, {Kriek} M, {Freeman} WR, {Sanders} RL, et~al. 2015.
\textit{\apj} 801:88

\bibitem[{{Shapley} et~al.(2003){Shapley}, {Steidel}, {Pettini} \&
  {Adelberger}}]{Shapley2003}
{Shapley} AE, {Steidel} CC, {Pettini} M, {Adelberger} KL. 2003.
\textit{\apj} 588:65--89

\bibitem[{{Sharma} et~al.(2014){Sharma}, {Roy}, {Nath} \&
  {Shchekinov}}]{Sharma2014}
{Sharma} P, {Roy} A, {Nath} BB, {Shchekinov} Y. 2014.
\textit{\mnras} 443:3463--3476

\bibitem[{{Shetty} \& {Ostriker}(2012)}]{Shetty2012_ostriker_maximalII}
{Shetty} R, {Ostriker} EC. 2012.
\textit{\apj} 754:2

\bibitem[{{Shopbell} \& {Bland-Hawthorn}(1998)}]{Shopbell1998}
{Shopbell} PL, {Bland-Hawthorn} J. 1998.
\textit{\apj} 493:129--153

\bibitem[{{Silich} et~al.(2003){Silich}, {Tenorio-Tagle} \&
  {Mu{\~n}oz-Tu{\~n}{\'o}n}}]{Silich2003}
{Silich} S, {Tenorio-Tagle} G, {Mu{\~n}oz-Tu{\~n}{\'o}n} C. 2003.
\textit{\apj} 590:791--796

\bibitem[{{Silich} et~al.(2004){Silich}, {Tenorio-Tagle} \&
  {Rodr{\'\i}guez-Gonz{\'a}lez}}]{Silich2004}
{Silich} S, {Tenorio-Tagle} G, {Rodr{\'\i}guez-Gonz{\'a}lez} A. 2004.
\textit{\apj} 610:226--232

\bibitem[{{Sirko} \& {Goodman}(2003)}]{Sirko2003}
{Sirko} E, {Goodman} J. 2003.
\textit{\mnras} 341:501--508

\bibitem[{{Skinner} \& {Ostriker}(2015)}]{Skinner2015}
{Skinner} MA, {Ostriker} EC. 2015.
\textit{\apj} 809:187

\bibitem[{{Smith} et~al.(2017){Smith}, {Bromm} \& {Loeb}}]{Smith2017}
{Smith} A, {Bromm} V, {Loeb} A. 2017.
\textit{\mnras} 464:2963--2978

\bibitem[{{Smith} et~al.(2020){Smith}, {Kannan}, {Tsang}, {Vogelsberger} \&
  {Pakmor}}]{Smith2020}
{Smith} A, {Kannan} R, {Tsang} BTH, {Vogelsberger} M, {Pakmor} R. 2020.
\textit{\apj} 905:27

\bibitem[{{Smith}(2013)}]{Smith2013}
{Smith} N. 2013.
\textit{\mnras} 434:102--113

\bibitem[{{Socrates} et~al.(2008){Socrates}, {Davis} \&
  {Ramirez-Ruiz}}]{Socrates2008}
{Socrates} A, {Davis} SW, {Ramirez-Ruiz} E. 2008.
\textit{\apj} 687:202--215

\bibitem[{{Somerville} \& {Dav{\'e}}(2015)}]{Somerville2015}
{Somerville} RS, {Dav{\'e}} R. 2015.
\textit{\araa} 53:51--113

\bibitem[{{Sparre} et~al.(2020){Sparre}, {Pfrommer} \& {Ehlert}}]{Sparre2020}
{Sparre} M, {Pfrommer} C, {Ehlert} K. 2020.
\textit{\mnras} 499:4261--4281

\bibitem[{{Sparre} et~al.(2019){Sparre}, {Pfrommer} \&
  {Vogelsberger}}]{Sparre2019}
{Sparre} M, {Pfrommer} C, {Vogelsberger} M. 2019.
\textit{\mnras} 482:5401--5421

\bibitem[{{Spruit}(2010)}]{Spruit2010}
{Spruit} HC. 2010.
\textit{{Theory of Magnetically Powered Jets}}. In \textit{Lecture Notes in
  Physics, Berlin Springer Verlag}, ed. T~{Belloni}, vol. 794.  233

\bibitem[{{Squire} et~al.(2022){Squire}, {Moroianu} \& {Hopkins}}]{Squire2022}
{Squire} J, {Moroianu} S, {Hopkins} PF. 2022.
\textit{\mnras} 510:110--130

\bibitem[{{Steidel} et~al.(2018){Steidel}, {Bogosavljevi{\'c}}, {Shapley},
  {Reddy}, {Rudie} et~al.}]{Steidel2018}
{Steidel} CC, {Bogosavljevi{\'c}} M, {Shapley} AE, {Reddy} NA, {Rudie} GC,
  et~al. 2018.
\textit{\apj} 869:123

\bibitem[{{Steidel} et~al.(2010){Steidel}, {Erb}, {Shapley}, {Pettini}, {Reddy}
  et~al.}]{Steidel2010}
{Steidel} CC, {Erb} DK, {Shapley} AE, {Pettini} M, {Reddy} N, et~al. 2010.
\textit{\apj} 717:289--322

\bibitem[{{Steinwandel} et~al.(2020){Steinwandel}, {Dolag}, {Lesch}, {Moster},
  {Burkert} \& {Prieto}}]{Steinwandel2020}
{Steinwandel} UP, {Dolag} K, {Lesch} H, {Moster} BP, {Burkert} A, {Prieto} A.
  2020.
\textit{\mnras} 494:4393--4412

\bibitem[{{Strickland} \& {Heckman}(2007)}]{Strickland2007}
{Strickland} DK, {Heckman} TM. 2007.
\textit{\apj} 658:258--281

\bibitem[{{Strickland} \& {Heckman}(2009)}]{Strickland2009}
{Strickland} DK, {Heckman} TM. 2009.
\textit{\apj} 697:2030--2056

\bibitem[{{Strickland} et~al.(2004){Strickland}, {Heckman}, {Colbert}, {Hoopes}
  \& {Weaver}}]{Strickland2004}
{Strickland} DK, {Heckman} TM, {Colbert} EJM, {Hoopes} CG, {Weaver} KA. 2004.
\textit{\apj} 606:829--852

\bibitem[{{Strickland} et~al.(2002){Strickland}, {Heckman}, {Weaver}, {Hoopes}
  \& {Dahlem}}]{Strickland2002}
{Strickland} DK, {Heckman} TM, {Weaver} KA, {Hoopes} CG, {Dahlem} M. 2002.
\textit{\apj} 568:689--716

\bibitem[{{Strickland} \& {Stevens}(2000)}]{Strickland2000}
{Strickland} DK, {Stevens} IR. 2000.
\textit{\mnras} 314:511--545

\bibitem[{{Strong} et~al.(2007){Strong}, {Moskalenko} \&
  {Ptuskin}}]{Strong2007}
{Strong} AW, {Moskalenko} IV, {Ptuskin} VS. 2007.
\textit{Annual Review of Nuclear and Particle Science} 57:285--327

\bibitem[{{Suchkov} et~al.(1996){Suchkov}, {Berman}, {Heckman} \&
  {Balsara}}]{Suchkov1996}
{Suchkov} AA, {Berman} VG, {Heckman} TM, {Balsara} DS. 1996.
\textit{\apj} 463:528--534

\bibitem[{{Sugahara} et~al.(2019){Sugahara}, {Ouchi}, {Harikane}, {Bouch{\'e}},
  {Mitchell} \& {Blaizot}}]{Sugahara2019}
{Sugahara} Y, {Ouchi} M, {Harikane} Y, {Bouch{\'e}} N, {Mitchell} PD, {Blaizot}
  J. 2019.
\textit{\apj} 886:29

\bibitem[{{Sugahara} et~al.(2017){Sugahara}, {Ouchi}, {Lin}, {Martin}, {Ono}
  et~al.}]{Sugahara2017}
{Sugahara} Y, {Ouchi} M, {Lin} L, {Martin} CL, {Ono} Y, et~al. 2017.
\textit{\apj} 850:51

\bibitem[{{Sukhbold} et~al.(2016){Sukhbold}, {Ertl}, {Woosley}, {Brown} \&
  {Janka}}]{Sukhbold2016}
{Sukhbold} T, {Ertl} T, {Woosley} SE, {Brown} JM, {Janka} HT. 2016.
\textit{\apj} 821:38

\bibitem[{{Sur} et~al.(2016){Sur}, {Scannapieco} \& {Ostriker}}]{Sur2016}
{Sur} S, {Scannapieco} E, {Ostriker} EC. 2016.
\textit{\apj} 818:28

\bibitem[{{Sutherland} \& {Dopita}(1993)}]{Sutherland1993}
{Sutherland} RS, {Dopita} MA. 1993.
\textit{\apjs} 88:253

\bibitem[{{Tacconi} et~al.(2020){Tacconi}, {Genzel} \&
  {Sternberg}}]{Tacconi2020}
{Tacconi} LJ, {Genzel} R, {Sternberg} A. 2020.
\textit{\araa} 58:157--203

\bibitem[{{Tan} \& {Fielding}(2024)}]{Tan2024}
{Tan} B, {Fielding} DB. 2024.
\textit{\mnras} 527:9683--9714

\bibitem[{{Tan} et~al.(2021){Tan}, {Oh} \& {Gronke}}]{Tan2021}
{Tan} B, {Oh} SP, {Gronke} M. 2021.
\textit{\mnras} 502:3179--3199

\bibitem[{{Thomas} et~al.(2023){Thomas}, {Pfrommer} \& {Pakmor}}]{Thomas2023}
{Thomas} T, {Pfrommer} C, {Pakmor} R. 2023.
\textit{\mnras} 521:3023--3042

\bibitem[{{Thompson} et~al.(2015){Thompson}, {Fabian}, {Quataert} \&
  {Murray}}]{Thompson2015}
{Thompson} TA, {Fabian} AC, {Quataert} E, {Murray} N. 2015.
\textit{\mnras} 449:147--161

\bibitem[{{Thompson} \& {Krumholz}(2016)}]{Thompson_Krumholz}
{Thompson} TA, {Krumholz} MR. 2016.
\textit{\mnras} 455:334--342

\bibitem[{{Thompson} et~al.(2009){Thompson}, {Prieto}, {Stanek}, {Kistler},
  {Beacom} \& {Kochanek}}]{Thompson2009}
{Thompson} TA, {Prieto} JL, {Stanek} KZ, {Kistler} MD, {Beacom} JF, {Kochanek}
  CS. 2009.
\textit{\apj} 705:1364--1384

\bibitem[{{Thompson} et~al.(2005){Thompson}, {Quataert} \&
  {Murray}}]{Thompson2005}
{Thompson} TA, {Quataert} E, {Murray} N. 2005.
\textit{\apj} 630:167--185

\bibitem[{{Thompson} et~al.(2006){Thompson}, {Quataert}, {Waxman}, {Murray} \&
  {Martin}}]{Thompson2006}
{Thompson} TA, {Quataert} E, {Waxman} E, {Murray} N, {Martin} CL. 2006.
\textit{\apj} 645:186--198

\bibitem[{{Thompson} et~al.(2016){Thompson}, {Quataert}, {Zhang} \&
  {Weinberg}}]{Thompson2016}
{Thompson} TA, {Quataert} E, {Zhang} D, {Weinberg} DH. 2016.
\textit{\mnras} 455:1830--1844

\bibitem[{{Tomaselli} \& {Ferrara}(2021)}]{Tomaselli2021}
{Tomaselli} GM, {Ferrara} A. 2021.
\textit{\mnras} 504:89--100

\bibitem[{{Tominaga} et~al.(2013){Tominaga}, {Blinnikov} \&
  {Nomoto}}]{Tominaga2013}
{Tominaga} N, {Blinnikov} SI, {Nomoto} K. 2013.
\textit{\apjl} 771:L12

\bibitem[{{Torres}(2004)}]{Torres2004}
{Torres} DF. 2004.
\textit{\apj} 617:966--986

\bibitem[{{Tremonti} et~al.(2007){Tremonti}, {Moustakas} \&
  {Diamond-Stanic}}]{Tremonti2007}
{Tremonti} CA, {Moustakas} J, {Diamond-Stanic} AM. 2007.
\textit{\apjl} 663:L77--L80

\bibitem[{{Tsang} \& {Milosavljevi{\'c}}(2015)}]{Tsang2015}
{Tsang} BTH, {Milosavljevi{\'c}} M. 2015.
\textit{\mnras} 453:1108--1120

\bibitem[{{Tsang} \& {Milosavljevi{\'c}}(2018)}]{Tsang2018}
{Tsang} BTH, {Milosavljevi{\'c}} M. 2018.
\textit{\mnras} 478:4142--4161

\bibitem[{{Tsung} et~al.(2022){Tsung}, {Oh} \& {Jiang}}]{Tsung2022}
{Tsung} THN, {Oh} SP, {Jiang} YF. 2022.
\textit{\mnras} 513:4464--4493

\bibitem[{{Tumlinson} et~al.(2017){Tumlinson}, {Peeples} \&
  {Werk}}]{Tumlinson2017}
{Tumlinson} J, {Peeples} MS, {Werk} JK. 2017.
\textit{\araa} 55:389--432

\bibitem[{{Ugliano} et~al.(2012){Ugliano}, {Janka}, {Marek} \&
  {Arcones}}]{Ugliano2012}
{Ugliano} M, {Janka} HT, {Marek} A, {Arcones} A. 2012.
\textit{\apj} 757:69

\bibitem[{{van Woerden} et~al.(1957){van Woerden}, {Rougoor} \&
  {Oort}}]{Vanwoerden1957}
{van Woerden} H, {Rougoor} GW, {Oort} JH. 1957.
\textit{Academie des Sciences Paris Comptes Rendus} 244:1691--1695

\bibitem[{{Veilleux} et~al.(1995){Veilleux}, {Kim}, {Sanders}, {Mazzarella} \&
  {Soifer}}]{Veilleux1995}
{Veilleux} S, {Kim} DC, {Sanders} DB, {Mazzarella} JM, {Soifer} BT. 1995.
\textit{\apjs} 98:171

\bibitem[{{Veilleux} et~al.(2020){Veilleux}, {Maiolino}, {Bolatto} \&
  {Aalto}}]{Veilleux2020}
{Veilleux} S, {Maiolino} R, {Bolatto} AD, {Aalto} S. 2020.
\textit{\aapr} 28:2

\bibitem[{{Veilleux} \& {Osterbrock}(1987)}]{Veilleux1987}
{Veilleux} S, {Osterbrock} DE. 1987.
\textit{\apjs} 63:295

\bibitem[{{Veilleux} et~al.(2009){Veilleux}, {Rupke} \&
  {Swaters}}]{Veilleux2009}
{Veilleux} S, {Rupke} DSN, {Swaters} R. 2009.
\textit{\apjl} 700:L149--L153

\bibitem[{{VERITAS Collaboration} et~al.(2009){VERITAS Collaboration},
  {Acciari}, {Aliu}, {Arlen}, {Aune} et~al.}]{veritas2009_m82}
{VERITAS Collaboration}, {Acciari} VA, {Aliu} E, {Arlen} T, {Aune} T, et~al.
  2009.
\textit{\nat} 462:770--772

\bibitem[{{Walch} \& {Naab}(2015)}]{Walch2015}
{Walch} S, {Naab} T. 2015.
\textit{\mnras} 451:2757--2771

\bibitem[{{Wang}(1995{\natexlab{a}})}]{Wang1995a}
{Wang} B. 1995{\natexlab{a}}.
\textit{\apj} 444:590

\bibitem[{{Wang}(1995{\natexlab{b}})}]{Wang1995b}
{Wang} B. 1995{\natexlab{b}}.
\textit{\apjl} 444:L17

\bibitem[{{Wang} et~al.(1997){Wang}, {Heckman}, {Weaver} \& {Armus}}]{Wang1997}
{Wang} J, {Heckman} TM, {Weaver} KA, {Armus} L. 1997.
\textit{\apj} 474:659--674

\bibitem[{{Wang} et~al.(2014){Wang}, {Nardini}, {Fabbiano}, {Karovska}, {Elvis}
  et~al.}]{Wang2014}
{Wang} J, {Nardini} E, {Fabbiano} G, {Karovska} M, {Elvis} M, et~al. 2014.
\textit{\apj} 781:55

\bibitem[{{Wang} et~al.(2022){Wang}, {Kassin}, {Faber}, {Koo}, {Cunningham}
  et~al.}]{Wang2022}
{Wang} W, {Kassin} SA, {Faber} SM, {Koo} DC, {Cunningham} EC, et~al. 2022.
\textit{\apj} 930:146

\bibitem[{{Weber} \& {Davis}(1967)}]{Weber1967}
{Weber} EJ, {Davis} Leverett J. 1967.
\textit{\apj} 148:217--227

\bibitem[{{Weinberg}(2023)}]{Weinberg2023}
{Weinberg} DH. 2023.
\textit{arXiv e-prints} :arXiv:2306.09133

\bibitem[{{Weiner} et~al.(2009){Weiner}, {Coil}, {Prochaska}, {Newman},
  {Cooper} et~al.}]{Weiner2009}
{Weiner} BJ, {Coil} AL, {Prochaska} JX, {Newman} JA, {Cooper} MC, et~al. 2009.
\textit{\apj} 692:187--211

\bibitem[{{Weingartner} \& {Draine}(2001)}]{Weingartner2001}
{Weingartner} JC, {Draine} BT. 2001.
\textit{\apj} 548:296--309

\bibitem[{{Wei{\ss}} et~al.(2001){Wei{\ss}}, {Neininger}, {H{\"u}ttemeister} \&
  {Klein}}]{Weiss2001}
{Wei{\ss}} A, {Neininger} N, {H{\"u}ttemeister} S, {Klein} U. 2001.
\textit{\aap} 365:571--587

\bibitem[{{Werhahn} et~al.(2021{\natexlab{a}}){Werhahn}, {Pfrommer} \&
  {Girichidis}}]{Werhahn2021_FRC}
{Werhahn} M, {Pfrommer} C, {Girichidis} P. 2021{\natexlab{a}}.
\textit{\mnras} 508:4072--4095

\bibitem[{{Werhahn} et~al.(2021{\natexlab{b}}){Werhahn}, {Pfrommer},
  {Girichidis}, {Puchwein} \& {Pakmor}}]{Werhahn_2021_voyager}
{Werhahn} M, {Pfrommer} C, {Girichidis} P, {Puchwein} E, {Pakmor} R.
  2021{\natexlab{b}}.
\textit{\mnras} 505:3273--3294

\bibitem[{{Werhahn} et~al.(2021{\natexlab{c}}){Werhahn}, {Pfrommer},
  {Girichidis} \& {Winner}}]{Werhahn2021_gamma}
{Werhahn} M, {Pfrommer} C, {Girichidis} P, {Winner} G. 2021{\natexlab{c}}.
\textit{\mnras} 505:3295--3313

\bibitem[{{Westmoquette} et~al.(2009){Westmoquette}, {Gallagher}, {Smith},
  {Trancho}, {Bastian} \& {Konstantopoulos}}]{Westmoquette2009}
{Westmoquette} MS, {Gallagher} JS, {Smith} LJ, {Trancho} G, {Bastian} N,
  {Konstantopoulos} IS. 2009.
\textit{\apj} 706:1571--1587

\bibitem[{{Westmoquette} et~al.(2007){Westmoquette}, {Smith}, {Gallagher},
  {O'Connell}, {Rosario} \& {de Grijs}}]{Westmoquette2007}
{Westmoquette} MS, {Smith} LJ, {Gallagher} J.~S. I, {O'Connell} RW, {Rosario}
  DJ, {de Grijs} R. 2007.
\textit{\apj} 671:358--373

\bibitem[{{Wibking} \& {Krumholz}(2023)}]{Wibking2023}
{Wibking} BD, {Krumholz} MR. 2023.
\textit{\mnras} 521:5972--5990

\bibitem[{{Wibking} et~al.(2018){Wibking}, {Thompson} \&
  {Krumholz}}]{Wibking2018}
{Wibking} BD, {Thompson} TA, {Krumholz} MR. 2018.
\textit{\mnras} 477:4665--4684

\bibitem[{{Wickramasinghe} et~al.(1966){Wickramasinghe}, {Donn} \&
  {Stecher}}]{Wickramasinghe1966}
{Wickramasinghe} NC, {Donn} BD, {Stecher} TP. 1966.
\textit{\apj} 146:590

\bibitem[{{Wiener} et~al.(2017){Wiener}, {Pfrommer} \& {Oh}}]{Wiener2017}
{Wiener} J, {Pfrommer} C, {Oh} SP. 2017.
\textit{\mnras} 467:906--921

\bibitem[{{Wu} et~al.(2019){Wu}, {Baker}, {Heckman}, {Hicks}, {Lutz} \&
  {Tacconi}}]{Wu2019}
{Wu} JF, {Baker} AJ, {Heckman} TM, {Hicks} EKS, {Lutz} D, {Tacconi} LJ. 2019.
\textit{\apj} 887:251

\bibitem[{{W{\"u}nsch} et~al.(2008){W{\"u}nsch}, {Tenorio-Tagle},
  {Palou{\v{s}}} \& {Silich}}]{Wunsch2008}
{W{\"u}nsch} R, {Tenorio-Tagle} G, {Palou{\v{s}}} J, {Silich} S. 2008.
\textit{\apj} 683:683--692

\bibitem[{{Wuyts} et~al.(2011{\natexlab{a}}){Wuyts}, {F{\"o}rster Schreiber},
  {Lutz}, {Nordon}, {Berta} et~al.}]{Wuyts2011a}
{Wuyts} S, {F{\"o}rster Schreiber} NM, {Lutz} D, {Nordon} R, {Berta} S, et~al.
  2011{\natexlab{a}}.
\textit{\apj} 738:106

\bibitem[{{Wuyts} et~al.(2011{\natexlab{b}}){Wuyts}, {F{\"o}rster Schreiber},
  {van der Wel}, {Magnelli}, {Guo} et~al.}]{Wuyts2011b}
{Wuyts} S, {F{\"o}rster Schreiber} NM, {van der Wel} A, {Magnelli} B, {Guo} Y,
  et~al. 2011{\natexlab{b}}.
\textit{\apj} 742:96

\bibitem[{{Wyse} \& {Silk}(1985)}]{Wyse1985}
{Wyse} RFG, {Silk} J. 1985.
\textit{\apjl} 296:L1--L5

\bibitem[{{XRISM Science Team}(2020)}]{XRISM2020}
{XRISM Science Team}. 2020.
\textit{arXiv e-prints} :arXiv:2003.04962

\bibitem[{{Xu} et~al.(2022){Xu}, {Heckman}, {Henry}, {Berg}, {Chisholm}
  et~al.}]{Xu2022}
{Xu} X, {Heckman} T, {Henry} A, {Berg} DA, {Chisholm} J, et~al. 2022.
\textit{\apj} 933:222

\bibitem[{{Xu} et~al.(2023{\natexlab{a}}){Xu}, {Heckman}, {Henry}, {Berg},
  {Chisholm} et~al.}]{Xu2023}
{Xu} X, {Heckman} T, {Henry} A, {Berg} DA, {Chisholm} J, et~al.
  2023{\natexlab{a}}.
\textit{\apj} 948:28

\bibitem[{{Xu} et~al.(2023{\natexlab{b}}){Xu}, {Heckman}, {Yoshida}, {Henry} \&
  {Ohyama}}]{Xu2023_M82}
{Xu} X, {Heckman} T, {Yoshida} M, {Henry} A, {Ohyama} Y. 2023{\natexlab{b}}.
\textit{\apj} 956:142

\bibitem[{{Yan} \& {Lazarian}(2002)}]{Yan2002}
{Yan} H, {Lazarian} A. 2002.
\textit{\prl} 89:281102

\bibitem[{{Yoast-Hull} et~al.(2013){Yoast-Hull}, {Everett}, {Gallagher} \&
  {Zweibel}}]{Yoast-Hull2013_M82}
{Yoast-Hull} TM, {Everett} JE, {Gallagher} J.~S. I, {Zweibel} EG. 2013.
\textit{\apj} 768:53

\bibitem[{{Yoast-Hull} et~al.(2015){Yoast-Hull}, {Gallagher} \&
  {Zweibel}}]{Yoast-Hull2015_Arp220}
{Yoast-Hull} TM, {Gallagher} JS, {Zweibel} EG. 2015.
\textit{\mnras} 453:222--228

\bibitem[{{Yoast-Hull} \& {Murray}(2019)}]{Yoast-Hull2019_Arp220}
{Yoast-Hull} TM, {Murray} N. 2019.
\textit{\mnras} 484:3665--3680

\bibitem[{{Yoshida} et~al.(2019){Yoshida}, {Kawabata}, {Ohyama}, {Itoh} \&
  {Hattori}}]{Yoshida2019}
{Yoshida} M, {Kawabata} KS, {Ohyama} Y, {Itoh} R, {Hattori} T. 2019.
\textit{\pasj} 71:87

\bibitem[{{Yoshida} et~al.(2016){Yoshida}, {Yagi}, {Ohyama}, {Komiyama},
  {Kashikawa} et~al.}]{Yoshida2016}
{Yoshida} M, {Yagi} M, {Ohyama} Y, {Komiyama} Y, {Kashikawa} N, et~al. 2016.
\textit{\apj} 820:48

\bibitem[{{Yuan} et~al.(2023){Yuan}, {Krumholz} \& {Martin}}]{Yuan2023}
{Yuan} Y, {Krumholz} MR, {Martin} CL. 2023.
\textit{\mnras} 518:4084--4105

\bibitem[{{Zabl} et~al.(2021){Zabl}, {Bouch{\'e}}, {Wisotzki}, {Schaye},
  {Leclercq} et~al.}]{Zabl2021}
{Zabl} J, {Bouch{\'e}} NF, {Wisotzki} L, {Schaye} J, {Leclercq} F, et~al. 2021.
\textit{\mnras} 507:4294--4315

\bibitem[{{Zhang} \& {Davis}(2017)}]{Zhang_Davis}
{Zhang} D, {Davis} SW. 2017.
\textit{\apj} 839:54

\bibitem[{{Zhang} et~al.(2018){Zhang}, {Davis}, {Jiang} \& {Stone}}]{Zhang2018}
{Zhang} D, {Davis} SW, {Jiang} YF, {Stone} JM. 2018.
\textit{\apj} 854:110

\bibitem[{{Zhang} \& {Thompson}(2012)}]{Zhang_Thompson}
{Zhang} D, {Thompson} TA. 2012.
\textit{\mnras} 424:1170--1178

\bibitem[{{Zhang} et~al.(2017){Zhang}, {Thompson}, {Quataert} \&
  {Murray}}]{Zhang2017}
{Zhang} D, {Thompson} TA, {Quataert} E, {Murray} N. 2017.
\textit{\mnras} 468:4801--4814

\bibitem[{{Zhang} et~al.(2014){Zhang}, {Wang}, {Ji}, {Smith}, {Foster} \&
  {Zhou}}]{Zhang2014}
{Zhang} S, {Wang} QD, {Ji} L, {Smith} RK, {Foster} AR, {Zhou} X. 2014.
\textit{\apj} 794:61

\bibitem[{{Zhu} et~al.(2015){Zhu}, {Comparat}, {Kneib}, {Delubac}, {Raichoor}
  et~al.}]{Zhu2015}
{Zhu} GB, {Comparat} J, {Kneib} JP, {Delubac} T, {Raichoor} A, et~al. 2015.
\textit{\apj} 815:48

\bibitem[{{Zweibel}(2017)}]{Zweibel2017}
{Zweibel} EG. 2017.
\textit{Physics of Plasmas} 24:055402

\end{thebibliography}
